\documentclass[paper]{aa}
\pdfoutput=1

\usepackage{amsmath}
\usepackage{natbib}
\usepackage{graphicx}
\usepackage{txfonts}
\usepackage{multirow}
\usepackage{textcomp}
\usepackage{subfigure}
\usepackage{lscape}
\usepackage{longtable}

\newcommand{\ume}{u_M}
\newcommand{\gme}{g_M}
\newcommand{\rme}{r_M}
\newcommand{\ime}{i_M}
\newcommand{\zme}{z_M}
\newcommand{\bdtruc}{BD~+17~4708}
\newcommand{\x}{{\mathbf x}}
\graphicspath{{eps/}}

\begin{document}

\title{Photometric Calibration of the Supernova Legacy Survey Fields\thanks{ Based on
observations obtained with MegaPrime/MegaCam, a joint project of CFHT
and CEA/DAPNIA, at the Canada-France-Hawaii Telescope (CFHT) which is
operated by the National Research Council (NRC) of Canada, the
Institut National des Sciences de l'Univers of the Centre National de
la Recherche Scientifique (CNRS) of France, and the University of
Hawaii. This work is based in part on data products produced at the
Canadian Astronomy Data Centre as part of the Canada-France-Hawaii
Telescope Legacy Survey, a collaborative project of NRC and CNRS.}~\thanks{Tables 13-22 are only 
available in electronic form at the CDS via anonymous ftp to {\tt cdsarc.u-strasbg.fr (130.79.128.5)}
or via {\tt http://cdsweb.u-strasbg.fr/cgi-bin/qcat?J/A+A/}}}

\author{
N.~Regnault\inst{1}
\and A.~Conley\inst{2}
\and J.~Guy\inst{1}
\and M.~Sullivan\inst{3}
\and J.-C. Cuillandre\inst{4}
\and P.~Astier\inst{1}
\and C.~Balland\inst{1,5}
\and S.~Basa\inst{6}
\and R.~G.~Carlberg\inst{2}
\and D.~Fouchez\inst{7}
\and D.~Hardin\inst{1}
\and I.~M.~Hook\inst{3,8}
\and D.~A.~Howell\inst{9,10}
\and R.~Pain\inst{1}
\and K.~Perrett\inst{2}
\and C.~J.~Pritchet\inst{11}
}

\institute{
LPNHE, CNRS-IN2P3 and Universit\'{e}s Paris 6 \& 7, 4 place Jussieu, F-75252 Paris Cedex 05, France
\and Department of Physics and Astronomy, University of Toronto, 50 St. George Street, Toronto, ON M5S 3H4, Canada
\and Department of Astrophysics, University of Oxford, Keble Road, Oxford OX1 3RH, UK
\and Canada-France-Hawaii Telescope Corp., Kamuela, HI 96743, USA
\and University Paris 11, F-91405 Orsay, France
\and LAM, CNRS, BP8, Traverse du Siphon, 13376 Marseille Cedex 12, France
\and CPPM, CNRS-IN2P3 and Universit\'e Aix-Marseille II, Case 907, 13288 Marseille Cedex 9, France
\and INAF -- Osservatorio Astronomico di Roma, via Frascati 33, 00040 Monteporzio (RM), Italy
\and Las Cumbres Observatory Global Telescope Network, 6740 Cortona Dr., Suite 102, Goleta, CA 93117
\and Department of Physics, University of California, Santa Barbara, Broida Hall, Mail Code 9530, Santa Barbara, CA 93106-9530
\and Department of Physics and Astronomy, University of Victoria, PO Box 3055, Victoria, BC V8W 3P6, Canada
}

\titlerunning{Photometric Calibration of the SNLS Fields}
\authorrunning{N.~Regnault et al.}

\offprints{nicolas.regnault\@@lpnhe.in2p3.fr}

\date{Received Mont DD, YYYY; accepted Mont DD, YYYY}

\abstract
{} {We present the photometric calibration of the Supernova Legacy
Survey (SNLS) fields. The SNLS aims at measuring the distances to
SNe Ia at ($0.3<z<1$) using MegaCam, the 1 deg$^2$ imager on the Canada-France-Hawaii Telescope (CFHT). The uncertainty affecting the photometric
calibration of the survey dominates the systematic uncertainty of the 
key measurement of the survey, namely the dark energy equation of state.
The photometric calibration of the SNLS requires obtaining a
uniform response across the imager, calibrating the science field
stars in each survey band (SDSS-like $ugriz$ bands) with respect to 
standards with known flux  in the same bands, and binding the calibration
to the $UBVRI$ Landolt standards used to calibrate the nearby SNe
from the literature necessary to produce cosmological constraints.}
{The spatial
non-uniformities of the imager photometric response
are mapped using dithered observations of dense stellar fields.
Photometric zero-points against Landolt standards are obtained.
The linearity of the instrument is studied.}
{We show that the imager filters
and photometric response are not uniform and publish correction
maps. We present models of the effective passbands of the instrument
as a function of the position on the focal plane. We define a natural magnitude 
system for MegaCam. We show that the systematics affecting the magnitude-to-flux 
relations can be reduced if we use the spectrophotometric standard star \bdtruc\ instead of Vega as 
a fundamental flux standard. We publish $ugriz$
catalogs of tertiary standards for all the SNLS fields. 
} {}


\maketitle

\section{Introduction}
\label{sec:introduction}

As we enter an era of precision supernova cosmology, photometric calibration
becomes an increasingly important contribution to the systematic error
budgets. All cosmology oriented surveys have therefore undertaken
ambitious calibration efforts in order to ``break the 1\% barrier''.
A notable example is the work published by the Sloan Digital Sky
Survey (SDSS) \citep{Ivezic07,Padmanabhan07}. Major future surveys are
also planning ambitious photometric calibration programs, relying on
stellar calibrator observations and in-situ laboratory measurements
\citep{Burke07,Magnier07,Keller07,Stubbs06,Tucker07}.

An example of a program demanding a better than 1\% photometric
precision is the measurement of the cosmological parameters using Type
Ia Supernovae (SNe~Ia) \citep{astier06, Wood-Vasey07, Riess07}. This
measurement is all about comparing the luminosity distances of a set
of distant SNe~Ia with those of their nearby counterparts. High-z ($z
\sim 0.7$) SN~Ia distances are determined from measurements taken with
the redder bands of the survey ($r$, $i$ and $z$), a model of the
SNe~Ia spectral energy distribution (SED) and a model of the survey
passbands.  Intermediate ($z \sim 0.4$) SN~Ia distances rely on the
same ingredients but measurements taken with the bluer bands of the
survey (typically $g$ and $r$). Nearby ($z \sim 0.05$) supernovae
usually come from datasets collected in the 1990s by dedicated
low-redshift surveys \citep{Hamuy96b,Riess99a,Jha06}, and calibrated in the
Landolt $UBVRI$ system. \citet{astier06} analyze in detail the
impact of the high-redshift survey calibration uncertainties on the
cosmological parameters. They show that a 1\% shift of the
high-redshift survey zero-points results in a variation of the
dark-energy equation of state parameter --$w$-- of 0.040; a 1 nm
precision of the survey passband central wavelength results in an
uncertainty of 0.013 on $w$; finally, an error 
of 1\% in the intercalibration of the $B$ and $R$ passbands 
(respectively $\lambda \sim 438\ {\rm nm}$ and $\lambda \sim 652\ {\rm nm}$) 
has also a sizeable impact on $w$, of 0.024.

Another example is the determination of photometric redshifts of
galaxies \citep{Ilbert06,Brodwin06}. The measurement involves
comparing the measured fluxes in several bands with synthetic
photometry computed from galaxy SED models and models of the survey
passbands. \citet{Ilbert06} demonstrate that photometric redshift
determination and galaxy type identification are sensitive to
1\%-level zero point changes. The analysis also requires an accurate
determination of the survey passbands. 

In many applications, notably the two presented above, the survey measurements must be converted
into physical fluxes at some point in the analysis, in order to be compared with predicted synthetic
fluxes, computed from a SED model and a model of the survey passbands.
A requirement of the photometric calibration is therefore that the
connection between the magnitudes and their broadband flux
counterparts is not broken. The most direct way to ensure this, is to
define the survey calibrated magnitudes as {\em natural magnitudes}.
In other words, the survey magnitude $m$ of an object whose measured broadband 
flux is $\phi_{ADU}$ must be defined as:
\begin{equation*}
  m = -2.5\log_{10} \phi_{ADU}  + \textrm{calibration coefficients}
\end{equation*}
and may not depend on the object's colors.

This paper presents the photometric calibration of the
Canada-France-Hawaii Telescope Supernova Legacy Survey
\citep[SNLS,][]{astier06} 3-year dataset, taken with the wide
field imager MegaCam. Our primary motivation is
the calibration of the luminosity distances to the type Ia supernovae
discovered by the SNLS. However, the results presented
here should be useful for all applications relying on photometric data
taken with MegaCam since the beginning of the CFHT Legacy Survey (CFHTLS) operations in 2003.

Many broadband magnitude systems have been defined over the last few
decades and implemented under the form of catalogs of standard star
magnitudes \citep{Landolt73, Landolt83, Landolt92, Menzies89,
  Stetson00, Stetson05, Smith02, Ivezic07}. The most widely used
standard star network is that of \citet{Landolt92}. Using it
to calibrate the SNLS dataset is not the most obvious choice, notably
because the Johnson-Kron-Cousins $UBVRI$ filters used by Landolt
differ significantly from the $ugriz$ filters which equip
MegaCam. Unfortunately, the magnitudes of the nearby supernovae used
to supplement the SNLS dataset are reported in the Landolt system.
Hence, adopting the same standard star network allows one to minimize
the systematic uncertainties which affect the comparison of the
(external) nearby and (SNLS) distant-SN~Ia luminosity distances.

The goal of the SNLS 3-year calibration was set by the results of the
systematic uncertainty analyzes presented in \citet{astier06}.
Improving the precision on $w$ requires pushing the
uncertainties on the calibrated supernova fluxes as close to 1\% as
possible. Attaining this kind of precision turned out to be
challenging.  The key problems that had to be solved were
(1) the control of the imager photometric response uniformity (2)
the modeling of the large and non-linear Landolt-to-MegaCam color
transformations (3) the choice of the optimal fundamental
spectrophotometric standard used to interpret the calibrated
magnitudes as physical fluxes minimizing the associated systematic
errors and (4) the modeling of the imager effective passbands.  All
these aspects are discussed in detail in this paper.

The main output of the calibration consists in catalogs of 
$g,r,i,z$ natural magnitudes for each of the four fields
surveyed by SNLS, along with a recipe to map these magnitudes to
fluxes.  Another key result is a model of the imager effective
passbands. The last important output of this work, the photometric response maps,
shall be released along with the next release of the survey images.

The calibration of the $u$-band data poses additional problems, and
deserves a paper of its own. {In particular, the mean wavelength of the
effective $u$ MegaCam passbands is extremely sensitive to the atmospheric
conditions and to the airmass. As an example, it varies by almost 2 nanometers 
between airmass 1 and 1.5. } Moreover, the DEEP field
$u$-band observations are not used by the SNLS project and are
therefore not time sequenced. As a consequence less epochs are
available, which makes the calibration much less robust. The
$u$-band magnitudes of the SNLS tertiaries are presented in
appendix. The precision on the calibrated $u$-band fluxes is of
about 5\%, much larger than the 1\% obtained in the other bands.

A plan of the paper follows. In \S \ref{sec:megaprime_instrument} and
\S \ref{sec:snls_survey} we present the MegaCam wide field imager and
the SNLS survey. 
{The photometry algorithms used to
derive the survey calibration are presented in \S
\ref{sec:photometric_reduction}. }
\S \ref{sec:elixir_pipeline} is devoted to the
presentation of Elixir, the image pre-processing pipeline developed at
CFHT. Given the stringent requirements on the calibration precision,
the Elixir results have been scrutinized by the SNLS collaboration,
and new maps of the imager photometric response have been
derived. This work is presented in \S
\ref{sec:the_photometric_grids}. An essential ingredient needed to
interpret the calibrated magnitudes as fluxes is a reliable model of
the imager effective passbands. The derivation of those is presented
in \S \ref{sec:megacam_passbands}. The last
sections of the paper are devoted to the main output of the
calibration: the derivation of the tertiary star catalogs. In \S
\ref{sec:landolt_stars} we discuss how we derive the survey
zero-points from the Landolt star observations. The next section, \S
\ref{sec:megacam_magnitudes}, is devoted to the definition of a natural MegaCam magnitude system. 
\S \ref{sec:flux_interpretation_of_megacam_magnitudes} discusses the conversion
of these natural magnitudes into physical fluxes, through the
use of a well chosen {\em fundamental flux standard}, namely the spectrophotometric standard star \bdtruc.
\S \ref{sec:tertiary_catalogs}
presents the derivation of the tertiary catalogs and the
selection of the photometric nights. Finally, we discuss 
in \S \ref{sec:systematic_uncertainties} the systematic uncertainties
affecting the calibrated magnitude measurements and their physical flux 
counterparts.

\section{MegaPrime and MegaCam}
\label{sec:megaprime_instrument}

MegaCam is a wide-field camera, hosted in the dedicated prime focus
environment MegaPrime on the Canada France Hawaii 3.6 m telescope
\citep{MegaCamPaper}. The camera images a field of
view $0.96 \times 0.94$ deg$^2$, using 36\footnote{In fact, the MegaCam focal plane 
is composed of 40 CCDs, but only the 36 central CCDs are actually read out, 
the 4 additional ones being almost completely vignetted by the filter frame. These four additional 
chips are kept as spares. } thinned E2V $2048\times4612$
CCDs, with pixels of 13.5 $\mu {\rm m}$ that subtend 0.185\arcsec on a
side ---the focal ratio being F/4.1. Each CCD is read out from two
amplifiers, which allows one to read out the 340 Mpixel focal plane in 35 seconds.
The output of each amplifier is sampled by a 16 bit ADC. The gains of the readout 
chain have been set to about 1.5 e$^-$ / ADU
with the consequence that 
only half of the MegaCam CCD full well ($\sim$ 200,000 e$^-$) is actually 
sampled by the readout electronics. 
The CFHT has an equatorial mount and the camera 
angle is fixed with respect to both the telescope and the sky.

{The linearity of the imager photometric response has been
carefully checked using images of a dense stellar field, observed
under photometric conditions, with increasing exposure times  (see
appendix \ref{sec:linearity} for details).  The linearity has been
found to be better than 1\% at the pixel level. The corresponding
upper limit for the non-linearities affecting the star fluxes is
much smaller, of the order of 0.1\%.}

The filter system is a juke box which holds up to 8 filters. The CFHT
Legacy Survey performs observations in five bands, labeled $\ume, \gme,
\rme, \ime, \zme$, similar to the SDSS $u,g,r,i,z$ bands \citep{fukugita96,Gunn98} and realized
using interference filters. The thinned MegaCam CCDs exhibit 
fringing, of about 6\% of the sky background peak to peak 
in $\ime$, and 15\% in $\zme$. The filters initially
mounted on MegaPrime were manufactured by Sagem/REOSC. 
In June 2007, the 
$i_M$ filter was accidentally broken and a new filter (labeled $i_{2M}$ hereafter), was 
procured from Barr Associates and installed on the camera in October 2007.
This new filter is not discussed in this work, and the calibration obtained
for the old filter cannot be applied to the new one.

\begin{figure}
\centering
\includegraphics[width=\linewidth]{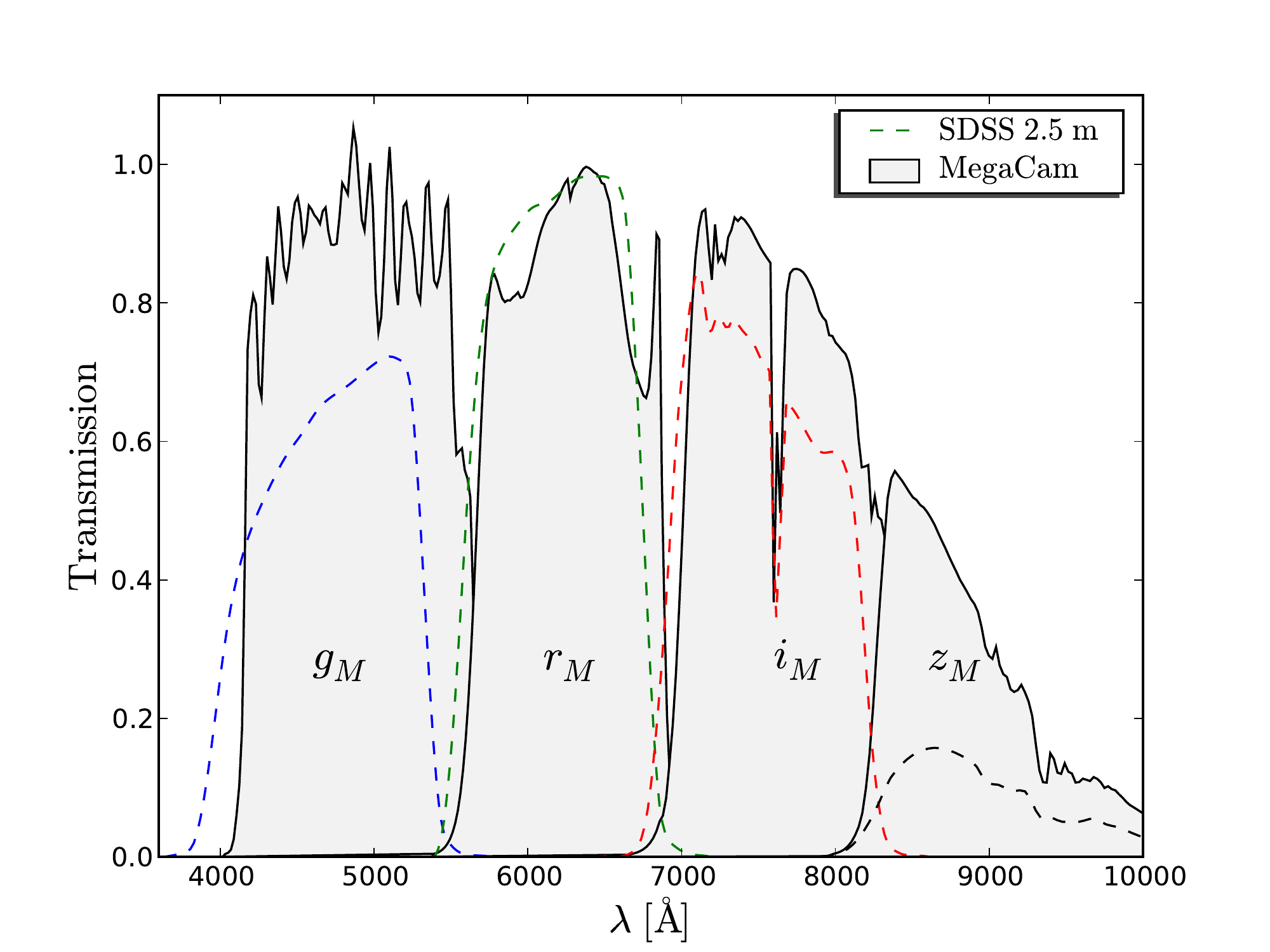}\\
\caption{SDSS 2.5-m effective passbands, and MegaCam effective
  passbands {\em at the center of the focal plane}. The $g'$ and
  $i'$ SDSS passbands are slightly bluer than the corresponding
  MegaCam passbands. The central value of the SDSS $r'$ and $z'$ are
  very similar to those of their MegaCam
  counterparts. \label{fig:megacam_sdss_passbands}}
\end{figure}

The transmission of the MegaPrime filters were characterized by 
their respective 
manufacturers 
and the CFHT team upon reception. Their most salient
feature is the significant spatial non-uniformity of
their transmission curves (namely of their central wavelength), which impacts 
significantly the observations (see \S \ref{sec:the_photometric_grids}). 
Interestingly, the new $i_{2M}$ filter exhibits a variation pattern
similar to the old one, although different manufacturers produced them.

Systematic
differences were found between the MegaCam and SDSS filters,
which translate into small color terms between both instruments (see
appendices \ref{sec:megacam_passband_tables} and \ref{sec:comparison_with_sdss}).
In particular, the $\ume$-band filter is significantly bluer than its 
SDSS counterpart. This is a deliberate design choice as the MegaCam E2V chips are
much more efficient in the blue than the SITe CCDs used in the SDSS imager.
For this reason, the $\ume$-filter is sometimes labeled $u^\star$, in order 
to distinguish it from a standard $u'$ filter.

The shutter is made of a half disk (1 meter in diameter) whose rotation
is controlled precisely, in order to ensure a
constant speed when the shutter crosses the CCD mosaic. The minimum 
exposure time that can be obtained is of 1 second, with uniformity 
of illumination of the large focal plane better than 10 ms.
In order to increase the accuracy of the short 
exposure times, the exact
duration of the exposure is measured with a 1 millisecond accuracy
using a dedicated system independent from the shutter motion
controller.

MegaPrime is operated and maintained by the CFHT team, which has
steadily improved the performance of the imager.
From the first light, MegaPrime images appeared to suffer from an image quality (IQ)
degradation of 0.25\arcsec\ (FWHM) from center to corner. For example, 
the mean image quality of the $\rme$-band exposures was 0.65\arcsec\ at the 
center of the focal plane, and almost 0.9\arcsec\ on the corners.
Despite an optimization of the vertical position of 
the four-lens wide field corrector (WFC), this situation prevailed until
November 2004, when the almost flat WFC L3 lens was flipped during an investigation. This resulted 
in a dramatic improvement of the IQ uniformity, with no degradation
of the median IQ. In July 2005, a small tilt was
applied to the camera itself resulting in an essentially uniform image quality.
Exposures with IQ varying from 0.4\arcsec to 0.48\arcsec across the field
of view have been routinely obtained since then.
A summary of the main modifications of the imager setup is presented in table
\ref{tab:grid_observations}.

\section{The SNLS Survey}
\label{sec:snls_survey}

The SNLS survey is a ground-based supernova
survey, aiming primarily at measuring the dark-energy equation of
state parameter $w$. It has been designed to build up a sample of 500 SNe~Ia
in the redshift range $0.3 < z < 1.0$. The SNLS project 
has two components: a large imaging survey using MegaPrime to detect
supernovae and monitor their lightcurves, and a spectroscopic survey
to confirm the nature of the candidates, and determine their redshift.
During its five years of operation (mid-2003 to mid-2008),
the survey delivered about 100 spectroscopically
identified SNe~Ia per year. The cosmological measurements based on the
first year dataset have been presented in \citet{astier06}.

SNLS exploits the DEEP component of the CFHT Legacy Survey (CFHTLS),
which targets four low Galactic extinction fields (see table \ref{table:deep_fields} for coordinates and extinction).
The data are time sequenced in $\gme,\rme,\ime$ and $\zme$ with observations
conducted every 3-4 nights in dark time, allowing the construction
of high-quality multicolor supernova light curves. Because SNLS is
highly vulnerable to gaps in the supernova light curves,
MegaPrime was mounted on the telescope at least 14 nights around 
every new moon during the course of the survey ---every such observing period is called a MegaCam (or MegaPrime) {run}.
The deep CFHTLS component comprises additional observations in the MegaCam
$\ume$ band, which are not time sequenced, and not directly used by the
SNLS. The calibration of the $\ume$ band data has however been included 
in the present work.

\begin{table}
\centering
\caption{Fields observed by the DEEP/SN component of the CFHTLS.\label{table:deep_fields}}
\begin{tabular}{ccc|c}
\hline
\hline
 Field &  RA          & Dec         &  E(B-V) $^{\mathrm{a}}$      \\
       & (2000)       & (2000)      & (MW)         \\
\hline
D1     &  02:26:00.00 & $-04$:30:00.0 & 0.027\\
D2     &  10:00:28.60 & $+02$:12:21.0 & 0.018\\
D3     &  14:19:28.01 & $+52$:40:41.0 & 0.010\\
D4     &  22:15:31.67 & $-17$:44:05.0 & 0.027\\
\hline
\end{tabular}
\begin{list}{}{}
  \item[$^\mathrm{a}$] \citep[from][]{Schlegel98}
\end{list}
\end{table}
The CFHT Legacy Survey observations are taken in ``queue service 
observing'' (QSO) mode by the CFHT staff \citep{Martin02}. For the SNLS, an image quality
better than 1\arcsec\ is required. Sometimes this constraint is relaxed to meet additional
constraints on the sequencing needed to ensure a regular sampling of
the supernova lightcurves. However, since the SNLS is a high-priority program, 
and the median seeing at CFHT (in the $\rme$-band) is 0.75'', 
the actual average IQ obtained over the four fields is about 
0.96\arcsec, 0.88\arcsec, 0.84\arcsec and 0.82\arcsec in the 
$\gme\ \rme\ \ime\ \zme$ bands. The cumulated exposure times acquired every four nights (1125 s, 1800 s, 1800 to
3600 s, 3600 s in $\gme, \rme, \ime$ and $\zme$ respectively) are
split into 5 to 10 individual exposures of 225 to 520 s each (see
\citet{Sullivan06-selection}). The science exposures are dithered, in
order to limit the impact of the dead areas (dead columns, gaps) on
the science observations. The dithering offsets are essentially determined by the 
size of the gaps between the CCDs, and reach 100 pixels in x
(or right ascension) and 500 pixels in y (or declination). 

{On average, the D1 and D2 fields are observed at an airmass of
  1.2, while the D3 and D4 fields are observed at a slightly higher
  airmass of about 1.3}.

Photometric calibration exposures are routinely taken, often at the beginning and/or
the end of every potentially photometric night. The science and
calibration observations can therefore be separated by a few hours,
and a careful detection of the non-photometric nights must be
implemented.  The standard calibration program consists in 5 band
observations of one \citet{Landolt92} field at a time. The calibration
fields also contain several standard stars from the
\citet{Smith02} catalog. Since these latter standards can be bright
(mag $\le 12$), the telescope is slightly defocused, by 0.1 mm, 
which still allows a Gaussian PSF profile, but reduces the maximum pixel flux 
by about a factor two in order to avoid saturation. The exposure time of
the calibration exposures is kept between two and three seconds. 

All the exposures taken with MegaCam are pre-processed at CFHT by the
Elixir pipeline (\citet{Magnier04} and \S \ref{sec:elixir_pipeline} of
this paper). This pre-processing stage comprises a bias removal, a flat field
correction, a photometric flat correction (see \S
\ref{sec:elixir_pipeline} and \S \ref{sec:the_photometric_grids}), and defringing of
the long exposures (over 10 seconds) taken in the $\ime$ and $\zme$ bands. No defringing
is applied to the calibration exposures.

\begin{table}
\centering
\caption{SNLS/DEEP and Landolt field observations\label{tab:epochs_and_seasons_per_field}}
\begin{tabular}{c|c|cccc}
\hline
\hline
 Field &  seasons     & \multicolumn{4}{c}{number of epochs}  \\
       &              & \multicolumn{4}{c}{photometric / total}  \\
       &              & $\gme$ & $\rme$ & $\ime$ & $\zme$     \\
\hline
D1      &  3.5  & 28 / 57 & 32 / 88 & 36 / 99  & 19 / 47 \\
D2      &  3    & 13 / 44 & 21 / 63 & 25 / 66  & 19 / 26 \\
D3      &  4    & 29 / 66 & 34 / 87 & 40 / 96  & 10 / 40 \\
D4      &  4    & 30 / 67 & 34 / 97 & 29 / 100 & 19 / 55 \\
\hline
Landolt $^{\mathrm{a}}$ & --    & 313 &  311 &  283  &  240    \\
\hline
\end{tabular}
\begin{list}{}{}
\item[$^\mathrm{a}$] all \citet{Landolt92} fields taken together (two to three Landolt fields may be observed each night).
\end{list}
\end{table}

This paper relies on the first 3.5 years of the survey -- from August 2003 to
December 2006.  Table \ref{tab:epochs_and_seasons_per_field} shows the
number of seasons and epochs for each field. It also presents the
number of photometric nights, as will be discussed in \S
\ref{sec:tertiary_catalogs}. As will be shown below, the large number of
observations allows us to implement a very robust calibration
procedure, and reach low levels of internal systematic uncertainties.
 
\section{Photometry}
\label{sec:photometric_reduction}

SNLS aims ultimately at assigning magnitudes to supernovae.  In
practice, this means that we have to measure ratios of supernovae
fluxes to standard star fluxes. The measurement of these ratios is
carried out in two steps : the ratio of (secondary) standards to
science field stars (i.e. tertiary standard candidates), and the
ratio of supernova to field stars. In order to be optimal, faint
supernovae should be measured using PSF photometry, and hence the
ratios of SN to tertiaries will be a ratio of PSF fluxes, in the same
exposure. This measurement is not in the scope of this paper,
and we refer interested readers to \cite{astier06,guy09}.

Because we have a large number of epochs and several images per epoch, 
the ratio of standard fluxes to science field star fluxes is not photon-noise
limited and can therefore be measured using aperture photometry. This avoids
the shortcomings of PSF flux ratios over different images, with
different PSFs. In this section, we describe the aperture photometry
algorithm and discuss systematics affecting the ratio of standards to
field star fluxes.

\subsection{Aperture Photometry}
\label{sec:aperture_photometry}

\begin{figure}
\begin{center}
\includegraphics[width=\linewidth]{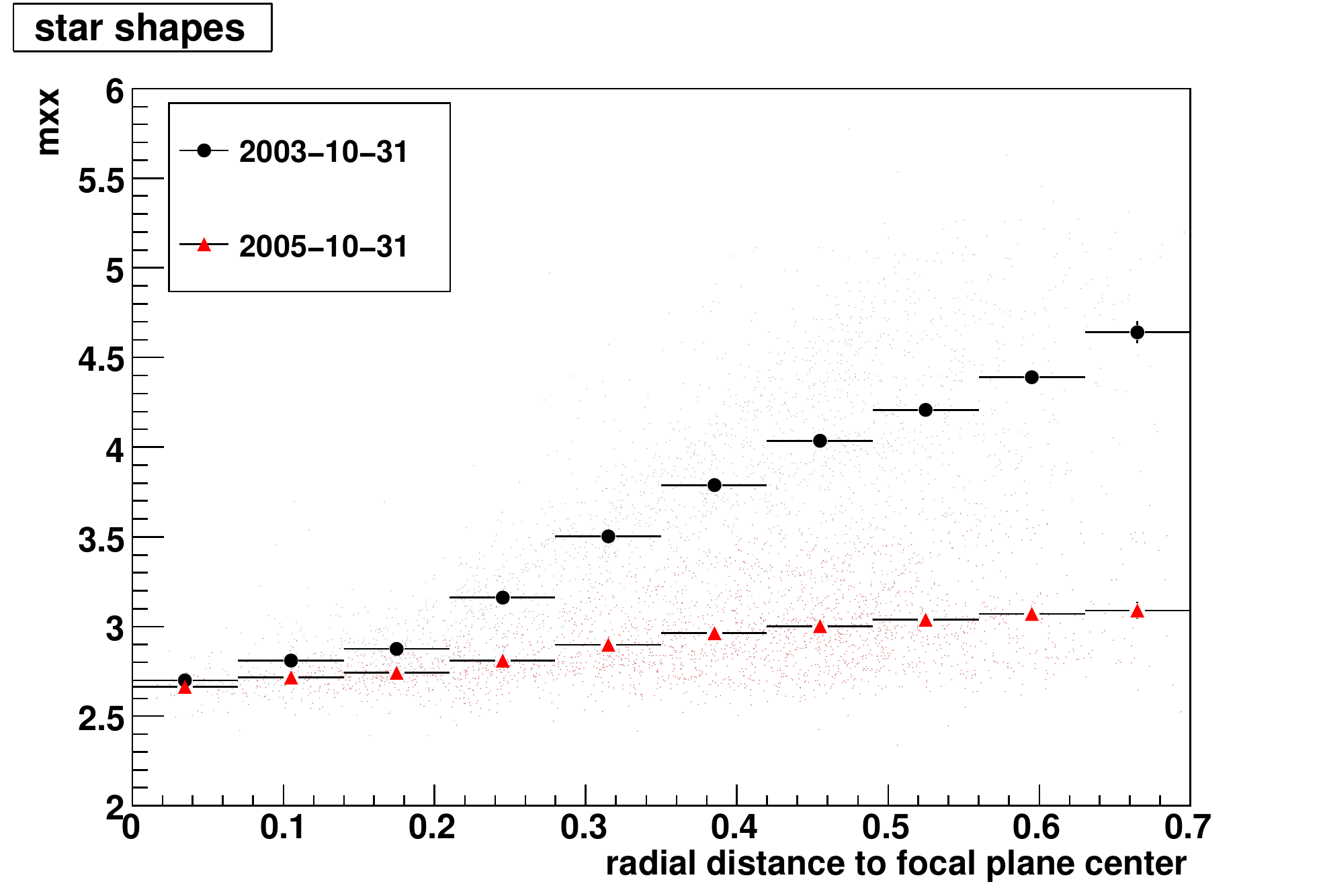}
\caption{Value of the bright star Gaussian moments as a function of
  the radial distance to the focal plane center. Black circles:
  Gaussian moments, computed on a $\rme$-band image, taken in
  2003-10-31 (before the L3-flip). Red triangles: Gaussian moments,
  computed on a $\rme$-band image of identical seeing (at the center
  of the focal plane), taken two years later on 2003-10-31.  As can be
  seen, the uniformity of the image quality has been noticeably
  improved.
  \label{fig:starshapes}}
\end{center}
\end{figure}

 The calibration and science data is reduced using the SNLS standard
 reduction software used in
 \citet{astier06}. Each exposure file is split into 36 smaller FITS
 images each corresponding to a CCD. On each smaller CCD-image, we
 detect the sources and model the sky background using the {\tt
 SExtractor} package \citep{Sex}. We recompute the sky background map,
 using only the pixels not affected by the detected object fluxes. We subtract
 this new background map from the image. We recompute the object
 Gaussian moments and aperture fluxes. 

 Figure \ref{fig:starshapes} show
 the Gaussian second moments of the field stars as a function of the
 position on the mosaic for a $r$-band image taken in 2003 and a
 $r$-band image taken in the same conditions, in 2005. It is apparent
 from this figure that the uniformity of the PSF has improved over the
 course of the survey. However, given the variations of the image
 quality across the mosaic, we scale the photometry apertures with an
 estimate of the local image quality, in order to try and minimize the spatial
 variations of the aperture corrections.
 
 The image quality on each CCD is estimated as the radius of the
 circle that has the same area as the ellipse defined by the average
 star second moments. We carry out the aperture photometry in a set of
 10 (IQ scaled) radii (typically from 5 to 30 pixels) of all the
 objects detected. The apertures are centered on the positions
 obtained from the Gaussian fit. Tests on synthetic data demonstrated
 that our aperture photometry code gives accurate estimates
 of fluxes and errors.
 
 Aperture pollution by neighbor objects can bias the flux estimations.
 We therefore chose to work with smaller apertures of radius equal to
 $7.5 \times$ HWHM (14 pixels on average). This leads to non-negligible
 aperture corrections, of at least 3\%. This is perfectly
 acceptable, as long as we check that these aperture corrections (1)
 are uniform across the focal plane and (2) are the same at the
 per mil level on the science and calibration exposures.
 
 To check that the scaling aperture strategy gives a uniform
 photometry, we studied the ratio of the flux within the standard
 aperture adopted for the calibration studies (7.5 HWHM) over the flux
 computed in a much larger aperture (20 HWHM)

\begin{figure}
\begin{center}
\includegraphics[width=\linewidth]{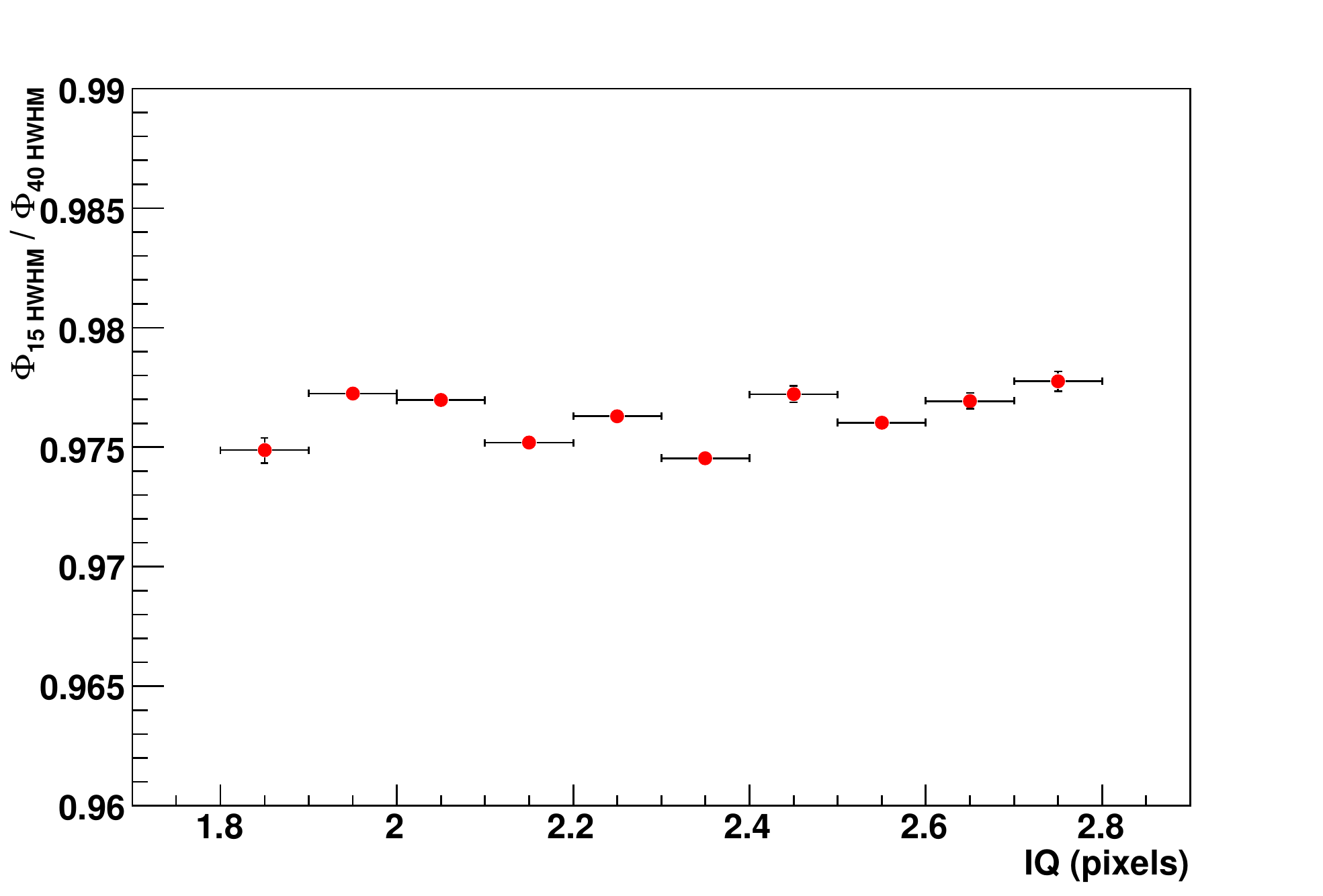}\\
\caption{Ratio of the flux computed in two apertures of different
  sizes (7.5 HWHM and 20 HWHM) as a function of the image quality
  computed for each CCD, for a set of $\rme$-band images taken on
  2003-10-01, when the PSF was not uniform over the focal plane. As
  can be seen, the ratio is not sensitive to the image quality, and
  hence, on the focal plane position. }
\label{fig:relative_aperture_corrections}
\end{center}
\end{figure}

 This shows that capturing a constant fraction of the flux of
 stars at the per mil level across the mosaic is not straightforward.
 There is a benefit for scaling apertures, but this approach is not
 sufficient for excellent quality photometry, in terms of systematic
 errors.  We however settled for scaling apertures, using an aperture
 of 7.5 seeings (RMS) in radius, which makes the aperture size an
 issue at the 0.2\% level.

\subsection{Background Estimation}

 The ``underlying background'' is the sum of all the underlying sources
 of photons which do not come from the star itself --sky background, mainly, 
 plus undetected objects. It is subtracted directly from
 the pixels using a background map computed by SExtractor. An
 incorrect background subtraction may bias the flux estimations, and
 this bias varies quadratically with the aperture radius. 
 In order to estimate the quality of the background measurements, we chose to study the variation of the aperture
 corrections in a large annulus around isolated stars as a function of
 the star fluxes, and measure how these corrections extrapolate to
 very low flux. The residual background measured is $+0.06(0.003)$,
 $-0.03(0.006)$, $-0.23(0.03)$ and $-0.04(0.02)$ ADU per pixel in the
 $\gme, \rme, \ime$- and $\zme$-bands respectively.
 In the $\gme$- and $\rme$-bands, 
 the impact of the residual background on the aperture
 magnitudes reported is smaller than 0.001 mag up to mag 21. Idem in the 
 $\zme$-band up to mag 19. In the $\ime$-band,
 however, it induces a sizeable magnitude dependent effect, reaching 0.005 mag
 at magnitude $21$. However, since the measurement of the
 residual background on a night basis is affected by large statistical
 uncertainties, we chose not to correct for this effect, and include
 it in the systematic uncertainty budget (\S \ref{sec:systematic_uncertainties}).

\subsection{Aperture Corrections}
\label{sec:aperture_corrections}

The tertiary stars in science frames are calibrated 
with respect to observations of 
standard star fields. Although in both kinds of frames
the fluxes are measured using aperture photometry, there is
a possibility that the fraction of the flux within the aperture
in both cases be different. Indeed, 
standard exposures are typically 2-3 seconds, without guiding, and with a telescope
slightly defocused, while science exposure
are integrated for 300 to 500 seconds, with guiding. In addition, there is evidence
that the telescope is still vibrating when taking the standard star exposures. As a 
consequence, the PSF on each dataset is expected to be different, leading to different 
aperture corrections.

   To address this issue, we computed the fractional increase of flux
of sufficiently isolated objects in an aperture twice as large (12 HWHM) as the 
one chosen for calibration purposes (7.5 HWHM).  This large aperture is 
considered here as an acceptable proxy for the total flux, because we expect
that the PSF shape at large radii becomes increasingly 
independent of atmospheric variability and guiding. This flux increase
is averaged over all measurements of a given night, separately 
for science and standard images, and separately for each band.
Only nights where both science and standards were observed enter
in these averages.
These averages are refered to as aperture corrections, although 
fluxes are not ``corrected''. We report in table \ref{table:aper_corr}
the average and r.m.s of these aperture corrections together with
statistics of the image quality. One can note that aperture corrections
differ only by 0.2\% between science and standard frames, and are well 
correlated over the same night. This difference is small and is probably
due to standards being observed on average
with a poorer image quality. These differential aperture corrections 
are about 0.2 \% for $g$,$r$,$i$ and $z$ bands, and compatible with
0 for $u$ band, where the image quality is on average identical for science
and standard fields. This effect is accounted for by correcting the 
tertiary standard calibrated magnitudes, as discussed in \S \ref{sec:systematic_uncertainties}.

\begin{table}[h]
\centering
\caption{Statistics of the average aperture corrections and image quality over
nights when both science and standards were observed. The correlation
coefficients (last column) are computed for data taken over the same night in the same band.
\label{table:aper_corr}}
\begin{tabular}{cc cc cc}
\hline
\hline
   & \multicolumn{2}{c}{science}  & \multicolumn{2}{c}{standards}  & \multirow{2}{*}{$\rho_{sci,std}$} \\
\cline{2-5}
 & average & r.m.s & average & r.m.s & \\
\hline 
AC & 2.34\% & 0.55\% & 2.10\% & 0.59\%  & 0.66 \\
IQ & 0.89\arcsec & 0.2\arcsec & 0.99\arcsec & 0.30\arcsec  & 0.42 \\
\hline
\end{tabular}
\end{table}

\section{The Elixir Pipeline}
\label{sec:elixir_pipeline}

 At the end of each MegaPrime run, {master twilight flat field} frames and {master fringe
frames} are built from all the exposures taken during the run,
including non-CFHTLS data (which represents about half of the total dataset). The Elixir pipeline, developed and operated by the
 CFHT team \citep{Magnier04}, 
builds master flat fields for each filter by stacking the individual flat field frames.
Individual frames inadvertently contaminated by clouds or nearby moon light
are rejected. They are identified by dividing each individual flat field exposure by the master flat, and 
inspecting the result visually. Typically, no more than one iteration is needed
to reject the outliers. In order to mitigate possible non-linearity residuals at 
the sub-percent level, individual flat field images are acquired in the 
10,000 to 15,000 ADU range. After two weeks, there are typically 40 to 60 
usable frames that can be stacked into a final normalized frame equivalent 
to a single 400,000 ADU counts per pixels, reducing to negligible photon noise levels (between one and two per mil).

If one measures the photometry of the same star on an image
flatfielded from the twilight flats, the flux varies by about $15\%$
in the $\ume$ band and $10\%$ in the $\gme$-, $\rme$-, $\ime$-, and $\zme$-bands when
moving the star from center to edges of the field of view.  The
variation is monotonic and essentially follows a circular pattern. 
A {\em photometric flat}, which ought to deliver uniform photometry across
the field of view (see below and \S \ref{sec:the_photometric_grids}) is then created by multiplying the master flat frame
by the maps of the imager photometric response non-uniformities. This frame is
the one used for flatfielding the science images of the entire run,
and allows for all multiplicative effects in the image to be corrected
at once. 
The
photometric flat was expected to correct this to within a percent.  \S
\ref{sec:the_photometric_grids} will reveal that this is not the case,
with 4\% peak-to-peak residuals remaining.

Fringe patterns are built by processing all $\ime$- and $\zme$-band
images corrected by the final flat. First the sky background is mapped
at a large scale (100 pixels) and subtracted. Then, the exposures are
scaled according to the fringe amplitudes measured on 100 peak-valley
pairs on each CCD. Since all CCDs see the same sky, a single scaling
factor is derived from the 36 CCDs. The scaled exposures are stacked,
and an iterative process similar to the one described above is carried
out, with a visual control allowing the rejection of frames containing
extended astrophysical sources such as large galaxies. Note that 
the fringe pattern contains the signature of the photometric grid.

After these steps, Elixir processes all the images of the run, and
derives an astrometric solution {\em per CCD} only, at the pixel scale level (0.2'') -- no global solution
over the mosaic is computed. The
goal is to provide the users with a first order astrometric
solution. Following this step, all the frames containing
\citet{Smith02} standards are identified and
processed, with the SExtractor package \citep{Sex}. The flux of the \citet{Smith02} standard stars 
using the SExtractor {\tt bestphot} algorithm. 
A median zero-point for the
entire run is derived for each filter, since it is not reasonable time wise to
obtain enough \citet{Smith02} standard star observations per night to derive solid zero-point
solutions. 
Again, the intention is to provide users with a photometric scaling, but 
SNLS uses its own procedures to calibrate the images, as the default calibration is 
not precise enough for our needs.

\subsection{Plate Scale Variations}
\label{sec:plate_scale_variations}

Plate scale variations cause a 
variation of the photometric response on images flat-fielded using
twilight images (or any kind of isotropic illumination), because
twilight images encode both sensitivity variations and variations
of the solid angle subtended by the instrument pixels. To correct
this photometric distortion introduced through flat fielding by plate 
scale variations,
it is common to resample images on a iso-area projection prior to photometry,
ignoring the Jacobian of the geometrical transformation in
the course of the resampling. As this approach assumes that
the only source of photometric non-uniformity of flatfielded images 
is the plate-scale variation,
we should compare plate-scale variations and photometric non-uniformity.
Note however that SNLS photometric reductions do not involve any resampling
on iso-area projections. 

\begin{figure}
\centering
\includegraphics[width=\linewidth]{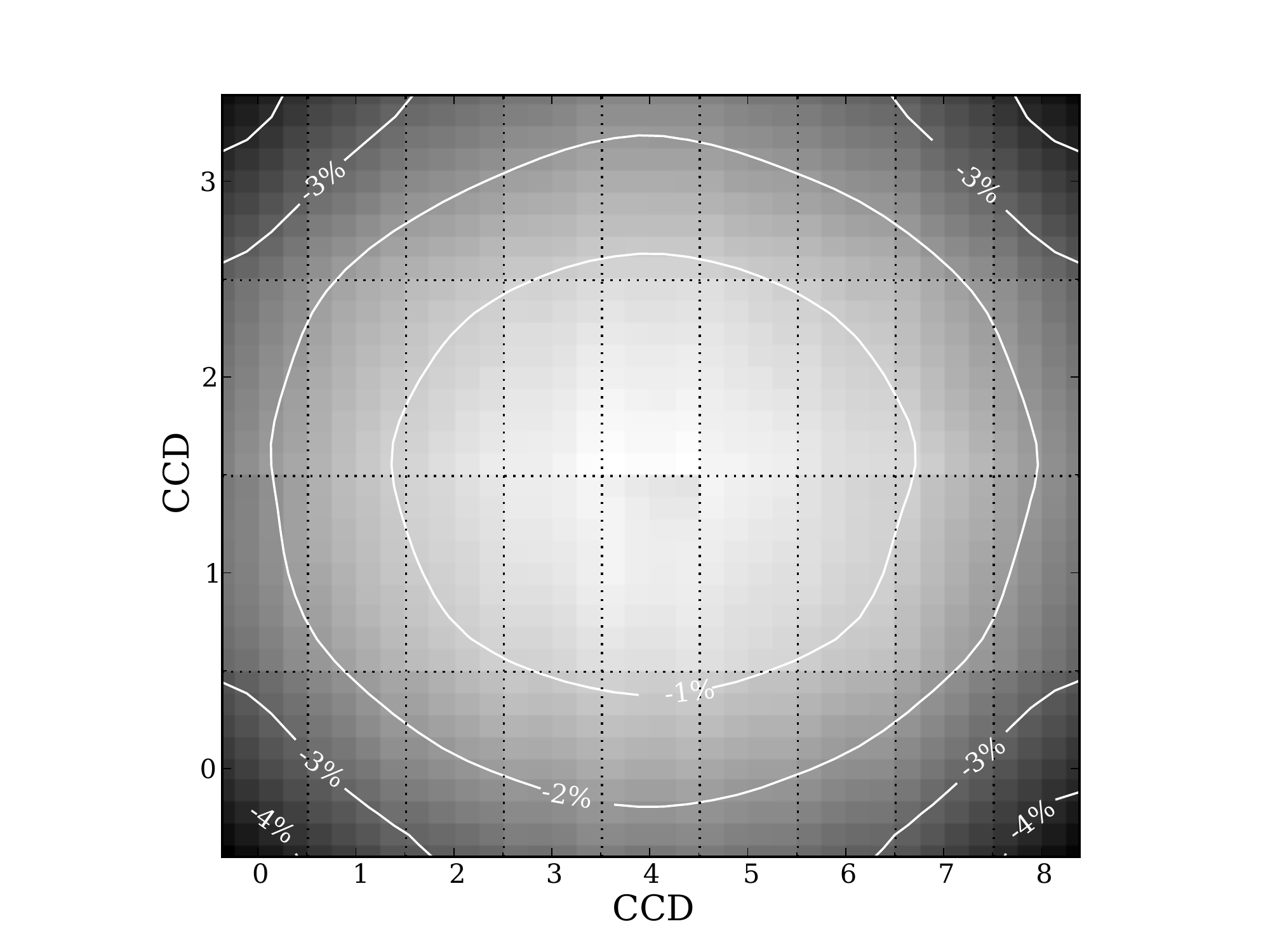}
\caption{Pixel area variations (i.e. square of the plate scale variations) 
computed on the MegaCam focal
 plane. Each grid rectangle corresponds to one CCD. CCD \#0 is up
 left. CCD \#35 is down right. Each CCD has been divided into
 $4\times9$ superpixels. The reference
 superpixel \ is located on CCD\#13, close to the center of the focal
 plane. On the edges of the focal plane, the solid angle subtended by
 a pixel can be up to 4\% smaller than on the center. Hence, due to
 this effect, the instrumental flux of a star measured on the edges,
 is at least 4\% {\em higher} than the instrumental flux of this same
 star, measured in the same conditions on the center of the
 mosaic. Note that the non-uniformities actually measured on dithered
 exposures of dense fields (\S \ref{sec:the_photometric_grids}
 and appendix \ref{sec:photometric_response_maps_details}) are about 3
 times larger. This indicates that the flatfield exposures taken
 during twilight are affected by stray light, absent from
the stars' light.}
\label{fig:plate_scale_variations}
\end{figure}

 The plate scale can be determined with a precision better than 1
 \textperthousand\ from the astrometry of dense stellar fields:
\begin{equation}
 \left|\frac{\partial \omega}{\partial x \partial y}\right| 
                                   = \frac{1}{\cos \delta} \times \left|\frac{\partial (\alpha \delta)}{\partial (x y)}\right|
\end{equation}
where $\partial \omega/\partial x \partial y$ is the solid angle subtended
by one pixel, $\alpha$ and $\delta$ refer to (e.g.) equatorial
coordinates on the sky, and $x$ and $y$ are pixel coordinates on CCDs.

Figure (\ref{fig:plate_scale_variations}) shows the variations of
$\left|{\partial xy}/\partial \omega\right|$ on the focal plane, relative to a point
located on CCD \#13, close to the center of the camera. The pixel area
scale can be up to 4\% smaller on the focal plane corners than on the
center. Since this variation is far smaller than the variation of
photometric response we measure ($\sim 10 - 15\%$), we should conclude that most of the
photometric non-uniformity is {\em not} due to plate scale
variations.

The measured plate scale variation pattern is similar in all bands and
seems to be extremely stable. It was not affected by the substantial 
changes of the imaging system : we have compared plate scales measured
on semester 2003B and 2006B and found them to be identical at the
1\textperthousand-level.

\subsection{Elixir strategy}
The residual non-uniformity of photometric response
is likely to be due to some source of light hitting the focal
plane that does not follow the normal light path that
forms a star image. This ``extra'' light
is called ``scattered light'' in Elixir parlance.
This name remains although the elimination of
genuine sources of scattered light
by extra baffling added in MegaPrime did not change
the overall shape of twilight images.

It is however not necessary to identify the source of photometric
response variations across the field of view to correct it.
The pragmatic approach of the Elixir team has been 
to measure as precisely as possible the non-uniformity
of the photometric response and include a correction in the flat-field images
applied to the raw images. This choice has two consequences. 
First, images flat-fielded this way have a spatially varying sky
background. Second, the 4\% response variation expected from plate-scale
variations is corrected for in the flat-fielding 
and users resampling images should 
account for the Jacobian of the resampling
transformation prior to photometry, in order to preserve fluxes
across the resampling.

The key data to measure the photometric response is a set of dithered
exposures on low Galactic latitude fields, also called the ``photometric
grid'' by the Elixir team. Elixir has implemented a reduction of these
data sets and the resulting correction maps are incorporated into the
flat-fields.  Since the goal of the SNLS is to calibrate the imager
with a sub percent precision, we have decided to put the Elixir
pipeline results under scrutiny and to redetermine independently the grid
correction maps applied to the data.  

\section{The Photometric Grids}
\label{sec:the_photometric_grids} 

The photometric response non-uniformity maps are derived
from dithered observations of the two dense fields listed in table
\ref{tab:grid_fields}. The dithered sequence starts with an exposure
followed by a series of 6 offset observations in the X direction and
6 in the Y direction. The steps increase logarithmically from
a few hundred pixels up to half a mosaic. Such observations are
performed almost each semester, and after each significant change in
the optical path. Table \ref{tab:grid_observations} lists all the
datasets taken since the beginning of the survey, along with the
improvements of the optical path performed by the CFHT team. In order 
to limit the shortcomings of extinction corrections, observations are carried out
in a single sequence and over a limited airmass range (typically 0.01 or less). This data 
is refered to as ``grid observations'' in the Elixir
parlance, and the corrections derived from them are called ``grid
corrections''. 

\begin{table}
\centering
\caption{The dense stellar fields observed to model the imager non-uniformities (Grid Fields).\label{tab:grid_fields}}
\begin{tabular}{lccc}
\hline
\hline
Name & RA & DEC & \\
\hline
 Grid-1 & 06:30:00.00 & 14:20:00.0 & winter field \\
 Grid-2 & 20:00:00.00 & 10:00:00.0 & summer field \\
\hline
\end{tabular}
\end{table}

\begin{table}
\centering
\caption{Grid field observations and MegaCam/MegaPrime optical
  improvements. \label{tab:grid_observations}}
\begin{tabular}{cccc}
\hline
\hline
Semester & Date       & Field  & band \\
\hline
\multirow{3}{0.75cm}{2003B} 
 & 2003-10-02 & Grid-2 & $\rme\ime$    \\
 & 2003-10-23 & Grid-2 & $\ume\gme\zme$   \\
 & 2003-10-24 & Grid-2 & $\ime \rme$    \\
\hline
\multirow{3}{0.75cm}{2004A} 
 & 2004-02-25 & \multicolumn{2}{c}{\em light baffle installed} \\
 & 2004-03-19 & Grid-1 & $\ume \gme \rme \ime \zme$ \\
\hline
\multirow{3}{0.75cm}{2004B} 
 & 2004-12-03 & \multicolumn{2}{c}{\em L3 Lens flipped upside-down} \\
 & 2004-12-03 & Grid-1 & $\gme \rme \ime$   \\
 & 2004-12-11 & Grid-1 & $\ume \gme \zme$   \\
\hline
\multirow{1}{0.75cm}{2004A} 
 & 2005-07-20 & \multicolumn{2}{c}{\em  Spacer moved (5.5 mm to 2.1 mm)} \\
\hline
\multirow{2}{0.75cm}{2005B} 
 & 2005-09-20 & \multicolumn{2}{c}{\em Focal plane tilt tweaked} \\
 & 2005-10-10 & Grid-2 & $\ume \gme \rme \ime \zme$ \\
\hline
\multirow{5}{0.75cm}{2006B} 
 & 2006-09-13 & Grid-2 & $\ime$ \\
 & 2006-09-14 & Grid-2 & $\rme$ \\
 & 2006-09-15 & Grid-2 & $\zme$ \\
 & 2006-09-16 & Grid-2 & $\ume$ \\
 & 2006-09-19 & Grid-2 & $\gme$ \\
\hline
\multirow{4}{0.75cm}{2007A} 
 & 2007-03-21 & Grid-1 & $\rme$ \\
 & 2007-03-22 & Grid-1 & $\ume \gme$ \\
 & 2007-03-23 & Grid-1 & $\ime$ \\
 & 2007-03-24 & Grid-1 & $\zme$ \\
\hline
\multirow{4}{0.75cm}{2007B} 
 & 2007-10-18 & Grid-1 & ${\ime}_2$ \\
 & 2007-11-08 & Grid-1 & ${\ime}_2$ \\
 & 2007-11-11 & Grid-1 & $\rme$ \\
 & 2007-11-12 & Grid-1 & $\gme$ \\
\hline
\end{tabular}
\end{table}

The grid exposures are processed with the standard Elixir procedure,
except that no photometric flat is applied to the pixels. The
instrumental fluxes of the grid stars are therefore affected by the
10\% to 15\% non-uniformities we are trying to model.

The data is reduced using the standard SNLS reduction procedure
described below (\ref {sec:aperture_photometry}). 
In particular, we filter the detected objects on each
individual exposure, selecting isolated star-like objects. In the
$\rme$-band, this leaves us with $\sim 1200$ usable flux measurements per
CCD. For a whole grid sequence (13 dithered observations), this represent about
100,000 stars measured twice or more, and about 600,000 flux
measurements.

\subsection{The Photometric Response Maps}
\label{sec:the_photometric_response_maps}

Our goal is to determine how the instrumental
magnitudes\footnote{i.e. $-2.5 \times \log_{10} \phi_{ADU}$, where $\phi_{ADU}$ is the instrumental flux of the object, 
  expressed in ADU per second.} of a star
vary as a function of the position on the focal plane. In practice, it
is convenient to choose a specific location as the {\em reference
  location}, $\x_0$, and the relation between the instrumental
magnitudes of an object, at positions $\x$ and $\x_0$, $m_{ADU|\x}$
and $m_{ADU|\x_0}$ can be parametrized as:
\begin{equation*}
  m_{ADU|\x} - m_{ADU|\x_0} = \delta zp(\x) + \delta k(\x) \times {\rm col}_{|\x_0}
\end{equation*}
where ${\rm col}$ is some color of the star.
The maps $\delta zp(\x)$ and $\delta k(\x)$ are determined from the
grid observations, and account for the non-uniformities of the
imager. By definition, these maps vanish at the reference location:
$\delta zp(\x_0) = 0$ and $\delta k(\x_0) = 0$. Physically
speaking, $\delta zp(\x)$ encodes the spatial variations of the overall integral of the bandpass,
while $\delta k(\x)$ encodes (at first order) the variations of its
central wavelength. In what follows, ``passband variations''
refer to variations of the passband shape irreducible to an overall constant.
If the imager passbands are uniform over the focal plane,
we expect $\delta k(\x)$ to be zero everywhere. 
Conversely, if the imager passbands do vary as a function of the position, we
should measure a non-zero color term between positions $\x$ and
$\x_0$.

The quantity ${\rm col}_{|\x_0}$ is the star color, measured in the
MegaCam passbands at the reference location. There is some degree of
arbitrariness in the definition of ${\rm col}_{|\x_0}$, since it is
always possible to redefine the color reference ${\rm col}
\rightarrow {\rm col} + \Delta {\rm col}$, and absorb the difference
in the $\delta zp(\x)$'s: $\delta zp(\x) \rightarrow \delta zp(\x) -
\delta k(\x) \times \Delta {\rm col}$. Hence, there is a color
convention associated with each $\delta zp(\x)$ map, which must be
made explicit.  We choose to define a conventional {\em grid
  reference color}, ${\rm col}_{grid}$, and parametrize the grid
corrections as:
\begin{equation}
  m_{ADU|\x} - m_{ADU|\x_0} = \delta zp(\x) + \delta k(\x) \times \Bigl[\ {\rm col}_{|\x_0} - {\rm col}_{grid} \Bigr]
\label{eqn:grid_correction_model}
\end{equation}
There are several ways to choose the grid reference colors. One
possibility is to define them as the mean color of the grid stars, in
order to minimize the statistical uncertainties carried by the $\delta
zp(\x)$ maps. Another way is to choose them close to the colors of the 
science objects under study. This way, the instrumental magnitudes 
of these objects, once corrected by the $\delta zp(\x)$ maps would 
be almost uniform on the focal plane. 
We must note however, that the grid reference colors are just internal
quantities, that do not affect the definition of the tertiary
magnitudes. 

In the next section, we detail how we extract the $\delta
zp(\x)$ and $\delta k(\x)$ maps from the grid data.

\subsection{Measuring the Photometric Response Maps}
\label{sec:measuring_the_photometric_response_maps}

As described in section \ref{sec:elixir_pipeline} the grid data in one
MegaCam band consists in 13 dithered observations of a dense stellar
field (about 100,000 isolated stars per exposure). Each field star
being observed at various locations on the focal plane, one can
compare the instrumental magnitudes at those locations and fit the
$\delta zp(\x)$ and $\delta k(\x)$ maps defined in the previous
section.

To parametrize the grid maps, we chose to develop them on basis
functions: $\delta zp(\x) = \sum_k \alpha_k p_k(\x)$ and $\delta k(\x)
= \sum_k \beta_k q_k(\x)$. Since the photometric response variations
may be sharp, especially when transitioning from one CCD to another,
we used a basis of independent superpixels (also called {\em cells}
thereafter), rather than smooth functions such as splines or
polynomials. To model the $\delta zp(\x)$ map, we divided each CCD
into $N_x \times N_y = 4 \times 9$ superpixels.  Each superpixel is
$512 \times 512$ pixels wide and contains about 70 bright, isolated grid
stars, which is enough to measure the $\delta zp(\x)$ with a precision better
than 0.001 mag.  We have found that determining the $\delta k(\x)$
requires more stars per superpixel, in order to have a sufficient 
color lever-arm in each cell. We therefore used larger cells,
dividing each CCD into $N_x \times N_y = 2 \times 3$ cells.

With such a parametrization, the number of grid parameters is $4 \times 9 \times 36  +
2 \times 3 \times 36 - 2 = 1510$, one cell being taken as a reference for each map. Fitting
such a model (i.e. building and inverting a $1510 \times 1510$
symmetric positive definite matrix) is routinely done on modern
desktop computers.

With only 13 dithered exposures, about 99\% of the grid
stars are never observed on the reference cells. Hence the
instrumental magnitudes of these stars at the reference location,
$m_{ADU|\x_0}$, are never directly measured and must be fitted along with
the grid map parameters. The same is true for the
star colors ${\rm col}_{|\x_0}$, which means that we must analyze the
grid data in two bands simultaneously. This adds about 200,000
(nuisance) parameters to the fit, turning it into a large
dimensionality, non-linear problem, which is much more difficult to solve.

In order to make the problem tractable, we have followed an iterative
approach, after checking with simulations that this is feasible.
In the first step, we fit the data independently in each band, ignoring the
$\delta k(\x) \times {\rm col}$ terms. We then update the star colors,
and refit the grid data in each band. We have found that it suffices
to iterate the procedure twice in order to retrieve the correct
colors. No such iterative procedure applies for the star magnitudes
though. Fitting 100,000 parameters is still possible, using approximate 
methods, which do not require to build the $\chi^2$ second derivative matrix, such as the 
conjugate gradient method. It is also possible to fit alternatively one set of parameters, keeping 
the other ones fixed. We have experienced however, that the conjugate gradient method
converges very slowly and can lead to wrong solutions, if one is not
ready to perform the $O(100,000)$ required iterations. Successive
optimizations on the grid parameters at fixed star magnitudes, and
vice versa also converge very slowly and remain far from the true
minimum.

Fortunately, we have found that the structure of the problem is
simple enough to allow us to obtain the true minimum of the $\chi^2$, and determine simultaneously the grid parameters,
the grid star magnitudes, along with their exact covariances. The details of the method are documented in details 
in appendix \ref{sec:photometric_response_maps_details}. The approach in fact applies to many
other calibration problems, such as least squares for astrometric
solutions as posed in e.g. \cite{Kaiser99,Padmanabhan07}. In the remaining 
of this section, we will present the results obtained with this technique.

\subsection{Monte-Carlo Checks}
\label{sec:monte_carlo_checks}

The extraction technique described in appendix \ref{sec:photometric_response_maps_details} was tested on simulated
data. Our main concern was to look for possible degeneracies
affecting the determination of the grid correction parameters, as well as
possible biases, coming from the fact that the fit is slightly
non-linear.

A realistic catalog of grid stars was built from real $\rme$- and
$\ime$-band measurements. Using this catalog, a photometric distortion
model and the dithering pattern applied to the grid sequence, we generated a hundred realizations
of a typical two-band grid run. From this set of data, the grid
corrections, $\delta{zp}(\x)$ and $\delta{k}(\x)$ were reconstructed from each
realization and compared to the photometric distorsion model.

\begin{figure}
\centering
\includegraphics[width=\linewidth]{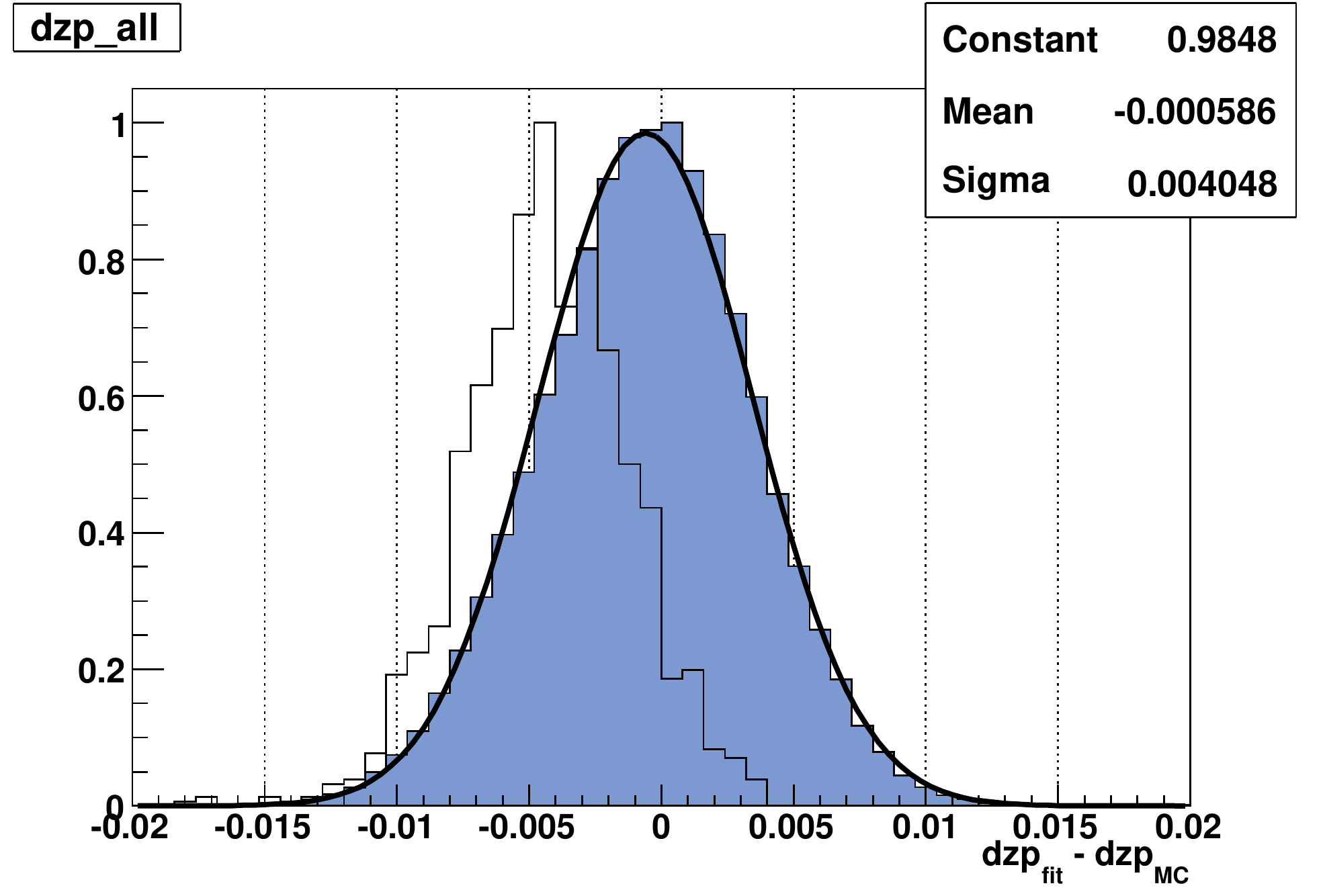}\\
\includegraphics[width=\linewidth]{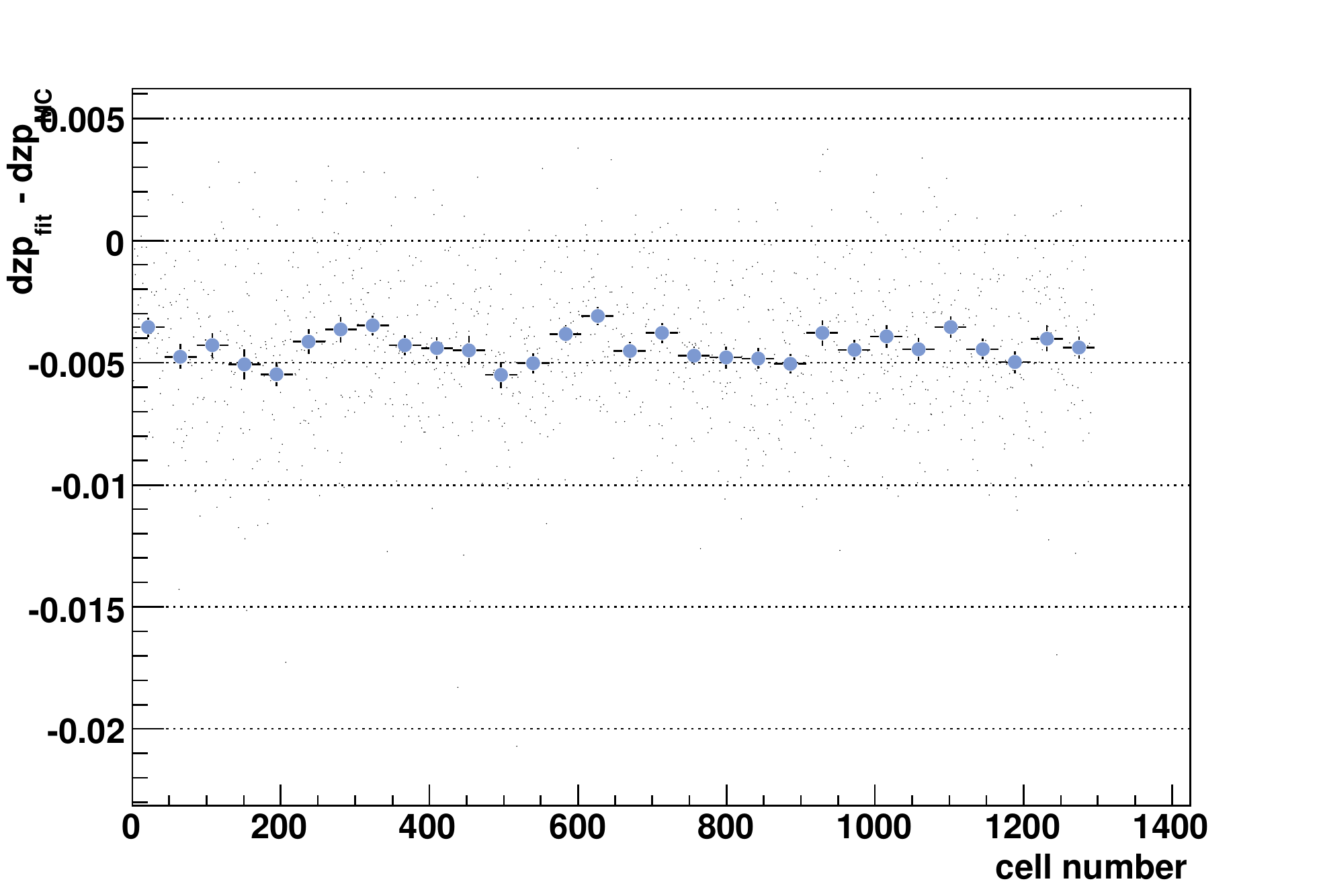}
\caption{Top panel: normalized distribution of the difference between 
 the reconstructed and simulated $\delta{zp}(\x)$ parameters, for 100
 realization (filled histogram) and 1 single realization (unfilled histogram). The
 reconstructed grid parameters are globally affected by the
 uncertainty of the reference cell zero point (see section
 \ref{sec:monte_carlo_checks}). Bottom panel: difference of the
 reconstructed and simulated grid parameters as a function of the cell
 number, for a single realization. No spatially dependent effect can
 be detected.}
\label{fig:grid_reconstruction_dzp}
\end{figure}

\begin{figure}
\centering
\includegraphics[width=\linewidth]{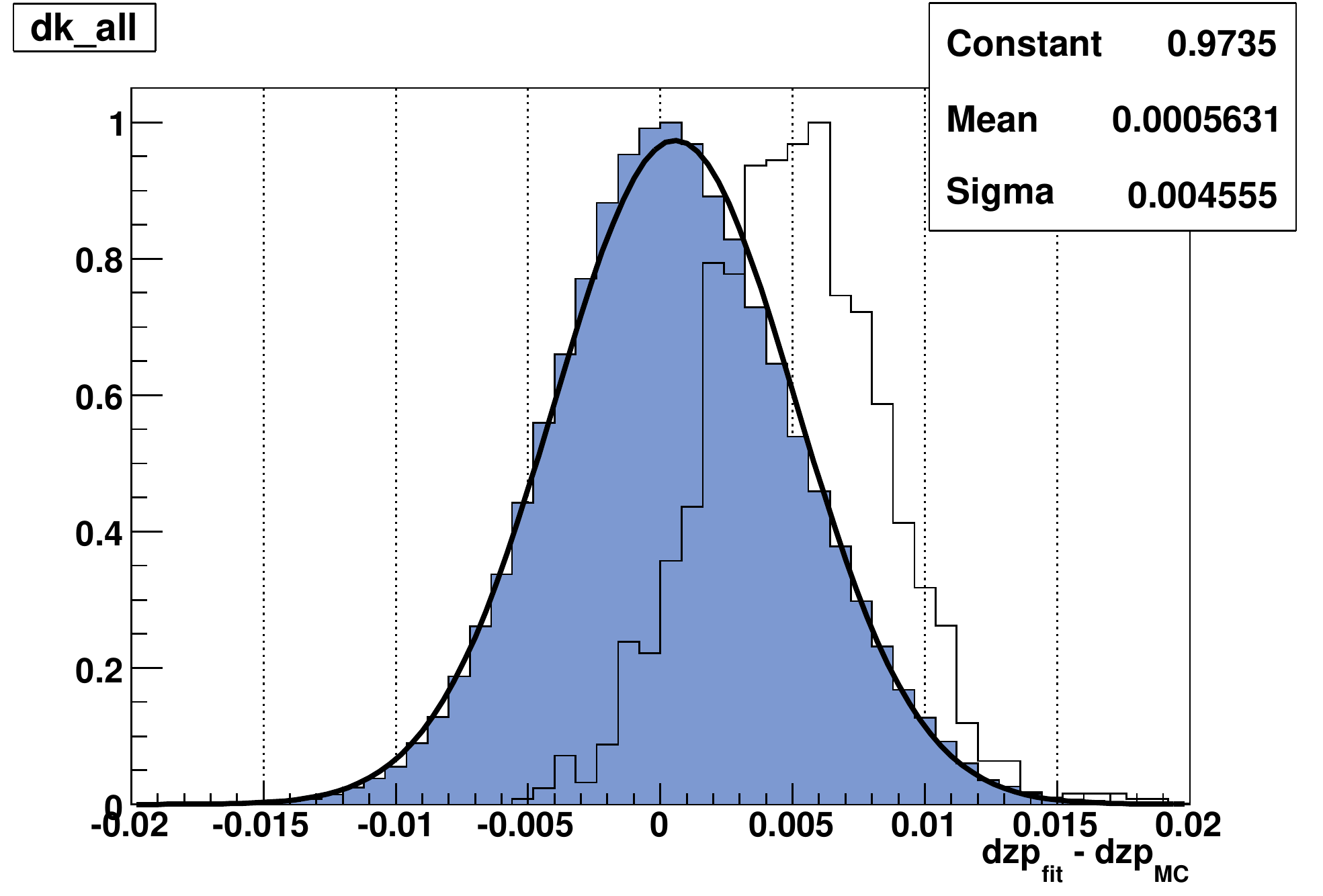}\\
\includegraphics[width=\linewidth]{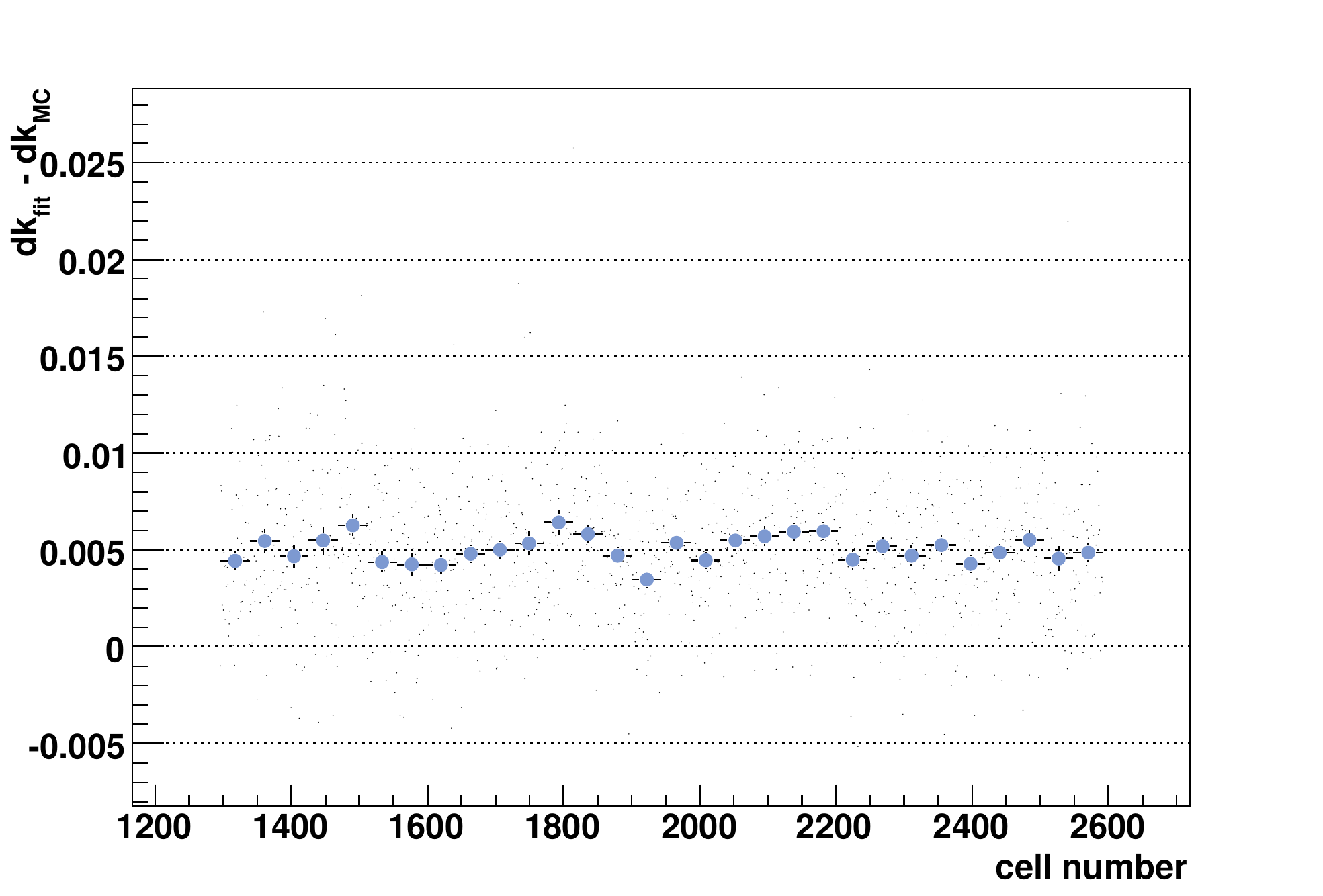}
\caption{Top panel: distribution of the difference between 
 the reconstructed and simulated $\delta k$ parameters, for 100
 realization (filled histogram) and 1 single realization (unfilled histogram). The
 reconstructed grid parameters are globally affected by the
 uncertainty of the reference cell zero point (see section
 \ref{sec:monte_carlo_checks}).  Bottom panel: difference of the
 reconstructed and simulated grid parameters as a function of the cell
 number, for a single realization. No spatially dependent effect can
 be detected.}
\label{fig:grid_reconstruction_dk}
\end{figure}

Figures (\ref{fig:grid_reconstruction_dzp}) and
(\ref{fig:grid_reconstruction_dk}) show the distribution of the
difference between the simulated photometric distorsion model and the
reconstructed grid calibration parameters, in the $\ime$-band, for all
the 100 realization processed. It is clear from those figures that the
extraction method is an unbiased estimator of the photometric
distorsion model, as could be expected from a (almost) linear fit.

On the same figure, we have represented the same difference, obtained
for one realization only. We see that the reconstructed grid
correction parameters are offset by a small amount, and the
value of the offset is almost exactly opposite for the
$\delta{zp}(\x)$ and the $\delta{\lambda}(\x)$. This effect comes
from the structure of the noise affecting the $\delta{zp}(\x)$ and
$\delta {k}(\x)$ map estimates. All the $\delta{zp}(\x)$ (resp. $\delta{
  k}(\x)$) are positively correlated with a correlation coefficient of
about $0.5$. These values can be easily explained if we consider that
each $\delta{zp}$ (resp. $\delta{k}_i$) cell value is the difference
between the the zero point (resp. color term) of cell number $i$ and
the zero point (resp. color term) of the reference cell. 

Robustification algorithms have been implemented in the fit procedure:
the fit was performed several times, and at each cycle, the
measurements with a partial $\chi^2$ above 4 sigmas of the median
$\chi^2$ were removed.
This is unavoidable, since least-square methods are
extremely vulnerable to outliers, and since we are dealing with $10^5 -
10^6$ measurements, potentially affected by undetected cosmics or bad
pixels. We have checked the robustification scheme by generating 2\%
of outliers in the simulated dataset. The error affecting the outliers
was chosen uniform in the $[-2{\rm mag};+2{\rm mag}]$ range. The
results presented in figures (\ref{fig:grid_reconstruction_dzp}) and
(\ref{fig:grid_reconstruction_dk}) show that the robustification
procedure correctly rejects the outliers.

The extraction method described above ignores possible variations of
the atmospheric absorption during one dithering scan. Since such
variations can bias the grid parameter reconstruction, it is tempting
to fit, along with the grid parameters, one global zero point per
exposure, in order to correct from such variations.  We have
implemented this model, and run it on simulated data. We have found,
inspecting the covariance matrix of the fit that there is an almost
total degeneracy between the grid parameters and the exposure
zero-points.

Not surprisingly, the uncertainties on the reconstructed grid
parameters, marginalized over the exposure zero points are much
higher, of the order of several percents. Hence, the only way to deal
with atmospheric absorption is to add a control exposure at the end of
the sequence, similar to the first exposure.  This would allow one to
assess whether the sequence was photometric or not. Unfortunately, no
such control exposure is currently available.  One may recover an
acceptable accuracy by fitting a coarser (typically one coefficient
per CCD) non-uniformity model to the data. One can still be reasonably confident that 
the atmospheric variations do not affect the grid data. Indeed, before taking a grid sequence, 
the observers ensured that the conditions were indeed photometric according to the 
SkyProbe\footnote{{\tt www.cfht.hawaii.edu/Instruments/Elixir/skyprobe}} monitor.

Finally, it is interesting to compare the minimum $\chi^2$ values
obtained when fitting either the grid model presented above or the
same model, with all $\delta {k}(\x)$ parameters fixed to zero. We notice that
the $\delta \chi^2$ between both models is of the order of 3000 (for
215 parameters added). This is a very significant improvement.  However the
fractional improvement on the fit residuals remains very modest: below
$10^{-2}$. This shows that when dealing with such a large
dimensionality problem, we can get a perfectly acceptable reduced
$\chi^2_{min}$ for a model which is obviously wrong. Hence, obtaining
the true minimum of the $\chi^2$ function using an exact technique
such as the one presented in appendix
\ref{sec:photometric_response_maps_details} is vital to
determine the optimal grid corrections.

\subsection{Results}
\label{sec:grid_results}

All the datasets presented in table \ref{tab:grid_observations} were
analyzed with the procedure described in \S
\ref{sec:measuring_the_photometric_response_maps}, 
\ref{sec:monte_carlo_checks} and appendix \ref{sec:photometric_response_maps_details}. We parametrize the
color corrections as follows:
\begin{eqnarray}
  g_{ADU|\x} & =  g_{ADU|\x_0} + \delta zp_g(\x) &+ \delta k_{ggr}(\x) \times \Bigl[ (g-r)_{|\x_0} - (g-r)_{grid} \Bigr] \nonumber \\
  r_{ADU|\x} & =  r_{ADU|\x_0} + \delta zp_r(\x) &+ \delta k_{rri}(\x) \times \Bigl[ (r-i)_{|\x_0} - (r-i)_{grid} \Bigr] \nonumber \\
  i_{ADU|\x} & =  i_{ADU|\x_0} + \delta zp_i(\x) &+ \delta k_{iri}(\x) \times \Bigl[ (r-i)_{|\x_0} - (r-i)_{grid} \Bigr] \nonumber \\
  z_{ADU|\x} & =  z_{ADU|\x_0} + \delta zp_z(\x) &+ \delta k_{ziz}(\x) \times \Bigl[ (i-z)_{|\x_0} - (i-z)_{grid} \Bigr] \nonumber 
\end{eqnarray}

\begin{figure*}
\centering
\mbox{\subfigure[$\delta {zp}_{g,g-r}(\x)$]{\includegraphics[width=0.45\linewidth]{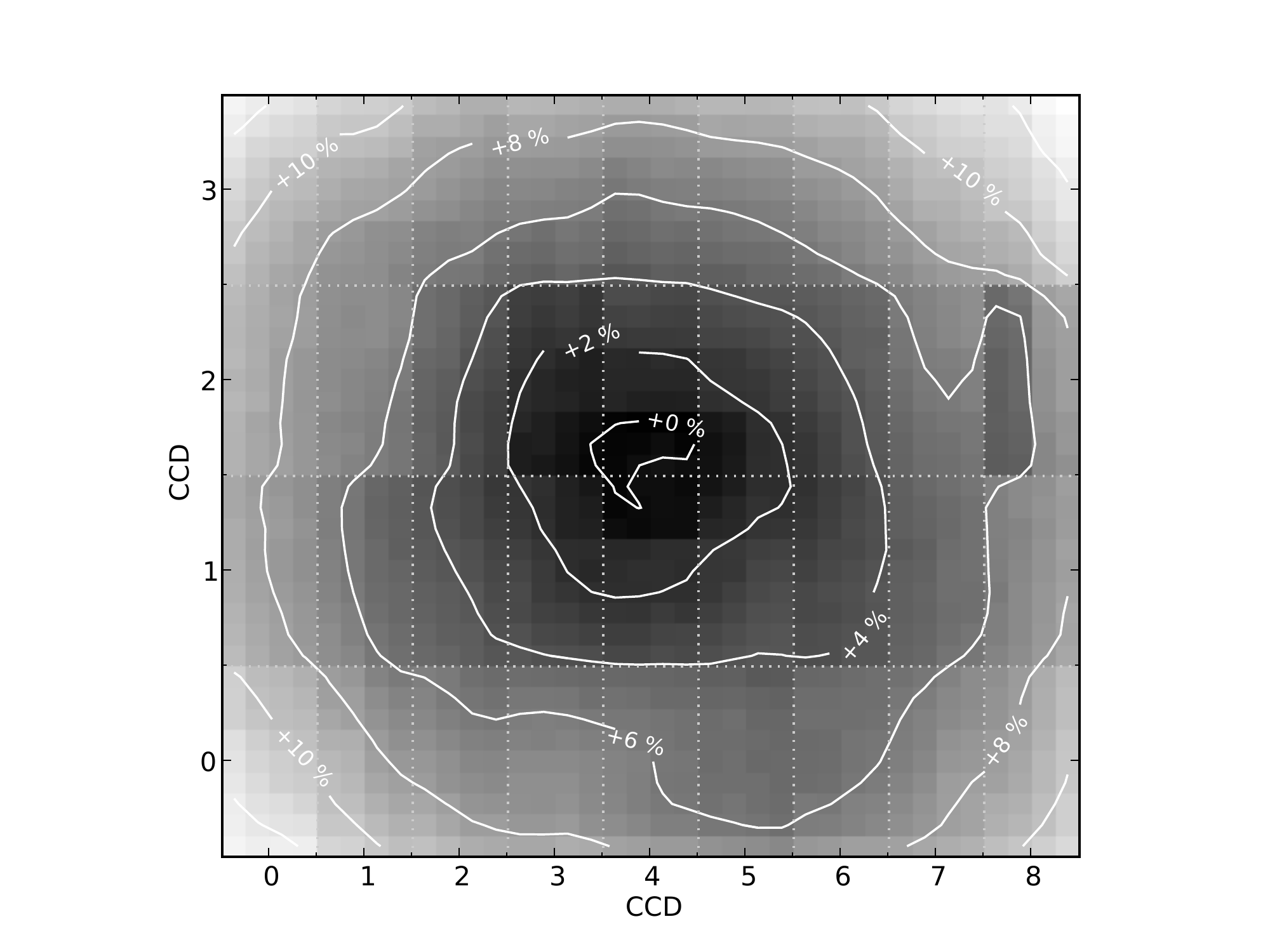}}
      \subfigure[$\delta {zp}_{r,r-i}(\x)$]{\includegraphics[width=0.45\linewidth]{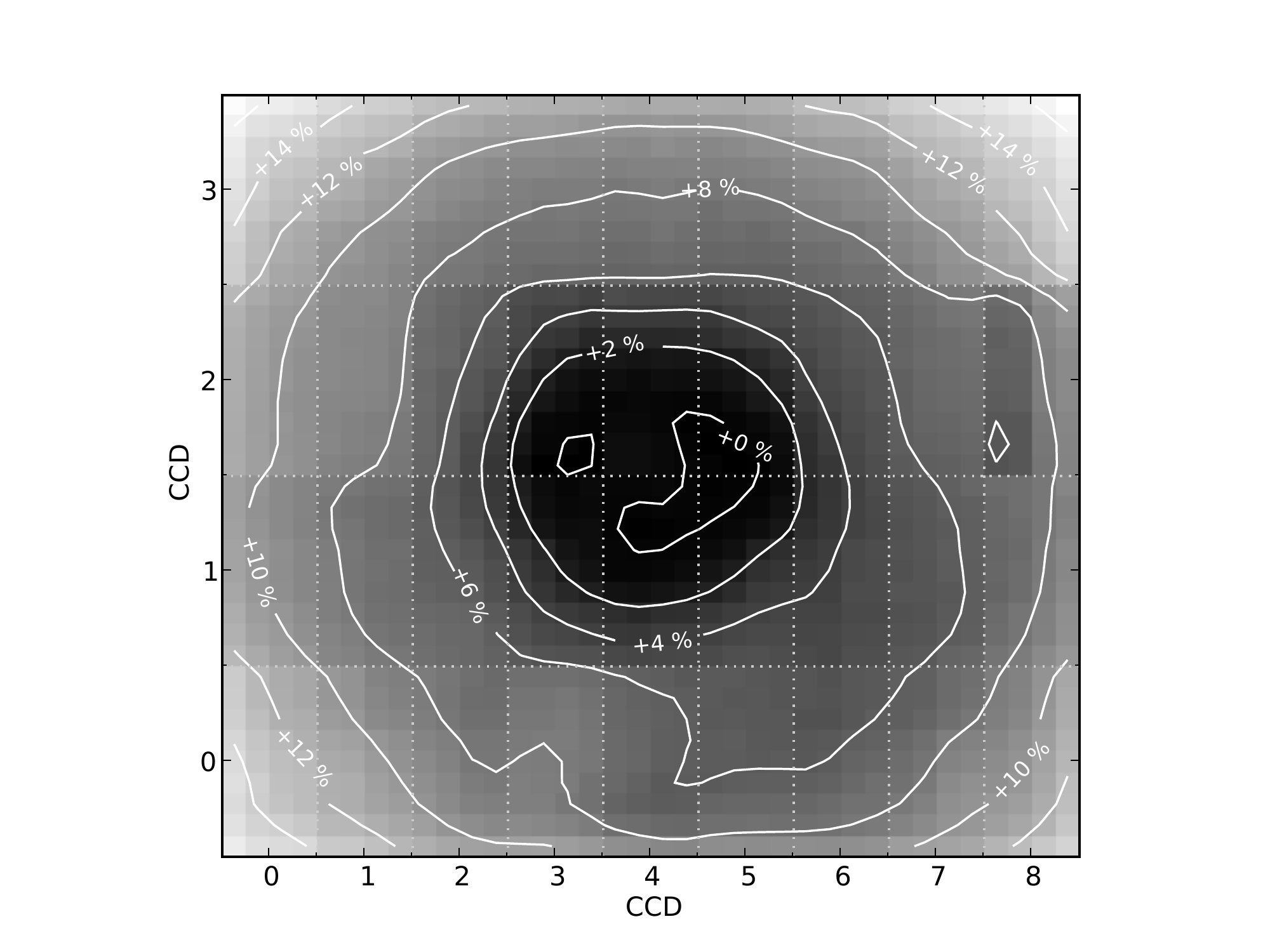}}}
\mbox{\subfigure[$\delta {zp}_{i,r-i}(\x)$]{\includegraphics[width=0.45\linewidth]{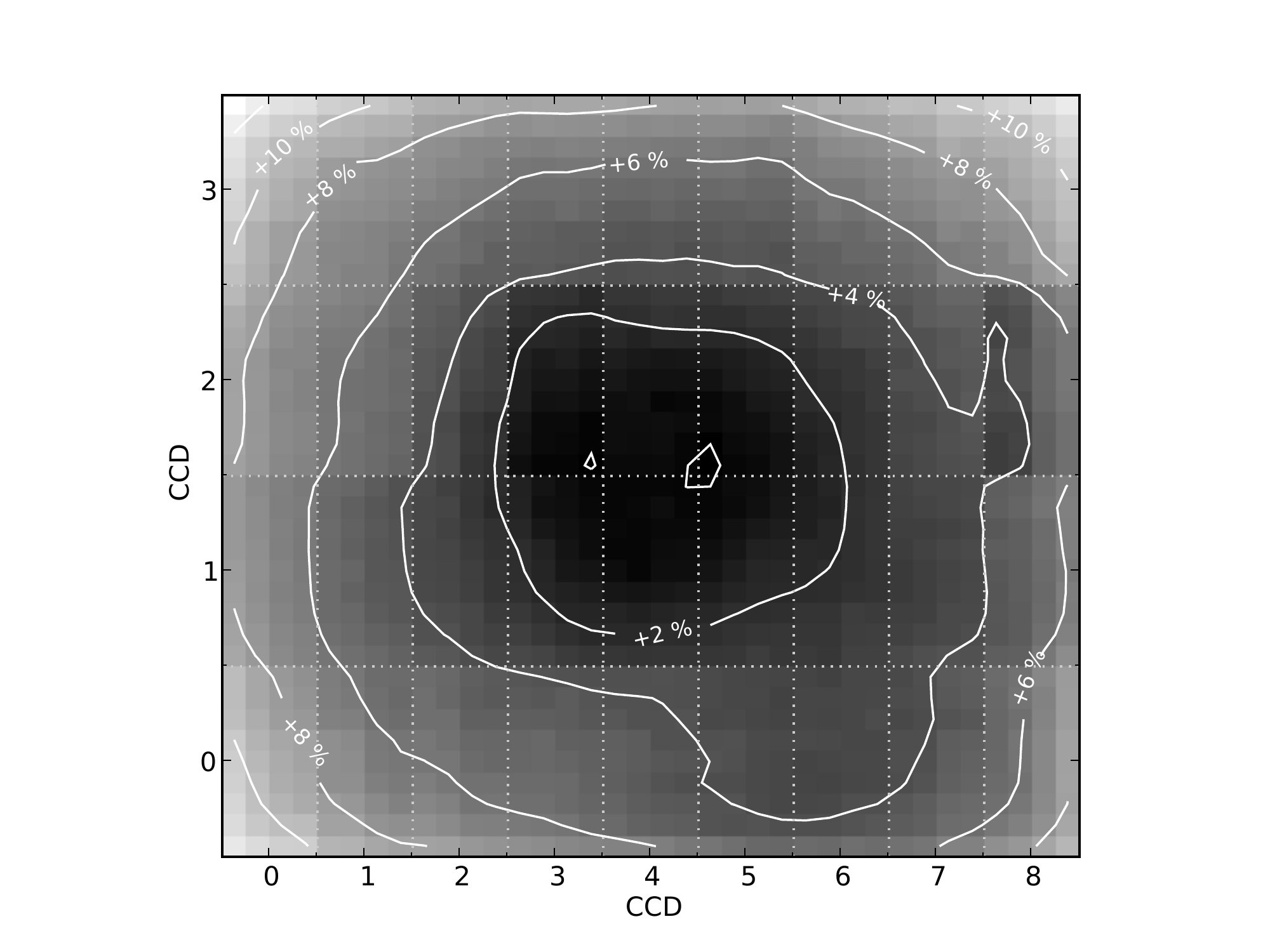}}
      \subfigure[$\delta {zp}_{z,i-z}(\x)$]{\includegraphics[width=0.45\linewidth]{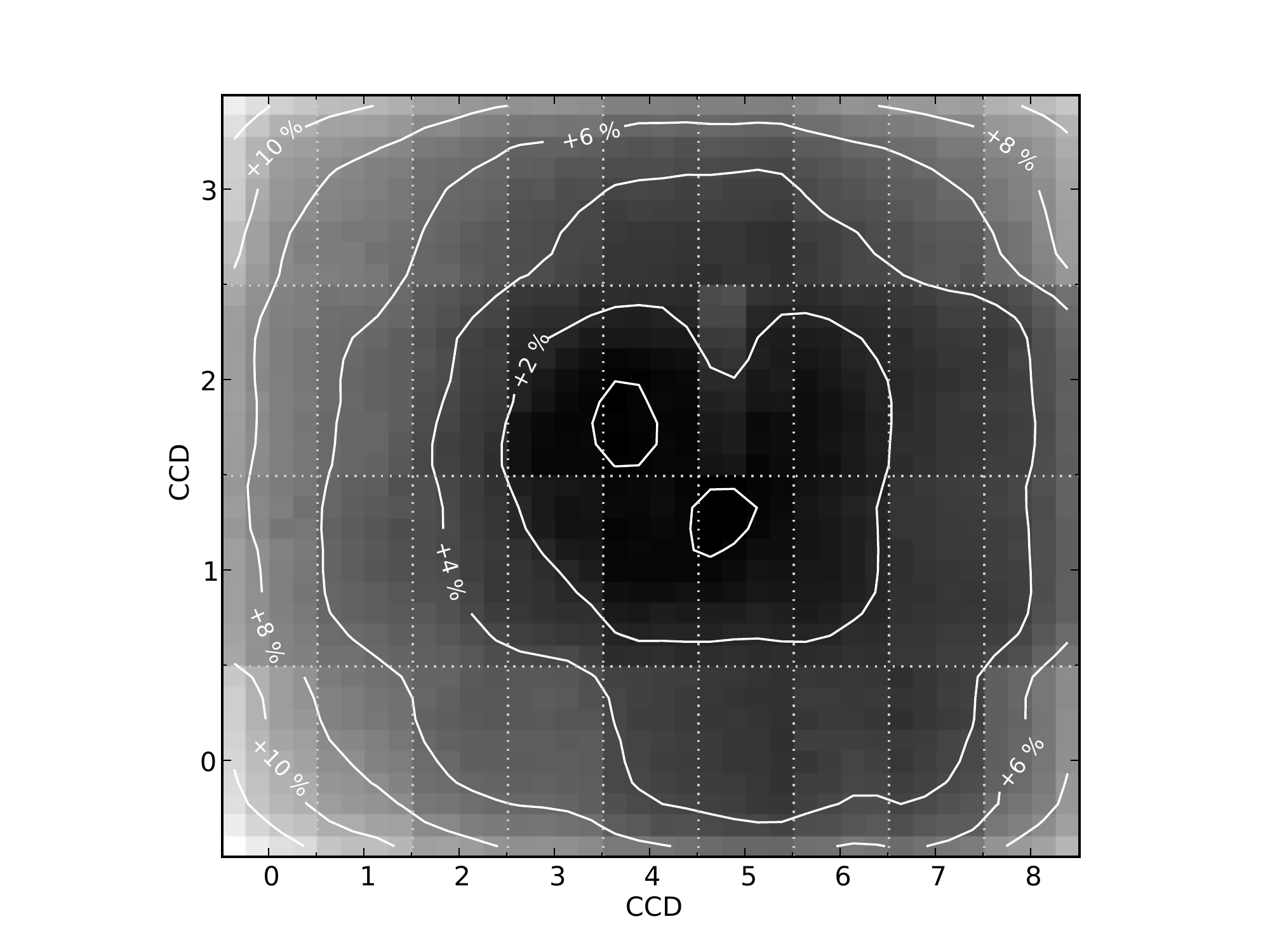}}}
\caption{$\gme, \rme, \ime$- and $\zme$-band $\delta {zp}(\x)$ maps
  determined from grid observations taken in 2005. In all bands, the
  instrumental flux of a star, measured on the corner a flat-field
  corrected image, will be 10 to 15\% higher than that of the same
  star, measured on the center of the mosaic. The plate scale
  variations account for about half of this effect. The other half is
  attributed to so-called ``scattered light'', i.e. light that does
  not follow the normal path that form a star.  In some bands, we see
  sharp steps between the two CCD amplifiers. These are due to the fact
  that the flat-field images are taken by combining all the images
  taken during a run, and that the gain of the amplifiers can vary by
  about 1\% during a run.
\label{fig:dzp_maps}}
\end{figure*}

\begin{figure*}
\centering
\mbox{\subfigure[$\delta {zp}_{r,r-i}(\x)\ \ (2003B)$]{\includegraphics[width=0.45\linewidth]{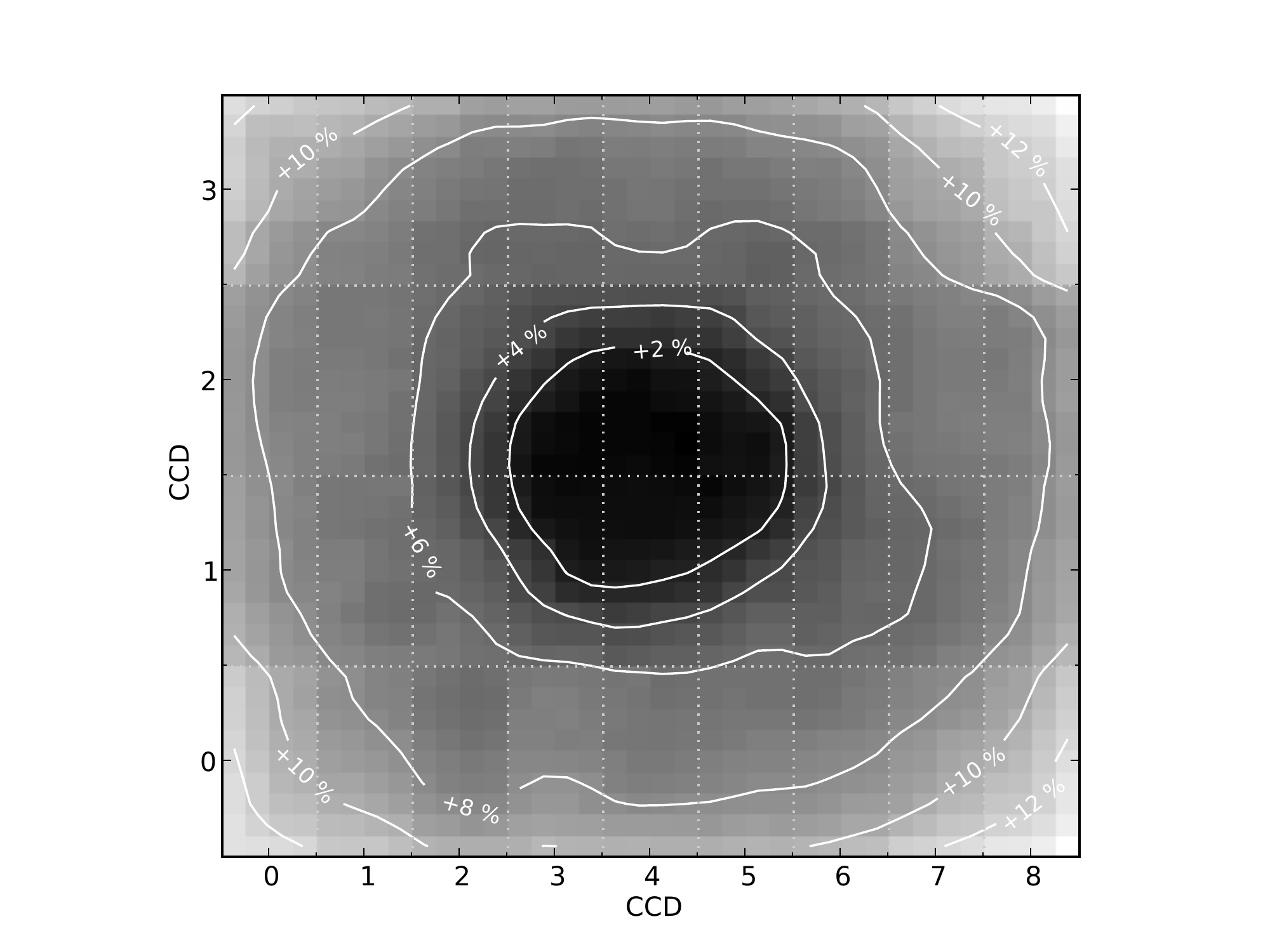}}
      \subfigure[$\delta {zp}_{r,r-i}(\x)\ \ (2004B)$]{\includegraphics[width=0.45\linewidth]{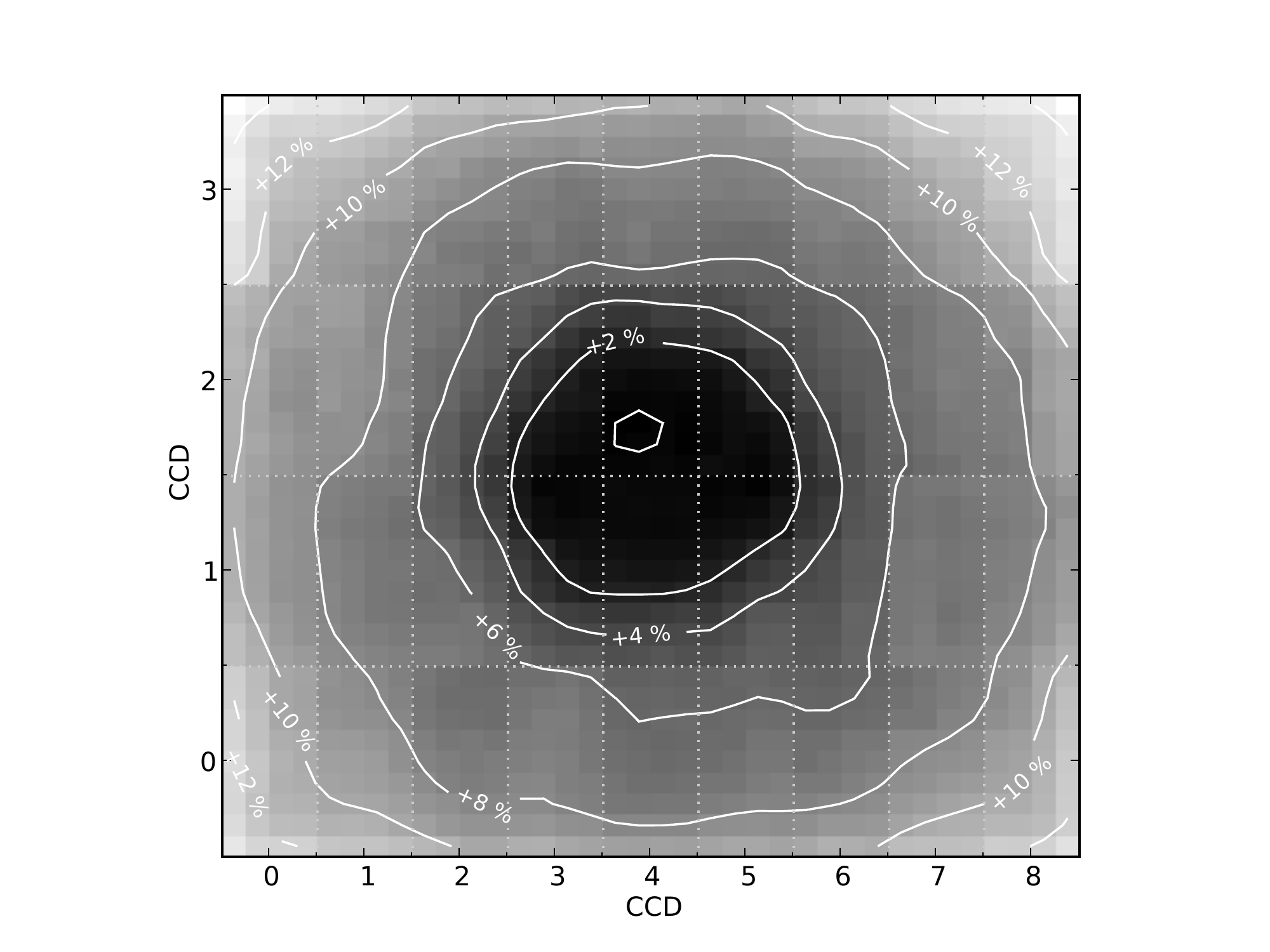}}}
\mbox{\subfigure[$\delta {zp}_{r,r-i}(\x)\ \ (2005B)$]{\includegraphics[width=0.45\linewidth]{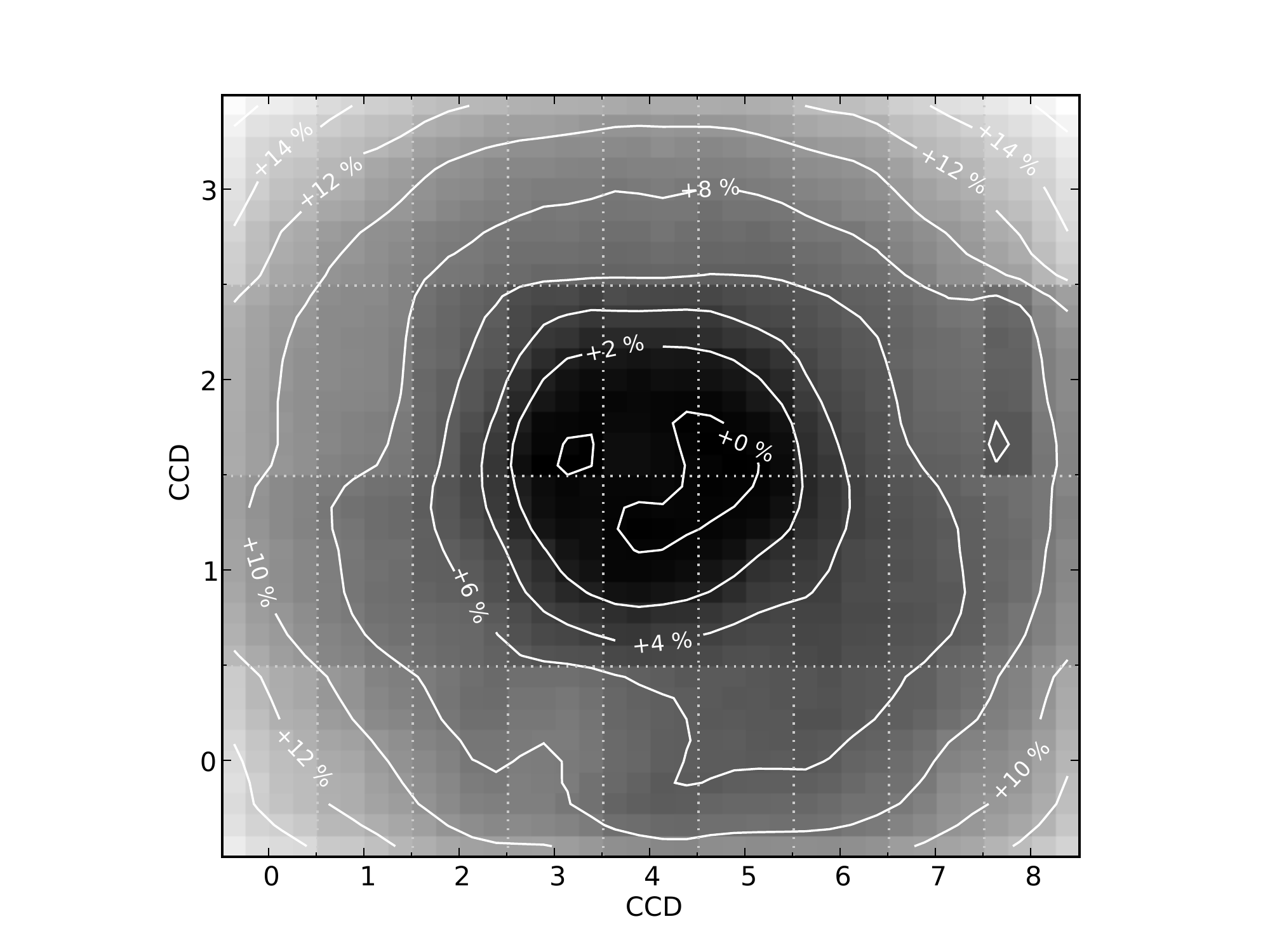}}
      \subfigure[$\delta {zp}_{r,r-i}(\x)\ \ (2006B)$]{\includegraphics[width=0.45\linewidth]{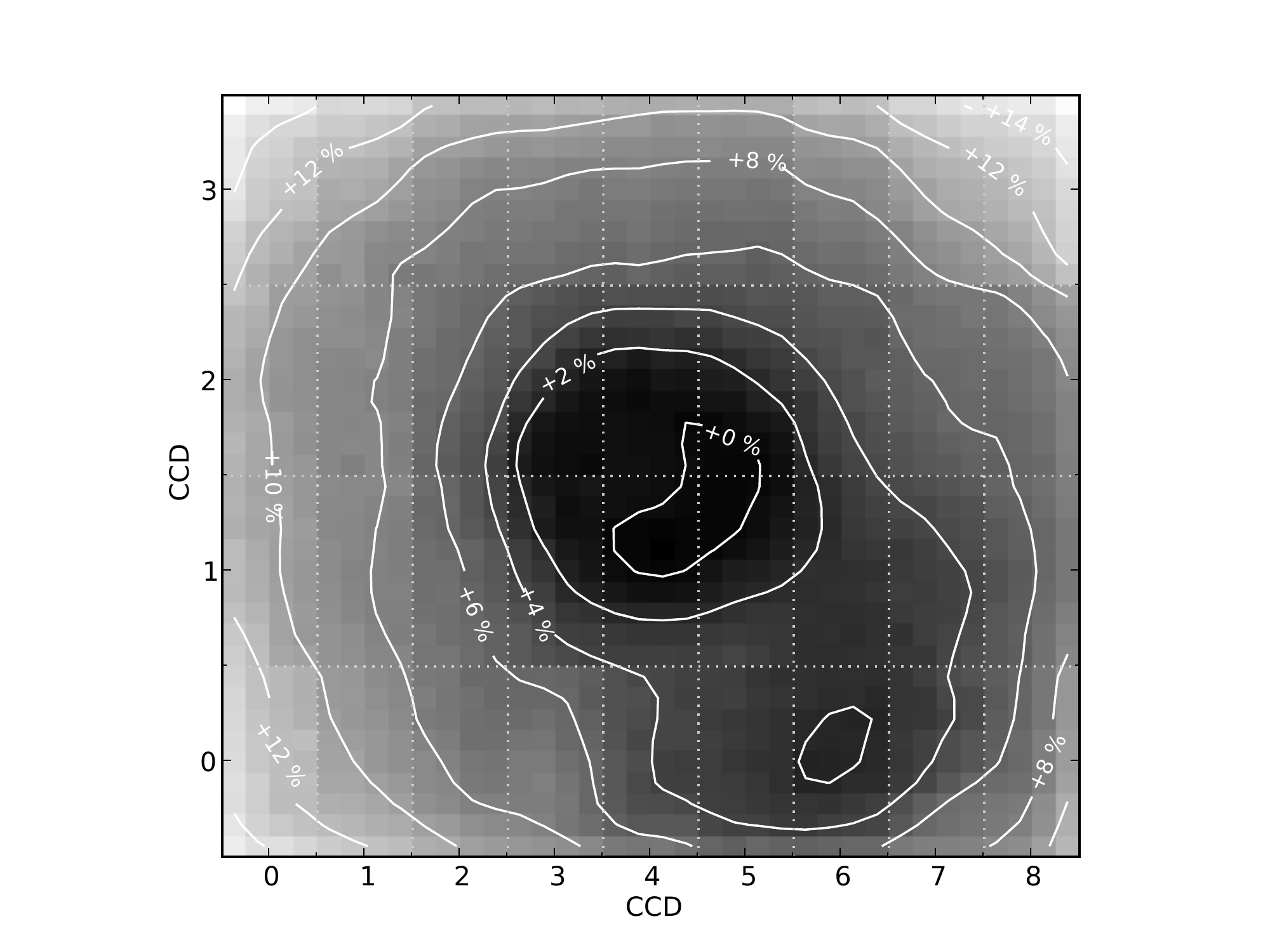}}}
\caption{$\rme-$band $\delta {zp}(\x)$ maps determined from grid
  observations taken from 2003 to 2007. The variations of the maps are due to metal shavings accumulating is
  the optical path, and modifying the ``scattered light''
  pattern. \label{fig:dzp_maps_2}}
\end{figure*}

Figure \ref{fig:dzp_maps} presents the $\gme, \rme, \ime$- and $\zme$-band $\delta
zp$ maps obtained on the 2005B dataset, using images flat-fielded with twilight flats. In all bands, a
same star will yield a higher instrumental flux if observed on the
edge of the camera than if observed on the center. 
These maps include the plate scale variations, but display larger
non-uniformities. \cite{sndice-spie08} present strong evidence
for internal reflections
in the MegaPrime wide-field corrector which add extra light in the center of 
the field of view, and hence tend to explain the need for correcting twilight flats.
The structure of the $\delta zp(\x)$
maps evolves slowly as a function of time, as shown on figure \ref{fig:dzp_maps_2}. This
evolution is mainly due to small changes in the optical path (see
table \ref{tab:grid_observations}) and to the accumulation of metal shavings
from the filter exchange mechanism on
the top optical surface below the filter mechanism\footnote{More precisely,
the tip-tilt plate, located above the wide-field corrector.}, due to the intense operation of the filter jukebox. 
This dust being somewhat shiny, it probably modified the internal 
reflection pattern within the optical path, corrupting the flatfields.
The dust was identified and removed in January 2007, and 
a preventive program to monitor the cleanliness of the optical path was set up.

\begin{figure*}
\centering
\mbox{\subfigure[$\delta k_{g,g-r}(\x)$]{\includegraphics[width=0.45\linewidth]{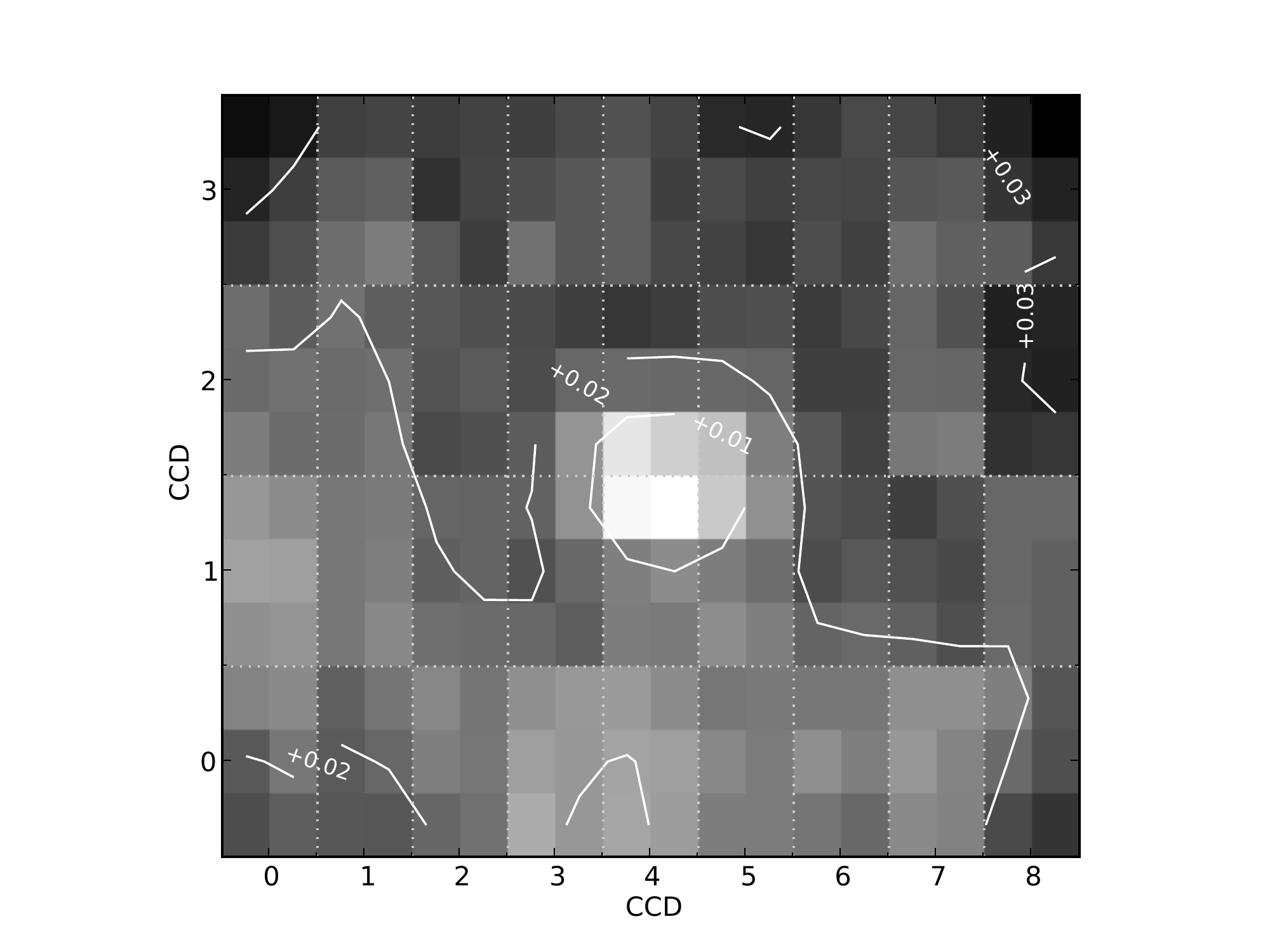}}\quad
      \subfigure[$\delta k_{r,r-i}(\x)$]{\includegraphics[width=0.45\linewidth]{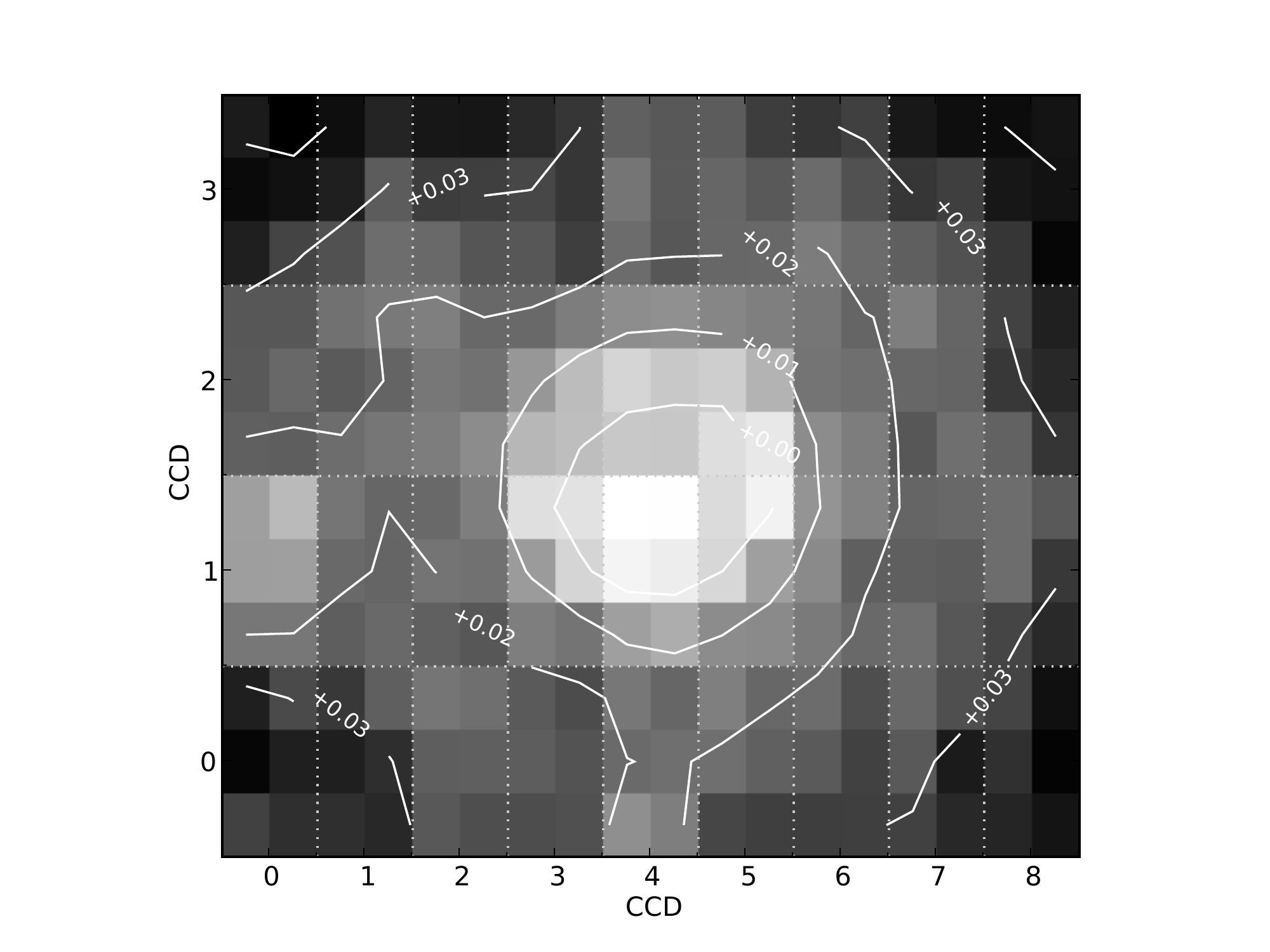}}}
\mbox{\subfigure[$\delta k_{i,r-i}(\x)$]{\includegraphics[width=0.45\linewidth]{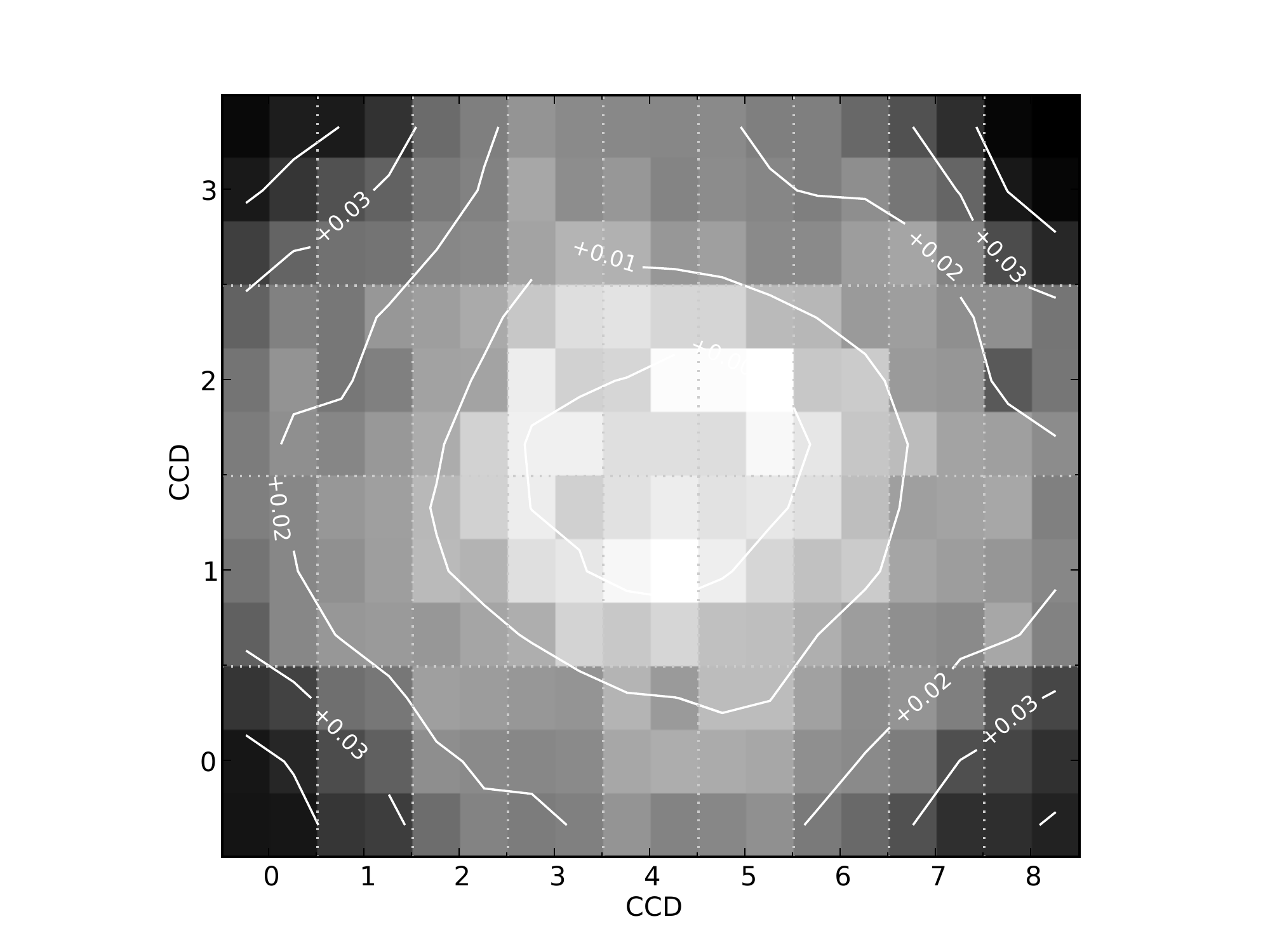}}\quad
      \subfigure[$\delta k_{z,i-z}(\x)$]{\includegraphics[width=0.45\linewidth]{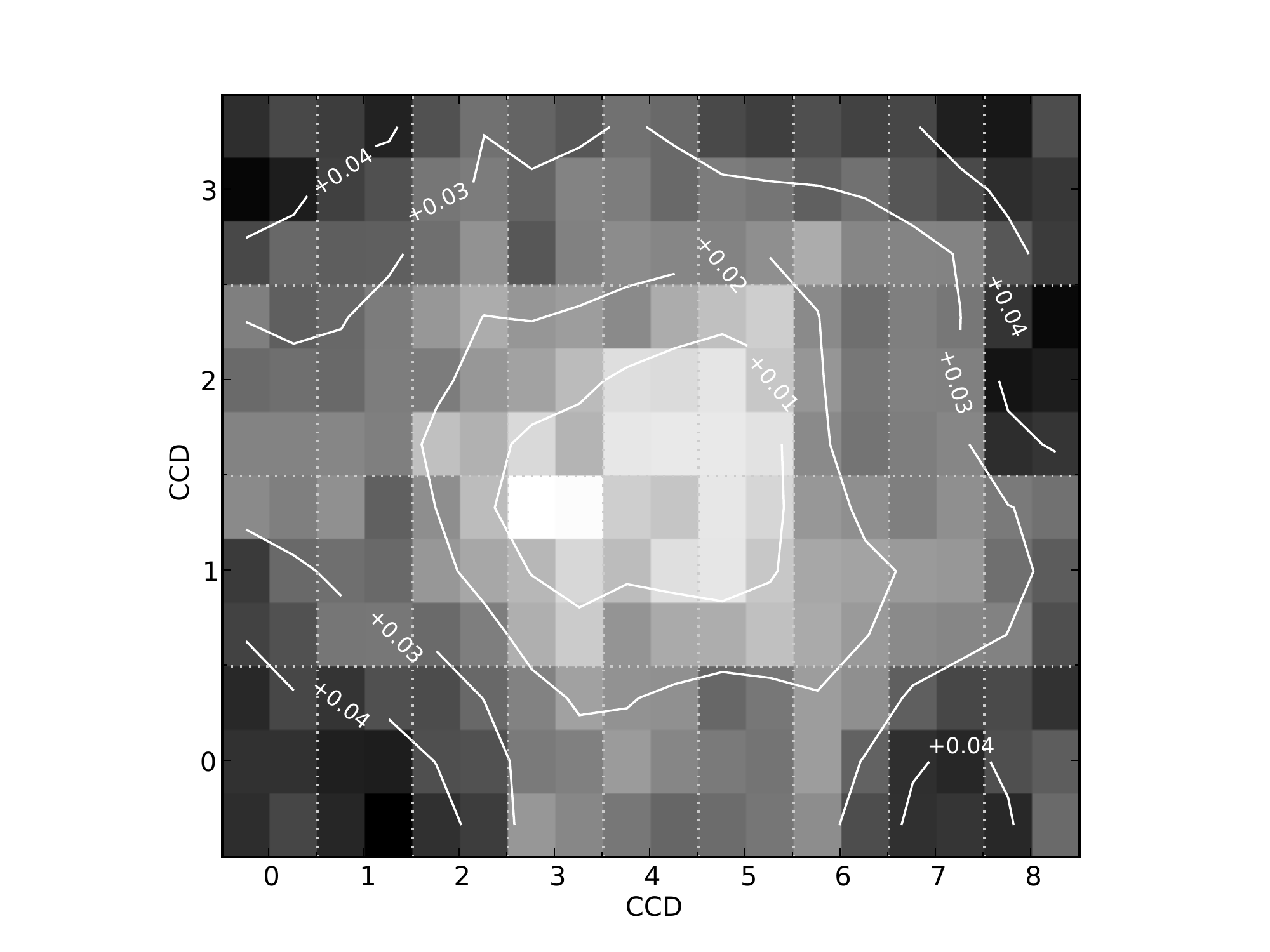}}}
\caption{$\gme$-, $\rme$, $\ime$- and $\zme$-band $\delta k(\x)$ maps
  determined from grid observations taken in 2005. In all bands, we
  observe a radial pattern, which is well reproduced using synthetic
  photometry and filter scans provided by the filter manufacturer
  (REOSC), as will be shown in \S\ref{sec:cmp_color_terms}. The contour lines are displayed to allow the reader to
  estimate the amplitude of the passband variations. }
\label{fig:dk_maps}
\end{figure*}

The $\delta k(\x)$ maps are presented in figure
\ref{fig:dk_maps}. We do measure non-zero color
terms between the reference location and any other focal plane
location. This came as a surprise, and was
later validated by the filter scans provided by the filter manufacturer
(see \S \ref{sec:megacam_passbands}). The non-uniformity pattern
is essentially invariant by rotation around the center of each filter. 
This is inherently the result of the manufacturer pushing the
technology when coating filters of that size ($300 {\rm mm} \times 300 {\rm mm}$) with
multiple layers. The coating chambers simply do have a projection beam
uniform enough to deliver the uniformity routinely achieved on smaller
filters.
The $\delta k(\x)$ maps were determined independently on each grid
set.  No significant variations of the maps were found between 2003
and 2006. 

The reduced $\chi^2_{\rm min}$ obtained from the fits is of about 4 in
the $\gme$-band and 3 in the $\rme, \ime$ and $\zme$-bands. Indeed, 
we have neglected the
contribution of the flat field errors and of the fringing when
evaluating the photometric measurement uncertainties. Since the fit is
linear, we have chosen to re-scale the photometric errors by the
appropriate amount, in order to obtain $\chi^2_{\rm min} / {\rm ndof}
\sim 1$, and to renormalize the covariance matrix of the grid
corrections accordingly. 

The statistical uncertainties affecting the $\delta zp(\x)$ are of
a little less than 0.001 mag in all bands. This uncertainty level is comparable to
the photon noise affecting the bright star instrumental magnitudes.
The statistical uncertainties on the $\delta k(\x)$ maps are of about
0.002 mags in all MegaCam bands. At the chosen grid reference color,
they are slightly correlated with the $\delta zp(\x)$ maps with a
correlation coefficient of $-0.25$. We have checked that, in all
bands, the statistical uncertainty introduced by the grid corrections
on the magnitudes transformed to the reference location is
never larger than 0.002 mag for stars of colors $\gme - \rme < 1.0$,
and of about 0.003 for stars of $\gme - \rme \sim 1.5$.

The fact that the imager passbands depend on the focal plane position
has important consequences on the calibration scheme. In particular, a
significant part (up to 1-2\%) of the uniformity corrections depend on
the color of each object.  We will detail how we account for the grid
corrections in section \ref{sec:applying_grid}.

\subsection{Comparison with Elixir findings}

\begin{figure}
\begin{center}
\includegraphics[width=\linewidth]{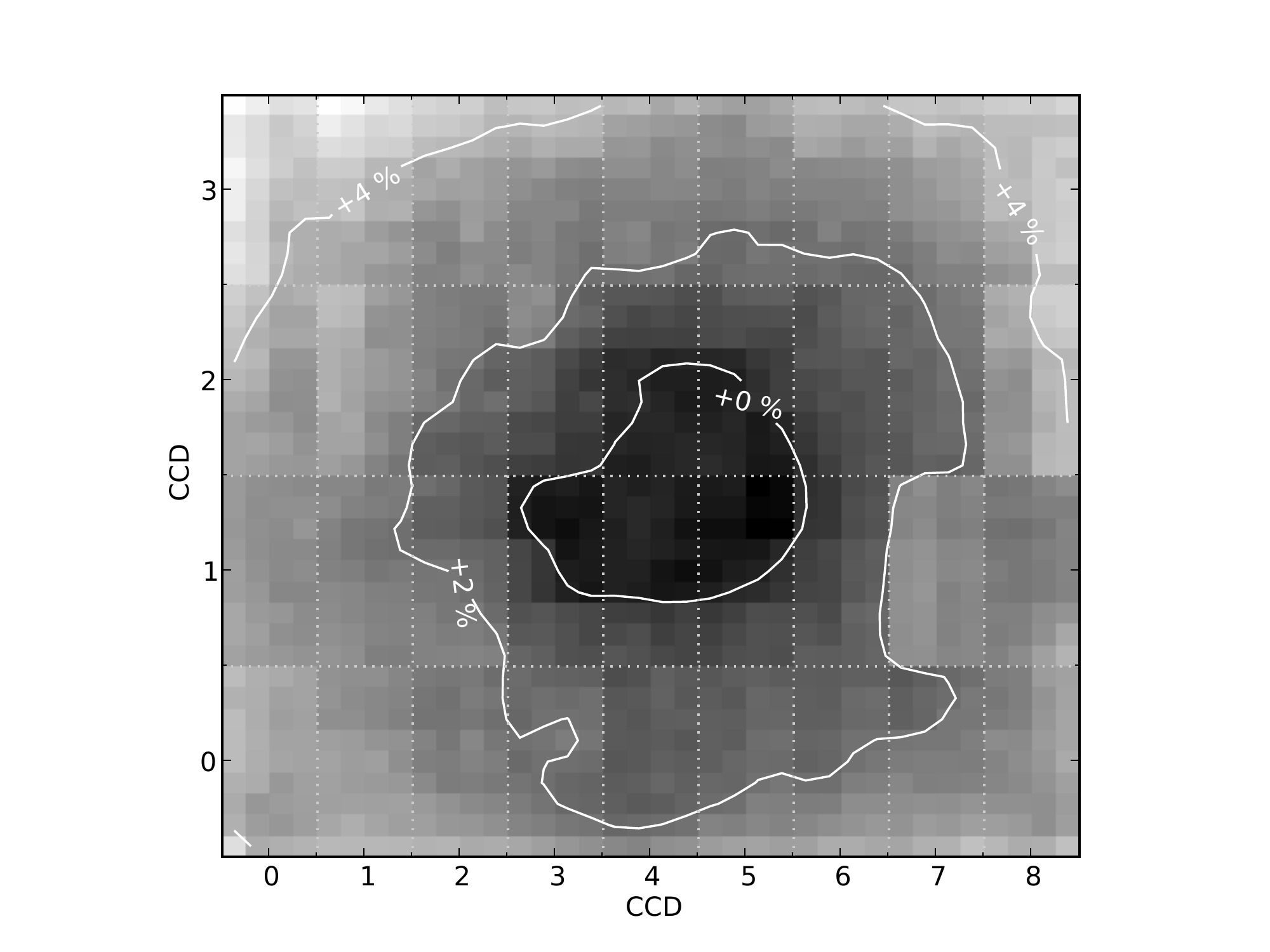}
\caption{Residual non-uniformities measured on a set of grid exposures
  taken during semester 2004B, and processed with Elixir, using the
  photometric correction maps in use for releases 
    T0004.\label{fig:residual_non_uniformities}}
\end{center}
\end{figure}

The Elixir analysis of the grid data did not find significant color terms
and solved for the $\delta zp(\x)$ without $\delta k(\x)$ terms and
did not solve for star magnitudes. The results are hence significantly
different : the $\delta zp(\x)$ maps re-analyzed by the SNLS collaboration span a range
larger by about 0.04 mag than the ones found by Elixir, as shown on
figure \ref{fig:residual_non_uniformities}. We found that the residual
non-uniformity pattern observed on the Elixir processed images 
is not constant and varies from one grid observation set to
another. The discovery of these non-uniformities triggered the
reanalysis of the grid dataset described above.

Up to the T0006 CFHTLS release (spring 2009), data were processed
using the standard Elixir recipe. The uniformity corrections presented
in the previous section will be integrated in 2009 after further
interactions with the Elixir team and the CFHTLS users community. This
is planned for the final release of the CFHTLS data T0007 (spring
2010).

\subsection{Applying the Grid Corrections to the Data}
\label{sec:applying_grid}

Because the grid corrections are color dependent, it is impossible to
obtain uniform instrumental magnitudes for all objects. In practice,
we have chosen to apply the $\delta zp(\x)$ maps directly to the
instrumental star magnitudes, defining so-called ``hat-magnitudes'':
\begin{eqnarray}
  \hat{g}_{ADU|\x} & = & g_{ADU|\x} - \delta zp_{g}(\x) \nonumber \\
                  & \ldots & \nonumber \\
  \hat{z}_{ADU|\x} & = & z_{ADU|\x} - \delta zp_{z}(\x) 
\end{eqnarray}
It would have been possible to apply them directly at the pixel
level. We have chosen to deal directly with the fluxes mainly for
practical reasons, as it did not imply a full reprocessing of the SNLS
data.

The hat magnitudes of stars of colors equal to the grid reference
colors are uniform across the focal plane. The hat magnitudes of all
other objects vary according to the relations:
\begin{eqnarray}
  \hat{g}_{ADU|\x} & = g_{ADU|\x_0} + \delta k_{ggr}(\x) \times \Bigl[ (g-r)_{|\x_0} - (g-r)_{grid}\Bigr] \nonumber\\
                  & \ldots  \nonumber\\
  \hat{z}_{ADU|\x} & = z_{ADU|\x_0} + \delta k_{ziz}(\x) \times \Bigl[ (i-z)_{|\x_0} - (i-z)_{grid}\Bigr]
\end{eqnarray}
It should be pointed out once again that the definition of the hat
magnitudes depend on the conventional grid reference colors, ${\rm col}_{grid}$.  
However, we
will see in \S \ref{sec:landolt_stars} and \S
\ref{sec:megacam_magnitudes} that the grid color conventions are
purely internal quantities and have no impact on the calibrated
tertiary magnitudes. 

The instrumental magnitudes entering the calibration pipeline were
measured on the survey images processed with the official Elixir
pipeline, which included a correction for the non-uniformities of the
imager. This correction was applied at the pixel level (see
above). Hence, what we measured were actually ``Elixir
hat-magnitudes'':
\begin{eqnarray}
  \hat{g}^{Elixir}_{ADU|\x} & = & g_{ADU|\x} - \delta zp^{Elixir}_{g}(\x) \nonumber \\
                  & \ldots & \nonumber \\
  \hat{z}^{Elixir}_{ADU|\x} & = & z_{ADU|\x} - \delta zp^{Elixir}_{z}(\x) 
\end{eqnarray}
As a consequence, the Elixir grid corrections had to be removed and
the new $\delta zp(\x)$ corrections applied:
\begin{eqnarray*}
  \hat{g}_{ADU|\x} & = & \hat{g}^{Elixir}_{ADU|\x} + \delta zp_g^{Elixir}(\x) - \delta zp_g(\x) \\
                  & \ldots & \\
  \hat{z}_{ADU|\x} & = & \hat{z}^{Elixir}_{ADU|\x} + \delta zp_z^{Elixir}(\x) - \delta zp_z(\x) 
\end{eqnarray*}

There are basically
two ways to handle the residual color corrections.
We can elect a specific focal plane position
(e.g. $\x_0$), transform all the hat magnitudes to this position,
using the $\delta k(\x)$ maps and choose to report all the MegaCam
magnitudes at this specific position. However, stars are complicated objects, 
and even main sequence stars do not follow linear color corrections over 
large color ranges, of one magnitude or more. Furthermore, non stellar objects such as
galaxies or supernovae do not obey the grid corrections. Instead, we 
chose to leave the science object instrumental magnitudes untouched,
and define a system of ``local natural magnitudes''. This is
possible because the ditherings applied to the survey images are
small, which ensures that each object is always observed with the same
effective filters\footnote{We call the combination of the filter passbands, 
the transmission of the optics, the reflectivity of the mirrors, the quantum 
efficiency of the detectors and the average transmission of the atmosphere {\em effective filters}.}. 
Working with natural magnitudes ensures that they 
can be directly converted into broadband
fluxes, provided that we have a model of the telescope effective
passbands at each position of the focal plane -- as well as a spectrum with known 
magnitudes. Building such a model is the subject of
\S \ref{sec:megacam_passbands}.

The average $\delta k(\x)$ maps may be obtained from the CDS\footnote{{\tt http://cdsweb.u-strasbg.fr/cgi-bin/qcat?J/A+A/}}.
The $\delta zp(\x)$ will be made available along with the next Elixir data release. 
Indeed, some further work 
with the Elixir team is needed, in order to decide the
optimal grid reference colors for each component of the CFHT Legacy Survey
and validate the new reduction procedure with a larger user base.

\section{MegaCam Passbands}
\label{sec:megacam_passbands}

\subsection{Transmission of the filters in the telescope beam}
MegaCam filters are interference filters and exhibit a transmission
depending on the crossing angle, as expected
for this type of filters. We describe in this paragraph how we synthesized
the transmission of the filters in the telescope beam from laboratory
measurements of the filter transmissions.

We have at our disposal two kinds of laboratory measurements: a set of transmission curves
measured at about a dozen of positions on the filter, all on the sides
of the filters because the equipment has a limited mechanical
clearance. Some of the positions were measured at 0, 2, 4 and 6 degrees
from normal incidence (Benedict, private communication). The other set of measurements was provided by
the filters manufacturer : it consists in a transmission curve 
at normal incidence at 10 positions along a radius for each filter.
From the first set of measurements, we can check that the circular
symmetry is an excellent approximation. We can also model the 
angular dependence. Assuming the circular symmetry, interpolating 
the other set of  measurements provides us with transmission curves 
at normal incidence anywhere on the filters. Since filters are located
at about 10 cm from the focal plane, we identify in what 
follows a position on the focal plane with a position on the filter.

  The angular dependence of interference filters transmission can be approximated by:
$$
T  \left( \lambda,\theta \right) = T \left(\lambda \left[ 1-\frac{\sin^2 \theta}{n^2} \right ]^{-1/2}, \theta=0 \right)
$$
where $n$ is the refracting index of the filter, and $\theta$ the incidence 
angle. This expression, exact for
a single Fabry-Perrot layer, is sufficiently accurate to describe the 
angular dependence measured on the MegaCam filters.
We find effective indices of  1.80, 1.70, 1.80, 1.60, 1.50 for $\ume, \gme, \rme, \ime$ and $\zme$
respectively, with an uncertainty of about 0.1.
  The angular dependence of the transmission has potentially two effects:
it induces a radial dependence of the transmission of a filter, even if
it were perfectly uniform; secondly, it shifts towards the blue the 
transmission in the telescope beam compared to the laboratory measurements at 
normal incidence.
The first effect is in fact very small for MegaCam: the mean angle of the beam
changes by about 10 milliradians between the center and the corner of the mosaic,
and the cut-on and cut-off wavelengths of a filter 
are shifted by less than a part in $10^4$, and 
cannot account for the position dependence of the 
transmission measured on the sky. The second effect turns out to be 
non-negligible in the f/4 beam at CFHT prime focus: it shifts
the filter central wavelengths by a few per mil.
  In order to synthesize the transmission of the filters, we integrated
the measurements provided by the manufacturer over the telescope beam
(with central occultation), assuming the above expression for the
angular dependence. If compared to normal incidence transmissions,
the effective wavelengths shift to the blue 
uniformly across the focal plane by amounts ranging from 4 \AA\ for $\ume$ band
to 14 \AA\ for $\zme$ band.

\subsection{Effective passbands}
The effective passbands combine many contributions,
among which (1) the CCD quantum efficiencies (2) the filter
transmissions (3) the transmission of the various lenses and windows
in the optical path (4) the mirror reflectivity (5) the average
atmospheric transmission at Mauna Kea as recently determined by the SuperNova Factory (SNF) Collaboration 
\citep{Buton09} and (6) the transmission spectrum of the telluric features 
including the strong ${\rm O_2}$, ${\rm OH}$ and ${\rm H_20}$ absorption features in the red and
near infrared \citep{Hinkle03}.

The effective passbands, along with their ingredients are listed in appendix
\ref{sec:megacam_passband_tables}. The full electronic version of
these tables can be retrieved from the CDS\footnote{{\tt http://cdsweb.u-strasbg.fr/cgi-bin/qcat?J/A+A/}}.

\subsection{Comparison of the observed and synthetic color terms}
\label{sec:cmp_color_terms}

\begin{figure}
\centering
\includegraphics[width=\linewidth]{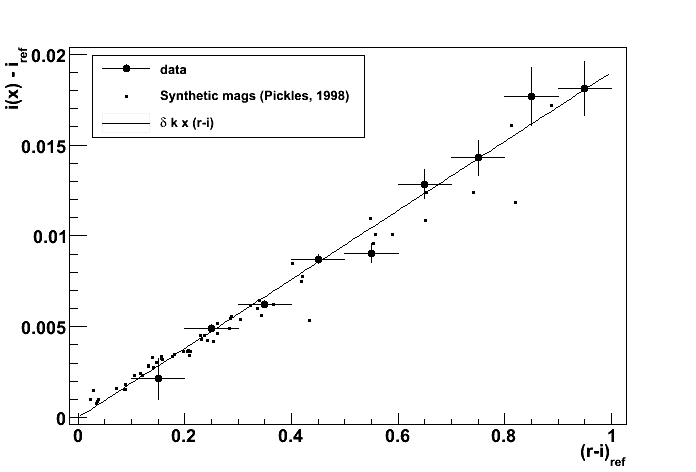}
\caption{Difference between the $\ime$ instrumental magnitude of stars located on
  the edge of the focal plane (ccd \# 9, 17, 18 \& 26) as a function
  of their $(\rme-\ime)$ color at the reference location. The large dots are
  the profile of the star measurements, corrected for the $\delta zp(\x)$ map, the small dots are the
  synthetic magnitudes, computed using the \citet{Pickles98} library
  and models of the passbands at the the center of the focal plane and
  at 11.5 cm from the center. The line shows the average linear grid color
  correction $\delta k(\x) \times (\rme-\ime)$. As can be seen, the grid
  color corrections are well approximated by linear relations.
\label{fig:grid_color_color}
}
\end{figure}

\begin{figure}
\centering
\includegraphics[width=\linewidth]{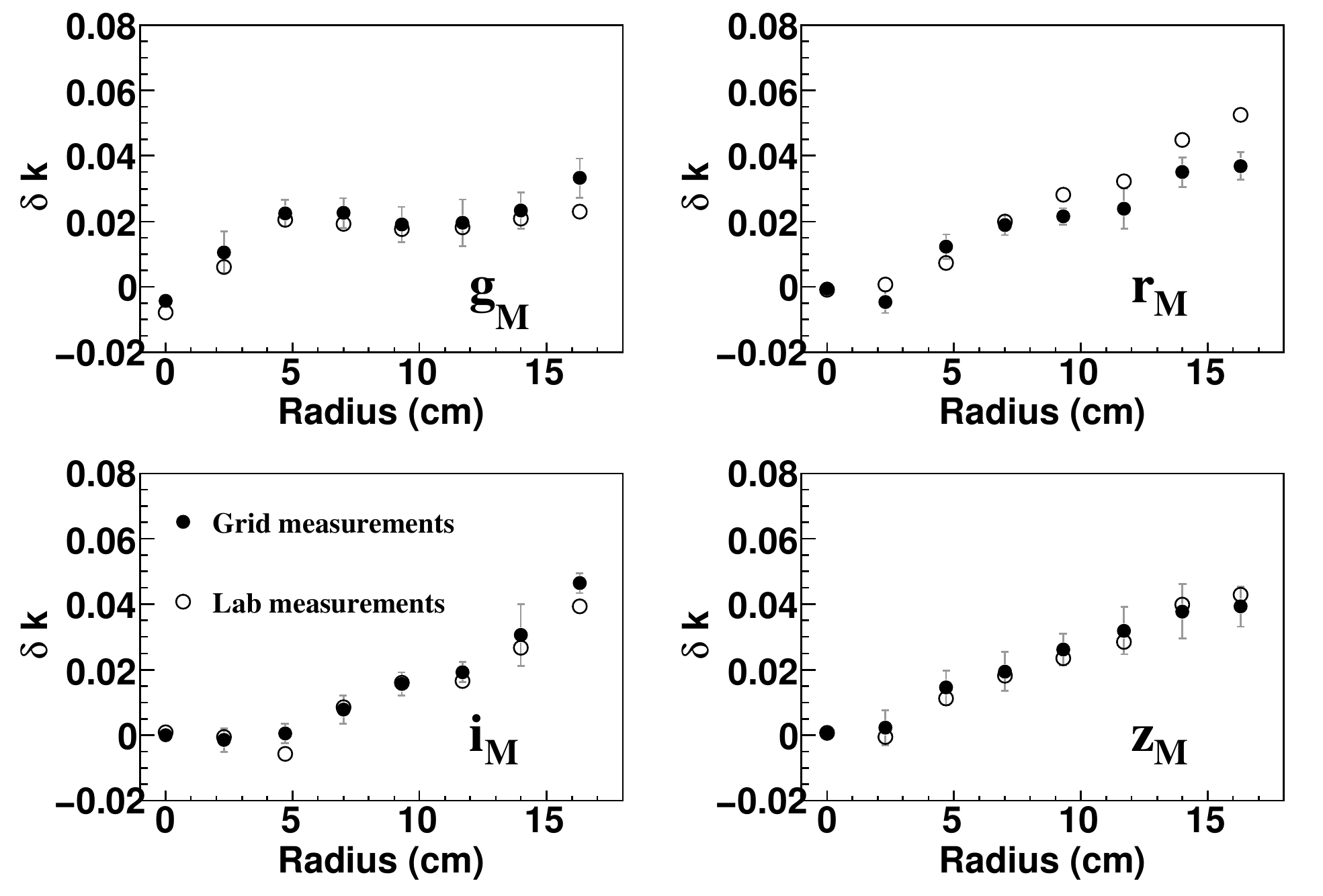}
\caption{Comparison of the color terms measured on the grid data, 
with synthetic ones as a function of the distance to field center.
The measurement points are averages at the quoted radius, with their error 
bars displaying the r.m.s over the circle.
The synthetic ones are computed by integrating the \citet{Pickles98} stellar
library in the synthetic transmissions. The synthetic color terms depend
only mildly on the color range and stellar types, except for the
$\rme$ band, where the disagreement worsens with the \citet{Stritzinger05}
library.
\label{fig:grid_synth_vs_observed}
}
\end{figure}

{Figure \ref{fig:grid_color_color} compares the $\ime$-band
  instrumental magnitudes of stars observed at the edge of the focal
  plane and at the reference location as a function of the star $(\rme
  - \ime)$ colors. As can be seen, there is a good agreement between
  the synthetic magnitudes computed using the effective passband
  models presented in this section and the grid measurements. Note
  also that the effect of the filter non-uniformities is small, and
  well described by a linear color term. } Figure
\ref{fig:grid_synth_vs_observed} compares the color terms across the
focal plane measured with the grid data with synthetic evaluations
from the \citet{Pickles98} stellar library. Except for the filter
itself, we used average transmissions, and in particular the average
quantum efficiency curve for the CCDs.

Similar results are obtained using the \citet{Stritzinger05} stellar
library, except for the $\rme$ band, where the agreement becomes worse
than shown here.  We suspect that stellar SEDs are poorly indexed by
$\rme-\ime$, and do not regard this mild disagreement (at the level of
0.01) as very serious. The fair agreement in the other bands, together with
its stability with a change of stellar library is indeed a good indication
of the quality of the grid solution. Taken at face value, the differences
between the two sets of curves cause shifts of the zero points below the millimagnitude
level for all bands (and is hence negligible) except for $\rme$ band where it 
would reach 2 millimagnitudes. We decided to ignore this potential contribution to 
the error budget because the synthetic $\rme$ color terms
are very sensitive to the assumed stellar population, making the 
result questionable.

\section{Landolt Stars}
\label{sec:landolt_stars}

\subsection{The Landolt Photometric System}

The photometric catalog published by \citet{Landolt92} is 
the most widely used standard star network. It contains $UBVRI$ Johnson-Kron-Cousins magnitudes of 526 stars 
centered on the celestial equator in the $11.5 < V < 16.0$ magnitude range. 
The repeatability is of about 1\% in all bands resulting in magnitude 
uncertainties of 0.003 mag. The uniformity of the catalog is believed 
to be excellent, of about 1\%. This standard star network was used to
calibrate the broadband observations of the nearby SNe~Ia used to
supplement the SNLS dataset. The systematic differences between
photometric systems are not well controlled and can amount to $2\% -
4\%$. In order to avoid this additional source of error in
the cosmological analyzes, we rely on the \citet{Landolt92} standard
star network to calibrate the SNLS survey.

The genealogy of the Landolt $UBV$ magnitudes can be traced back to
the pioneering work of
\citet{Johnson51,Johnson53,Johnson54,Johnson55,Johnson63}. \citet{Landolt73}
transferred the \citet{Johnson63} photometric system to a set of 642
bright ($6.0 < V < 12.5$) equatorial stars, selected for their
stability. This program was conducted at Kitt Peak National
Observatory (KPNO), using the 16- and 36-inches telescopes, and a
refrigerated 1P21 photomultiplier similar to that used by Johnson and
collaborators. $R$- and $I$-band magnitudes were added about ten years
later.  The catalog published in \citep{Landolt83} gives $UBVRI$
magnitudes of 223 $7.0 < V < 12.5$ stars. This work was based on
observations made on the Cerro Tololo Inter-American Observatory
(CTIO) 0.4- and 0.9-m telescopes, using a different type of
photomultiplier (RCA 31034) and slightly different filters. The $UBV$
magnitudes were tied to the \citet{Landolt73} system, the $RI$ band
observations were tied to the system defined by
\citet{Cousins78}. Finally, \citet{Landolt92} published $UBVRI$
magnitudes of 526 fainter ($11.5<V<16.0$) equatorial stars, tied to
the 1983 catalog. This catalog was built from observations taken with
the CTIO 0.9-m and 1.5-m telescopes, RCA 31034A and Hamamatsu R943-02
photomultipliers and the same $UBVRI$ filters used to develop the 1983
catalog. Additional papers containing photometry of particularly well-studied
stars, mainly spectrophotometric standards were also published \citep{Landolt92b,Landolt07b}. In particular, 
\citet{Landolt07b} contains Landolt magnitudes of 31 stars, among which 
spectrophotometric standards used to calibrate the Hubble Space Telescope. Most 
\citet{Landolt92} standards are red objects with an average $B-V$ color of 
$0.81 \pm 0.52 (rms)$. The Landolt stars observed by MegaCam are mainly located
in the Selected Area (SA) fields, which contain very few blue stars. The average 
$B-V$ color of the Landolt stars observed with MegaCam is $0.77 \pm 0.31 (rms)$.
As a consequence there are very few observed stars in the color range $0 < B-V < 0.25$.

Contrary to statements found in the literature, the Landolt system is
{\em not} defined in terms of any particular magnitude of Vega, nor is the
Johnson system. The absolute ``gray'' zero-point of the Johnson system
is linked to the former ``International Photovisual System'' via 9
stars in the so-called ``North Polar Sequence'', reobserved by
\citet{Johnson51}. The color zero-points were set from 6 A0V stars of
which Vega is one, by the condition that the average $B-V$ and $U-B$
color index of these stars is exactly zero
\citep{Johnson53}. Vega and the North Polar Sequence objects were too 
bright for Landolt to observe, even on small telescopes. As noted above, the Landolt $R$ and $I$
magnitudes are tied to the system defined by \citet{Cousins76}. The zero-points of this 
latter system are also defined so that the colors of a ``typical'' A0V star are all zero. However, 
no mention of Vega appears in \citet{Cousins76}. 
Hence, there are
large uncertainties on the magnitudes of Vega in the Landolt system.

It must also be noted that the Landolt system is {\em not} a natural system. The
reduction procedure used to derive the calibrated magnitudes from the
raw observations is described in great detail in \citet{Landolt07c}
and can be summarized as follows. First, airmass corrections are
applied to the data in order to obtain magnitudes above the
atmosphere. Second order
(i.e. color dependent) corrections are applied to the $B-V, U-B, V-R,
R-I, V-I$ color indexes, while only a first order correction was
applied to the $V$ band magnitudes. 
Then, zero-points {\em as well as (unpublished) linear color corrections} are determined,
for each night using standard stars. This procedure is not
discussed in the \citet{Landolt73,Landolt83,Landolt92} papers. The zero-points and color corrections 
are parametrized as:
\begin{align*}
  V     &=& V_{|X=0}     &+& z &+& f \times (B-V)_{|X=0} \\
  (B-V) &=& (B-V)_{|X=0} &+& a &+& b \times (B-V)_{|X=0} \\
  (U-B) &=& (U-B)_{|X=0} &+& c &+& d \times (U-B)_{|X=0} \\
  (V-R) &=& (V-R)_{|X=0} &+& p &+& q \times (V-R)_{|X=0} \\
  (R-I) &=& (R-I)_{|X=0} &+& r &+& s \times (R-I)_{|X=0} \\
  (V-I) &=& (V-I)_{|X=0} &+& t &+& u \times (V-I)_{|X=0} 
\end{align*}
where the $(X=0)$ quantities are the instrumental magnitudes and
colors extrapolated to an airmass of zero. The color correction
parameters vary from night to night, and can be large. For example,
while building the \citet{Landolt07b} catalog, typical values of
$+0.026$, $+1.036$, $+0.913$, $+1.033$ and $+1.093$ and $+1.069$ were
derived for the $f, b, d, q, s$ and $u$ parameters respectively
(Landolt, private communication).  Finally, at the end of the reduction procedure, smaller color corrections
are applied to the calibrated magnitudes, in order to account for the
changes of instrumentation throughout each program. These corrections
are discussed in the Landolt papers. They are
non-linear and parametrized using (non-continuous) piecewise-linear
functions. They are generally about a factor two smaller than the
unpublished color corrections discussed above.

An obvious difficulty when trying to calibrate MegaCam against the
Landolt system, is that the $UBVRI$ and $\gme, \rme,
\ime, \zme$ passbands are extremely different, leading to large and
non-linear color transformations between both systems. 
Furthermore, these color transformation are extremely
difficult to model, especially for blue stars ($0 < B-V < 0.4$) given
the scarcity of the Landolt stars in this region.

Note also that the reddest Landolt band, $I$-band, is significantly
bluer than the MegaCam $\zme$-band. Hence, calibrating the latter band
requires extrapolating the Landolt calibration to redder wavelengths.

In \citet{astier06}, we attempted to model these color
transformations, using synthetic photometry and the library of spectra
published by \citet{Pickles98}. The color transformations were
interpolated from the synthetic MegaCam-Landolt $\gme-V\ vs.\ B-V \ldots
\zme-I\ vs.\ R-I$ color-color diagrams.  The synthetic magnitudes of
each spectrum were computed using (1) models of the Landolt and MegaCam
passbands and (2) the measurement of the Vega spectral energy distribution
published by \citet{Bohlin04}. The MegaCam filter model was not as
sophisticated as the one presented in \S \ref{sec:megacam_passbands}
and in appendix \ref{sec:megacam_passband_tables}. It was an average
of the Sagem / REOSC measurements. Hence it was significantly bluer than the
filters at the center of the focal plane. To approximate the Landolt
passbands, we used the determinations published in \citet{Bessel90},
each one being shifted in wavelength by a quantity $\delta \lambda$
that had to be determined. These shifts were estimated by comparing
the synthetic and observed Landolt magnitudes of the Baldwin Stone
Southern Hemisphere spectrophotometric standard, observed by
\citet{Landolt92b, Hamuy92, Hamuy94}. We found at that time that the
$B, V, R$ and $I$ Bessell filters had to be {\em blueshifted} by 41,
27, 21 and 25 \AA\ respectively, in order to match the Landolt
effective filters. The Landolt-to-MegaCam synthetic color
transformations computed using these blueshifted filters
reproduced the measurements well.

Redoing this study using the new determination of the MegaCam filters,
and alternate libraries of spectrophotometric standards with known
Landolt magnitudes, \citep[][aka CALSPEC]{Stritzinger05, Bohlin07} we
were not able to reproduce the blueshifts listed above. Both libraries
gave compatible results, namely small redshifts of less than 10 \AA\ 
for the $B$, $V$ and $R$-bands, and a blueshift of about $-40$
\AA\ for the $I$-band.  The reasons for the discrepancy between the
results relying on the Hamuy data, on one hand, and the Stritzinger or
CALSPEC data on the other hand are not yet well understood. It may
point to a wavelength, or flux calibration problem affecting the
\citet{Hamuy92,Hamuy94} spectra. By an unhappy coincidence, the
(probably) incorrect Bessell filter blueshifts derived in
\citet{astier06} allowed us nevertheless to obtain the correct
synthetic color transformations, because the MegaCam filter models we
used at that time were also slightly bluer than our current model at
the center of the focal plane, where most Landolt stars are observed
(see figure \ref{fig:nb_stars_per_ccd_and_night}).

Because of the large differences between the Landolt and MegaCam 
filters, the modeling of the large Landolt-to-MegaCam color relations 
turned out to be delicate. 
In addition to the broken linear relations used in this paper (see below), we also
considered a more physical model based on SED libraries and
manipulating filter passbands.  To decide between the two approaches,
we did a series of blind tests on fake standard star observations
based on physical models and including realistic noise.  We found that
the broken linear method used here was considerably more robust to
uncertainties in the stellar library, although at the cost of
not providing as direct a physical interpretation to our measurements.

Further attempts to refine the analysis presented in \citet{astier06}
using synthetic photometry proved unsuccessful. Eventually, we came to
the following conclusions.
First, it is illusory to seek a description of the Landolt system as
a natural system of some ``effective'' hypothetical instrument.  Using
shifted Bessell filters to describe the Landolt passbands is not
accurate given how the {\em shape} of these filters differ from the
shape of the filters used by Landolt. For example, using a refined
modeling of the Landolt instrument, we obtained an estimate of the
$B-V$ magnitude of Vega which differed by 0.02 mag from the estimate
obtained with shifted Bessell filters.

Second, the concept of Landolt-to-MegaCam color transformation is not
well-defined. These transformations depend on the mean properties
(metallicity and $\log g$) of the stellar population observed by
Landolt. In particular, since there are less than a dozen of Landolt
stars observed in the $0 < B-V < 0.25$ region --where most A0V stars lie, including Vega-- there are large
systematic uncertainties affecting these transformations in this
region. 

Finally, it is possible to reduce significantly the impact of the
systematics affecting the Landolt-to-MegaCam color transformations by
choosing a fundamental standard whose colors are close to the mean
color of the Landolt stars. It is then enough to model roughly the
color transformations using piecewise-linear functions, as described
in the next section.  The choice of an optimal primary standard is
discussed in \S \ref{sec:flux_interpretation_of_megacam_magnitudes}.

\subsection{Calibration Model}

\begin{figure}
\centering
\includegraphics[width=\linewidth]{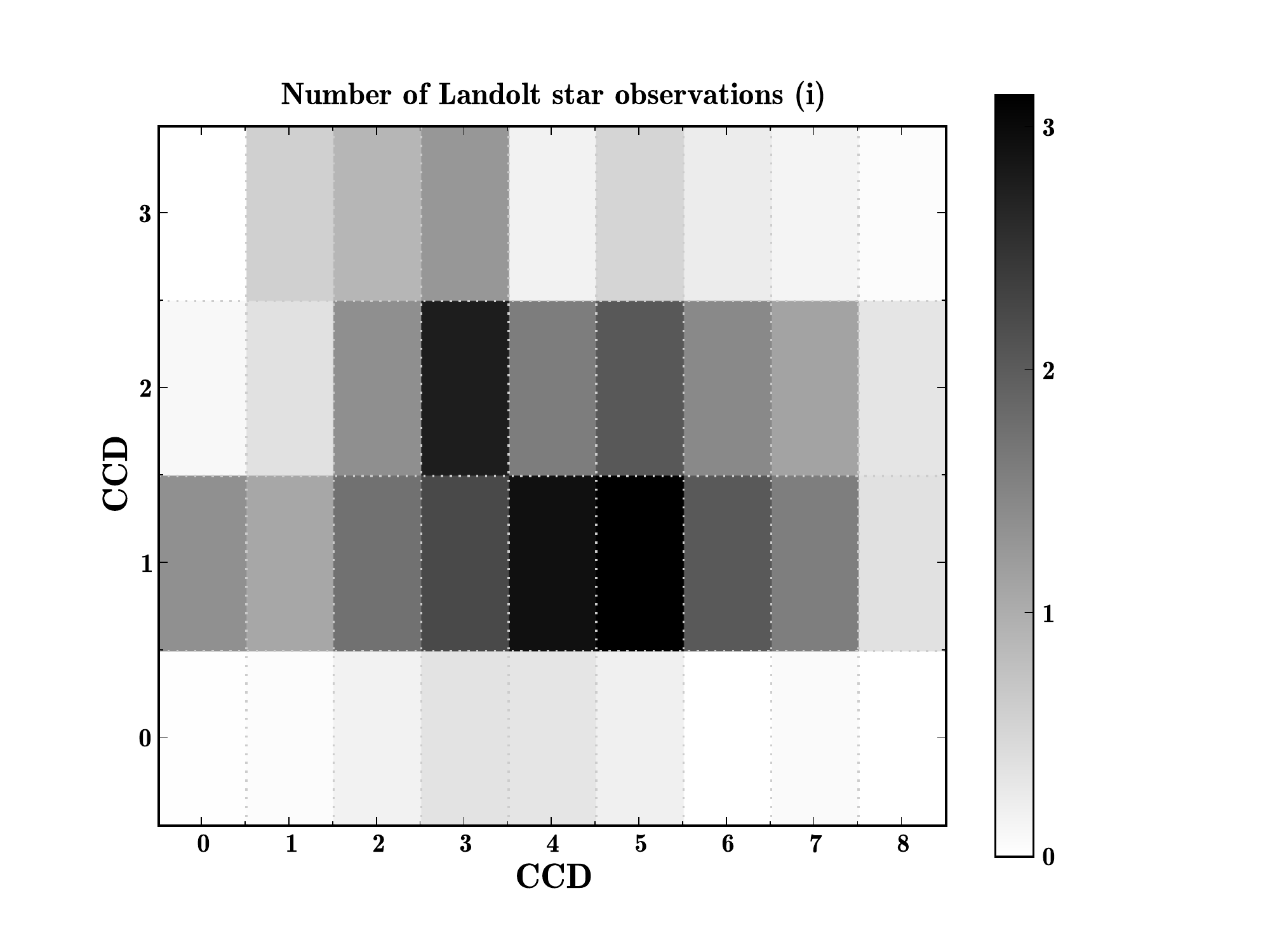}\\
\caption{Average number of Landolt star observations per CCD, in the
  $\ime$-band. CCD number 0 is top left, while CCD number 35 bottom
  right.  Two to three stars per night and per band are observed on
  the central CCDs. On the other hand, often less than 0.5 (and even
  less than 0.1 for CCDs 27 to 35) star are observed on the CCDs
  located on the side of the focal plane. Hence, each CCD cannot be
  calibrated independently, and we have to rely on the uniformity
  maps to propagate the calibration to the entire focal
  plane. \label{fig:nb_stars_per_ccd_and_night}}
\end{figure}

Landolt fields are observed almost each potentially photometric night.
The flux of the Landolt stars is measured using the aperture
photometry algorithm described in section
\ref{sec:photometric_reduction}. On average, between 2 and 3 stars are
observed each night, in each band, on each central CCD (figure
\ref{fig:nb_stars_per_ccd_and_night}). On the other hand, less than
0.5 Landolt star per night are observed on the CCDs which are on the sides
of the focal plane.
The Landolt star observations
allow us to determine the zero-points of each night, in each band,
along with additional (nuisance) parameters, such as the coefficients
describing the Landolt to MegaCam transformations and the mean airmass
coefficients.

Due to the significant differences between the Landolt
Johnson-Kron-Cousins-$UBVRI$ and MegaCam $g_M,r_M,i_M,z_M$
filter-sets, the Landolt-to-MegaCam color transformations were found
to be large and even non-linear. We model them using piecewise-linear
functions of the form:
\begin{equation*}
  {C}({\rm color}; \alpha, \beta) =
  \begin{cases}
    \ \alpha \times {\rm color}, \ \ \ \  \text{if ${\rm color} < {\rm color}_{break}$} \\
    \ \alpha \times {\rm color}_{break} + \beta \times \Bigl( {\rm color} - {\rm color}_{break}\Bigr), \\ 
    \ \ \ \ \ \ \ \ \ \ \ \ \ \ \ \ \ \ \ \ \ \ \ \ \ \ \ \ \ \ \ \ \ \ \ \ \ \ \ \ {\text{otherwise}} \\
  \end{cases}
\end{equation*}
where ${\rm color}$ is a Landolt color. The equation above has 
three unknown parameters: 2 slopes, $\alpha$ and $\beta$, and a ``color break''
marking the transition between the two slopes. The slopes $\alpha$ and
$\beta$ are fitted along with the zero-points.  The parameter ${\rm
  color}_{break}$ is adjusted by studying the fit residuals, and
fixed. We have found that the sensitivity of the zero points to the
break position is small: in all bands they vary by less than 
0.0015 mag if the break varies by 0.1 mag. As discussed above, if we choose a fundamental 
standard whose colors are similar to the Landolt star colors, the 
impact of the break position will be even smaller than 0.001 mag.

Since the MegaCam passbands are not uniform, there are, in practice as
many photometric systems as there are locations on the focal plane. 
The density of Landolt stars in the calibration fields does not allow us to
calibrate independently each location, 
and we must rely on the grid maps to propagate the
calibration to the whole focal plane. To establish the calibration
equations, let's first assume that the Landolt stars are all observed
at the reference location, $\x_0$. The relations between the MegaCam
instrumental ``hat magnitudes'' (corrected for ``gray'' non-uniformities) as defined in \S \ref{sec:the_photometric_grids}
and the magnitudes reported by Landolt ($UBVRI$) can be parametrized as:
\begin{align*}
  \hat {g}_{ADU|\x_0} & = & V &- k_g  (X-1) &+ C(B-V; \alpha_{g}, \beta_{g}) &+ ZP_g \\
  \hat{r}_{ADU|\x_0} & = & R &- k_r  (X-1) &+ C(V-R; \alpha_{r}, \beta_{r}) &+ ZP_r \\
  \hat{i}_{ADU|\x_0} & = & I &- k_i  (X-1) &+ C(R-I; \alpha_{i}, \beta_{i}) &+ ZP_i \\
  \hat{z}_{ADU|\x_0} & = & I &- k_z  (X-1) &+ C(R-I; \alpha_{z}, \beta_{z}) &+ ZP_z 
\end{align*}
$X$ is the airmass of the observation, $k_u, \ldots k_z$ are the
airmass coefficients and $ZP_u, \ldots ZP_z$ the zero-points.  The
free parameters of the calibration relations above are the five
airmass terms $k_{ugriz}$, the ten color transformations slopes
$\alpha_{ugriz}$ and $\beta_{ugriz}$ and about 1600 zero points ---one
zero-point per night and per band. All these parameters are fit
simultaneously on the whole calibration dataset.

The airmass range of the calibration data taken each night is
extremely variable and does not allow one to fit an airmass
coefficient per night and per band. For this reason, we have chosen to
fit one global airmass coefficient per band. This has no consequence
on the tertiary magnitudes which are built by averaging the data
coming from many different epochs. 

Quite often, second order terms of the form $k' \times (X-1) \times
{\rm color}$ enter the airmass parametrization and are neglected here.
This will be discussed in more detail in \S
\ref{sec:systematic_uncertainties}. We have estimated the magnitude of
these contributions using the passband models discussed in \S
\ref{sec:megacam_passbands} and synthetic photometry and found them to
be extremely small ($< 0.001$ mag).

In reality, the Landolt stars are not observed at the focal plane reference
location, but at many random locations. We
therefore rely on the grid transformations determined in \S
\ref{sec:the_photometric_grids} in order to relate the instrumental
magnitudes of the Landolt stars at any position with the same
magnitudes at the reference position.  The calibration equations
actually implemented in the zero-point fit are therefore a little more 
complex:
\begin{align*}
  \hat{g}_{ADU|\x} & = & \delta k_{ggr}(\x)\ \Bigl((g-r)_{|\x_0} - (g-r)_{grid}\Bigr) + V &- k_g  (X-1) &+ \\
                  &   &  C(B-V; \alpha_{g}, \beta_{g}) + ZP_g \\
                  & \ldots & \\
  \hat{z}_{ADU|\x} & = & \delta k_{ziz}(\x)\ \Bigl((i-z)_{|\x_0} - (i-z)_{grid}\Bigr) + I &- k_z  (X-1) &+ \\
                  &   &  C(R-I; \alpha_{z}, \beta_{z}) + ZP_z 
\end{align*}
These equations imply that the MegaCam colors of the Landolt stars,
measured at the reference location: $(g-r)_{|\x_0}$, \ldots
$(i-z)_{|\x_0}$ must be determined in the course of the calibration
fit. They are determined iteratively from the MegaCam observations of
the Landolt stars: in a first pass, we ignore the color grid corrections and
obtain a first approximation of the MegaCam colors of the Landolt
stars.  These colors are then injected into the fit, and we iterate.

The fact that the grid reference colors appear explicitly in the
calibration equations may seem a little odd. As mentioned above, these
colors are internal quantities, and should not impact the calibration.
Indeed, it can be verified that changing the grid reference colors and
changing the $\delta zp(\x)$ maps entering in the definition of the hat magnitudes accordingly has
no impact on the value of the zero points.

Finally, a few words must be said about the uncertainty model.  One
must recall that the Landolt stars do not form an exact one-parameter
sequence, and cannot be perfectly indexed by one color only. 
In practice, the Landolt-MegaCam color-color diagrams are affected by an intrinsic
star-to-star dispersion partly of astrophysical origin, of about 1\% in the
$\gme \rme \ime$-bands and 2\% in the $\zme$-band.
The dispersion of the fit residuals is dominated by this star-to-star
dispersion and not by the star measurement uncertainties (closer to
0.1\%). 
Since each star is measured several times, this means that the
error model has a very specific structure which must be modeled. The
star-to-star dispersion, in each band, is estimated from the fit
residuals. The correlations between the same band contributions of
each star are accounted for by fitting one additional parameter per
star and per band, $\delta {\rm mag}$, this parameter being
constrained by adding terms in the $\chi^2$ of the form $\left(\delta
{\rm mag} / \sigma_{star-to-star}\right)^2$.

\subsection{Results}
\label{sec:landolt_stars_results}

\begin{table*}
\begin{center}
\caption{Landolt to MegaCam color transformation slopes and airmass correction terms.
  \label{tab:zpfit_results}}
\begin{tabular}{ccccccc}
\hline
\hline
band & Landolt            & color & $\alpha$ & $\beta$ & $k$ & $k_{Elixir}$ $^\mathrm{a}$ \\
     & color index        & break &          &         &     &              \\
\hline
$g_M$  & $B-V$  & $+0.45$   & $+0.4957 \pm 0.0153$  &  $+0.4583 \pm 0.0026$  &  $-0.1830\pm 0.0017$ &  $-0.15$ \\
$r_M$  & $V-R$  & $+0.65$   & $+0.1654 \pm 0.0049$  &  $+0.2079 \pm 0.0248$  &  $-0.1346\pm 0.0017$ &  $-0.10$ \\
$i_M$  & $R-I$  & $+0.40$   & $+0.2069 \pm 0.0093$  &  $+0.1702 \pm 0.0056$  &  $-0.0467\pm 0.0017$ &  $-0.04$ \\
$z_M$  & $R-I$  & $+0.35$   & $-0.1949 \pm 0.0301$  &  $-0.4420 \pm 0.0133$  &  $-0.0585\pm 0.0034$ &  $-0.03$ \\
\hline 
\end{tabular}
\end{center}
\begin{list}{}{}
  \item[$^\mathrm{a}$] canonical airmass coefficients reported by the Elixir pipeline in the image headers.
\end{list}
\end{table*}

\begin{figure*}
\centering
\mbox{\subfigure[$\gme - V\ vs.\ B-V$]{\includegraphics[width=0.45\linewidth]{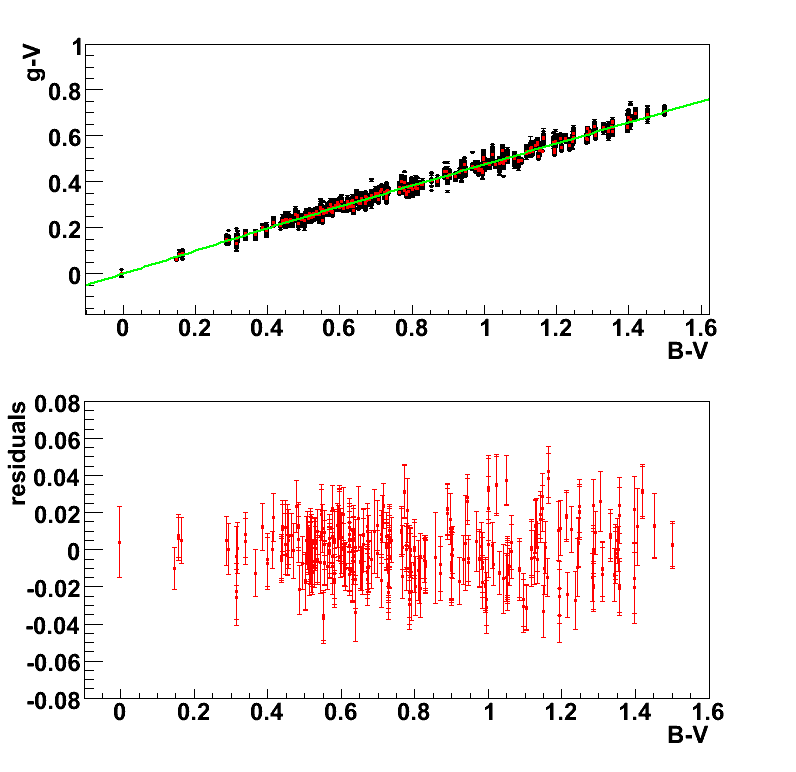}}
      \subfigure[$\rme - R\ vs.\ V-R$]{\includegraphics[width=0.45\linewidth]{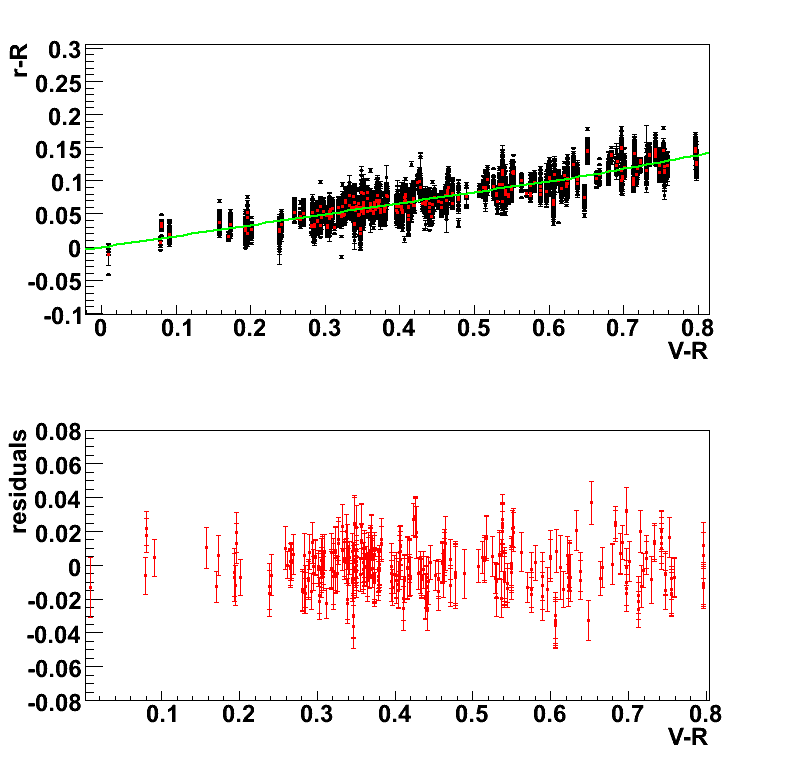}}}
\mbox{\subfigure[$\ime - I\ vs.\ R-I$]{\includegraphics[width=0.45\linewidth]{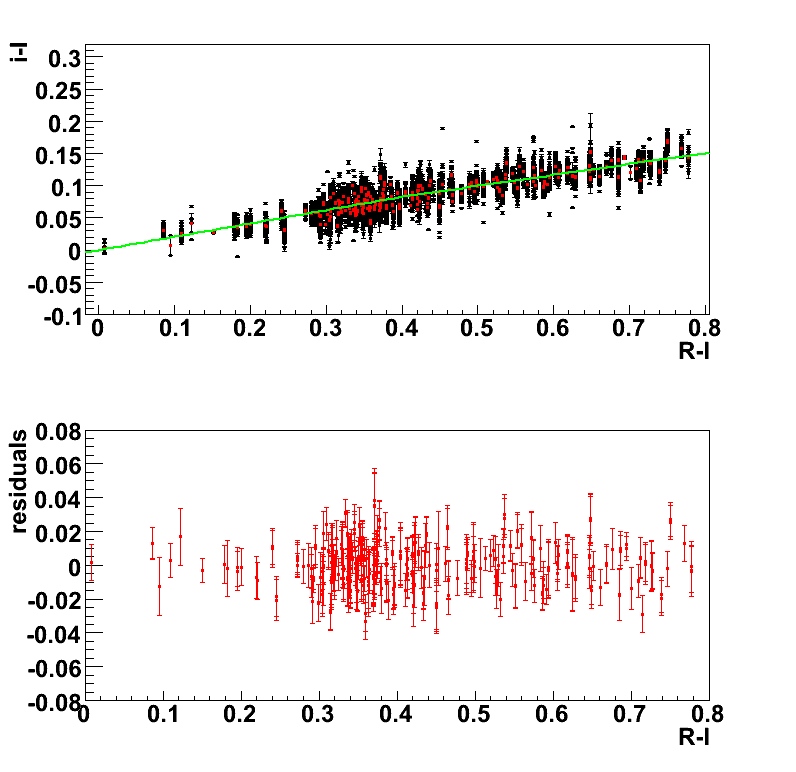}}
      \subfigure[$\zme - I\ vs.\ R-I$]{\includegraphics[width=0.45\linewidth]{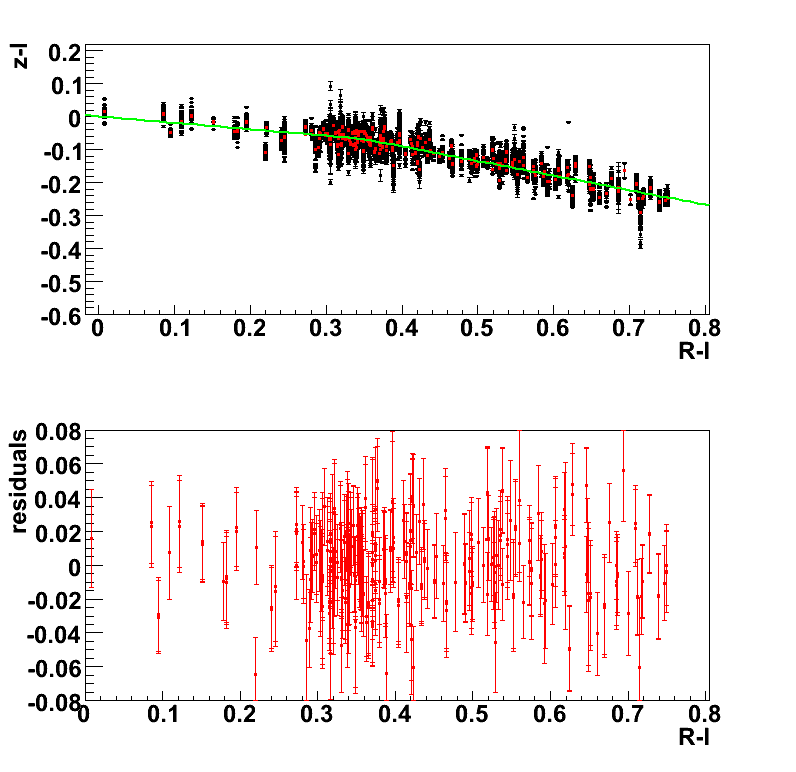}}}
\caption{Upper panels: color-color plots considered in the $\gme$,
  $\rme$, $\ime$ and $\zme$ zero-point fits. The black points
  correspond to individual Landolt star measurements. The red points
  are the average of the calibrated measurements of a same star.  The
  color-color transformations are modeled as piecewise-linear
  functions, with breaks at $B-V = +0.45$, $V-R=+0.65$, $R-I=+0.40$
  and $R-I=+0.35$ in the $\gme$, $\rme$, $\ime$ and $\zme$ bands
  respectively.  Lower panels: color-color plots residuals (average of
  each Landolt star's calibrated measurements only).
\label{fig:zp_color_color}}
\end{figure*}

Figure \ref{fig:zp_color_color} presents the color-color diagrams
considered in the zero-point fit described above. The parameters of
the Landolt-to-MegaCam color transformations are summarized in table
\ref{tab:zpfit_results}. These slopes are measured with precisions
of about 1\% on average, this error budget being dominated by the
star-to-star dispersion.

\begin{figure*}
\centering
\mbox{\subfigure[$\gme$]{\includegraphics[width=0.45\linewidth]{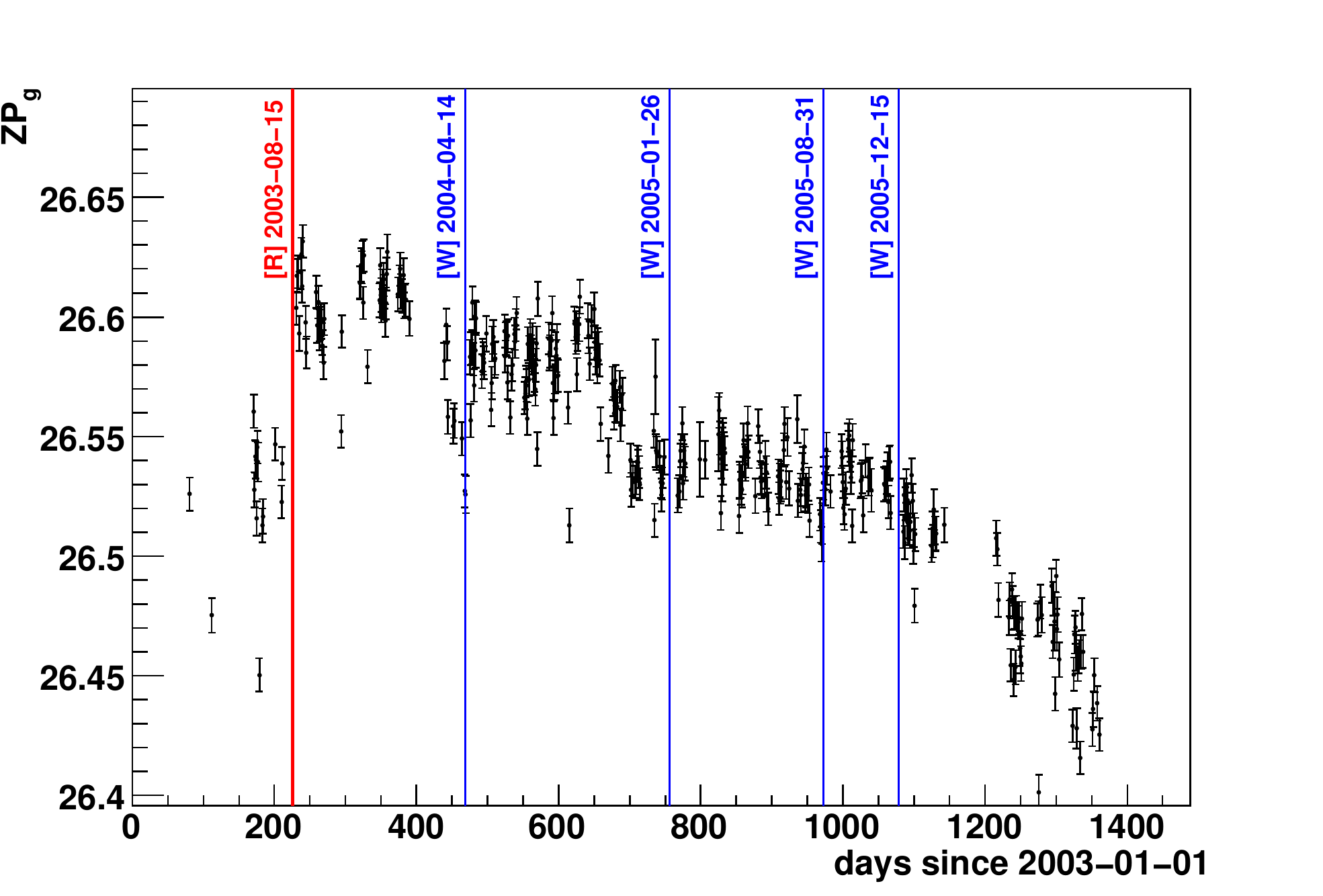}}
      \subfigure[$\rme$]{\includegraphics[width=0.45\linewidth]{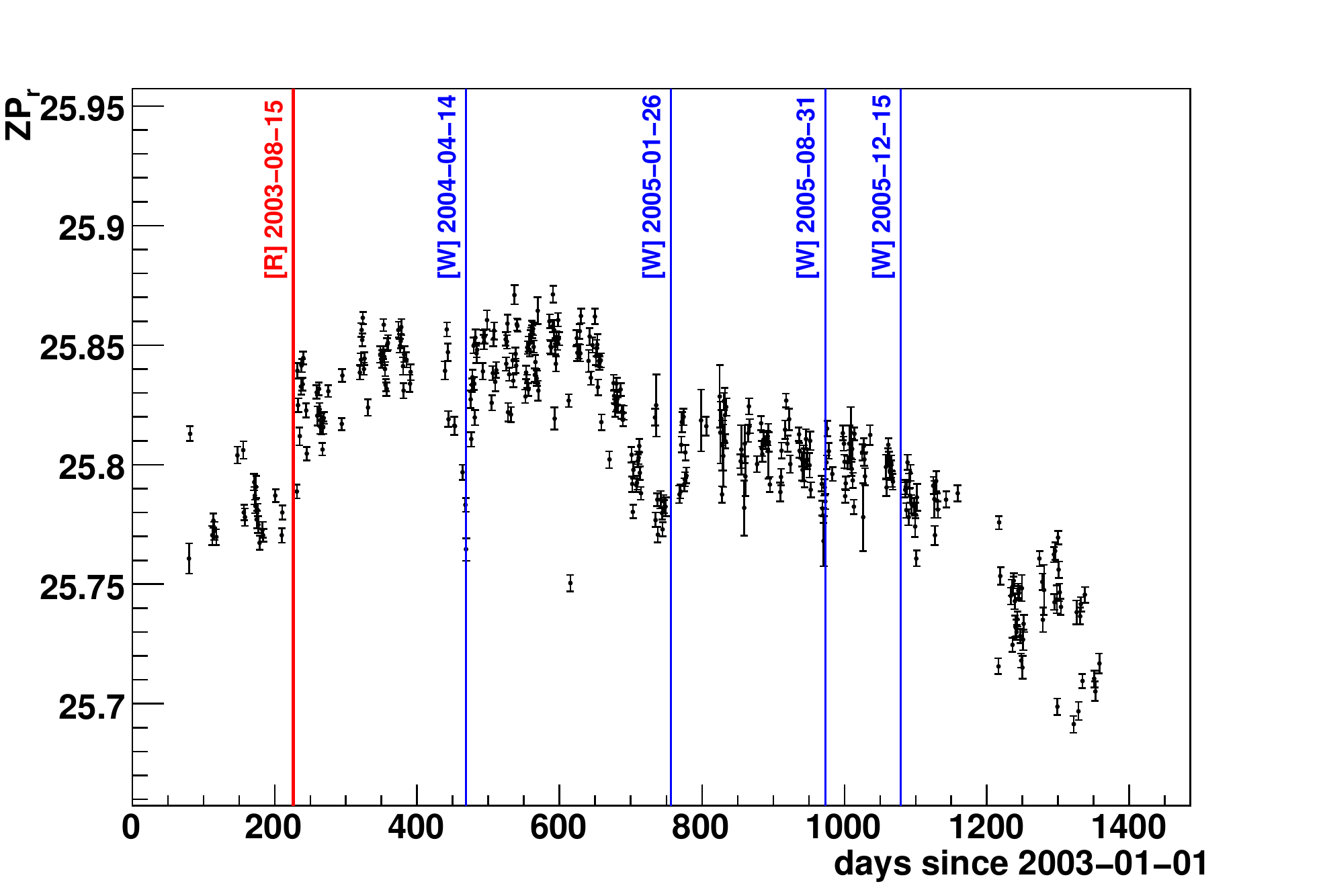}}}
\mbox{\subfigure[$\ime$]{\includegraphics[width=0.45\linewidth]{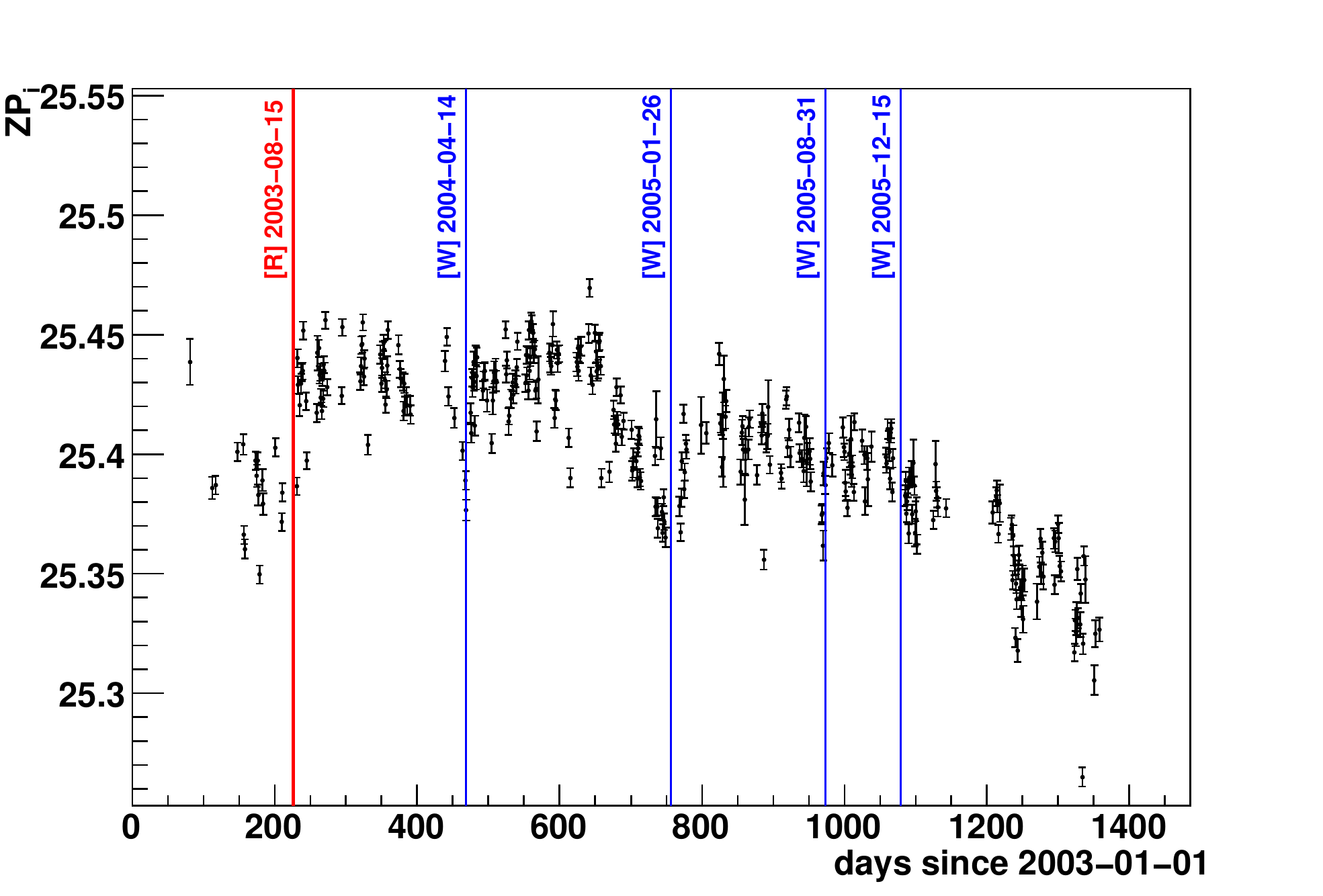}}
      \subfigure[$\zme$]{\includegraphics[width=0.45\linewidth]{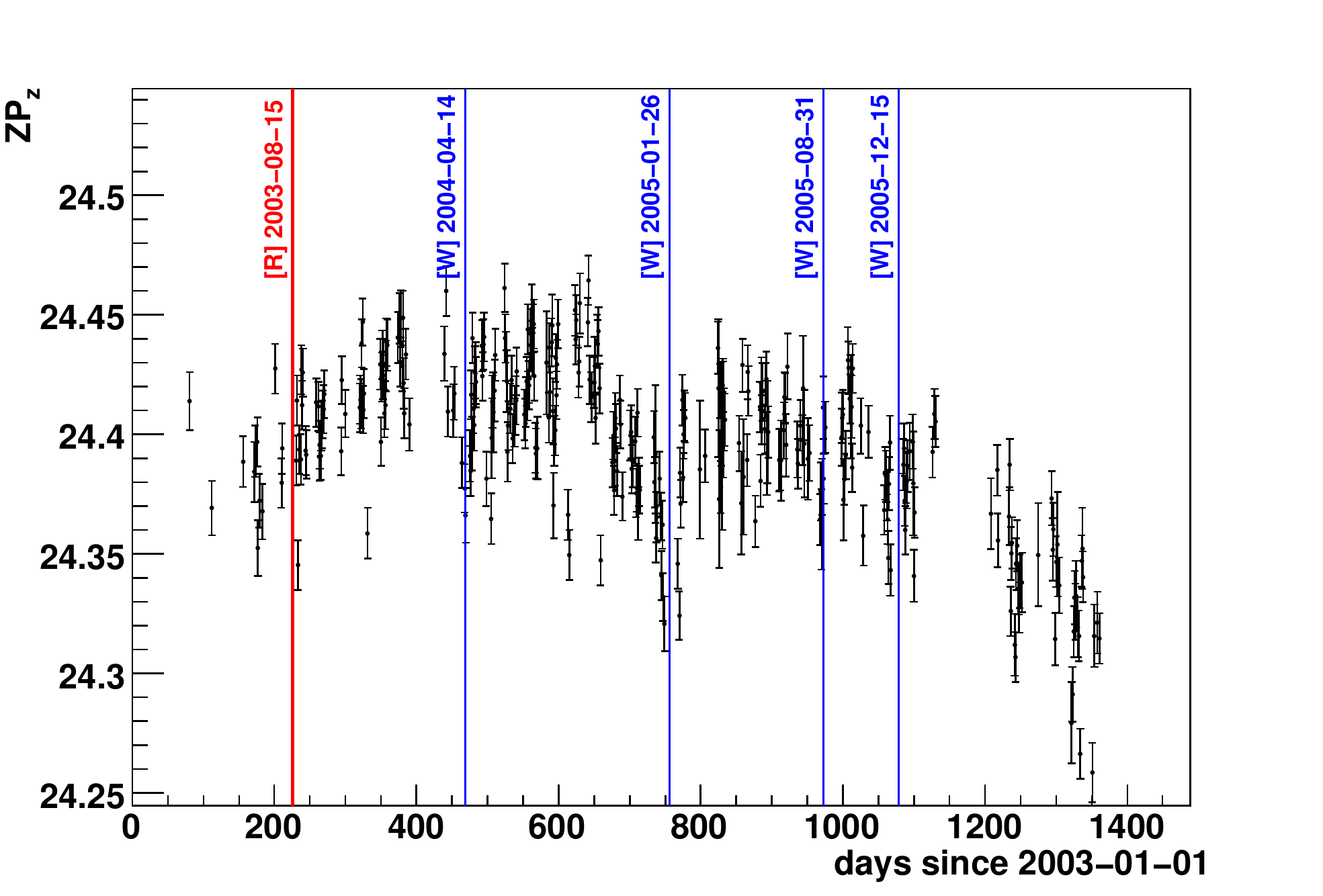}}}
\caption{$\gme$, $\rme$, $\ime$ and $\zme$ zero-point as a function of
  time. The zero-point evolution is classically due to dust
  accumulating in the optical path and the natural degradation of the mirror coating over time. We also indicate the main events 
 (recoating, labeled ``[R]'', and washings ``[W]'') which affected the primary mirror during 
  the three first years of the survey. The significant improvement
  observed around +200 days is due a mirror recoating 
  (August 15th, 2003). The effect of two additional mirror
  cleanings, which took place on April 14th, 2004 and January 26th,
  2005 are also clearly visible. The full MegaCam / MegaPrime history log is
  available at {\tt http://www.cfht.hawaii.edu/Instruments/Imaging/MegaPrime/megaprimehistory.html}.
\label{fig:zero_point_versus_time}}
\end{figure*}

The airmass coefficients are summarized in table
\ref{tab:zpfit_results}. As can be seen, they do differ from the
mean coefficients reported by Elixir, and these differences induce
sizeable differences on the calibrated magnitudes. 

Figure \ref{fig:zero_point_versus_time} presents the variation of the
$\gme, \rme, \ime$ and $\zme$ zero-points as a function of time. The
effect of dust accumulating on the optical path is clearly visible.
The recoveries observed at various points of the survey can be
explained with the full MegaCam / MegaPrime history log\footnote{{\tt
    http://www.cfht.hawaii.edu/Instruments/Imaging\-/MegaPrime\-/megaprimehistory.html}}. The
spectacular improvement observed at the beginning of the survey is due
to the recoating of the mirror (August 15th, 2003). Two additional
significant improvements were due to mirror
cleanings (April 14th, 2004 and January 26th, 2005). Note that an
additional mirror recoating took place on August, 2007, during the
fourth year of the survey. It allowed one to recover the efficiency levels 
observed at the beginning of the survey.

The zero-point uncertainties determined at this stage are of about
0.007, 0.003, 0.004 and 0.011 mag in the $\gme, \rme, \ime$- and
$\zme$-bands respectively. These uncertainties, are dominated by the
uncertainties on the color transformation parameters discussed
above. Indeed the average Landolt star colors are very far from zero: $B-V
\sim 0.77$, $V-R \sim 0.46$ and $R-I \sim 0.42$. Since the color
transformations are global, this also means that the zero-points in a
given band are correlated. If we remove the color
transformation contributions, the intrinsic zero-point uncertainties
are of about 0.003 mag in the $\gme, \rme, \ime$-bands and 0.005 mag
in $\zme$.

We will see in the next sections that the impact of the zero-point
uncertainties on the flux interpretation of the tertiary standard
actually depends on the colors of the fundamental standard used to
interpret the fluxes. If we choose a standard such as Vega, the
color transformation uncertainties will dominate. If we choose a star whose
colors are closer to the average Landolt star colors, 
the zero-point contributions to the calibrated flux 
error budget will be close to the intrinsic uncertainties.

\section{MegaCam Magnitudes}
\label{sec:megacam_magnitudes}

We are now ready to define the MegaCam magnitude system. As discussed
in the introduction of this paper, a requirement of the SNLS
calibration is that the MegaCam magnitudes must be easily interpreted
as physical fluxes. This requires that the MegaCam
magnitudes be defined as natural magnitudes, i.e. their definition should not 
integrate any term depending on each star's color.

\subsection{Uniform Magnitudes}

First, let's assume that the MegaCam passbands are spatially uniform. Or
equivalently, that all our tertiary standards are observed at the
focal plane reference location. In this case, the definition of the
MegaCam magnitudes is:
\begin{eqnarray*}
  g_{|\x_0} & \equiv & \hat{g}_{ADU|\x_0} + ZP_g \nonumber \\
       & \ldots & \nonumber \\
  z_{|\x_0} & \equiv & \hat{z}_{ADU|\x_0} + ZP_z 
\end{eqnarray*}
In reality, the tertiary standards are observed at various reference
locations. But since we know how the hat magnitudes transform all over
the focal plane, we can define the MegaCam magnitudes as:
\begin{eqnarray}
  g_{|\x_0} & \equiv & \hat{g}_{ADU|\x} - \delta k_{ggr}(\x) \times \Bigl[ (g-r)_{|\x_0} - (g-r)_{grid} \Bigr] + ZP_g \nonumber \\
      & \ldots & \nonumber \\
  z_{|\x_0} & \equiv & \hat{z}_{ADU|\x} - \delta k_{ziz}(\x) \times \Bigl[ (i-z)_{|\x_0} - (i-z)_{grid} \Bigr] + ZP_z 
\end{eqnarray}
With this definition, the calibrated magnitudes do not depend on the
focal plane position where the stars have been observed. We call 
them {\em Uniform Magnitudes}. These
are typically the kind of magnitudes we would like to report to the
end users, since they hide the complexity of the MegaCam imager to
the external user. However, the Uniform Magnitudes are not natural
magnitudes: indeed, they include a color term in their definition.

\subsection{Local Natural Magnitudes}

The definition above tells us how to define the MegaCam natural
magnitudes. They must incorporate the grid color corrections. 
We define these magnitudes as:
\begin{eqnarray}
  g_{|\x} & \equiv & {g}_{|\x_0} + \delta k_{ggr}(\x) \times \Bigl[ (g-r)_{|\x_0} - (g-r)_{ref} \Bigr] \nonumber \\
          & \ldots & \nonumber \\
  z_{|\x} & \equiv & {z}_{|\x_0} + \delta k_{ziz}(\x) \times \Bigl[ (i-z)_{|\x_0} - (i-z)_{ref} \Bigr] 
\end{eqnarray}
where $g_{|\x_0}$ \ldots $z_{|\x_0}$ are the magnitudes defined above,
and $(g-r)_{ref}$ \ldots $(i-z)_{ref}$ arbitrary color zero
points. With this definition, the relation between the calibrated 
magnitudes and the instrumental magnitudes is:
\begin{eqnarray}
  g_{|\x} & = & \hat{g}_{ADU|\x} + ZP_g - \delta k_{ggr}(\x) \times \Bigl[ (g-r)_{ref} - (g-r)_{grid} \Bigr] \nonumber \\
         & \ldots & \nonumber \\
  z_{|\x} & = & \hat{z}_{ADU|\x} + ZP_z - \delta k_{ziz}(\x) \times \Bigl[ (i-z)_{ref} - (i-z)_{grid} \Bigr] 
\end{eqnarray}
As we can see, the $g_{|\x}$, \ldots $z_{|\x}$ form a natural
magnitude system. We call them {\em Local Natural Magnitudes}. On the
other hand, the definition of these magnitudes explicitly depends on
where each star was observed on the focal plane.

Once again, it is easy to verify that the calibrated magnitudes do not
depend on the choice of the grid reference colors, since the quantities
$\hat{m}_{ADU|\x} + \delta k(\x) \times ({\rm color}_{grid}) $ are
themselves independent of those reference colors.
We also have introduced another set of reference colors: $(g-r)_{ref},
\ldots (i-z)_{ref}$. These colors have the same status as the grid
reference colors: they have been introduced to make explicit that
there is some degree of arbitrariness in the Local Magnitude
definition. In practice, it is useful to take these colors equal the
colors of the fundamental standard at the center of the focal plane:
\begin{eqnarray*}
  (g-r)_{ref} & = & (g-r)_{|\x_0}^{ref}\nonumber \\
  (r-i)_{ref} & = & (r-i)_{|\x_0}^{ref}\nonumber \\
  (i-z)_{ref} & = & (i-z)_{|\x_0}^{ref}
\end{eqnarray*}

\section{Flux Interpretation of the MegaCam Magnitudes}
\label{sec:flux_interpretation_of_megacam_magnitudes}

Now that we have defined a system of natural magnitudes, we still have
to explicit the magnitude to flux conversion. To do this, we rely on a
``fundamental spectrophotometric standard'', i.e. a star of known SED,
$S_{ref}(\lambda)$ and known MegaCam magnitudes in the system defined
above. The calibrated broadband flux, $F_{|\x}$ of an object of magnitude
$m_{|\x}$ is then:
\begin{equation}
  \text{F}_{|\x} = 10^{-0.4\ \left(m_{|\x} - m_{ref}\right)} \times \int S_{ref}(\lambda) T(\lambda;\x) d\lambda
  \label{eqn:magnitude_to_flux_transformation}
\end{equation}
where $T(\lambda;\x)$ is the effective passband of the imager at
location $\x$ on the focal plane.

Systematic uncertainties affect this mapping function. First, the
MegaCam passbands are not known perfectly. Furthermore, the SED of the
fundamental standard is not perfectly measured.  Finally, the MegaCam
magnitudes of the fundamental standard, in the system defined in the
previous section are not known perfectly. At best, the fundamental
standard is directly observed with the survey telescope, and the
measurement uncertainties must be taken into account. Most often, the
fundamental standard is too bright to be directly observed, and its
magnitudes must be inferred in some way, introducing additional
uncertainties.

It must also be noted that the quantities relevant for the cosmology
studies are not exactly physical fluxes, but rather the {\em ratio} of
physical fluxes, measured in different passbands. In other words, the
cosmological measurements are completely insensitive to any change of
the absolute flux scale. For this reason, we report the
uncertainties of the fluxes and magnitudes relative to a reference
band, namely the Landolt $V$-band. In particular, the relevant
uncertainties are those affecting (1) the ratios $\int
S_{ref}(\lambda) T(\lambda;\x)d\lambda / \int S_{ref}(\lambda)
V(\lambda) d\lambda$ and (2) the colors $g_{ref}-V$, $r_{ref}-V$,
$i_{ref} - V$ and $z_{ref} - V$.

\subsection{Selecting a Fundamental Standard}
\label{sec:selecting_a_fundamental_standard}

No spectrophotometric standard usable as a fundamental
standard has reliable magnitudes established from MegaCam observations. Such a program is
underway, but has not been completed yet. The MegaCam magnitudes of
the fundamental standard must therefore be infered from its Landolt
magnitudes, using the Landolt to MegaCam color transformations. For
example, if the Landolt colors of the star are bluer than the color
break:
\begin{eqnarray}
  g_{|\x_0}^{ref} & = & V^{ref} + \alpha_g \times (B-V)^{ref} + \Delta g^{ref}_{|\x_0} \nonumber \\
                & \ldots & \nonumber \\
  z_{|\x_0}^{ref} & = & I^{ref} + \alpha_z \times (R-I)^{ref} + \Delta z^{ref}_{|\x_0}
\label{eqn:megacam_magnitudes_fundamental_standard}
\end{eqnarray}
where the quantities $\Delta m^{ref}_{|\x_0}$ account for the fact
that the fundamental standard departs slightly from the
Landolt-to-MegaCam color-color law. The uncertainties on the MegaCam
magnitudes of the fundamental standard account directly as systematic
errors on the calibrated fluxes. The $\Delta m^{ref}_{|\x_0}$
offsets can be as large as one percent, given the standard deviation of the residuals 
to the Landolt-to-MegaCam color laws. Hence, they must be
evaluated, using synthetic photometry. Furthermore, we point out that
the uncertainty on the Landolt-to-MegaCam color transformations also
affect the estimates of $g_{|\x_0}$, \ldots $z_{|\x_0}$. Our concern
in this section, is therefore to choose a fundamental standard, which
would allow us to minimize this error budget.

A standard choice is the star $\alpha$-Lyr{\ae}, also called
Vega. Vega is one of the six A0V stars that define the zero-points of
the historical \citet{Johnson51, Johnson53} UBV system. The Landolt
magnitudes of this star being all close to zero (although not exactly
zero), the Landolt system is often (improperly) referred to as a ``Vega based
system'' and the SED of Vega is generally used to convert the Landolt
magnitudes into fluxes. This was the approach used in
\citet{astier06}. The ``canonical'' magnitudes of Vega
reported, for example in \citet{fukugita96} were used to twist the SED of Vega
measured by \citet{Bohlin04} and to get an approximation of the SED of the 
hypothetical fundamental standard of the Landolt system.

However, a close study of the systematic error budget shows that Vega
is not a wise choice if we seek a 1\% precision or better on the
calibrated fluxes. First, Landolt did not actually
observe Vega directly while building his catalogs due to its brightness. An estimate of the Landolt
Vega $U-B$ and $B-V$ colors can be obtained by propagating the initial
measurements of \citet{Johnson51} through the color transformations
cited by \citet{Landolt73, Landolt83, Landolt92}. The $R-I$ color
index can traced back from the original \citet{Cousins78} papers. However, 
values of $R-I$ that differ by about 
0.05 mag  are cited in \citet{Taylor86,fukugita96}.
Furthermore, no measurement of the $V-R$ color index could be found in the 
literature. In any case, it is not possible to guarantee a precision of 0.01 mag
on the Vega magnitudes reconstructed in such a way.

Another problem with Vega, is that it is significantly bluer than the
average Landolt star ($B-V \sim 0.77$) and the average tertiary
standards. As discussed in \S \ref{sec:landolt_stars}, there are 
large uncertainties associated with the modeling of the Landolt-to-MegaCam 
color transformations. 
In particular, given the low number of blue Landolt stars, 
it is nearly impossible to tell anything about the 
linearity of the color transformations in the bluer parts of the color-color diagrams.
Furthermore, as noted in \S \ref{sec:landolt_stars}, the blue side of the
Landolt-to-MegaCam transformations (the $\alpha$ parameters) are
determined with a precision not better than 1.5\% in $g$ and $3\%$ in
$z$. This induces an additional uncertainty of about 1\% on the
MegaCam colors of Vega infered with equation \ref{eqn:megacam_magnitudes_fundamental_standard}. 
This problem disappears if we use, as a fundamental standard, 
a star of colors close to the average color of the Landolt stars, 
and if possible, a star directly observed by Landolt.

Few Landolt stars have measured spectral energy distributions,
covering the wavelength-range $3000\AA\ - 11000\AA$ covered by
MegaCam.  The HST community has put considerable effort into building
a database of high-quality spectrophotometric standards ---the
so-called CALSPEC database \citep{CALSPEC}. An absolute flux scale was
defined, based on NLTE models of three pure hydrogen white-dwarfs: G191-B2B, GD~153 and GD~71. The SEDs
of several key standards such as Vega and \bdtruc\ have been
re-observed and re-calibrated with the {\em Space Telescope Imaging Spectrograph} (STIS) 
and the {\em Near Infrared Camera and Multi-Object Spectrometer} (NICMOS) instruments
\citep{Bohlin00, Bohlin04, Bohlin07}. More recently,
\citet[][hereafter LU07]{Landolt07b} published magnitudes of some of the CALSPEC HST
spectrophotometric standards. Combining these two sources, we found
that six stars have simultaneously known Landolt magnitudes and known SEDs, 
published by the CALSPEC project and 
measured exclusively with the HST STIS and NICMOS instruments:
AGK~+81~266, \bdtruc, G~191-B2B, GD~71, GRW~705824 and LDS~749B. Most of
these objects are very blue stars, except \bdtruc, whose colors are
close to the average Landolt colors.

We have therefore selected \bdtruc\ as a fundamental standard. This
F8-type star has been chosen as a fundamental standard for many photometric
systems, notably that of the Sloan Digital Sky Survey (SDSS)
\citep{fukugita96, Smith02, Gunn98, Gunn06, Ivezic07} and consequently has been
studied by many groups, which have derived estimates of its extinction, effective
temperature, metallicity and surface gravity \citep[see][and references
  therein]{Ramirez06}. \citet{Bohlin04b} have measured the absolute
spectral energy distribution of \bdtruc\ in the wavelength range
$1700\ \AA < \lambda < 10000\ \AA$, with an accuracy of less than
0.5\% in the transfer of the flux calibration of the three white
dwarfs primary standards, and an accuracy of the relative flux distribution 
of about 2\%\footnote{In this analysis we use the latest
  version posted on the CALSPEC ftp server: {\tt
    ftp://ftp.stsci.edu/cdbs/current\_calspec/} and labeled {\tt
    bd17d4708\_stisnic\_002.fits}}. 

On the other hand, it has been pointed out that \bdtruc\ may be a
binary system, with a faint late-M companion of mass $\sim 0.15
M_\odot$ revolving around the main star in about 220 days 
\citep{Latham88, Ramirez06}. Furthermore, with a $V$-band
magnitude of 9.464, \bdtruc\ is a little bright to be observed
directly with MegaCam with a good accuracy. The indirect determination
of its MegaCam magnitudes is presented in the next section.

In addition to \bdtruc, LU07 present magnitudes for the fundamental
CALSPEC white dwarf calibrators.  We also considered using these
stars, but the fact that they are even bluer than Vega
exacerbates the extrapolation problem when calculating their MegaCam magnitudes.
Furthermore, there is some evidence that a similar problem affects the
Landolt magnitudes themselves.  As noted earlier, the original Landolt
telescope/detector system no longer exists, so the magnitudes
tabulated in LU07 have been transformed from some natural system to
the \citet{Landolt92} system using a similar method as that described
in 9.1.  These transformations were calculated using ``typical''
Landolt stars, which are much redder than the white dwarfs.
Therefore, these transformations may not be accurate for the very blue
white dwarfs, especially in $B$ and $U$ where the absence of a Balmer
break makes the SEDs very different.

Without knowing the exact Landolt catalog system passbands we have no
way of precisely calculating the amount of bias present or correcting
for it, but we can check for the plausibility of this issue by
simulating a similar set of observations and reduction procedures as
actually used by LU07.  We start with an assumed model for the Landolt
catalog passbands \citet{Bessel90} and use the natural system
passbands given in LU07, folding in mirror reflectivity and
photomultiplier response, then carry out synthetic photometry using a
library of SEDs \citet{Pickles98} in both systems.  We then calculate
multi-step linear transformations following the prescription of
\citet{Landolt92} using the redder stars and compare the transformed
magnitudes of the CALSPEC WDs to the actual synthetic magnitudes in
the natural system.  We find that the transformations are biased at
the 1-2\% level, particularly in $B$ where the LU07 $B$ filter has a
``notch'' near the peak transmission.  The exact amount of bias is
sensitive to the assumed Landolt catalog filters, but its existence
is not.  We conclude that there is reason to be cautious when using
the WDs for photometric calibration.  Note that these concerns have no
effect on the CALSPEC spectroscopic calibration.

\subsection{The MegaCam Magnitudes of \bdtruc}
\label{sec:megacam_magnitudes_bdtruc}

At first order, the MegaCam magnitudes of \bdtruc\ can be derived from
the magnitudes and colors reported by Landolt, and the
Landolt-to-MegaCam color transformations determined in \S
\ref{sec:landolt_stars} (see equation \ref{eqn:megacam_magnitudes_fundamental_standard}). 
In table \ref{tab:magnitudes_of_bd17}, we report the
$g_M-V$, $r_M-V$, $i_M-V$ and $z_M-V$ colors of \bdtruc. The
uncertainties quoted by Landolt are propagated, assuming that the
magnitudes and colors $V$, $B-V$, $V-R$ and $R-I$ reported by Landolt
are essentially independent. Note that the $i_M-V$ and $z_M-V$ are
strongly correlated, with a correlation coefficient of 0.96. The
statistical uncertainties affecting the Landolt-to-MegaCam
color transformations do also affect the MegaCam colors reported in
table \ref{tab:magnitudes_of_bd17}. However, we will see in \S
\ref{sec:tertiary_catalogs} that the impact of the color
transformations on the calibrated tertiary fluxes is actually much
smaller. As a consequence, we do not include them in the final
uncertainty budget listed in table \ref{tab:magnitudes_of_bd17}.

The determination of the offsets $\Delta g_{|\x_0}^{ref}$ \ldots
$\Delta z_{|\x_0}^{ref}$ defined in equation \ref{eqn:megacam_magnitudes_fundamental_standard} is a little more complex. These quantities
account for how the \bdtruc\ magnitudes differ, on average, from those
of the Landolt stars whose colors are close to \bdtruc. To estimate
them, we will rely on (1) estimates of the extinction, temperature,
metallicity and surface gravity of \bdtruc, (2) rough estimates of the
same quantities for Landolt stars of colors similar to those of
\bdtruc\ and (3) the Phoenix library of synthetic star SED models
\citep[][and references therein]{Hauschildt97, Baron98, Hauschildt99}\footnote{The study presented here relies on version 2.6.1 of the Phoenix / GAIA spectral library, that can be retrieved from the Phoenix ftp server: {\tt ftp://ftp.hs.uni-hamburg.de/pub/outgoing/phoenix/GAIA}}.

The extinction, temperature, surface gravity and metallicity of
\bdtruc\ are estimated in \citet{Ramirez06}: $E(B-V) \simeq 0.010 \pm
0.003$, $T_{eff} \simeq 6141 \pm 50 K$, $\log g = 3.87 \pm 0.08$ and
$[{\rm M/H}] = -1.74 \pm 0.09$.  \citep{Ramirez06} also gives rough
estimates of the type (late-M), mass ($\sim 0.15 M_\odot$) and
effective temperature ($\sim 3000 K$) of its faint companion.

On the other hand, not much is known about the Landolt stars. We
assume them to be nearby disk stars. Their mean metallicity may be
derived from \citet{Ramirez07}: $-0.5 < [{\rm M/H}] < -0.3$,
substantially higher than the metallicity of \bdtruc. Landolt stars of
colors similar to those of \bdtruc\ have an effective temperature
$T_{eff} \sim 6200 K$ and a surface gravity $\log g \sim 4.3$ 
\citep{Allen76}. In what follows, we estimate
the magnitude offsets induced by (1) the fact that \bdtruc\ is a
likely binary system (2) the metallicity differences between \bdtruc\ and the
mean Landolt star (3) the surface gravity differences and (4) the
extinction differences.

\paragraph{Binarity} We can estimate the impact of the likely faint companion
with synthetic photometry. We select the Phoenix stellar models whose
parameters are as close as possible to those of \bdtruc\ and its
companion, respectively $T_{eff} = 6000 K$, $\log g = 3.5$, $[M/H] =
-2$ and $T_{eff} = 3000 K$, $\log g = 4.5$ and $[M/H] = -2$ \citep{Ramirez06}. The
impact on the calibration is given by the difference of residuals to
the Landolt-to-MegaCam color transformation, with and without the
contribution of the faint companion. We obtain (in the sense with
companion {\em minus} without companion):
\begin{eqnarray}
  \Delta g \simeq +0.001 \nonumber \\
  \Delta r \simeq +0.004 \nonumber \\
  \Delta i \simeq -0.002 \nonumber \\
  \Delta z \simeq -0.015 
\end{eqnarray}
There is a large uncertainty, of about 50\% on those numbers. 
We account for them as additional systematic uncertainties,
as we do not known the fraction of Landolt stars that are also in
binary systems.

\paragraph{Metallicity} 
{As noted above the metallicity of \bdtruc\ is significantly lower
than that of Landolt stars of similar colors.  The impact of this
difference was estimated by computing the offsets between the MegaCam
synthetic magnitudes of GAIA / Phoenix stars of metallicities and
colors close to that of \bdtruc, and the synthetic magnitudes of
Phoenix SEDs of similar colors but metallicity close to that of
Landolt stars. We obtained the following offsets (in the sense
\bdtruc\ mag {\em minus} Landolt mag):}
\begin{eqnarray}
  \Delta g = +0.007 \pm 0.002 \nonumber \\
  \Delta r = +0.003 \pm 0.001 \nonumber \\
  \Delta i = +0.002 \pm 0.001 \nonumber \\
  \Delta z = -0.009 \pm 0.001 
\end{eqnarray}
These offsets
are applied to the first order estimates of the magnitudes of \bdtruc.

\paragraph{Surface Gravity} The impact of the surface gravity differences 
between the Landolt stars and \bdtruc\ can be evaluated in a similar
fashion. We find that the corrections are smaller than 0.001 mag in
all bands except in $\zme$ {(in the sense \bdtruc\ mag {\em minus} Landolt mag)}:
\begin{eqnarray}
  \Delta z =  -0.001 \pm 0.003 
\end{eqnarray}
We apply this offset to the \bdtruc\ magnitude estimates, and 
retain a systematic uncertainty of 0.003 mag in the systematic error
budget.

\paragraph{Extinction} The mean reddening affecting the Landolt stars is
poorly known. It can be constrained by the locus of the Landolt stars,
in the $V-R$ vs. $R-I$ color-color diagram. Indeed, the $V-R$ and
$R-I$ colors of the Landolt stars are very well correlated in the
color region of \bdtruc. More precisely, the quantity $\Delta_{VRI} =
(R-I) - 0.7 (V-R)$, computed in the same color region, has an RMS of
0.019 and an average value of $0.995 \pm 0.002$. For \bdtruc,
$\Delta_{VRI} = 0.111$. This quantity can be related to unmeasured
quantities such as the reddening using synthetic photometry:
\begin{eqnarray*}
  \Delta_{VRI}^{\bdtruc} - \Delta_{VRI}^{Landolt} & = & \frac{\partial \Delta_{VRI}}{\partial E(B-V)}\ \delta E(B-V) + \\
                                              &   & \frac{\partial \Delta_{VRI}}{\partial [M/H]}\ \delta [M/H] + \\
                                              &   & \frac{\partial \Delta_{VRI}}{\partial \log g}\ \delta\log g
\end{eqnarray*}
Using the Phoenix models, we find that the derivatives of
$\Delta_{VRI}$ w.r.t. the extinction and metallicity in the color
region of \bdtruc\ to be, respectively: $0.22$, $0.02$,
the effect of the surface gravity being essentially negligible in this
color range. This can be translated into a constraint on
the extinction difference between \bdtruc\ and the Landolt stars of
similar colors:
\begin{equation*}
  \Delta E(B-V) = E(B-V)(BD\ +17) - E(B-V)({\rm Landolt}) \simeq 0.045
\end{equation*}
Again, using synthetic photometry and the \citet{Cardelli89} law, this
can be translated into calibration offsets:
\begin{eqnarray}
  \Delta \gme & = -0.01 \times \Delta E(B-V) & = -0.0005 \nonumber \\
  \Delta \rme & = +0.02 \times \Delta E(B-V) & = +0.0009 \nonumber \\
  \Delta \ime & = +0.01 \times \Delta E(B-V) & = +0.0005 \nonumber \\
  \Delta \zme & = -0.19 \times \Delta E(B-V) & = -0.0085 
\end{eqnarray}
The effect of the reddening differences between
\bdtruc\ and the Landolt stars impacts essentially the $\zme$-band
magnitudes, because this band requires an extrapolation from 
$I$ to $\zme$. We will consider these offsets as additional systematic
uncertainties.

To summarize: we have derived MegaCam Natural colors $g-V$,
$r-V$, $i-V$ and $z-V$ for \bdtruc, relying on the
Landolt-to-MegaCam color transformations, and on synthetic photometry
computed from the Phoenix / GAIA spectral library. The results along
with their uncertainties are summarized in table
\ref{tab:magnitudes_of_bd17}. The Landolt magnitudes of \bdtruc, as well
as our estimates of the \bdtruc\ MegaCam magnitudes are all correlated.
The full covariance matrix, ${\mathbf V}_{\bdtruc}$, reported in table
\ref{tab:megacam_mags_fundamental_standard_covmat} (appendix
\ref{sec:uncertainties_and_covariance_matrices}) is the sum of two $9
\times 9$ components:
\begin{equation*}
  {\mathbf V}_{\bdtruc} = {\mathbf V}_{Landolt} + {\mathbf V}_{\Delta m}
\end{equation*}
${\mathbf V}_{\Delta m}$ contains the uncertainties on the deviations
to the linear Landolt-to-MegaCam color corrections derived in this
section and assumed to be independent.  ${\mathbf V}_{Landolt}$
accounts for the Landolt uncertainties on the $g-V$, $r-V$, $i - V$,
$z-V$, $U-V$, $B-V$, $R-V$ and $I-V$-colors. Note that since
\citet{Landolt07b} do not discuss the correlations between their
measurement uncertainties, we had to make an assumption on their error
budget: we have chosen to assume that their $V$, $U-B$, $B-V$, $V-R$,
$R-I$ and $V-I$ measurements are essentially independent. This is
equivalent to assuming that the individual magnitudes measurements are
all affected by an overall ``gray'' uncertainty, that correlate them
positively. This ``gray'' uncertainty is reflected in the higher $V$-band
uncertainty reported by \citet{Landolt07b}. The uncertainties on the
MegaCam $X-V$ colors are obtained by propagating the Landolt errors
using equation \ref{eqn:megacam_magnitudes_fundamental_standard}.

To be complete, we should also add the contribution of the
uncertainties that affect the Landolt-to-MegaCam transformation
coefficients themselves.  However, as will be discussed later, this
contribution is itself strongly correlated with the zero-point
uncertainties, hence with the tertiary star magnitudes. Since the
uncertainties we are ultimately interested in are those which affect
the {\em differences} between the tertiary star magnitudes and the
corresponding magnitudes of \bdtruc, we will discuss this term later
in the analysis, once we have discussed the tertiary star uncertainty
budget. Note however that this contribution depends on the {\em
  difference} between the \bdtruc\ colors and the average Landolt star
colors. Since \bdtruc\ is much redder than Vega, this contribution is
greatly reduced by using this star instead of Vega.

Finally, the MegaCam magnitudes of \bdtruc, along with their
covariance matrix can be computed from the quantities listed in table
\ref{tab:magnitudes_of_bd17}. We find $g_M = 9.6906 \pm 0.0035$, $r_M
= 9.2183 \pm 0.0050$, $i_M = 8.9142 \pm 0.0037$ and $z_M = 8.7736 \pm
0.0180$.  The full covariance matrix (without the color transformation
uncertainties) is reported in appendix
\ref{sec:uncertainties_and_covariance_matrices} (table
\ref{tab:megacam_mags_fundamental_standard_covmat}).

\begin{table*}
\begin{center}
\caption[]{Landolt and MegaCam Magnitudes and colors of \bdtruc\ with their uncertainties.\label{tab:magnitudes_of_bd17}}
\begin{tabular}{l|ccc|cccccc|c}
\hline
\hline
      &  1st order        &  offset              & final    & Landolt          & Color            & $\log g$      & $[M/H]$ & $E(B-V)$ & binarity & total \\
      &                   &                      & value    & uncertainties    & transformations$\dagger$ &       &         &          &          & uncertainty \\
\hline 
$V$   &  --               & --                   & $+9.464$ & $\pm 0.0026$     &   --           &  --             &    --   &  --      & --       & $\pm 0.0026$ \\
$U-V$ &  --               & --                   & $+0.260$ & $\pm 0.0026$     &   --           &  --             &    --   &  --      & --       & $\pm 0.0026$ \\
$B-V$ &  --               & --                   & $+0.443$ & $\pm 0.0015$     &   --           &  --             &    --   &  --      & --       & $\pm 0.0015$ \\
$R-V$ &  --               & --                   & $-0.298$ & $\pm 0.0011$     &   --           &  --             &    --   &  --      & --       & $\pm 0.0011$ \\
$I-V$ &  --               & --                   & $-0.618$ & $\pm 0.0013$     &   --           &  --             &    --   &  --      & --       & $\pm 0.0013$ \\
\hline
$g_M-V$ & $+0.2196$         &       $+0.007$     &  $+0.2266$  & $\pm 0.0007$     &  $\pm 0.0067$  &    $<0.001$             & $\pm 0.002$  & $<0.001$     &  $\pm 0.001$   & $\pm 0.0023$ \\
$r_M-V$ & $-0.2487$         &       $+0.003$     &  $-0.2457$  & $\pm 0.0009$     &  $\pm 0.0015$  &    $<0.001$             & $\pm 0.001$  & $\pm 0.001$  &  $\pm 0.004$   & $\pm 0.0043$ \\
$i_M-V$ & $-0.5518$         &       $+0.002$     &  $-0.5498$  & $\pm 0.0013$     &  $\pm 0.0029$  &    $<0.001$             & $\pm 0.001$  & $<0.001$     &  $\pm 0.002$   & $\pm 0.0026$ \\
$z_M-V$ & $-0.6804$         &       $-0.010$     &  $-0.6904$  & $\pm 0.0013$     &  $\pm 0.0185$  &    $\pm 0.003$          & $\pm 0.001$  & $\pm 0.009$  &  $\pm 0.015$   & $\pm 0.0178$ \\
\hline
\end{tabular}
\end{center}
\begin{list}{}{}
\item[$^\dagger$] Not included in the total uncertainty budget
  reported in the last column (total uncertainty). The total impact of
  the color transformation uncertainties on the calibrated tertiary
  fluxes is much smaller. It is discussed in section
  \ref{sec:tertiary_catalogs}.
\end{list}
\end{table*}

\section{Tertiary Catalogs}
\label{sec:tertiary_catalogs}

We now turn to the production of the tertiary standard catalogs.  One
such catalog is produced for each of the four SNLS fields, using
science and calibration exposures taken under photometric
conditions. Once established for each of the four SNLS fields, the
tertiary standard catalogs allow any user to propagate the calibration
to any SNLS exposure. In this section, we discuss the general
procedure to derive tertiary standard catalogs from the science and
calibration exposures.

\subsection{Tertiary Star Selection}

The tertiary standard catalogs should only contain well measured,
isolated, non-variable stars. The star identification is carried out
as follows. The objects are detected on deep stacks of the SNLS
fields, using the SExtractor package \citep{Sex}. The second moments
($m_{xx}$, $m_{yy}$, $m_{xy}$) of the sources detected on each CCD are
estimated from a 2-D Gaussian fit. The star locus in the $m_{xx}$
versus $m_{yy}$ diagram is then identified. A first list of tertiary
standard candidates is established. It contains all the isolated
objects belonging to the star locus and measured with a
signal-to-noise ratio better than 10. By ``isolated'', we mean that
the aperture flux pollution due to the closest neighbor must be less
that 0.1\% of the star flux.
 
\begin{figure}
\begin{center}
\includegraphics[width=\linewidth]{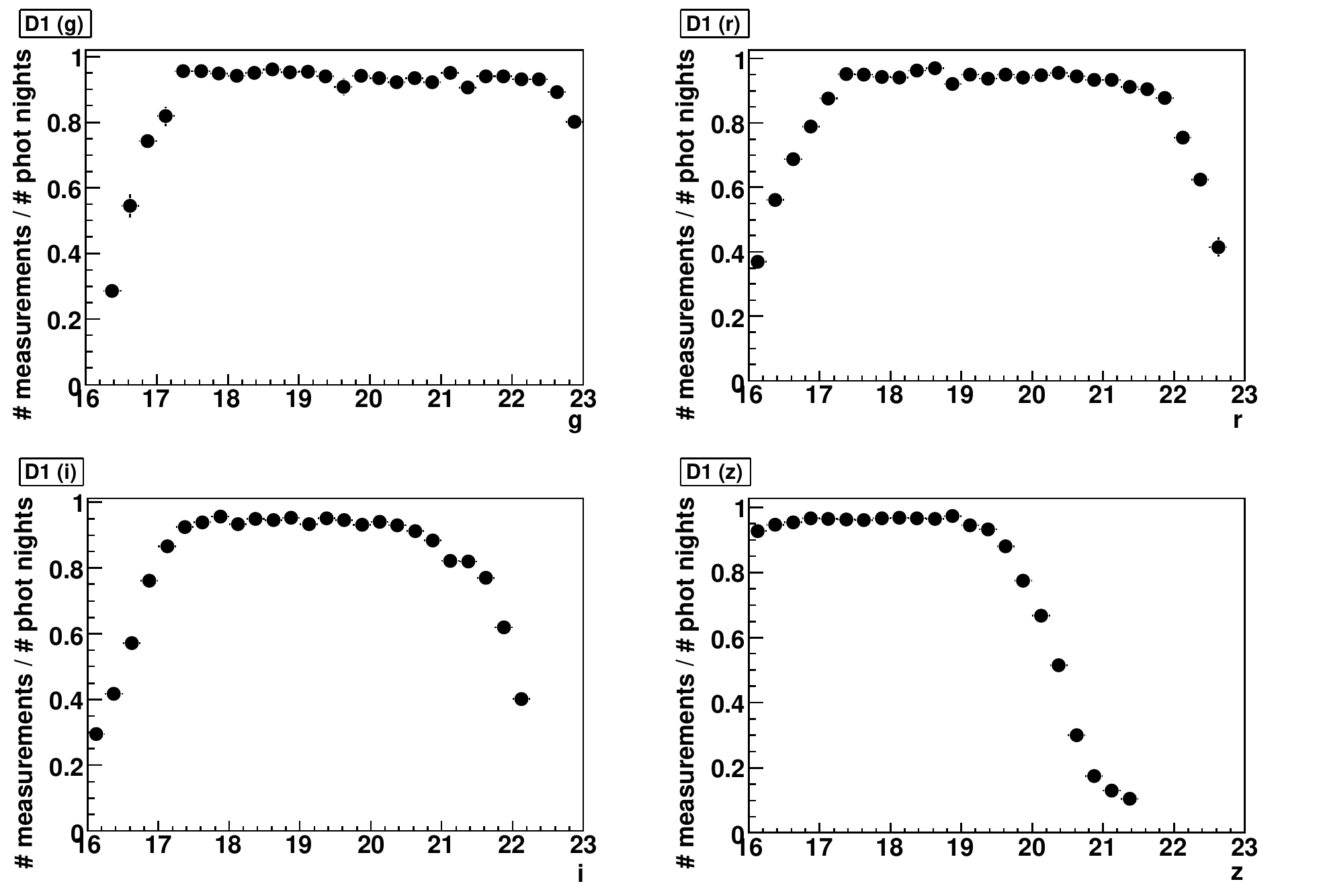}\\
\caption{Average number of measurements per tertiary standard (in percentage of the total number of good epochs) in the
 $\gme, \rme, \ime$- and $\zme$-bands, as a function of the tertiary standard
 magnitude. The efficiency drops we observe depend on the
 signal-to-noise cut applied during the object detection. These drops
 set the magnitude limits of the tertiary catalogs. The drop observed
 for small magnitude stars is mainly due to saturation. We keep all
 the bright stars, provided that they have more than 3 non-saturated
 measurements. \label{fig:nbmeasurements_per_tertiary_standard}}
\end{center}
\end{figure}

The selection is then refined by examining the contents of each star
aperture. If the aperture contains pixels flagged as bad in the dead
pixel maps, identified cosmics or pixels whose value is below 10 sigma
below the sky background, the star measurement is rejected.  At the
end of this selection process, we end up with approximately 100 to 200
tertiary standard candidates per CCD, depending on the field.

The master lists of tertiary star candidates are then matched with the
detection catalogs produced for each individual frame. The
completeness of these catalogs varies from one epoch to another,
depending on the observing conditions. Figure
\ref{fig:nbmeasurements_per_tertiary_standard} shows the number of
measurements as a function of the star magnitude for the D1 field.  We
see that our catalogs are complete below mag 22, 21.5, 21, and 19 for
bands $\gme$, $\rme$, $\ime$ and $\zme$ respectively. Above theses magnitudes, an
increasing fraction of the detections does not pass the
signal-to-noise cuts ($> 5$) and the calibrated magnitudes we may
compute from these measurements will be biased. We therefore choose
not to report calibrated magnitudes fainter than the thresholds listed
above.

The last cut applied to the tertiary star candidates is based on their
variability: at the end of the calibration process, we examine the
lightcurves of each candidate (actually, we compute the partial
$\chi^2$ of each individual calibrated measurement around the average
calibrated magnitude) and reject all the objects which display a
dispersion above a given threshold.

\subsection{Building the Tertiary Catalogs}

The flux of the field stars are measured using the photometry
algorithms described in \S \ref{sec:photometric_reduction}. We 
use the exact same algorithms that were used to measured the standard star fluxes. 
In practice, we use adaptive apertures of 7.5 seeing (14
pixels on average). The measured fluxes are divided by the exposure time reported in the image
headers with a precision of 3 milliseconds, and then transposed to an
airmass of 1 using the extinction coefficients reported
in table \ref{tab:zpfit_results}.  {The effect of varying airmass on the 
$\gme, \rme, \ime$  and $\zme$ effective passbands is very small (below 5\AA\ in 
the 1.0 - 1.6 airmass range). Therefore, no second-order airmass-color correction was applied.
No evidence for such an effect was found in the calibration residuals. } 
Finally, we compute the so-called
``hat instrumental magnitudes'' de-applying the Elixir uniformity maps, 
and applying the $\delta zp(\x)$ as described in \S \ref{sec:the_photometric_grids}.

The science exposures are taken in sequences of 5 to 10 exposures,
depending on the band. These measurements are merged in order to
produce one single list of averaged magnitudes per night. The merging
procedure selects one exposure of the sequence as a reference, and
allows for one free photometric alignment factor for 
each exposure. This permits to account for the
atmospheric extinction variations during the sequence, as well as
additional effects, such as fluctuations of the image quality,
inducing variations of the aperture corrections.
We found that on most nights, the fluctuations from one exposure to another do
not exceed 0.3\%.

The Local Natural Magnitudes described in the previous section are
then computed using the night and band zero-points, the photometric
grid maps $\delta k(\x)$, the grid reference colors and the MegaCam
colors of the fundamental standard (namely \bdtruc). At
this stage, we have a lightcurve for each selected tertiary star, with
observations taken at 20 epochs on average, spanning a time range of 3 years. All the
calibrated measurements of each selected tertiary star, in each band
are then averaged, in order to produce the tertiary catalogs. This
averaging process is iterative: it comprises several outlier rejection
steps, and attempts to identify the non-photometric nights, as
described below. In the following of this section, we detail the most
important points of the procedure.

\subsection{Photometric Error Model}

The uncertainties estimated by the photometric algorithm, reflect only
the Poissonian fluctuations of the background and star photon counts. 
They do not account for fluctuations of the
seeing and atmospheric transmission from one exposure to another. As a
consequence, the uncertainties affecting the bright stars are
underestimated. Since we do perform an
outlier rejection while averaging the night fluxes, it is essential to
build a realistic error model.

The uncertainties $\sigma_\phi$ affecting each measurement $\phi$ can
be classically parametrized as:
\begin{equation}
 \sigma_\phi = a\ \sqrt{\phi}\ \oplus\ b\ \phi\ \oplus\ c
\end{equation}
where $\oplus$ is the quadratic summation symbol. The first term is
the stochastic Poisson noise.  The second term describes all the
multiplicative fluctuations from one exposure to another, such as the
atmospheric extinction, the aperture correction variations and the flat-field noise. The last term
accounts for all the fluctuations which are independent of the flux,
primarily the background subtraction residuals. The $a,b$
and $c$ coefficients are fitted on the data for each night, in each
band. On photometric
nights, the typical values of the $b$ coefficient is $0.002$. This
gives an estimate of the photometric repeatability on short time
scales, in photometric conditions.

\subsection{Outlier Rejection}

The calibrated measurements being averaged are polluted by a small
fraction of outliers, usually due to CCD defects or
cosmics.  The outlier rejection algorithm is based on the comparison
of the partial $\chi^2$ of each individual measurement, object or
night, compared to the $\chi^2$ value we could expect from a
5-$\sigma$ measurement, object or night. The expected 5-$\sigma$ cut
is computed in a robust way, from the median value of the individual
measurements.

Additional procedures were implemented, in order to identify the
``non-photometric nights'', as well as the variable stars.

\subsection{Photometric Night Selection}

As noted in \S \ref{sec:snls_survey}, the calibration exposures
are often taken several hours before or after the science exposures. The
atmospheric transparency may therefore vary significantly between both
sets of exposures. However the SNLS dataset is exceptional in the
sense that each field was observed on a very large number of epochs,
with the calibration and science images taken in many different
configurations (different times in the night, variable time intervals
between the science and calibration exposures). This ensures that the
atmospheric transparency variations affect similarly the calibration
and science exposures. Therefore, these variations can be treated just
like an additional source of noise ---once we have identified the
pathological nights, displaying variations of more than 10\%, or
suffering from a large amount of absorption.

The night selection was performed as follows: first, the nights which
are obviously affected by a large absorption, i.e.  whose fitted
zero-points depart from the average zero-points measured on the
neighboring nights by more than 0.1 mag are identified and removed.
Then, the nights containing a science exposure sequence which is non
photometric, i.e. presenting an exposure-to-exposure variation greater than 1\%
are also rejected.

After having applied those two cuts, we are left with two sets of
nights. The first set contains the nights during which a large number
of calibration exposures was taken, over a large time range ($> 3$
hours). The second set contains the nights with calibration exposures
concentrated at a given time.  The long term photometric stability of
the first set of nights can be estimated directly by measuring the
dispersion of the zero-points determined on each calibration
exposure. The photometric stability of the second set cannot be
estimated likewise; instead, since the dispersion of the calibrated
magnitudes around their mean can be measured, we can identify and
reject the non-stable nights from science data only.

Tertiary catalogs were built using (1) the first set of nights,
rejecting those which display a zero-point stability worse than 1\%
and (2) the same nights plus the nights from the second set, the
pathological nights being identified and removed, using the
calibration residuals. Table \ref{tab:night_to_night_disp} summarizes
the number of nights flagged as photometric for both sets, and the
measured dispersion of the calibration residuals. This dispersion of
about 1\% is mainly due to the atmospheric variations.  As we can see,
the dispersion is about the same in both sets of nights. With the
larger set however, the statistical gain on the precision on the mean
calibrated magnitudes is significant. Moreover, the mean differences between 
the magnitudes computed with each set of nights are compatible with the 
expected dispersion, which shows that no non-photometric night has been 
accidentally left in the larger set. 
As a consequence, the tertiary
catalogs released with this paper are the ones built with the larger
set. In all fields, the uncertainty due to the atmospheric absorption
variations is of about 0.002 mag in $\gme, \rme, \ime$ and 0.003 mag
in $\zme$.

\begin{table*}
\begin{center}
\caption{Night-to-night dispersion, number of nights and resulting
  statistical uncertainty for night sets 1 \& 2.
  \label{tab:night_to_night_disp}}
\begin{tabular}{lc|ccc|ccc}
\hline
\hline
      &      & \multicolumn{3}{c|}{Set \#1}               & \multicolumn{3}{c}{Set \#2} \\
      &      &          &           &                    &           &           &                     \\
\hline
field & band & $\sigma$ & \# nights & $\sigma / \sqrt{N}$&  $\sigma$ & \# nights & $\sigma / \sqrt{N}$ \\
      &      &          &           &                    &           &           &                     \\
\hline
D1    & $g$  & 0.007    &   13        & 0.002            &  0.009 & 28 & 0.002 \\ 
      & $r$  & 0.009    &   14        & 0.002            &  0.007 & 32 & 0.001 \\ 
      & $i$  & 0.007    &   11        & 0.002            &  0.009 & 36 & 0.002 \\ 
      & $z$  & 0.014    &   11        & 0.004            &  0.014 & 19 & 0.003 \\ 
\hline
D2    & $g$  & 0.007    &    8        & 0.003            &  0.007 & 13 & 0.002 \\ 
      & $r$  & 0.006    &   12        & 0.002            &  0.007 & 21 & 0.002 \\ 
      & $i$  & 0.004    &    7        & 0.002            &  0.008 & 25 & 0.002 \\ 
      & $z$  & 0.011    &    3        & 0.006            &  0.010 & 10 & 0.003 \\ 
\hline
D3    & $g$  & 0.002    &   15        & 0.002            &  0.009 & 29 & 0.002 \\ 
      & $r$  & 0.007    &   18        & 0.002            &  0.009 & 34 & 0.002 \\ 
      & $i$  & 0.006    &   14        & 0.002            &  0.013 & 40 & 0.002 \\ 
      & $z$  & 0.009    &    3        & 0.005            &  0.011 & 10 & 0.003 \\ 
\hline
D4    & $g$  & 0.008    &   18        & 0.002            &  0.009 & 30 & 0.002 \\ 
      & $r$  & 0.006    &   16        & 0.002            &  0.007 & 34 & 0.001 \\ 
      & $i$  & 0.006    &   11        & 0.002            &  0.008 & 29 & 0.002 \\ 
      & $z$  & 0.007    &    7        & 0.003            &  0.013 & 19 & 0.003 \\ 
\hline
\end{tabular}
\end{center}
\end{table*}

\subsection{Results}
\label{sec:tertiary_catalogs_results}

The Local Natural Magnitudes of the tertiary standard are listed in
appendix in tables \ref{tab:d1_tertiaries}, \ref{tab:d2_tertiaries},
\ref{tab:d3_tertiaries} and \ref{tab:d4_tertiaries}. We also report
the corresponding $\delta k(\x)$ grid coefficients at the mean
focal plane position where the star was observed, so that the Uniform Magnitudes
can be computed from the Local Magnitudes.

\begin{figure*}
\centering
\mbox{\subfigure[$\rme-\ime\ vs. \gme-\rme$]{\includegraphics[width=0.45\linewidth]{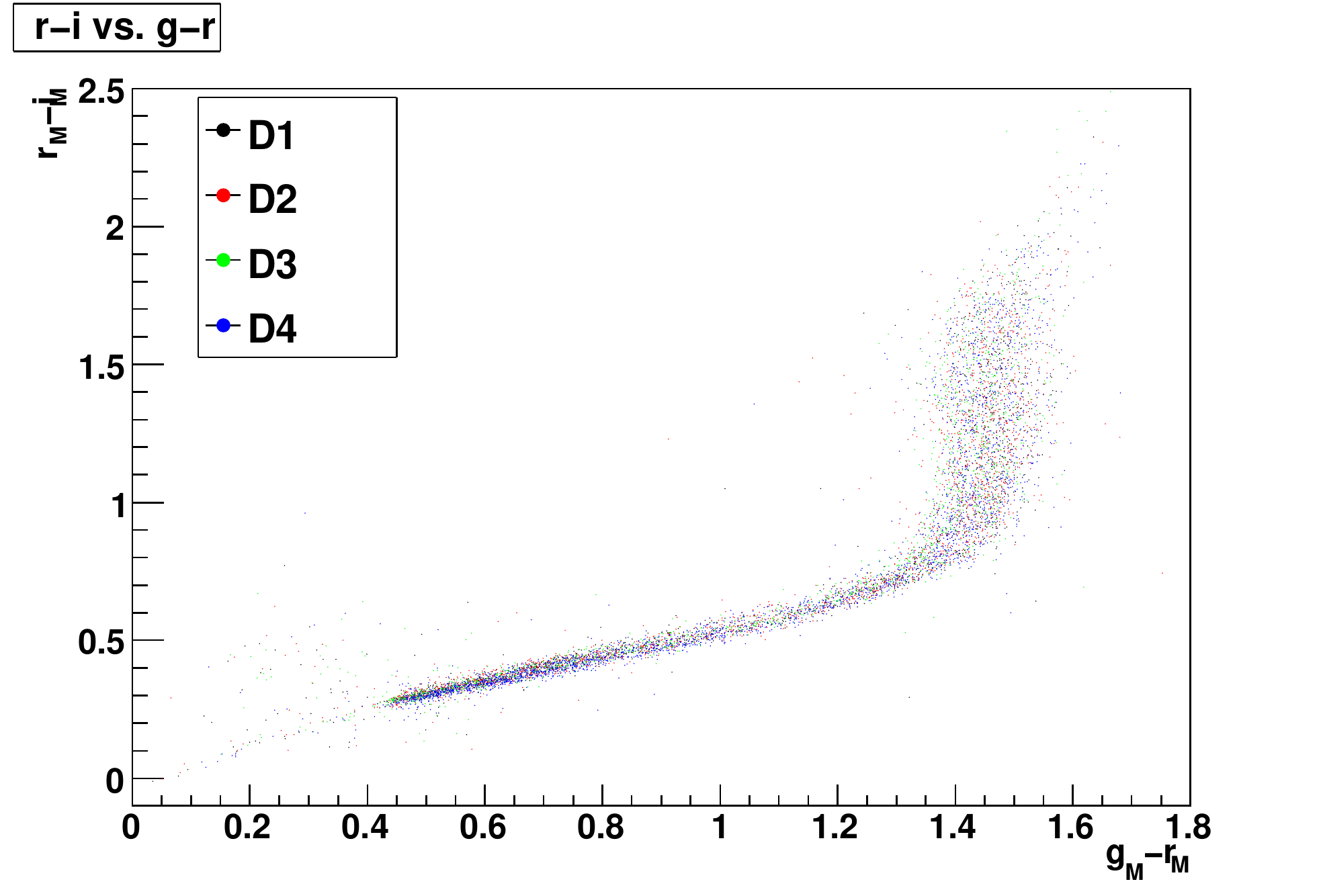}}
      \subfigure[$\rme-\ime\ vs. \gme-\rme$ (profile)]{\includegraphics[width=0.45\linewidth]{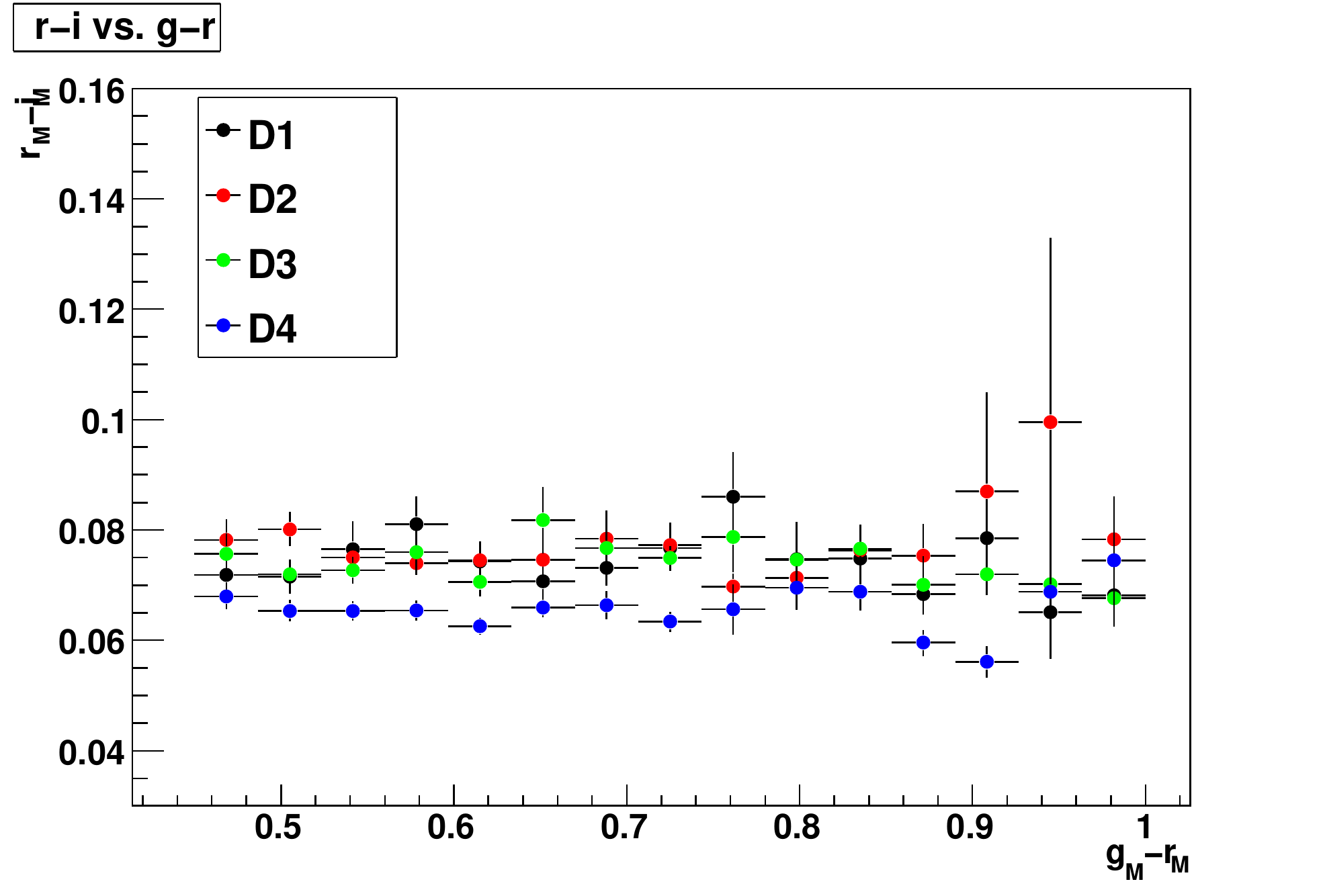}}}
\caption{Left: $\rme - \ime$ versus $\gme - \rme$ color-color plot of
  the tertiary standards {(Uniform Magnitudes)}. The agreement between all fields is good,
  and the contamination of non-stellar objects seems to be very low.
  Right: a zoom on the blue branch of the same color-color plot, after
  subtraction of the mean slope. We see that the colors of all four
  fields agree at the percent level.\label{fig:color_color_plots}}
\end{figure*}

Figure \ref{fig:color_color_plots} shows the $\rme-\ime vs. \gme-\rme$
color-color plot for all four SNLS fields. {Since we are comparing
objects over the whole focal plane, this figure was produced using
Uniform Magnitudes instead of Local Natural Magnitudes. }
As we can see, the color distributions of the four SNLS fields are compatible at the 1\%-level.
{Note that larger field-to-field differences in the stellar loci would
not necessarily suggest a drift of the calibration. Indeed, we have
verified using the Phoenix / GAIA library that larger effects, up to 4
to 6\% may be observed due, for example, to systematic differences in
the star metallicities. }

{ 
As can be seen on figure \ref{fig:color_color_plots}, the MegaCam
magnitude system defined in this paper is close to a Vega-based
system. Indeed, it is tied to Landolt through quasi-linear color
relations. Note however that our magnitudes probably depart from true
Vega magnitudes by a few percents, and the amplitude of this departure
is not known with precision. Other surveys, such as the Sloan Digital
Sky Survey (SDSS) use a different calibration path and report
magnitudes in an almost AB-system. Hence, the MegaCam magnitudes
differ from SDSS magnitudes by (1) a small color term due to small
differences in the effective passbands and (2) a constant term, which
accounts for the differences between a AB- and a Vega-like system (see
appendix \ref{sec:comparison_with_sdss} for more details).  }

\subsection{Statistical Uncertainties}
\label{sec:tertiary_catalogs_statistical_uncertainties}

\begin{figure*}
\begin{center}
\includegraphics[width=\linewidth]{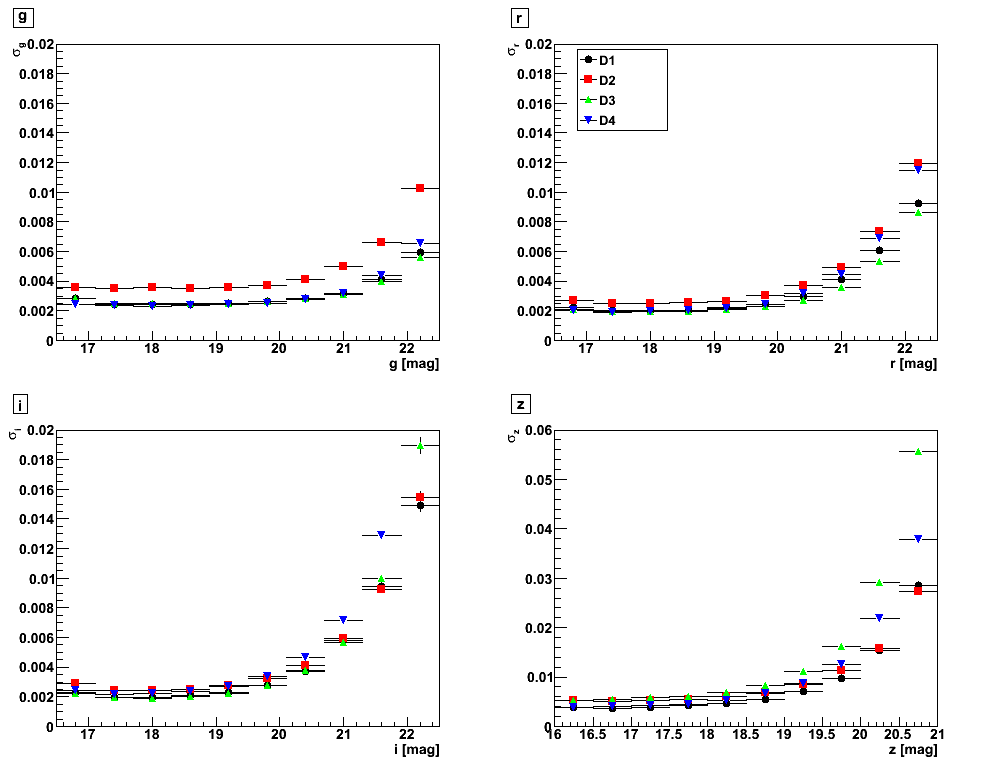}\\
\caption{Average random uncertainties as a function of the magnitudes
  for the SNLS DEEP fields, in the $\gme, \rme, \ime$ and
  $\zme$-bands.
  \label{fig:random_uncertainties}}
\end{center}
\end{figure*}

The statistical uncertainties affecting the tertiary standard
calibrated fluxes comprise four independent components: (1) the
tertiary standard flux measurement uncertainties ---photon noise,
flatfield noise and readout noise--- (2) the atmospheric transparency
variations between the science and calibration exposures (3) 
the intrinsic zero-points uncertainties and (4) the Landolt-to-MegaCam 
color transformation uncertainties. In this section, we shall
discuss the impact of each contribution.

As discussed in the previous sections, we are interested in the
uncertainties affecting the tertiary standard relative {\em
  fluxes}. Inspecting equation
\ref{eqn:magnitude_to_flux_transformation}, we see that they are
essentially equal to the statistical uncertainties on the {\em
  differences} between the tertiary standard calibrated magnitudes and
the MegaCam magnitudes of \bdtruc: $g_{|\x} - g_{ref}$, \ldots
$z_{|\x} - z_{ref}$.

Table \ref{tab:night_to_night_disp} summarizes the
night-to-night dispersion affecting the calibrated tertiary fluxes.
These dispersions are mainly attributed to the atmospheric
transparency variations between the calibration and science exposures.
We have looked for systematic variations of the atmospheric
transparency as a function of the time of the night (\S
\ref{sec:systematic_uncertainties}) and concluded that we can safely
assume these variations to be independent from night to night. Hence,
we derive a statistical uncertainty by dividing the measured
dispersion by the square root of the number of photometric nights.
The resulting uncertainties are of about 0.002 mag in $\gme, \rme, \ime$
and 0.003 in $\zme$.

The intrinsic zero-point uncertainties are of about 0.003 in the
$\gme, \rme,$ and $\ime$-bands and 0.005 in the $\zme$-band. Given the
large number of epochs, the resulting statistical error is lower than
0.001 mag in the $\gme, \rme, \ime$-bands and of about 0.001 mag in
the $\zme$-band.

The uncertainties reported along with the calibrated magnitudes of the
tertiaries (see tables \ref{tab:d1_tertiaries},
\ref{tab:d2_tertiaries}, \ref{tab:d3_tertiaries} and
\ref{tab:d4_tertiaries} in appendix \ref{sec:tertiary_catalog_tables})
combine the uncertainties related to the photon-noise, zero-point and
atmospheric variations. {They are shown on figure
\ref{fig:random_uncertainties}.  In the $\gme, \rme$ and $
\ime$-bands, for stars brighter than 20, they are dominated by the
night-to-night atmospheric dispersion reported in table \ref{tab:night_to_night_disp}, 
and then by photon noise. In the $\zme$-band, the photon noise dominates for stars 
fainter than about 18.5. This is expected given the low quantum efficiency 
of the MegaCam detectors in the red. }

The last cause of statistical uncertainties are the
Landolt-to-MegaCam color transformations which impact the $g_{|\x} -
g_{\bdtruc}$, \ldots $z_{|\x} - z_{\bdtruc}$ magnitude
differences. Since we fit global color transformations, these
uncertainties are correlated and do not average out, whatever the
number of epochs. We found them to be equal to $0.002$ mag in $\gme$,
$0.0015$ mag in $\rme$, $0.0012$ in $\ime$ and $0.005$ in $\zme$. This
is the dominant contribution to the statistical uncertainty budget.

\section{Systematic Uncertainties}
\label{sec:systematic_uncertainties}

The systematic uncertainties that affect the calibrated fluxes are
summarized in table \ref{tab:systematic_uncertainties}.

\subsection{Photometry Related Errors}

As discussed in \S \ref{sec:photometric_reduction}, we use variable
size apertures to measure the flux of the calibration and tertiary
stars.  We do not expect the PSF to be identical on the science
exposures, which are long guided exposures, and calibration exposures
of much shorter exposure time, not guided, and slightly defocused. We
found that on average, the seeing measured on the calibration
exposures is higher than the seeing measured on the science exposures
by about 0.13 arcseconds. We also have evidence that the shape of the
PSF is different. As described in \S \ref{sec:aperture_corrections},
we also have computed the fractional increase of flux between our
standard apertures and apertures twice as large. The differences
between these science image aperture corrections {\em minus} the
calibration image aperture corrections are summarized in table
\ref{tab:aperture_corrections}. On average, we collect 0.2\% more
photons on the science images than on the calibration images. The
tertiary star magnitudes are corrected for this effect. The residual
systematic uncertainties are well below 0.001 mag.

\begin{table}
\begin{center}
\caption{Aperture correction differences between the science and calibration exposures, 
  in the sense science {\em minus} calibration exposures. 
  \label{tab:aperture_corrections}}
\begin{tabular}{lc}
\hline
\hline
band   & $\delta {\rm aper}$  \\
\hline
$g_M$  & $+0.0027 \pm 0.0003$ \\
$r_M$  & $+0.0026 \pm 0.0004$ \\
$i_M$  & $+0.0020 \pm 0.0003$ \\
$z_M$  & $+0.0020 \pm 0.0005$ \\
\hline
\end{tabular}
\end{center}
\end{table}

\subsection{The Airmass Corrections}

As described in \S \ref{sec:landolt_stars}, the airmass dependent term
is modeled as a linear correction, neglecting the second order
corrections, of the form: $k' \times (X-1) \times {\rm color}$. The
second order coefficients can be modeled using synthetic
photometry. Their theoretical values are summarized in table
\ref{tab:second_order_airmass_corrections}. Neglecting the second order
correction result in a bias equal to: 
\begin{equation*}
  k' \times (X-1) \times \Bigl[ <{\rm color}_{science}> - <{\rm color}_{Landolt}> \Bigr]
\end{equation*}
The values of these biases are listed in table
\ref{tab:second_order_airmass_corrections}. As we can see, they are
smaller than 0.001 mag in all bands. We therefore ignore them in the
systematic error budget.

\begin{table}
\begin{center}
\caption{Synthetic second order airmass corrections.\label{tab:second_order_airmass_corrections}}
\begin{tabular}{cccccc}
\hline
\hline
band  & color  & $k'$    & $<{\rm color}>$    & $<{\rm color}>$   & bias $^\mathrm{a}$\\
      & index  & (synth) & (Landolt) & (tertiaries) &      \\
\hline 
$g_M$ &  $g-r$      & $-0.014$  &   $+0.735$  &  $+0.950$    &  $<0.001$    \\
$r_M$ &  $r-i$      & $-0.005$  &   $+0.408$  &  $+0.618$    &  $<0.001$    \\
$i_M$ &  $r-i$      & $<0.001$  &   $+0.408$  &  $+0.618$    &  $<0.001$    \\
$z_M$ &  $i-z$      & $<0.001$  &   $+0.192$  &  $+0.272$    &  $<0.001$    \\
\hline
\end{tabular}
\end{center}
\begin{list}{}{}
\item[$^\mathrm{a}$] Effects on the tertiary magnitudes of neglecting
  the second order correction. The resulting bias is smaller than
  0.001 mag in all bands.
\end{list}
\end{table}

\subsection{The Shutter Precision}

As described in \S \ref{sec:megaprime_instrument}, the actual exposure
time is measured using a dedicated system with a precision
of 0.001 second. The shutter
ballistics was investigated by the CFHT team using sequences of
exposures of the same field, taken with increasing exposure
times. \citet{Cuillandre05} showed that the specification of 3 ms was
actually met. The study could not measure a systematic shutter
uncertainty. In order to be conservative, we assign an error of 3 ms
to all the exposure times. This error is negligible for the science
exposures, but not for the calibration exposures, of mean exposure
time $T_{exp} \sim 2$ s. Being conservative, we consider that this induces a systematic error on the
tertiary magnitudes, of $0.0015$ mag (correlated) in all bands.

\subsection{Landolt Catalog Internal Dispersion}

The \citet{Landolt92} catalog is known to be remarkably uniform, with
a reported internal dispersion of 0.003 mag. We have checked this
assertion, and estimated the resulting error on the tertiary
magnitudes.  Indeed, we observe regularly 8 Landolt fields, containing
about $250$ Landolt stars and representing a little less than half of
the whole Landolt catalog. Since each SNLS field is itself calibrated
with a subset of no more than the three or four Landolt fields that
are observable along with it, we can expect to measure a sizeable
error due to the Landolt catalog internal dispersion.

The field-to-field dispersion of the zero-points was measured on the
stable nights during which more than one Landolt field was observed.
We have measured it to be of 0.002 mag in $gri$ and 0.004 in
$z$. Since each SNLS field is calibrated with about 4
different Landolt fields on average, we estimate the uncertainty to be
of about 0.001 mag in the $gri$-bands and 0.002 mag in the $z$-band.

\subsection{Grid Reference Colors}

As discussed in \S \ref{sec:the_photometric_grids}, there is a hidden
color reference associated to each $\delta zp(\x)$ map. The grid
sequences were calibrated using the average run zero-points provided
by the Elixir pipeline. These zero-points are valid for an entire run
(about two weeks), and do not account for the night-to-night
variations. We have evaluated the night-to-night dispersion of the
zero-points to be of about $\sigma_{n-to-n} \sim 0.01 - 0.02\ {\rm
  mag}$. The grid reference colors are known to this precision, and
each $\delta zp(\x)$ map pixel is therefore affected by an error,
equal to $\delta k(\x) \times \sigma_{n-to-n}$. Table
\ref{tab:grid_reference_colors} summarizes the largest associated 
uncertainty.

\begin{table}
\begin{center}
\caption{Uncertainties on the grid reference colors and the associated
  max error affecting the grid
  pixels.\label{tab:grid_reference_colors}}
\begin{tabular}{lcccc}
\hline
\hline
band  & color & $\max_\x \left[\delta k(\x)\right]$  & $\sigma_{\rm color}$ & max grid error \\
      &       &                                      & &                \\
\hline
$g_M$  & $g-r$ & 0.04                                & 0.021 &  $< 0.001$              \\
$r_M$  & $r-i$ & 0.04                                & 0.013 &  $< 0.001$              \\
$i_M$  & $r-i$ & 0.05                                & 0.013 &  $< 0.001$                \\
$z_M$  & $i-z$ & 0.06                                & 0.019 &  $< 0.001$                \\
\hline
\end{tabular}
\end{center}
\end{table}

\subsection{Adequacy of the Grid Color Corrections}

The agreement of the MegaCam passband model and the grid corrections
has been discussed in \ref{sec:cmp_color_terms}. We have found that
there is an excellent agreement between the synthetic color terms
determined using the \citet{Pickles98} library and the color terms
measured from the grid corrections, except in the $\rme$-band, where
there might be a slight disagreement. We found, using the real spatial
distribution of the Landolt star measurements, that the differences
between the model predictions and grid measurements have an impact of
less than 0.001 mag on the zero-points in the $\gme$, $\ime$ and
$\zme$-bands, and of 0.002 mag in the $\rme$-band.

\subsection{The Magnitudes of \bdtruc}

\bdtruc\ is too bright to be observed directly with MegaCam. Its
MegaCam magnitudes were determined indirectly in \S
\ref{sec:megacam_magnitudes}. The uncertainties on these estimates are
dominated by astrophysical considerations, especially the impact of
the fact that \bdtruc\ is possibly a binary system.

\subsection{The SED of \bdtruc}

The last source of systematic uncertainties that must be discussed is
related to the measurement of the SED of \bdtruc. We use version 2 of the 
determination published by \citet{Bohlin04b}, available on the CALSPEC web site.
The uncertainties affecting this measurement come from two
sources. First, the repeatability of the STIS observations, which is
routinely monitored with repeated observations of the star
AGK~+81~266. According to figure 1 of \citet{Bohlin04b}
this repeatability is of about 0.3\% in the $\gme$-, $\rme$-, $\ime$-
and and 0.6\% in the $\zme$-bands respectively (1 $\sigma$ errors).
There are hints that there are correlations between these
uncertainties. We have computed our own repeatability model using the
original monitoring spectra of AGK~+81~266, integrated in the MegaCam
passbands.  We found a repeatability of 0.2\%, 0.3\%, 0.3\% and 0.6\%
in the $\gme$-, $\rme$-, $\ime$- and $\zme$-bands respectively, with
strong correlations between the neighboring bands.

The other main source of uncertainty come from the model flux
distributions of the three primary standards themselves. This includes
internal uncertainties of the NLTE white dwarf atmosphere modeling, as
well as uncertainties in the determination of the star metallicities,
surface gravity and effective temperature, from the observation of the
Balmer line profiles. From the results presented in \citet{Bohlin02},
we adopt a 0.5\% uncertainty (1 $\sigma$) over the range 3000\AA -
10000\AA. This translates into uncertainties on the $\gme$, $\rme$,
$\ime$ and $\zme$ synthetic broadband fluxes of \bdtruc, {\em relative
  to the synthetic $V$-band flux}, which are essentially negligible,
except for the $\ime$- and $\zme$-bands (0.15\% and 0.24\%
respectively).

The sum of the two contributions is dominated by the repeatability of
the STIS instrument.  Combining both sources of uncertainties, we
obtain the covariance matrix listed in appendix
\ref{sec:uncertainties_and_covariance_matrices}, table
\ref{tab:bd17_sed_covmat}.

\begin{table*}
\begin{center}
\caption{Summary of the systematic uncertainties affecting the
  calibrated magnitudes and fluxes. \label{tab:systematic_uncertainties}}
\begin{tabular}{lcccc}
\hline
\hline 
                         & $g_M$  & $r_M$  & $i_M$ & $z_M$ \\
                         &        &        &       &      \\
\hline 
Aperture corrections                   & $<0.001$  & $<0.001$ &  $<0.001$    &   $<0.001$       \\
Background subtraction                 & $<0.001$  & $<0.001$ &  $\pm 0.005$ &   $<0.001$       \\
Shutter precision                      & $\pm 0.0015$ & $\pm 0.0015$ & $\pm 0.0015$  & $\pm 0.0015$ \\
Linearity                              & $<0.001$  & $<0.001$ &  $<0.001$    &   $<0.001$       \\
Second order airmass corrections       & $<0.001$  & $<0.001$  & $<0.001$ &  $<0.001$           \\
Grid Reference Colors                  & $<0.001$  & $<0.001$  & $<0.001$ &  $<0.001$           \\
Grid Color Corrections                 & $<0.001$  & $<0.001$  & $\pm 0.002$ & $<0.001$         \\
Landolt catalog                        & $\pm 0.001$ & $\pm 0.001$ & $\pm 0.001$ & $\pm 0.002$  \\
Magnitudes of \bdtruc                  & $\pm 0.002$ & $\pm 0.004$ & $\pm 0.003$ & $\pm 0.018$  \\
\hline
                                       &           &        &       &      \\
Total                                  & $\pm 0.003$ & $\pm 0.004$ &  $\pm 0.006$  &  $\pm 0.018$    \\
\hline
                                       &             &             &               &      \\
SED of \bdtruc                            & $\pm 0.001$ & $\pm 0.002$ &  $\pm 0.004$  &  $\pm 0.007$    \\
\hline 
Total                                  & $\pm 0.003$ & $\pm 0.005$ &  $\pm 0.007$  &  $\pm 0.019$    \\
\hline 
\end{tabular}
\end{center}
\end{table*}

\subsection{Summary: the Full Uncertainty Budget}

We obtain the full uncertainty budget by combining the statistical and
systematic uncertainties. The full covariance matrix is listed in
table \ref{tab:final_covariance_matrix} (appendix
\ref{sec:uncertainties_and_covariance_matrices}).  As discussed above,
we have chosen to report the uncertainties that affect the quantities
$\gme - g_{ref}$, $\rme - r_{ref}$, $\ime -i_{ref} $ and $\zme -
z_{ref}$, i.e. the differences between the tertiary magnitudes
\bdtruc\ magnitudes. Note that these quantities are correlated with
the Landolt magnitudes of \bdtruc: $U_{ref}$, $B_{ref}$, $V_{ref}$,
$R_{ref}$ and $I_{ref}$. Since the SNLS cosmology analysis does
compare nearby supernova $UBVRI$ magnitudes with the MegaCam magnitudes
of more distant supernovae, we cannot ignore these correlations.
Hence, we report a $9 \times 9$ covariance matrix, containing all
these correlations.

The contribution of the uncertainties affecting the measurement of the
SED of \bdtruc\ are listed in table
\ref{tab:bd17_sed_covmat}. Similarly, we report a $9 \times 9$
covariance matrix, including the 5 Landolt bands in addition to the
MegaCam bands.

\section{Discussion}
\label{sec:discussion}

In this paper, we have characterized the MegaCam focal plane
photometric response, built a model of the MegaCam passbands, and
defined a system of {\em Local Natural Magnitudes}, depending on the
focal plane position. This system is implemented as four catalogs of
tertiary standard stars, one for each of the four CFHTLS DEEP
field. The relations between these magnitudes and their physical flux
counterparts have been explicited. They rely on a specific star,
\bdtruc, with known Landolt $UBVRI$ and MegaCam $\gme, \rme, \ime$,
$\zme$-magnitudes, and a known Spectral Energy Distribution measured
and calibrated independently by \citet{Bohlin04b}.

{
The statistical and systematic uncertainties affecting the tertiary
magnitudes and their flux counterparts have been discussed in \S
\ref{sec:megacam_magnitudes_bdtruc}, \S
\ref{sec:tertiary_catalogs_statistical_uncertainties} and \S
\ref{sec:systematic_uncertainties}. The uncertainties affecting the
MegaCam magnitudes of each tertiary standard may be classically split
into two different contributions. First, a statistical uncertainty
which accounts for the flux measurement shot noise, the number of
epochs averaged and the average photometric conditions under which
this star was observed. This contribution is summarized in figure
\ref{fig:random_uncertainties}. In the $\gme, \rme$ and $\ime$-bands,
it amounts to about 0.0025 mag for a mag 18 star and stays below 0.005
mag up to mag 21. In the $\zme$-band, it is higher, of about 0.005 mag
for a mag 18 star, and reaches 0.01 mag for a mag 19.5 star. The
second contribution is a systematic uncertainty which is summarized in
the first section of table \ref{tab:systematic_uncertainties} and
characterizes how well the MegaCam system is tied to the Landolt
system: 0.002 in $\gme$ and $\rme$, 0.006 in $\ime$ and 0.003 in
$\zme$. Finally, the magnitude-to-flux conversion introduces
additional uncertainties, of 0.005 mag or better in the $\gme, \rme, \ime$-bands and
0.019 mag in $\zme$. The final systematic uncertainty budget is
detailed in table \ref{tab:systematic_uncertainties} and in the
covariance matrices \ref{tab:final_covariance_matrix} and
\ref{tab:bd17_sed_covmat} listed in appendix
\ref{sec:uncertainties_and_covariance_matrices}.

In order to fully quantify how well this system is tied to Landolt, we
should also add to this budget the impact of the Landolt-to-MegaCam
color term uncertainties. If one considers the full relevant color
range from Vega ($B-V \sim$ 0) to the mean colors of the Landolt stars
($B-V \sim 0.77$), one finds that this contribution is large, between 
0.005 and 0.015 mag, depending on the band. \bdtruc\ was chosen as a fundamental standard
partly because its colors are closer to the mean color of the Landolt
stars. This allows one to reduce the final impact of the
Landolt-to-MegaCam color term uncertainties down to less than $0.002$
mag in $\gme, \rme, \ime$ and $0.005$ mag in $\zme$.

The statistical and systematic error budgets are extremely similar
from one field to another, mainly because each field was treated on an
equal footing, with the same observation strategy. The only systematic
difference from one field to another is the mean effective airmass
$\sim 1.2$ for D1 and D2 and $\sim 1.3$ for D3 and D4. Propagating
the uncertainties affecting the airmass coefficients determined in \S
\ref{sec:landolt_stars}, one finds that the impact of these
differences on the final tertiary magnitudes is very small, below 0.001 mag.

The MegaCam magnitudes defined in this paper are tied to the Landolt
system using (quasi-)linear color relations. Hence Vega (like any A0V star)
should have MegaCam colors close to zero. Note however that
the Landolt and MegaCam magnitudes of Vega are affected with large
uncertainties of a few percents at least. We have not been able
to quantify with precision the departure of this system from a
strictly Vega-based system. This is the reason why we have discarded
Vega as a fundamental standard, and used \bdtruc\ instead.

We would like to stress again, that the photometric system
presented in this paper has been designed in such a way that no
color correction (or grid color correction) ever has to be applied
to the science objects' magnitudes. Indeed, our main goal is to keep
the connection between the magnitudes reported by the survey and the
underlying physical fluxes. This is especially important when
dealing with non-stellar objects, such as supernovae or galaxies.

The main application of this work is the calibration of the
supernova lightcurves obtained during the 3 first years of the SNLS
survey. Nevertheless, we release all the necessary information so
that any MegaCam dataset can be anchored to this system, provided
that one of the SNLS DEEP fields is regularly observed during the
program along with the science data.

The calibration of the SNLS 3-year supernova lightcurves from the
tertiary catalogs presented in this paper is discussed in detail in
\citet[][in prep]{guy09} and anyone interested in tying a dataset to our
system should study this paper as a first step. The recommended
procedure may be briefly outlined as follows: 
\begin{enumerate}
\item first, the instrumental magnitudes of the SNLS tertiary stars used
as calibrators must be measured using the exact same photometry
algorithm as the one used to estimate the flux of the science objects,

\item then, the zero-points are derived by comparing the instrumental
magnitudes of the tertiaries with the calibrated magnitudes released
with this paper (see appendix \ref{sec:tertiary_catalog_tables}),

\item the Local Natural Magnitudes can then be obtained
by applying those zero-points to the science objects' instrumental
magnitudes,

\item finally physical broadband fluxes may be derived from
the Local Natural Magnitudes using (a) equation
\ref{eqn:magnitude_to_flux_transformation} 
(b) the MegaCam magnitudes of \bdtruc\ reported in table \ref{tab:magnitudes_of_bd17}
(c) the SED of \bdtruc\ discussed in \S \ref{sec:selecting_a_fundamental_standard}  and 
(d) the MegaCam passband model detailed in appendix \ref{sec:megacam_passband_tables}. The filter model at the exact focal
plane location of the science target may be obtained by interpolating
between the filter scans detailed in tables \ref{tab:u_band_table} to
\ref{tab:z_band_table}.
\end{enumerate}
}

Very recently, \citet{Landolt09} published an extension of the catalog
used in this paper. Bluer as well as fainter stars were added to the
\citet{Landolt92} catalog. The author tied the measurements to his
former magnitude system. Note that for a small subset of the
\citet{Landolt92} catalog (about 40-50 stars), revised magnitudes were
published. In general the magnitude changes are small, of about 0.003
mag, except for a handful of stars (less than 10), for which they can
amount to 0.01 to 0.1 mag. We have checked that, for the stars used in
this analysis, the average magnitude difference between both catalogs
is smaller than 0.001 mag.

The SDSS-II first year dataset \citep{Holtzman08} relies on a similar
calibration scheme. The calibration of the SDSS SN~Ia lightcurves
relies on the catalog of tertiary standard stars published by
\citet{Ivezic07}. This latter catalog contains about 1.01 million
non-variable stars located on the SDSS equatorial stripe 82
($|\delta_{J2000}| < 1.266, 20h34 < \alpha < 4h$) and is believed to
be uniform at the 1\% level. Although the reduction of drift scan
imaging data poses different problems, it is interesting to note that
\citet{Ivezic07} had to solve very similar problems, such as the
uniformity of the photometric response along a drift scan, or the
small differences between the filters which equip each column of the
camera. 

The \citet{Ivezic07} catalog is calibrated to the natural SDSS 2.5-m
photometric system, which deviates from a perfect AB system by about
4\% in the $u$-band, and 1 to 2\% in the $griz$-bands. Therefore,
\citet{Holtzman08} chose to tie their magnitudes to the HST
white-dwarf scale. Several potential CALSPEC primary standards were
observed with the smaller SDSS Photometric Telescope (PT) and their
magnitudes were transferred to the SDSS 2.5-m photometric system using
linear color transformations, supplemented by synthetic photometry in
order to determine star-specific offsets with respect to the color
transformations. As in our case, the color transformations are not
well-defined in the white-dwarf color range. Hence, they chose to use
the three CALSPEC red, faint, solar analogs P330E, P177D and P041C as
primary standards. The calibration path adopted by \citet{Holtzman08}
is extremely similar to that presented in this paper. Its main
advantage is that the SDSS Photometric Telescope response is well
characterized, and that the color transformations between the PT and
the science telescope are smaller.

There are several shortcomings affecting the calibration presented
is this paper.  First, \bdtruc\ has not been observed directly with MegaCam.
We therefore rely on (1) the Landolt measurements of \bdtruc, (2) the
Landolt-to-MegaCam color transformations, (3) on the Phoenix / GAIA
synthetic libraries and combined with estimates of \bdtruc's
metallicity, surface gravity and extinction to estimate its MegaCam
magnitudes. Then, the time interval between the calibration and
science measurements is of several hours. We are therefore sensitive
to the variations of the atmospheric conditions on such long periods.
The large number of epochs allows one to reduce this source of
uncertainty to 2 mmags on average. However, a more robust calibration
would be obtained if calibration and science observations could be
separated by intervals of a few minutes only, and if repeated
exposures of the same field through the observation sequence could
allow one to estimate the photometricity of the observing condition,
during the science and calibration observations. Finally, the current
grid dataset, with observations every 6 months, does not allow one to
monitor precisely the run-to-run variations of the imager photometric
response.

In order to overcome these shortcomings, we have designed a dedicated
calibration program, called MAPC (MegaCam Absolute Photometric
Calibration). MAPC combines observations of the DEEP fields, HST
spectrophotometric standards which are a little fainter than \bdtruc, and
dithered observations of the SDSS equatorial stripe 82, all taken in a
little less than 30 minutes. This program will allow us to redefine
the MegaCam Local Natural Magnitudes independently of the Landolt
magnitudes. We also will be able to explicit the connections of the
MegaCam magnitudes with the widely used SDSS magnitude system.

A second article dealing with the precise photometric calibration of
MegaCam data will follow the current effort. It should be noted that
the photometric precision delivered since first light by the Elixir
pipeline (4\%) falls within the typical scientific requirements and 
has not been a limitation for most users. We expect, however, that
the new level of precision provided by our current work will enable 
science currently unforeseen with the MegaCam data set.

\begin{acknowledgements}
The authors wish to thank R. Bohlin for helpful advice. We are
grateful to A. Landolt for giving us details about his instrumentation
and analysis pipeline. We thank C. Buton and the SNF Collaboration for
letting us use their atmospheric transmission curve prior to
publication. We are grateful to S. Bailey for his careful rereading 
of the manuscript and to V. Ruhlmann-Kleider for useful
comments and remarks.  Canadian collaboration members acknowledge
support from NSERC and CIAR; French collaboration members from
CNRS/IN2P3, CNRS/INSU, PNCG. Most of the data reduction was carried
out at the Centre de Calcul de l'IN2P3 (CCIN2P3, Lyon, France).
\end{acknowledgements}

\bibliographystyle{aa}


\begin{appendix}
\section{Measuring the Photometric Response Maps (details)}
\label{sec:photometric_response_maps_details}

We have shown in \S \ref{sec:the_photometric_grids} that fitting the
uniformity maps $\delta zp(\x)$ and $\delta k(\x)$
from the dithered grid field observations is a very large
dimensionality problem, involving about 200,000 parameters. Of these
parameters, most are the grid stars instrumental magnitudes
(i.e. nuisance parameters). The grid maps are developed on independent
superpixels $\delta zp(\x) = \sum_{k=1}^N \alpha_k p_k(\x)$, $\delta
k(\x) = \sum_{k=1}^{N'} \beta_k q_k(\x)$. The $\delta zp(\x)$ are fit
on $512 \times 512$ superpixels, with 1296 superpixels on the focal
plane.  The precision obtained on the $\delta k(\x)$ parameters
depends on the color lever arm. In order to maximize it, the $\delta
k(\x)$ maps are developed on larger ($1024 \times 1537$) superpixels,
with only 216 superpixels on the focal plane. Since both maps have a
reference cell, there are $N = 1295$ and $N' = 215$ $\alpha_k$ and
$\beta_k$ coefficients respectively, representing 1510 parameters in
total.
We have checked that first order methods, which do not require one to
build the second derivatives of the $\chi^2$ converge very slowly and
do not allow one to obtain the exact solution in a reasonable amount
of iterations.  In this section, we show that the structure of the
problem is such that the dimensionality of the normal equations can be
in fact greatly reduced. This allows us to obtain the exact solution
of the problem in a single step.

First, the model fitted on the instrumental
magnitudes of the grid stars is as follows. This model connects the expectation of the $j$th measured instrumental
magnitude of star $i$, $m_{ADU|\x}(i,j)$ with the fit parameters, i.e.
the grid map parameters $\alpha_k$ and $\beta_k$ and the magnitudes of
the grid stars at the reference location, $m_{|\x_0}(i)$. The star
colors are fixed, and recomputed iteratively until they do not vary 
by more than 0.0001 mag. The model can be written:
\begin{eqnarray*}
  E \left [ m_{ADU|\x}(i,j) \right ] &=& m_{|\x_0}(i) + \sum_k \alpha_k p_k(\x_{ij}) + \nonumber\\
                  && \sum_k \beta_k q_k(\x_{ij}) \times \Bigl({\rm col}_{|\x_0}(i) - {\rm col}_{grid}\Bigr)
\end{eqnarray*}
Grouping together the grid map parameters $\alpha_k$ and $\beta_k$ in
a single vector ${\mathbf p}$, and the star magnitudes in another
(much larger) vector ${\mathbf M}$, the equation above can be
rewritten in the more compact form:
\begin{equation}
  E \left [ m_{ADU|\x}(i,j) \right ] = {\mathbf D}_{ij}^T {\mathbf p} + {\mathbf S}_i^T {\mathbf M}
\end{equation}
${\mathbf D}_{ij}$ and
${\mathbf S}_i$ are one-dimensional vectors, defined for each
measurement $j$, and each star $i$. The large vector ${\mathbf S}_{i}$ contains a single
non-zero element:
\begin{eqnarray}
  {\mathbf D}_{ij}[k]   &=& p_k(x_{ij}) \\
  {\mathbf D}_{ij}[k+N] &=& q_k(x_{ij}) \times {\rm color}_{i} \\
  {\mathbf S}_{i}[j]    &=& \delta_{ij}
\end{eqnarray}

The $\chi^2$ to minimize can be written
as:
\begin{equation}
 \chi^2({\mathbf p}, {\mathbf M}) = \sum_{ij} w_{ij} \left({\mathbf D}_{ij}^T {\mathbf p} + {\mathbf S}_i^T {\mathbf M} - {m}_{ADU|\x}(i,j)\right)^2
 \label{eqn:model_chi2}
\end{equation}
where $w_{ij}$ is the weight of the measurement $ {m}_{ADU|\x}(i,j)$.
 The associated normal equations are:
\begin{equation}
 \begin{pmatrix}
  {\mathbf W}_p & {\mathbf A}^T \\
  {\mathbf A}   & {\mathbf W}_M \\
 \end{pmatrix}
 \times
 \begin{pmatrix}
  {\mathbf p} \\
  {\mathbf M} \\
 \end{pmatrix}
 = 
 \begin{pmatrix}
  {\mathbf B}_p \\
  {\mathbf B}_M \\
 \end{pmatrix}
\label{eqn:normal_equation}
\end{equation}
where 
\begin{eqnarray}
{\mathbf W}_p  & = & \sum_{ij}\ w_{ij}\ {\mathbf D}_{ij}\ {\mathbf D}_{ij}^T \\
{\mathbf W}_M  & = & \sum_{ij}\ w_{ij}\ {\mathbf S}_i\ {\mathbf S}_i^T \\
{\mathbf A}    & = & \sum_{ij}\ w_{ij}\ {\mathbf S}_i\ {\mathbf D}_{ij}^T 
\end{eqnarray}
and 
\begin{eqnarray}
{\mathbf B}_p  & = & \sum_{ij}\ w_{ij}\ m_{ADU|\x}(i,j)\ {\mathbf D}_{ij} \\
{\mathbf B}_M  & = & \sum_{ij}\ w_{ij}\ m_{ADU|\x}(i,j)\ {\mathbf S}_i
\end{eqnarray}
Since all coordinates but one of each
 vector ${\mathbf S}_i$ are zero, all the matrices ${\mathbf S}_i
 {\mathbf S}_i^T$ are diagonal. Hence ${\mathbf W}_M$ is diagonal. We
 can eliminate ${\mathbf M}$ from the above normal equations:
\begin{equation}
 \left({\mathbf W}_p - {\mathbf A}^T {\mathbf W}_M^{-1} {\mathbf A}\right) {\mathbf p} 
 = 
 {\mathbf B}_p - {\mathbf A}^T {\mathbf W}_M^{-1} {\mathbf B}_M
 \label{eqn:model_solution_p}
\end{equation}
and get a computationally tractable equation because ${\mathbf W}_M^{-1}$ is diagonal.
This linear equation has the dimensionality of ${\mathbf p}$ and can be solved
for  ${\mathbf p}$.
 ${\mathbf M}$ can then be determined from:
\begin{equation}
{\mathbf M} = {\mathbf W}_M^{-1} \Bigl( {\mathbf B}_M-{\mathbf A} {\mathbf p} \Bigr)
 \label{eqn:model_solution_M}
\end{equation}
using again that ${\mathbf W}_M$ is diagonal. 

Equations \ref{eqn:model_solution_p} and \ref{eqn:model_solution_M}
yield the exact solution of the linear least squares problem
(\ref{eqn:model_chi2}). Besides reaching the least squares minimum,
the method described here provides us with the true
covariance matrix of the calibration parameters. Indeed, one can
show from equation (\ref{eqn:normal_equation}) that the full
covariance matrix, marginalized over the grid star magnitudes is given
by:
\begin{equation}
{\mathbf C}
 = \left({\mathbf W}_p - {\mathbf A}^T {\mathbf W}_M^{-1} {\mathbf
 A}\right)^{-1}
\end{equation}

What allowed us to reduce the dimensionality of the problem and obtain
the exact solution is the very specific structure of the $\chi^2$
second derivative matrix, which contained a very large diagonal
sub-block. Such a structure seems to be very common in calibration
problems, such as the ones described in \citet{Kaiser99,
  Padmanabhan07}. This structure can be observed when we can distinguish a set of
global parameters (here, the grid map parameters), combined with
another much larger set of local parameters (here, the magnitudes). 
The key feature
of the least-squares problem that makes it eligible for this
technique is that no $\chi^2$ term contains two elements
of  ${\mathbf M}$. The same procedure still applies if the $m_i$ are vectors,
with the difference that ${\mathbf W}_M$ is block-diagonal. 
In those cases, the
factorization described here can be used, making the problem tractable or 
at least saving large
amounts of computing time.

\end{appendix}
\begin{appendix}
\section{MegaCam Passbands}
\label{sec:megacam_passband_tables}

\begin{figure}
\centering
\includegraphics[width=\linewidth]{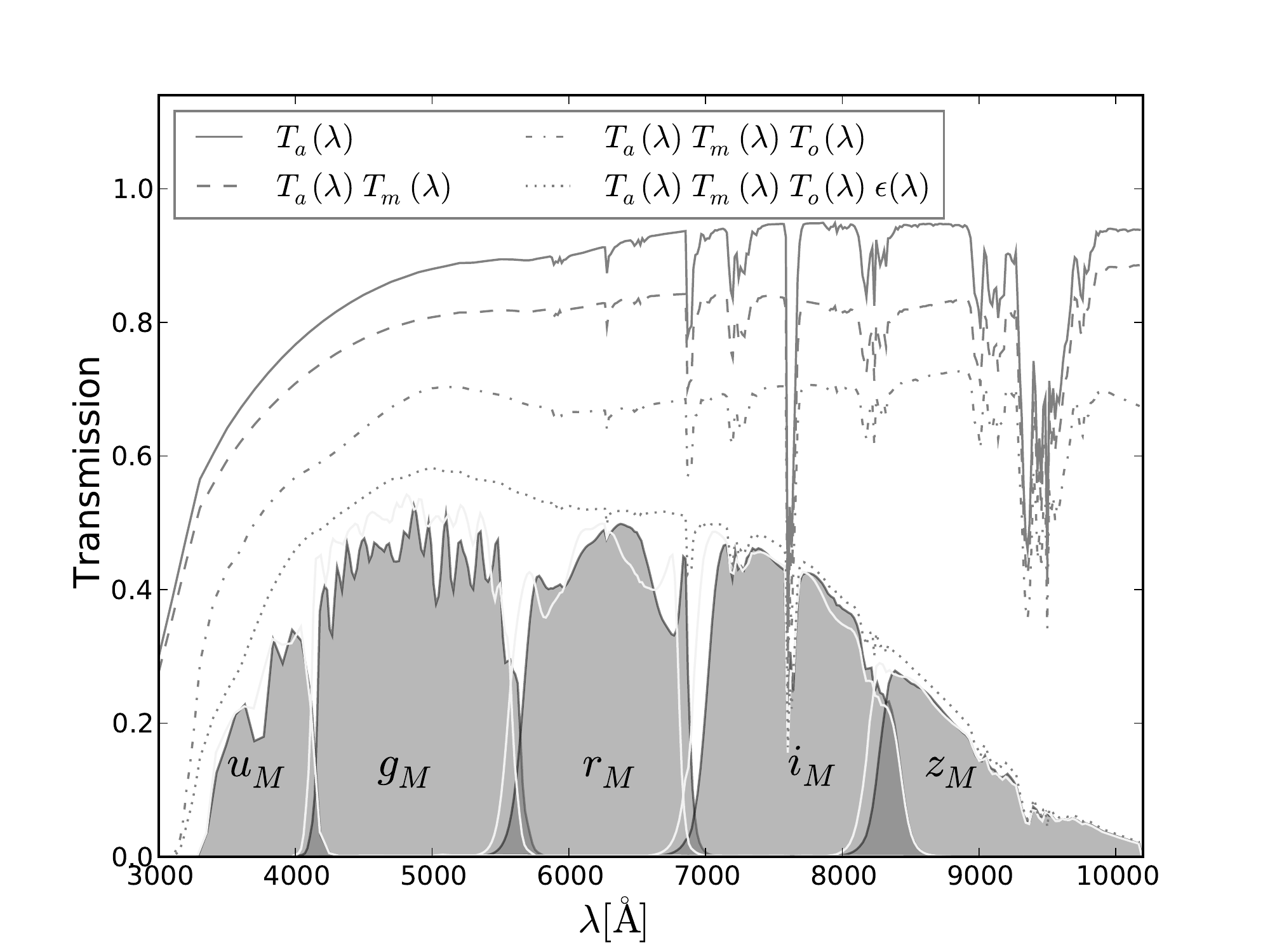}\\
\caption{MegaCam effective passbands at the center of the focal plane
  (filled) and close to the sides of the focal plane (solid gray lines). 
  As discussed in \S \ref{sec:the_photometric_grids}, the
  filters are bluer on the edges than at the center of the camera. We
  also display the cumulative effect of the main ingredients of the effective passbands:
  average quantum efficiency of the MegaCam CCDs $\epsilon(\lambda)$, mirror reflectivity $T_m(\lambda)$,
  transmission of the wide field adapter optics (including the camera
  window) $T_o(\lambda)$, and average atmospheric transmission $T_a(\lambda)$. The mirror
  reflectivity is essentially flat, and does not impact the passband
  shape. In the $\zme$-band, the red cutoff is determined by the
  quantum efficiency curve. The blue side of the $\ume$-band is shaped
  by the quantum efficiency curve as well as the optics and
  atmospheric transmissions, however the cutoff itself seems to be
  determined by the filter transmission.
  \label{fig:megacam_passband_ingredients}}
\end{figure}

In this section, we detail our model of the ingredients of the MegaCam
effective passbands, presented in \S \ref{sec:megacam_passbands}
and summarized on figure \ref{fig:megacam_passband_ingredients}. 
For all
bands, the model consists in the product of five components:
\begin{equation}
  T(\lambda; \x) = T_f(\lambda; \x) \times T_o(\lambda) \times R_m(\lambda) \times T_a(\lambda) \times \varepsilon(\lambda)
\end{equation}
$T_f(\lambda; \x)$ is the position dependent transmission of the
interference filters. $T_o(\lambda)$ is the transmission of the four
lens optical system which equip MegaPrime. It also includes the
transmission of the camera window.  $R_m(\lambda)$ refers to the
reflectivity of the primary mirror. $T_a(\lambda)$ is the average
transmission of the atmosphere above Mauna Kea. Finally,
$\varepsilon(\lambda)$, is the mean quantum efficiency of the E2V CCDs
which equip the focal plane of MegaCam.

The transmission of the optical system, $T_o(\lambda)$ and the
reflectivity of the primary mirror, $R_m(\lambda)$ were obtained from
the CFHT team.
The quantum efficiency $\varepsilon(\lambda)$ of the camera was obtained from the CEA
team.
It is actually an
average model, derived from the measured quantum efficiencies of the
chips which equip MegaCam.
The Mauna Kea atmospheric transmission $T_a(\lambda)$ is being measured by the Nearby
Factory Collaboration using the SuperNova Integral Field Spectrometer
\citep{Buton09}. We use a preliminary version of this measurement. Finally, 
the $O_2$ and $OH$ absorption lines have a sizeable impact on the $\zme$ passband. 
We use the determination presented in \citet{Hinkle03}\footnote{{\tt ftp://ftp.noao.edu/catalogs/atmospheric\_transmission/}}. 
The resolution of the original determined being of about 0.15 \AA, we have rebinned it, 
with to reach a bin size of about 3\AA. Table \ref{tab:megacam_open_transmission} displays the product of all the 
components listed above.

Finally, the filter transmissions, $T_f(\lambda, \x)$, were characterized by their
manufacturer (Sagem~/~REOSC). For each filter, ten scans were
performed at ten different locations --- namely, at the center of the
filter, and then, at 23, 47, 70, 93, 117, 140, 163, 186 and 210
millimeters from the center, along a diagonal. The transmissions
reported by Sagem / REOSC were blueshifted as described in \S
\ref{sec:megacam_passbands}, in order to account for the fact that the
f/4 beam does not cross the filters at a normal incidence on average.
The resulting blueshifted transmissions $\ume, \gme, \rme, \ime, $ and
$\zme$ are reported in tables \ref{tab:u_band_table},
\ref{tab:g_band_table}, \ref{tab:r_band_table}, \ref{tab:i_band_table}
and \ref{tab:z_band_table} respectively. In order to build a passband
model which is continuous as a function of the position, the scans
presented in those tables were interpolated, assuming a central
symmetry around the focal plane center, as indicated by the $\delta
k(\x)$ maps.

The sidereal positions of the objects can be mapped to ``filter
coordinates'' (in millimeters), using the following formula:
\begin{eqnarray}
  x_{f} &=& F \times \Bigl[ \cos(\delta)\ \sin(\alpha - \alpha_0) \Bigr] \nonumber \\
  y_{f} &=& F \times \Bigl[ -\cos(\delta) \sin(\delta_0) \cos(\alpha - \alpha_0) + \sin(\delta) \cos(\delta_0) \Bigr]
\end{eqnarray}
where $(x_f, y_f)$ are the point coordinates in the filter frame
(expressed in millimeters), $F$ is the MegaPrime focal length (14,890
millimeters), $(\alpha,\delta)$ and $(\alpha_0, \delta_0)$ are the
sidereal position of the object and the field center respectively.  In
the formula above, we use the fact that the distance between the
filters and the focal plane is negligible compared to the focal
length.

\onltab{13}{

\begin{table}
\caption{Open Transmission of the MegaCam imager (no filters) \label{tab:megacam_open_transmission}}
\begin{center}
\begin{tabular}{cc}
\hline
\hline
$\lambda$ & transmission \\
$(\AA)$   &              \\
\hline 
\ldots & \ldots \\
3100.0 & 0.0013 \\
\ldots & \ldots \\
4100.0 & 0.4817 \\
\ldots & \ldots \\
5100.0 & 0.5764\\
\ldots & \ldots \\
6100.0 & 0.5190\\
\ldots & \ldots \\
7100.0 & 0.4976\\
\ldots & \ldots \\
8100.0 & 0.3705\\
\ldots & \ldots \\
9100.0 & 0.1394\\
\ldots & \ldots \\
10100.0 & 0.0272\\
\ldots & \ldots \\
\hline 
\end{tabular}
\end{center}
\end{table}
}

\onllongtab{14}{
\begin{landscape}
\begin{longtable}{ccccccccccc}
\caption{$\ume$ filter scan \label{tab:u_band_table}} \\
\hline
\hline 
$\lambda$    & $T_{u_M}(\lambda; r)$ 
              & $T_{u_M}(\lambda; r)$ 
              & $T_{u_M}(\lambda; r)$ 
              & $T_{u_M}(\lambda; r)$ 
              & $T_{u_M}(\lambda; r)$ 
              & $T_{u_M}(\lambda; r)$ 
              & $T_{u_M}(\lambda; r)$ 
              & $T_{u_M}(\lambda; r)$ 
              & $T_{u_M}(\lambda; r)$ 
              & $T_{u_M}(\lambda; r)$ \\

  $(\AA)$    & $r = 0\ \rm{mm}$
              & $r = 23\ \rm{mm}$
              & $r = 47\ \rm{mm}$
              & $r = 70\ \rm{mm}$
              & $r = 93\ \rm{mm}$
              & $r = 117\ \rm{mm}$
              & $r = 140\ \rm{mm}$
              & $r = 163\ \rm{mm}$
              & $r = 186\ \rm{mm}$
              & $r = 210\ \rm{mm}$ \\
\hline 
\endfirsthead
\caption{continued.} \\
\hline 
\hline 
$\lambda$    & $T_{u_M}(\lambda; r)$ 
              & $T_{u_M}(\lambda; r)$ 
              & $T_{u_M}(\lambda; r)$ 
              & $T_{u_M}(\lambda; r)$ 
              & $T_{u_M}(\lambda; r)$ 
              & $T_{u_M}(\lambda; r)$ 
              & $T_{u_M}(\lambda; r)$ 
              & $T_{u_M}(\lambda; r)$ 
              & $T_{u_M}(\lambda; r)$ 
              & $T_{u_M}(\lambda; r)$ \\

  $(\AA)$    & $r = 0\ \rm{mm}$
              & $r = 23\ \rm{mm}$
              & $r = 47\ \rm{mm}$
              & $r = 70\ \rm{mm}$
              & $r = 93\ \rm{mm}$
              & $r = 117\ \rm{mm}$
              & $r = 140\ \rm{mm}$
              & $r = 163\ \rm{mm}$
              & $r = 186\ \rm{mm}$
              & $r = 210\ \rm{mm}$ \\

\endhead
\hline 
\endfoot
  3275 & 0.0000 & 0.0000 & 0.0020 & 0.0000 & 0.0000 & 0.0000 & 0.0000 & 0.0000 & 0.0000 & 0.0000 \\ 
  3280 & 0.0000 & 0.0000 & 0.0041 & 0.0023 & 0.0020 & 0.0000 & 0.0000 & 0.0001 & 0.0001 & 0.0002 \\ 
  3285 & 0.0000 & 0.0020 & 0.0052 & 0.0048 & 0.0040 & 0.0021 & 0.0019 & 0.0019 & 0.0023 & 0.0024 \\ 
  3290 & 0.0019 & 0.0043 & 0.0068 & 0.0062 & 0.0052 & 0.0044 & 0.0039 & 0.0040 & 0.0048 & 0.0052 \\ 
  3295 & 0.0041 & 0.0056 & 0.0091 & 0.0083 & 0.0068 & 0.0057 & 0.0051 & 0.0052 & 0.0063 & 0.0069 \\ 
  3300 & 0.0054 & 0.0075 & 0.0125 & 0.0113 & 0.0091 & 0.0076 & 0.0067 & 0.0069 & 0.0086 & 0.0094 \\ 
  3305 & 0.0073 & 0.0103 & 0.0174 & 0.0156 & 0.0124 & 0.0103 & 0.0090 & 0.0093 & 0.0118 & 0.0130 \\ 
  3310 & 0.0100 & 0.0144 & 0.0247 & 0.0221 & 0.0173 & 0.0142 & 0.0123 & 0.0129 & 0.0164 & 0.0181 \\ 
  3315 & 0.0139 & 0.0203 & 0.0355 & 0.0317 & 0.0247 & 0.0200 & 0.0173 & 0.0180 & 0.0229 & 0.0252 \\ 
  3320 & 0.0196 & 0.0290 & 0.0512 & 0.0458 & 0.0354 & 0.0285 & 0.0244 & 0.0254 & 0.0321 & 0.0351 \\ 
  3325 & 0.0280 & 0.0417 & 0.0740 & 0.0661 & 0.0510 & 0.0408 & 0.0347 & 0.0360 & 0.0449 & 0.0488 \\ 
  3330 & 0.0400 & 0.0598 & 0.1055 & 0.0944 & 0.0730 & 0.0582 & 0.0492 & 0.0506 & 0.0624 & 0.0671 \\ 
  3335 & 0.0574 & 0.0856 & 0.1479 & 0.1331 & 0.1038 & 0.0828 & 0.0696 & 0.0708 & 0.0862 & 0.0921 \\ 
  3340 & 0.0822 & 0.1211 & 0.2022 & 0.1834 & 0.1450 & 0.1163 & 0.0975 & 0.0981 & 0.1182 & 0.1257 \\ 
  3345 & 0.1158 & 0.1671 & 0.2651 & 0.2429 & 0.1961 & 0.1589 & 0.1336 & 0.1338 & 0.1596 & 0.1687 \\ 
  3350 & 0.1605 & 0.2242 & 0.3341 & 0.3096 & 0.2564 & 0.2116 & 0.1797 & 0.1796 & 0.2123 & 0.2233 \\ 
  3355 & 0.2168 & 0.2907 & 0.4043 & 0.3795 & 0.3234 & 0.2735 & 0.2359 & 0.2364 & 0.2762 & 0.2887 \\ 
  3360 & 0.2829 & 0.3627 & 0.4702 & 0.4474 & 0.3933 & 0.3415 & 0.3010 & 0.3031 & 0.3485 & 0.3623 \\ 
  3365 & 0.3558 & 0.4362 & 0.5287 & 0.5099 & 0.4631 & 0.4142 & 0.3740 & 0.3781 & 0.4268 & 0.4409 \\ 
  3370 & 0.4292 & 0.5027 & 0.5757 & 0.5630 & 0.5270 & 0.4860 & 0.4494 & 0.4554 & 0.5040 & 0.5171 \\ 
  3375 & 0.4964 & 0.5567 & 0.6092 & 0.6038 & 0.5812 & 0.5510 & 0.5217 & 0.5300 & 0.5739 & 0.5848 \\ 
  3380 & 0.5519 & 0.5962 & 0.6298 & 0.6319 & 0.6233 & 0.6059 & 0.5862 & 0.5963 & 0.6319 & 0.6400 \\ 
  3385 & 0.5923 & 0.6211 & 0.6404 & 0.6484 & 0.6514 & 0.6469 & 0.6369 & 0.6481 & 0.6732 & 0.6794 \\ 
  3390 & 0.6169 & 0.6341 & 0.6440 & 0.6557 & 0.6670 & 0.6719 & 0.6700 & 0.6806 & 0.6979 & 0.7042 \\ 
  3395 & 0.6300 & 0.6391 & 0.6434 & 0.6572 & 0.6740 & 0.6850 & 0.6891 & 0.6981 & 0.7130 & 0.7206 \\ 
  3400 & 0.6316 & 0.6365 & 0.6382 & 0.6528 & 0.6726 & 0.6877 & 0.6950 & 0.7042 & 0.7223 & 0.7311 \\ 
  3405 & 0.6237 & 0.6275 & 0.6282 & 0.6437 & 0.6650 & 0.6827 & 0.6925 & 0.7049 & 0.7281 & 0.7363 \\ 
  3410 & 0.6100 & 0.6145 & 0.6161 & 0.6318 & 0.6541 & 0.6732 & 0.6868 & 0.7043 & 0.7294 & 0.7355 \\ 
  3415 & 0.5954 & 0.6017 & 0.6053 & 0.6209 & 0.6430 & 0.6641 & 0.6815 & 0.7036 & 0.7250 & 0.7305 \\ 
  3420 & 0.5841 & 0.5923 & 0.5975 & 0.6132 & 0.6354 & 0.6581 & 0.6778 & 0.7010 & 0.7185 & 0.7269 \\ 
  3425 & 0.5761 & 0.5850 & 0.5919 & 0.6076 & 0.6298 & 0.6529 & 0.6734 & 0.6942 & 0.7156 & 0.7303 \\ 
  3430 & 0.5694 & 0.5777 & 0.5853 & 0.6008 & 0.6228 & 0.6459 & 0.6655 & 0.6863 & 0.7219 & 0.7432 \\ 
  3435 & 0.5636 & 0.5705 & 0.5778 & 0.5925 & 0.6140 & 0.6364 & 0.6565 & 0.6832 & 0.7390 & 0.7634 \\ 
  3440 & 0.5617 & 0.5668 & 0.5739 & 0.5868 & 0.6064 & 0.6286 & 0.6528 & 0.6910 & 0.7622 & 0.7838 \\ 
  3445 & 0.5694 & 0.5735 & 0.5804 & 0.5906 & 0.6073 & 0.6305 & 0.6611 & 0.7123 & 0.7835 & 0.7985 \\ 
  3450 & 0.5915 & 0.5959 & 0.6032 & 0.6100 & 0.6238 & 0.6470 & 0.6841 & 0.7409 & 0.7964 & 0.8044 \\ 
  3455 & 0.6271 & 0.6338 & 0.6413 & 0.6456 & 0.6562 & 0.6788 & 0.7170 & 0.7678 & 0.7999 & 0.8023 \\ 
\ldots &  \ldots &  \ldots &  \ldots &  \ldots &  \ldots &  \ldots &  \ldots &  \ldots &  \ldots &  \ldots \\ 
\end{longtable}
\end{landscape}
}
\onllongtab{15}{
\begin{landscape}
\begin{longtable}{ccccccccccc}
\caption{$\gme$ filter scan \label{tab:g_band_table}} \\
\hline
\hline 
$\lambda$    & $T_{g_M}(\lambda; r)$ 
              & $T_{g_M}(\lambda; r)$ 
              & $T_{g_M}(\lambda; r)$ 
              & $T_{g_M}(\lambda; r)$ 
              & $T_{g_M}(\lambda; r)$ 
              & $T_{g_M}(\lambda; r)$ 
              & $T_{g_M}(\lambda; r)$ 
              & $T_{g_M}(\lambda; r)$ 
              & $T_{g_M}(\lambda; r)$ 
              & $T_{g_M}(\lambda; r)$ \\

  $(\AA)$    & $r = 0\ \rm{mm}$
              & $r = 23\ \rm{mm}$
              & $r = 47\ \rm{mm}$
              & $r = 70\ \rm{mm}$
              & $r = 93\ \rm{mm}$
              & $r = 117\ \rm{mm}$
              & $r = 140\ \rm{mm}$
              & $r = 163\ \rm{mm}$
              & $r = 186\ \rm{mm}$
              & $r = 210\ \rm{mm}$ \\
\hline 
\endfirsthead
\caption{continued.} \\
\hline 
\hline 
$\lambda$    & $T_{g_M}(\lambda; r)$ 
              & $T_{g_M}(\lambda; r)$ 
              & $T_{g_M}(\lambda; r)$ 
              & $T_{g_M}(\lambda; r)$ 
              & $T_{g_M}(\lambda; r)$ 
              & $T_{g_M}(\lambda; r)$ 
              & $T_{g_M}(\lambda; r)$ 
              & $T_{g_M}(\lambda; r)$ 
              & $T_{g_M}(\lambda; r)$ 
              & $T_{g_M}(\lambda; r)$ \\

  $(\AA)$    & $r = 0\ \rm{mm}$
              & $r = 23\ \rm{mm}$
              & $r = 47\ \rm{mm}$
              & $r = 70\ \rm{mm}$
              & $r = 93\ \rm{mm}$
              & $r = 117\ \rm{mm}$
              & $r = 140\ \rm{mm}$
              & $r = 163\ \rm{mm}$
              & $r = 186\ \rm{mm}$
              & $r = 210\ \rm{mm}$ \\

\endhead
\hline 
\endfoot
  3975 & 0.0000 & 0.0000 & 0.0000 & 0.0000 & 0.0000 & 0.0000 & 0.0000 & 0.0000 & 0.0007 & 0.0009 \\ 
  3980 & 0.0000 & 0.0000 & 0.0000 & 0.0000 & 0.0000 & 0.0000 & 0.0000 & 0.0000 & 0.0027 & 0.0031 \\ 
  3985 & 0.0000 & 0.0000 & 0.0000 & 0.0000 & 0.0000 & 0.0000 & 0.0000 & 0.0000 & 0.0042 & 0.0048 \\ 
  3990 & 0.0000 & 0.0000 & 0.0000 & 0.0000 & 0.0000 & 0.0000 & 0.0000 & 0.0007 & 0.0050 & 0.0056 \\ 
  3995 & 0.0000 & 0.0000 & 0.0007 & 0.0007 & 0.0007 & 0.0007 & 0.0008 & 0.0028 & 0.0058 & 0.0065 \\ 
  4000 & 0.0000 & 0.0000 & 0.0026 & 0.0026 & 0.0026 & 0.0027 & 0.0030 & 0.0044 & 0.0067 & 0.0073 \\ 
  4005 & 0.0000 & 0.0000 & 0.0040 & 0.0040 & 0.0040 & 0.0041 & 0.0046 & 0.0051 & 0.0075 & 0.0082 \\ 
  4010 & 0.0000 & 0.0000 & 0.0045 & 0.0046 & 0.0046 & 0.0048 & 0.0053 & 0.0058 & 0.0084 & 0.0092 \\ 
  4015 & 0.0000 & 0.0000 & 0.0051 & 0.0052 & 0.0052 & 0.0054 & 0.0060 & 0.0065 & 0.0094 & 0.0104 \\ 
  4020 & 0.0008 & 0.0007 & 0.0056 & 0.0057 & 0.0058 & 0.0061 & 0.0067 & 0.0073 & 0.0107 & 0.0120 \\ 
  4025 & 0.0028 & 0.0026 & 0.0062 & 0.0063 & 0.0064 & 0.0067 & 0.0074 & 0.0081 & 0.0123 & 0.0142 \\ 
  4030 & 0.0041 & 0.0037 & 0.0068 & 0.0069 & 0.0071 & 0.0074 & 0.0083 & 0.0091 & 0.0147 & 0.0173 \\ 
  4035 & 0.0046 & 0.0041 & 0.0076 & 0.0077 & 0.0079 & 0.0083 & 0.0093 & 0.0104 & 0.0181 & 0.0219 \\ 
  4040 & 0.0050 & 0.0046 & 0.0086 & 0.0088 & 0.0090 & 0.0094 & 0.0108 & 0.0121 & 0.0230 & 0.0285 \\ 
  4045 & 0.0055 & 0.0052 & 0.0102 & 0.0104 & 0.0105 & 0.0111 & 0.0128 & 0.0147 & 0.0302 & 0.0380 \\ 
  4050 & 0.0060 & 0.0061 & 0.0125 & 0.0127 & 0.0127 & 0.0134 & 0.0158 & 0.0184 & 0.0403 & 0.0511 \\ 
  4055 & 0.0067 & 0.0073 & 0.0158 & 0.0160 & 0.0159 & 0.0168 & 0.0201 & 0.0238 & 0.0541 & 0.0683 \\ 
  4060 & 0.0077 & 0.0090 & 0.0204 & 0.0207 & 0.0205 & 0.0216 & 0.0262 & 0.0313 & 0.0718 & 0.0893 \\ 
  4065 & 0.0091 & 0.0115 & 0.0270 & 0.0273 & 0.0270 & 0.0284 & 0.0347 & 0.0416 & 0.0931 & 0.1134 \\ 
  4070 & 0.0111 & 0.0152 & 0.0361 & 0.0365 & 0.0359 & 0.0378 & 0.0462 & 0.0554 & 0.1174 & 0.1398 \\ 
  4075 & 0.0139 & 0.0202 & 0.0478 & 0.0484 & 0.0477 & 0.0501 & 0.0610 & 0.0725 & 0.1435 & 0.1671 \\ 
  4080 & 0.0178 & 0.0270 & 0.0622 & 0.0630 & 0.0623 & 0.0656 & 0.0790 & 0.0927 & 0.1703 & 0.1950 \\ 
  4085 & 0.0231 & 0.0356 & 0.0784 & 0.0796 & 0.0793 & 0.0835 & 0.0992 & 0.1147 & 0.1974 & 0.2241 \\ 
  4090 & 0.0301 & 0.0459 & 0.0952 & 0.0970 & 0.0975 & 0.1028 & 0.1204 & 0.1373 & 0.2257 & 0.2564 \\ 
  4095 & 0.0389 & 0.0574 & 0.1121 & 0.1144 & 0.1161 & 0.1226 & 0.1420 & 0.1601 & 0.2578 & 0.2950 \\ 
  4100 & 0.0492 & 0.0697 & 0.1284 & 0.1313 & 0.1343 & 0.1421 & 0.1634 & 0.1832 & 0.2965 & 0.3432 \\ 
  4105 & 0.0606 & 0.0826 & 0.1448 & 0.1484 & 0.1525 & 0.1618 & 0.1857 & 0.2082 & 0.3456 & 0.4042 \\ 
  4110 & 0.0726 & 0.0963 & 0.1632 & 0.1672 & 0.1722 & 0.1832 & 0.2111 & 0.2379 & 0.4078 & 0.4789 \\ 
  4115 & 0.0845 & 0.1118 & 0.1859 & 0.1902 & 0.1957 & 0.2086 & 0.2424 & 0.2753 & 0.4832 & 0.5645 \\ 
  4120 & 0.0967 & 0.1312 & 0.2164 & 0.2209 & 0.2263 & 0.2415 & 0.2832 & 0.3241 & 0.5693 & 0.6553 \\ 
  4125 & 0.1098 & 0.1572 & 0.2579 & 0.2625 & 0.2673 & 0.2850 & 0.3363 & 0.3863 & 0.6593 & 0.7421 \\ 
  4130 & 0.1255 & 0.1937 & 0.3137 & 0.3183 & 0.3219 & 0.3422 & 0.4037 & 0.4624 & 0.7445 & 0.8159 \\ 
  4135 & 0.1467 & 0.2458 & 0.3865 & 0.3908 & 0.3928 & 0.4153 & 0.4860 & 0.5506 & 0.8173 & 0.8710 \\ 
  4140 & 0.1762 & 0.3164 & 0.4746 & 0.4785 & 0.4785 & 0.5022 & 0.5784 & 0.6439 & 0.8709 & 0.9050 \\ 
  4145 & 0.2186 & 0.4066 & 0.5731 & 0.5764 & 0.5745 & 0.5980 & 0.6731 & 0.7333 & 0.9034 & 0.9205 \\ 
  4150 & 0.2775 & 0.5115 & 0.6733 & 0.6758 & 0.6730 & 0.6942 & 0.7611 & 0.8101 & 0.9181 & 0.9239 \\ 
  4155 & 0.3547 & 0.6195 & 0.7631 & 0.7649 & 0.7624 & 0.7799 & 0.8323 & 0.8668 & 0.9209 & 0.9221 \\ 
\ldots &  \ldots &  \ldots &  \ldots &  \ldots &  \ldots &  \ldots &  \ldots &  \ldots &  \ldots &  \ldots \\ 
\end{longtable}
\end{landscape}
}
\onllongtab{16}{
\begin{landscape}
\begin{longtable}{ccccccccccc}
\caption{$\rme$ filter scan \label{tab:r_band_table}} \\
\hline
\hline 
$\lambda$    & $T_{r_M}(\lambda; r)$ 
              & $T_{r_M}(\lambda; r)$ 
              & $T_{r_M}(\lambda; r)$ 
              & $T_{r_M}(\lambda; r)$ 
              & $T_{r_M}(\lambda; r)$ 
              & $T_{r_M}(\lambda; r)$ 
              & $T_{r_M}(\lambda; r)$ 
              & $T_{r_M}(\lambda; r)$ 
              & $T_{r_M}(\lambda; r)$ 
              & $T_{r_M}(\lambda; r)$ \\

  $(\AA)$    & $r = 0\ \rm{mm}$
              & $r = 23\ \rm{mm}$
              & $r = 47\ \rm{mm}$
              & $r = 70\ \rm{mm}$
              & $r = 93\ \rm{mm}$
              & $r = 117\ \rm{mm}$
              & $r = 140\ \rm{mm}$
              & $r = 163\ \rm{mm}$
              & $r = 186\ \rm{mm}$
              & $r = 210\ \rm{mm}$ \\
\hline 
\endfirsthead
\caption{continued.} \\
\hline 
\hline 
$\lambda$    & $T_{r_M}(\lambda; r)$ 
              & $T_{r_M}(\lambda; r)$ 
              & $T_{r_M}(\lambda; r)$ 
              & $T_{r_M}(\lambda; r)$ 
              & $T_{r_M}(\lambda; r)$ 
              & $T_{r_M}(\lambda; r)$ 
              & $T_{r_M}(\lambda; r)$ 
              & $T_{r_M}(\lambda; r)$ 
              & $T_{r_M}(\lambda; r)$ 
              & $T_{r_M}(\lambda; r)$ \\

  $(\AA)$    & $r = 0\ \rm{mm}$
              & $r = 23\ \rm{mm}$
              & $r = 47\ \rm{mm}$
              & $r = 70\ \rm{mm}$
              & $r = 93\ \rm{mm}$
              & $r = 117\ \rm{mm}$
              & $r = 140\ \rm{mm}$
              & $r = 163\ \rm{mm}$
              & $r = 186\ \rm{mm}$
              & $r = 210\ \rm{mm}$ \\

\endhead
\hline 
\endfoot
  5295 & 0.0000 & 0.0000 & 0.0000 & 0.0000 & 0.0000 & 0.0000 & 0.0000 & 0.0001 & 0.0000 & 0.0000 \\ 
  5300 & 0.0000 & 0.0000 & 0.0000 & 0.0000 & 0.0000 & 0.0000 & 0.0000 & 0.0012 & 0.0002 & 0.0000 \\ 
  5305 & 0.0000 & 0.0000 & 0.0000 & 0.0000 & 0.0000 & 0.0000 & 0.0000 & 0.0027 & 0.0012 & 0.0000 \\ 
  5310 & 0.0000 & 0.0000 & 0.0000 & 0.0000 & 0.0000 & 0.0000 & 0.0001 & 0.0035 & 0.0028 & 0.0002 \\ 
  5315 & 0.0000 & 0.0000 & 0.0000 & 0.0000 & 0.0000 & 0.0000 & 0.0013 & 0.0038 & 0.0037 & 0.0012 \\ 
  5320 & 0.0000 & 0.0000 & 0.0000 & 0.0000 & 0.0000 & 0.0000 & 0.0029 & 0.0042 & 0.0040 & 0.0027 \\ 
  5325 & 0.0000 & 0.0000 & 0.0000 & 0.0000 & 0.0000 & 0.0001 & 0.0038 & 0.0046 & 0.0044 & 0.0035 \\ 
  5330 & 0.0000 & 0.0000 & 0.0000 & 0.0000 & 0.0000 & 0.0012 & 0.0041 & 0.0050 & 0.0048 & 0.0039 \\ 
  5335 & 0.0000 & 0.0000 & 0.0000 & 0.0000 & 0.0000 & 0.0027 & 0.0045 & 0.0055 & 0.0052 & 0.0042 \\ 
  5340 & 0.0000 & 0.0000 & 0.0000 & 0.0000 & 0.0012 & 0.0035 & 0.0049 & 0.0061 & 0.0058 & 0.0046 \\ 
  5345 & 0.0000 & 0.0000 & 0.0000 & 0.0000 & 0.0027 & 0.0038 & 0.0054 & 0.0067 & 0.0064 & 0.0051 \\ 
  5350 & 0.0000 & 0.0000 & 0.0000 & 0.0012 & 0.0036 & 0.0042 & 0.0059 & 0.0073 & 0.0070 & 0.0056 \\ 
  5355 & 0.0000 & 0.0000 & 0.0000 & 0.0027 & 0.0039 & 0.0046 & 0.0065 & 0.0081 & 0.0077 & 0.0061 \\ 
  5360 & 0.0000 & 0.0000 & 0.0000 & 0.0035 & 0.0043 & 0.0050 & 0.0071 & 0.0089 & 0.0085 & 0.0068 \\ 
  5365 & 0.0000 & 0.0000 & 0.0013 & 0.0038 & 0.0047 & 0.0055 & 0.0078 & 0.0098 & 0.0094 & 0.0074 \\ 
  5370 & 0.0000 & 0.0000 & 0.0028 & 0.0042 & 0.0051 & 0.0060 & 0.0086 & 0.0108 & 0.0104 & 0.0082 \\ 
  5375 & 0.0000 & 0.0000 & 0.0036 & 0.0045 & 0.0055 & 0.0065 & 0.0094 & 0.0119 & 0.0115 & 0.0090 \\ 
  5380 & 0.0012 & 0.0013 & 0.0040 & 0.0050 & 0.0061 & 0.0071 & 0.0104 & 0.0131 & 0.0127 & 0.0100 \\ 
  5385 & 0.0028 & 0.0028 & 0.0043 & 0.0054 & 0.0066 & 0.0078 & 0.0114 & 0.0145 & 0.0140 & 0.0110 \\ 
  5390 & 0.0036 & 0.0037 & 0.0047 & 0.0059 & 0.0073 & 0.0086 & 0.0125 & 0.0159 & 0.0155 & 0.0121 \\ 
  5395 & 0.0039 & 0.0040 & 0.0051 & 0.0064 & 0.0079 & 0.0094 & 0.0138 & 0.0176 & 0.0171 & 0.0134 \\ 
  5400 & 0.0042 & 0.0044 & 0.0056 & 0.0070 & 0.0087 & 0.0103 & 0.0152 & 0.0194 & 0.0189 & 0.0148 \\ 
  5405 & 0.0046 & 0.0048 & 0.0061 & 0.0077 & 0.0095 & 0.0113 & 0.0167 & 0.0214 & 0.0208 & 0.0163 \\ 
  5410 & 0.0050 & 0.0052 & 0.0066 & 0.0084 & 0.0104 & 0.0125 & 0.0184 & 0.0237 & 0.0231 & 0.0180 \\ 
  5415 & 0.0055 & 0.0057 & 0.0072 & 0.0092 & 0.0115 & 0.0137 & 0.0203 & 0.0262 & 0.0255 & 0.0199 \\ 
  5420 & 0.0060 & 0.0062 & 0.0079 & 0.0101 & 0.0126 & 0.0151 & 0.0224 & 0.0290 & 0.0283 & 0.0221 \\ 
  5425 & 0.0065 & 0.0068 & 0.0087 & 0.0111 & 0.0138 & 0.0166 & 0.0248 & 0.0321 & 0.0313 & 0.0244 \\ 
  5430 & 0.0071 & 0.0074 & 0.0095 & 0.0121 & 0.0152 & 0.0182 & 0.0273 & 0.0356 & 0.0347 & 0.0270 \\ 
  5435 & 0.0078 & 0.0081 & 0.0104 & 0.0133 & 0.0167 & 0.0201 & 0.0302 & 0.0394 & 0.0385 & 0.0299 \\ 
  5440 & 0.0085 & 0.0088 & 0.0114 & 0.0146 & 0.0183 & 0.0221 & 0.0334 & 0.0436 & 0.0428 & 0.0332 \\ 
  5445 & 0.0093 & 0.0096 & 0.0125 & 0.0160 & 0.0202 & 0.0244 & 0.0369 & 0.0484 & 0.0475 & 0.0368 \\ 
  5450 & 0.0102 & 0.0105 & 0.0137 & 0.0176 & 0.0222 & 0.0269 & 0.0409 & 0.0537 & 0.0528 & 0.0409 \\ 
  5455 & 0.0111 & 0.0116 & 0.0150 & 0.0194 & 0.0245 & 0.0297 & 0.0453 & 0.0596 & 0.0587 & 0.0454 \\ 
  5460 & 0.0122 & 0.0127 & 0.0165 & 0.0213 & 0.0270 & 0.0328 & 0.0501 & 0.0662 & 0.0653 & 0.0505 \\ 
  5465 & 0.0134 & 0.0139 & 0.0181 & 0.0235 & 0.0298 & 0.0362 & 0.0555 & 0.0734 & 0.0726 & 0.0562 \\ 
  5470 & 0.0147 & 0.0152 & 0.0199 & 0.0258 & 0.0329 & 0.0400 & 0.0614 & 0.0814 & 0.0807 & 0.0625 \\ 
  5475 & 0.0161 & 0.0167 & 0.0219 & 0.0285 & 0.0363 & 0.0442 & 0.0680 & 0.0901 & 0.0896 & 0.0695 \\ 
\ldots &  \ldots &  \ldots &  \ldots &  \ldots &  \ldots &  \ldots &  \ldots &  \ldots &  \ldots &  \ldots \\ 
\end{longtable}
\end{landscape}
}
\onllongtab{17}{
\begin{landscape}
\begin{longtable}{ccccccccccc}
\caption{$\ime$ filter scan \label{tab:i_band_table}} \\
\hline
\hline 
$\lambda$    & $T_{i_M}(\lambda; r)$ 
              & $T_{i_M}(\lambda; r)$ 
              & $T_{i_M}(\lambda; r)$ 
              & $T_{i_M}(\lambda; r)$ 
              & $T_{i_M}(\lambda; r)$ 
              & $T_{i_M}(\lambda; r)$ 
              & $T_{i_M}(\lambda; r)$ 
              & $T_{i_M}(\lambda; r)$ 
              & $T_{i_M}(\lambda; r)$ 
              & $T_{i_M}(\lambda; r)$ \\

  $(\AA)$    & $r = 0\ \rm{mm}$
              & $r = 23\ \rm{mm}$
              & $r = 47\ \rm{mm}$
              & $r = 70\ \rm{mm}$
              & $r = 93\ \rm{mm}$
              & $r = 117\ \rm{mm}$
              & $r = 140\ \rm{mm}$
              & $r = 163\ \rm{mm}$
              & $r = 186\ \rm{mm}$
              & $r = 210\ \rm{mm}$ \\
\hline 
\endfirsthead
\caption{continued.} \\
\hline 
\hline 
$\lambda$    & $T_{i_M}(\lambda; r)$ 
              & $T_{i_M}(\lambda; r)$ 
              & $T_{i_M}(\lambda; r)$ 
              & $T_{i_M}(\lambda; r)$ 
              & $T_{i_M}(\lambda; r)$ 
              & $T_{i_M}(\lambda; r)$ 
              & $T_{i_M}(\lambda; r)$ 
              & $T_{i_M}(\lambda; r)$ 
              & $T_{i_M}(\lambda; r)$ 
              & $T_{i_M}(\lambda; r)$ \\

  $(\AA)$    & $r = 0\ \rm{mm}$
              & $r = 23\ \rm{mm}$
              & $r = 47\ \rm{mm}$
              & $r = 70\ \rm{mm}$
              & $r = 93\ \rm{mm}$
              & $r = 117\ \rm{mm}$
              & $r = 140\ \rm{mm}$
              & $r = 163\ \rm{mm}$
              & $r = 186\ \rm{mm}$
              & $r = 210\ \rm{mm}$ \\

\endhead
\hline 
\endfoot
  6565 & 0.0000 & 0.0000 & 0.0000 & 0.0000 & 0.0000 & 0.0000 & 0.0000 & 0.0000 & 0.0001 & 0.0001 \\ 
  6570 & 0.0000 & 0.0000 & 0.0000 & 0.0000 & 0.0000 & 0.0000 & 0.0000 & 0.0005 & 0.0005 & 0.0005 \\ 
  6575 & 0.0000 & 0.0000 & 0.0000 & 0.0000 & 0.0000 & 0.0000 & 0.0000 & 0.0013 & 0.0014 & 0.0013 \\ 
  6580 & 0.0000 & 0.0000 & 0.0000 & 0.0000 & 0.0000 & 0.0000 & 0.0000 & 0.0023 & 0.0024 & 0.0023 \\ 
  6585 & 0.0000 & 0.0000 & 0.0000 & 0.0000 & 0.0000 & 0.0000 & 0.0000 & 0.0033 & 0.0034 & 0.0032 \\ 
  6590 & 0.0000 & 0.0000 & 0.0000 & 0.0000 & 0.0000 & 0.0000 & 0.0000 & 0.0037 & 0.0038 & 0.0036 \\ 
  6595 & 0.0000 & 0.0000 & 0.0000 & 0.0000 & 0.0000 & 0.0000 & 0.0005 & 0.0039 & 0.0041 & 0.0039 \\ 
  6600 & 0.0000 & 0.0000 & 0.0000 & 0.0000 & 0.0000 & 0.0000 & 0.0014 & 0.0042 & 0.0044 & 0.0042 \\ 
  6605 & 0.0000 & 0.0000 & 0.0000 & 0.0000 & 0.0000 & 0.0000 & 0.0024 & 0.0045 & 0.0047 & 0.0044 \\ 
  6610 & 0.0000 & 0.0000 & 0.0000 & 0.0000 & 0.0000 & 0.0004 & 0.0034 & 0.0048 & 0.0050 & 0.0047 \\ 
  6615 & 0.0000 & 0.0000 & 0.0000 & 0.0000 & 0.0004 & 0.0013 & 0.0038 & 0.0051 & 0.0053 & 0.0051 \\ 
  6620 & 0.0000 & 0.0000 & 0.0000 & 0.0000 & 0.0014 & 0.0023 & 0.0041 & 0.0055 & 0.0057 & 0.0054 \\ 
  6625 & 0.0000 & 0.0000 & 0.0000 & 0.0000 & 0.0023 & 0.0033 & 0.0043 & 0.0059 & 0.0061 & 0.0058 \\ 
  6630 & 0.0000 & 0.0000 & 0.0000 & 0.0004 & 0.0033 & 0.0036 & 0.0046 & 0.0063 & 0.0066 & 0.0062 \\ 
  6635 & 0.0000 & 0.0000 & 0.0000 & 0.0014 & 0.0037 & 0.0038 & 0.0049 & 0.0067 & 0.0070 & 0.0067 \\ 
  6640 & 0.0000 & 0.0000 & 0.0000 & 0.0023 & 0.0039 & 0.0041 & 0.0053 & 0.0072 & 0.0076 & 0.0072 \\ 
  6645 & 0.0000 & 0.0000 & 0.0000 & 0.0033 & 0.0042 & 0.0043 & 0.0056 & 0.0077 & 0.0081 & 0.0077 \\ 
  6650 & 0.0005 & 0.0005 & 0.0000 & 0.0036 & 0.0044 & 0.0046 & 0.0060 & 0.0083 & 0.0087 & 0.0083 \\ 
  6655 & 0.0014 & 0.0014 & 0.0000 & 0.0039 & 0.0047 & 0.0049 & 0.0064 & 0.0089 & 0.0094 & 0.0089 \\ 
  6660 & 0.0024 & 0.0023 & 0.0005 & 0.0041 & 0.0050 & 0.0053 & 0.0069 & 0.0095 & 0.0101 & 0.0095 \\ 
  6665 & 0.0034 & 0.0033 & 0.0014 & 0.0044 & 0.0054 & 0.0056 & 0.0074 & 0.0102 & 0.0108 & 0.0103 \\ 
  6670 & 0.0037 & 0.0036 & 0.0024 & 0.0047 & 0.0057 & 0.0060 & 0.0079 & 0.0110 & 0.0116 & 0.0110 \\ 
  6675 & 0.0039 & 0.0039 & 0.0034 & 0.0050 & 0.0061 & 0.0064 & 0.0084 & 0.0118 & 0.0125 & 0.0119 \\ 
  6680 & 0.0042 & 0.0041 & 0.0037 & 0.0053 & 0.0065 & 0.0068 & 0.0090 & 0.0127 & 0.0135 & 0.0128 \\ 
  6685 & 0.0044 & 0.0044 & 0.0040 & 0.0056 & 0.0070 & 0.0073 & 0.0097 & 0.0137 & 0.0145 & 0.0138 \\ 
  6690 & 0.0047 & 0.0047 & 0.0042 & 0.0060 & 0.0074 & 0.0078 & 0.0104 & 0.0147 & 0.0156 & 0.0148 \\ 
  6695 & 0.0050 & 0.0050 & 0.0045 & 0.0064 & 0.0080 & 0.0084 & 0.0111 & 0.0158 & 0.0169 & 0.0160 \\ 
  6700 & 0.0054 & 0.0053 & 0.0048 & 0.0069 & 0.0085 & 0.0090 & 0.0119 & 0.0171 & 0.0182 & 0.0173 \\ 
  6705 & 0.0057 & 0.0056 & 0.0051 & 0.0073 & 0.0091 & 0.0096 & 0.0128 & 0.0184 & 0.0197 & 0.0187 \\ 
  6710 & 0.0061 & 0.0060 & 0.0054 & 0.0078 & 0.0098 & 0.0103 & 0.0138 & 0.0199 & 0.0213 & 0.0202 \\ 
  6715 & 0.0065 & 0.0064 & 0.0058 & 0.0084 & 0.0105 & 0.0110 & 0.0148 & 0.0214 & 0.0230 & 0.0218 \\ 
  6720 & 0.0069 & 0.0068 & 0.0062 & 0.0090 & 0.0112 & 0.0118 & 0.0160 & 0.0232 & 0.0249 & 0.0236 \\ 
  6725 & 0.0074 & 0.0073 & 0.0066 & 0.0096 & 0.0120 & 0.0127 & 0.0172 & 0.0250 & 0.0269 & 0.0256 \\ 
  6730 & 0.0079 & 0.0078 & 0.0071 & 0.0103 & 0.0129 & 0.0136 & 0.0185 & 0.0270 & 0.0292 & 0.0277 \\ 
  6735 & 0.0085 & 0.0083 & 0.0075 & 0.0110 & 0.0138 & 0.0146 & 0.0199 & 0.0293 & 0.0316 & 0.0300 \\ 
  6740 & 0.0091 & 0.0089 & 0.0081 & 0.0118 & 0.0149 & 0.0157 & 0.0215 & 0.0317 & 0.0343 & 0.0326 \\ 
  6745 & 0.0097 & 0.0096 & 0.0086 & 0.0127 & 0.0160 & 0.0169 & 0.0232 & 0.0343 & 0.0372 & 0.0354 \\ 
\ldots &  \ldots &  \ldots &  \ldots &  \ldots &  \ldots &  \ldots &  \ldots &  \ldots &  \ldots &  \ldots \\ 
\end{longtable}
\end{landscape}
}
\onllongtab{18}{
\begin{landscape}
\begin{longtable}{ccccccccccc}
\caption{$\zme$ filter scan \label{tab:z_band_table}} \\
\hline
\hline 
$\lambda$    & $T_{z_M}(\lambda; r)$ 
              & $T_{z_M}(\lambda; r)$ 
              & $T_{z_M}(\lambda; r)$ 
              & $T_{z_M}(\lambda; r)$ 
              & $T_{z_M}(\lambda; r)$ 
              & $T_{z_M}(\lambda; r)$ 
              & $T_{z_M}(\lambda; r)$ 
              & $T_{z_M}(\lambda; r)$ 
              & $T_{z_M}(\lambda; r)$ 
              & $T_{z_M}(\lambda; r)$ \\

  $(\AA)$    & $r = 0\ \rm{mm}$
              & $r = 23\ \rm{mm}$
              & $r = 47\ \rm{mm}$
              & $r = 70\ \rm{mm}$
              & $r = 93\ \rm{mm}$
              & $r = 117\ \rm{mm}$
              & $r = 140\ \rm{mm}$
              & $r = 163\ \rm{mm}$
              & $r = 186\ \rm{mm}$
              & $r = 210\ \rm{mm}$ \\
\hline 
\endfirsthead
\caption{continued.} \\
\hline 
\hline 
$\lambda$    & $T_{z_M}(\lambda; r)$ 
              & $T_{z_M}(\lambda; r)$ 
              & $T_{z_M}(\lambda; r)$ 
              & $T_{z_M}(\lambda; r)$ 
              & $T_{z_M}(\lambda; r)$ 
              & $T_{z_M}(\lambda; r)$ 
              & $T_{z_M}(\lambda; r)$ 
              & $T_{z_M}(\lambda; r)$ 
              & $T_{z_M}(\lambda; r)$ 
              & $T_{z_M}(\lambda; r)$ \\

  $(\AA)$    & $r = 0\ \rm{mm}$
              & $r = 23\ \rm{mm}$
              & $r = 47\ \rm{mm}$
              & $r = 70\ \rm{mm}$
              & $r = 93\ \rm{mm}$
              & $r = 117\ \rm{mm}$
              & $r = 140\ \rm{mm}$
              & $r = 163\ \rm{mm}$
              & $r = 186\ \rm{mm}$
              & $r = 210\ \rm{mm}$ \\

\endhead
\hline 
\endfoot
  7840 & 0.0000 & 0.0000 & 0.0000 & 0.0000 & 0.0000 & 0.0000 & 0.0000 & 0.0000 & 0.0000 & 0.0000 \\ 
  7845 & 0.0000 & 0.0000 & 0.0000 & 0.0000 & 0.0000 & 0.0000 & 0.0000 & 0.0003 & 0.0000 & 0.0000 \\ 
  7850 & 0.0000 & 0.0000 & 0.0000 & 0.0000 & 0.0000 & 0.0000 & 0.0002 & 0.0007 & 0.0000 & 0.0000 \\ 
  7855 & 0.0000 & 0.0000 & 0.0000 & 0.0000 & 0.0000 & 0.0000 & 0.0007 & 0.0015 & 0.0000 & 0.0000 \\ 
  7860 & 0.0000 & 0.0000 & 0.0000 & 0.0000 & 0.0000 & 0.0000 & 0.0015 & 0.0022 & 0.0000 & 0.0000 \\ 
  7865 & 0.0000 & 0.0000 & 0.0000 & 0.0000 & 0.0000 & 0.0000 & 0.0022 & 0.0031 & 0.0003 & 0.0000 \\ 
  7870 & 0.0000 & 0.0000 & 0.0000 & 0.0000 & 0.0000 & 0.0000 & 0.0031 & 0.0038 & 0.0007 & 0.0000 \\ 
  7875 & 0.0000 & 0.0000 & 0.0000 & 0.0000 & 0.0000 & 0.0002 & 0.0038 & 0.0040 & 0.0014 & 0.0000 \\ 
  7880 & 0.0000 & 0.0000 & 0.0000 & 0.0000 & 0.0000 & 0.0007 & 0.0040 & 0.0043 & 0.0021 & 0.0000 \\ 
  7885 & 0.0000 & 0.0000 & 0.0000 & 0.0000 & 0.0000 & 0.0014 & 0.0043 & 0.0046 & 0.0029 & 0.0001 \\ 
  7890 & 0.0000 & 0.0000 & 0.0000 & 0.0000 & 0.0002 & 0.0022 & 0.0046 & 0.0049 & 0.0036 & 0.0003 \\ 
  7895 & 0.0000 & 0.0000 & 0.0000 & 0.0000 & 0.0008 & 0.0030 & 0.0049 & 0.0053 & 0.0038 & 0.0008 \\ 
  7900 & 0.0000 & 0.0000 & 0.0000 & 0.0001 & 0.0015 & 0.0036 & 0.0053 & 0.0056 & 0.0041 & 0.0015 \\ 
  7905 & 0.0000 & 0.0000 & 0.0000 & 0.0007 & 0.0022 & 0.0039 & 0.0056 & 0.0060 & 0.0044 & 0.0022 \\ 
  7910 & 0.0000 & 0.0000 & 0.0000 & 0.0014 & 0.0031 & 0.0041 & 0.0060 & 0.0065 & 0.0047 & 0.0030 \\ 
  7915 & 0.0000 & 0.0000 & 0.0000 & 0.0022 & 0.0038 & 0.0044 & 0.0065 & 0.0069 & 0.0050 & 0.0036 \\ 
  7920 & 0.0000 & 0.0000 & 0.0001 & 0.0029 & 0.0040 & 0.0047 & 0.0069 & 0.0075 & 0.0053 & 0.0039 \\ 
  7925 & 0.0000 & 0.0000 & 0.0008 & 0.0036 & 0.0043 & 0.0051 & 0.0074 & 0.0080 & 0.0057 & 0.0042 \\ 
  7930 & 0.0000 & 0.0000 & 0.0015 & 0.0039 & 0.0046 & 0.0054 & 0.0080 & 0.0086 & 0.0061 & 0.0044 \\ 
  7935 & 0.0000 & 0.0000 & 0.0022 & 0.0041 & 0.0049 & 0.0058 & 0.0086 & 0.0093 & 0.0066 & 0.0047 \\ 
  7940 & 0.0000 & 0.0000 & 0.0030 & 0.0044 & 0.0053 & 0.0062 & 0.0093 & 0.0100 & 0.0071 & 0.0050 \\ 
  7945 & 0.0001 & 0.0000 & 0.0037 & 0.0047 & 0.0056 & 0.0067 & 0.0100 & 0.0108 & 0.0076 & 0.0054 \\ 
  7950 & 0.0008 & 0.0001 & 0.0040 & 0.0051 & 0.0060 & 0.0072 & 0.0107 & 0.0116 & 0.0082 & 0.0057 \\ 
  7955 & 0.0014 & 0.0008 & 0.0043 & 0.0054 & 0.0064 & 0.0077 & 0.0116 & 0.0126 & 0.0088 & 0.0061 \\ 
  7960 & 0.0022 & 0.0015 & 0.0046 & 0.0058 & 0.0069 & 0.0083 & 0.0126 & 0.0137 & 0.0096 & 0.0066 \\ 
  7965 & 0.0031 & 0.0023 & 0.0050 & 0.0064 & 0.0077 & 0.0092 & 0.0141 & 0.0153 & 0.0106 & 0.0072 \\ 
  7970 & 0.0039 & 0.0033 & 0.0056 & 0.0072 & 0.0086 & 0.0104 & 0.0159 & 0.0173 & 0.0119 & 0.0080 \\ 
  7975 & 0.0044 & 0.0042 & 0.0063 & 0.0080 & 0.0096 & 0.0116 & 0.0180 & 0.0196 & 0.0133 & 0.0089 \\ 
  7980 & 0.0048 & 0.0046 & 0.0070 & 0.0089 & 0.0107 & 0.0129 & 0.0202 & 0.0221 & 0.0149 & 0.0098 \\ 
  7985 & 0.0053 & 0.0050 & 0.0077 & 0.0098 & 0.0118 & 0.0143 & 0.0225 & 0.0246 & 0.0165 & 0.0108 \\ 
  7990 & 0.0056 & 0.0054 & 0.0082 & 0.0106 & 0.0128 & 0.0155 & 0.0246 & 0.0269 & 0.0180 & 0.0117 \\ 
  7995 & 0.0060 & 0.0057 & 0.0088 & 0.0114 & 0.0137 & 0.0167 & 0.0266 & 0.0292 & 0.0194 & 0.0125 \\ 
  8000 & 0.0064 & 0.0061 & 0.0094 & 0.0122 & 0.0148 & 0.0180 & 0.0288 & 0.0316 & 0.0209 & 0.0134 \\ 
  8005 & 0.0068 & 0.0065 & 0.0101 & 0.0131 & 0.0159 & 0.0194 & 0.0313 & 0.0344 & 0.0227 & 0.0144 \\ 
  8010 & 0.0073 & 0.0069 & 0.0108 & 0.0141 & 0.0171 & 0.0210 & 0.0340 & 0.0374 & 0.0246 & 0.0155 \\ 
  8015 & 0.0078 & 0.0074 & 0.0116 & 0.0152 & 0.0185 & 0.0227 & 0.0370 & 0.0408 & 0.0267 & 0.0167 \\ 
  8020 & 0.0083 & 0.0079 & 0.0125 & 0.0163 & 0.0199 & 0.0246 & 0.0402 & 0.0444 & 0.0289 & 0.0181 \\ 
\ldots &  \ldots &  \ldots &  \ldots &  \ldots &  \ldots &  \ldots &  \ldots &  \ldots &  \ldots &  \ldots \\ 
\end{longtable}
\end{landscape}
}

\end{appendix}

\begin{appendix}
\section{Linearity Checks}
\label{sec:linearity}

Photometric standards are bright stars. Even when observed with short
exposure times, their brightest pixel is commonly close to saturation.
On the contrary, most if not all supernovae represent a small increase 
over the sky level. Since any amplifying electronic system is bound 
to become non-linear when approaching saturation, we tried to measure
or bound non-linearities of MegaCam's photometric response.

For this purpose, we did not use the light emitting diodes built
in the MegaPrime setup, although they enable in principle to inject
controlled amounts of light in the imager. These diodes illuminate
the detectors almost uniformly, and the response could be 
different than to localized astronomical sources. Using genuine
astronomical observations is obviously less flexible, and we had
to restrict our linearity test to checking if star measurements
are altered at the high end of the dynamic range.

\begin{figure}
\centering
\includegraphics[width=\linewidth]{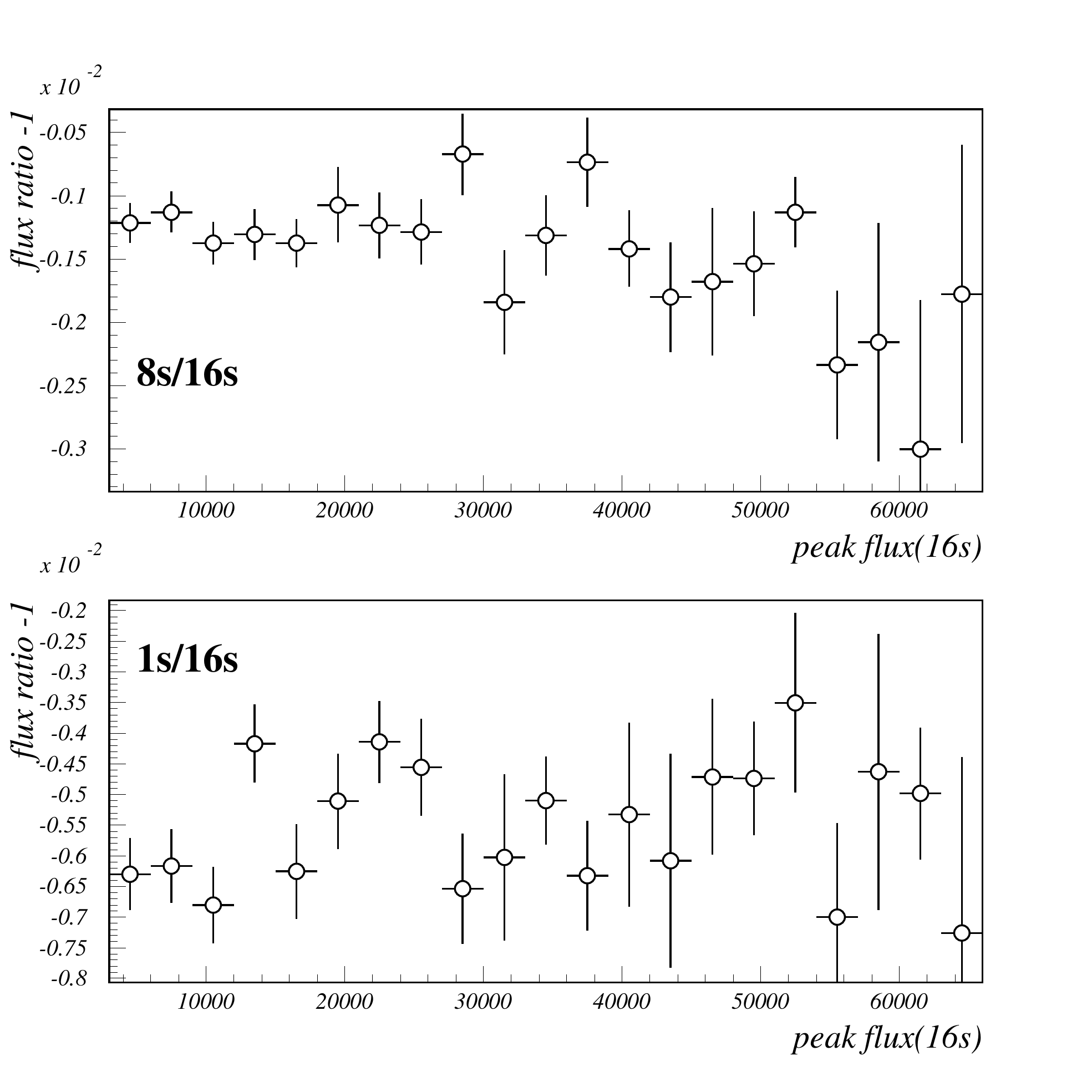}
\caption{Difference to unity of ratios of fluxes measured in different 
duration exposures, as a function of peak flux. We deduce from this plot
that star fluxes measured at the high end of the dynamic range are on average
not altered by more than about one part in a thousand. The flux independent
offsets can be attributed to e.g. unaccounted for aperture corrections or
errors in the reported exposure times.}
\label{fig:flux_ratios}
\end{figure}

   Images of a low Galactic latitude field were observed on the same
night, within ten minutes, with exposure times of 1,2,4,8 and 16 s, at an almost constant 
airmass of 1.01 and without ditherings between exposures.
These images were flat-fielded using standard flats, but this is essentially
irrelevant to what follows. We measured fluxes of stars
in a 16 pixels radius aperture ($f_{16}$), after a
thorough estimation of the background level. Namely, we first detected sources
down to a S/N of about 3, generously masked pixels attributed to these sources
and used the remaining pixels to compute a local background average level 
below each source. For these low background levels, it is important
to use a mean rather than a median, because the Poisson distribution 
describing the pixel statistics is skewed. We also computed the aperture flux
in a 27 pixels radius ($f_{27}$) and fitted a linear relation to 
$f_{27}-f_{16}$ vs $f_{16}$ of the isolated measurements in each CCD. 
The slope provides us with an average aperture correction to 27 pixels
radius and the offset indicates the quality of the background subtraction.
We averaged the aperture corrections per exposure and applied them to the 
$f_{16}$ measurements, and checked that our background residuals (mostly below 
0.1 ADU per pixel) do not affect the fluxes we are considering by more than
one part in a thousand. Fig. \ref{fig:flux_ratios} displays ratios of fluxes
(per unit time) measured with different exposure times as a function of 
object brightest pixel value, averaged over the mosaic. Average values
of these ratios differ slightly from unity, and the difference
is compatible with the expected differences of aperture corrections
beyond 27 pixels in radius. A lower response at the high end of the range
would cause a rise of this ratio with peak flux, and we do not see
such a trend.
We assume that the system is linear on the lower half of the peak flux 
range and compare
the average ratio at peak flux (in the 16 s exposure) below 
32,000 ADUs to the same quantity above 50,000 ADUs: we find differences 
below the per mil with uncertainties below the per mil as well. We hence 
conclude that possible average departures from linearity
affect bright star fluxes by less than one part in a thousand 
(which roughly corresponds to 1 \% at the pixel level). Possible non-linearities are therefore ignored.

\end{appendix}
\begin{appendix}
\section{Uncertainties and Covariance Matrices}
\label{sec:uncertainties_and_covariance_matrices}

We now report the various covariance matrices discussed in the
paper. All of them (tables
\ref{tab:megacam_mags_fundamental_standard_covmat}, 
\ref{tab:final_covariance_matrix} and 
\ref{tab:bd17_sed_covmat}) are available online.

First, we report the covariance matrix of the uncertainties affecting
the estimated MegaCam $g_{ref}$, $r_{ref}$, $i_{ref}$, and $z_{ref}$
and the Landolt $U_{ref}$, $B_{ref}$, $V_{ref}$, $R_{ref}$ and
$I_{ref}$ \bdtruc\ magnitudes. It is listed in table
\ref{tab:megacam_mags_fundamental_standard_covmat}. Note that this
matrix was established assuming that the Landolt $V_{ref}$,
$U-B_{ref}$, $B-V_{ref}$, $V-R_{ref}$, $R-I_{ref}$ and $V-I_{ref}$
measurements are all independent. As discussed in \S
\ref{sec:flux_interpretation_of_megacam_magnitudes}, this is
equivalent to assuming that the Landolt calibrated magnitudes are all
affected by an overall ``gray scale'' uncertainty, which correlate
them positively and that the Landolt $V$-band uncertainty is a good
estimate of it. This gray scale uncertainty does not affect the
Landolt colors though.

\begin{table*}
\caption{Full covariance matrix of the \bdtruc\ MegaCam and Landolt
  magnitude uncertainties. See \S
  \ref{sec:flux_interpretation_of_megacam_magnitudes} for
  details. \label{tab:megacam_mags_fundamental_standard_covmat}}
\begin{equation*}  
  {\mathbf V}_{\bdtruc} = 10^{-6} \times \begin{pmatrix}
 12.3128 &  6.7600 &  6.7600 &  6.7600 &  7.8753 &  7.8753 &  6.7600 &  6.7600 &  6.7600 \\ 
      &  25.6028 &  6.7600 &  6.7600 &  6.7600 &  6.7600 &  6.7600 &  7.7698 &  6.7600 \\ 
      &        &  13.4847 &  8.4173 &  6.7600 &  6.7600 &  6.7600 &  6.7600 &  8.4500 \\ 
      &        &        &  324.4808 &  6.7600 &  6.7600 &  6.7600 &  6.7600 &  8.4500 \\ 
      &        &        &        &  13.4200 &  9.0100 &  6.7600 &  6.7600 &  6.7600 \\ 
      &        &        &        &        &  9.0100 &  6.7600 &  6.7600 &  6.7600 \\ 
      &        &        &        &        &        &  6.7600 &  6.7600 &  6.7600 \\ 
      &        &        &        &        &        &        &  7.9700 &  6.7600 \\ 
      &        &        &        &        &        &        &        &  8.4500 \\    
  \end{pmatrix}  
  \ \ \ \ \ \ \ \ \ 
  \begin{matrix} 
    g_{ref} \\
    r_{ref} \\
    i_{ref} \\
    z_{ref} \\
    U_{ref} \\
    B_{ref} \\
    V_{ref} \\
    R_{ref} \\
    I_{ref} \\
  \end{matrix}
\end{equation*}
\end{table*}

Then, we also report the covariance matrix of the uncertainties
affecting the MegaCam tertiary magnitudes. As discussed in \S
\ref{sec:flux_interpretation_of_megacam_magnitudes}, the relevant
quantities used to map calibrated magnitudes into fluxes are the
differences between the tertiary standard magnitudes and the
magnitudes of \bdtruc: $\gme - g_{ref} \ldots \zme - z_{ref}$. By
construction, these uncertainties are correlated to the Landolt
magnitudes of \bdtruc. Hence, we report in table
\ref{tab:final_covariance_matrix} the covariance matrix of the $\gme -
g_{ref}$, \ldots $\zme - z_{ref}$, $U_{ref}$, \ldots $I_{ref}$
magnitude uncertainties. This matrix contains (1) the statistical
uncertainties affecting the zero points and the Landolt-to-MegaCam
color transformation slopes and (2) the systematic uncertainties
listed in the upper part of table
\ref{tab:systematic_uncertainties}. As discussed in \S
\ref{sec:tertiary_catalogs}, the influence of the color transformation
slopes is greatly reduced thanks to the fact that the colors of
\bdtruc\ is much closer to the average colors of the Landolt stars.

\begin{figure}
\begin{center}
\includegraphics[width=0.95\linewidth]{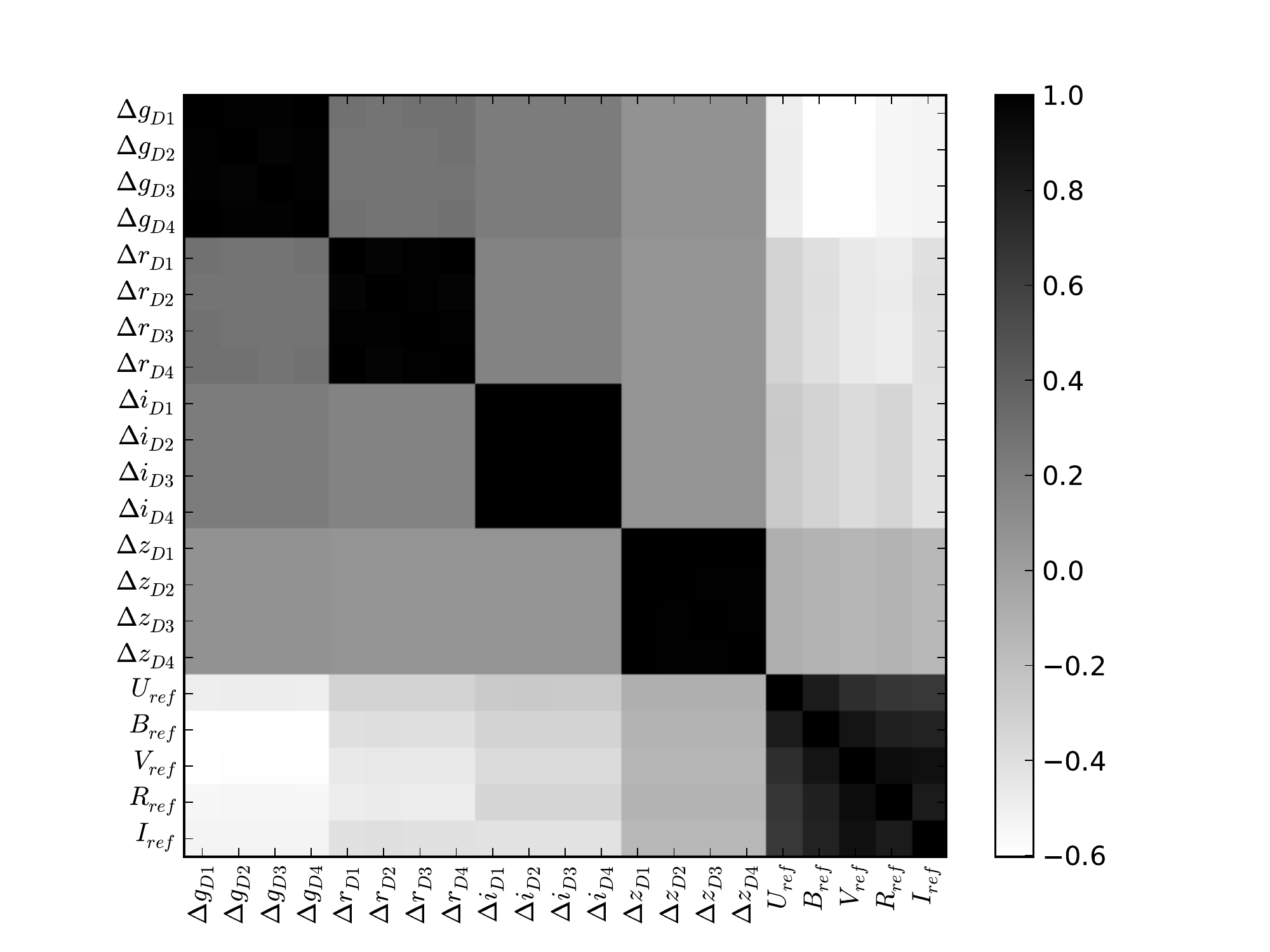}
\caption{Correlation matrix of the (statistical and systematic)
  uncertainties affecting the $\Delta g_{D1} = g - g_{ref} (D1)$,
  \ldots $\Delta g_{D4} = g - g_{ref} (D4)$ \ldots\ $\Delta z(D1) = z
  - z_{ref}(D1)$ \ldots\ $\Delta z_{D4} = z - z_{ref}(D4)$, $U_{ref}$
  \ldots\ $I_{ref}$ magnitudes. As can be seen, the uncertainties on
  each field's tertiary star magnitudes are almost fully correlated in
  each band. Hence, we report only an average $9 \times 9$ matrix,
  valid for each field. The non-zero correlations between the bands
  come from the overall ``gray scale uncertainty'' that affect the
  Landolt magnitudes. \label{fig:full_covmat_split_by_fields}}
\end{center}
\end{figure}

In principle, we could have separately reported the
uncertainties affecting each field's tertiary standards. Indeed, some
SNLS fields are better observed that others. However, we have noticed
that the total uncertainty budget is dominated by the
systematics. Hence, the $\gme - g_{ref}$ \ldots $\zme - z_{ref}$
uncertainties are almost fully correlated, as can be seen on figure
\ref{fig:full_covmat_split_by_fields}. Hence, we have decided to
report only an ``average'' $9 \times 9$ matrix, valid for each SNLS
field, instead of the full $21 \times 21$ matrix.

\begin{table*}
\caption{Final covariance matrix of the uncertainties affecting (1)
  the {\em differences} between the tertiary MegaCam magnitudes and
  the MegaCam magnitudes of \bdtruc\ and (2) the Landolt magnitudes of
  \bdtruc. \label{tab:final_covariance_matrix} }
\begin{equation*}
{\mathbf V}_{final} = 10^{-6} \times \begin{pmatrix}
21.2013 &   6.7600 &   6.7600 &    6.7600 &  -7.8753 &  -7.8753 &  -6.7600 &  -6.7600 &  -6.7600 \\ 
        &  34.0033 &   6.7600 &    6.7600 &  -6.7600 &  -6.7600 &  -6.7600 &  -7.7698 &  -6.7600 \\ 
        &          &  50.5137 &    8.4173 &  -6.7600 &  -6.7600 &  -6.7600 &  -6.7600 &  -8.4500 \\ 
        &          &          &  345.0753 &  -6.7600 &  -6.7600 &  -6.7600 &  -6.7600 &  -8.4500 \\ 
        &          &          &           &  13.4200 &   9.0100 &   6.7600 &   6.7600 &   6.7600 \\ 
        &          &          &           &          &   9.0100 &   6.7600 &   6.7600 &   6.7600 \\ 
        &          &          &           &          &          &   6.7600 &   6.7600 &   6.7600 \\ 
        &          &          &           &          &          &          &   7.9700 &   6.7600 \\ 
        &          &          &           &          &          &          &          &   8.4500 \\ 
\end{pmatrix}
  \ \ \ \ \ \ \ \ \ 
  \begin{matrix} 
    \gme - g_{ref} \\
    \rme - r_{ref} \\
    \ime - i_{ref} \\
    \zme - z_{ref} \\
    U_{ref} \\
    B_{ref} \\
    V_{ref} \\
    R_{ref} \\
    I_{ref} \\
  \end{matrix}
\end{equation*}
\end{table*}

Finally, there is a last set of (systematic) uncertainties, which
affects only the calibrated {\em fluxes} derived from the tertiary
standard magnitudes: the errors related to the measurement of the SED
of \bdtruc, including the repeatability of the STIS measurements, and
the absolute calibration of the STIS instrument. These uncertainties
are discussed in \S \ref{sec:systematic_uncertainties}. The associated 
covariance matrix is reported in table \ref{tab:bd17_sed_covmat}.

\begin{table*}
  \caption{Covariance matrix of the uncertainties affecting the
    synthetic instrumental magnitudes of
    \bdtruc. \label{tab:bd17_sed_covmat}}
  \begin{equation*}
    {\mathbf V}_{SED_{BD +17}} = 10^{-6} \times \begin{pmatrix}
5.2760 &  1.1170 &  -2.0222  &  -6.1807 &   6.2336 &   6.0589 &  3.2774 &   0.6793 &  -3.2182 \\ 
       &  8.9793 &   7.8773  &   9.9974 &   2.3035 &   1.7390 &  3.6392 &   8.9824 &   8.3187 \\ 
       &         &   14.2570 &  23.3526 &  -3.5755 &  -2.5778 &  1.7173 &   8.8588 &  16.9508 \\ 
       &         &           &  46.0064 &  -9.7314 &  -7.8124 &  0.5591 &  11.9371 &  29.7526 \\ 
       &         &           &          &  11.0517 &   8.4008 &  3.6545 &   1.4931 &  -5.4785 \\ 
       &         &           &          &          &   7.5107 &  3.6710 &   1.1458 &  -4.1358 \\ 
       &         &           &          &          &          &  3.2572 &   3.4286 &   1.3518 \\ 
       &         &           &          &          &          &         &   9.1231 &   9.6018 \\ 
       &         &           &          &          &          &         &          &  20.7291 \\ 
    \end{pmatrix}
\ \ \ \ \ \ 
  \begin{matrix} 
    g_{ref} \\
    r_{ref} \\
    i_{ref} \\
    z_{ref} \\
    U_{ref} \\
    B_{ref} \\
    V_{ref} \\
    R_{ref} \\
    I_{ref} \\
  \end{matrix}
  \end{equation*}
\end{table*}

\end{appendix}

\begin{appendix}
\section{Tertiary Catalogs}
\label{sec:tertiary_catalog_tables}

The tertiary catalogs built for the SNLS fields D1, D2, D3 and D4 are
listed in tables \ref{tab:d1_tertiaries}, \ref{tab:d2_tertiaries},
\ref{tab:d3_tertiaries} and \ref{tab:d4_tertiaries} respectively.

They include the $\ume$ band calibrated magnitudes, derived as
discussed in the previous section. For each star, we report the Local
Natural Magnitudes, along with the $\delta k(\x)$ grid coefficient at
the focal plane position where the star was observed. The Uniform
Magnitudes can then be derived using the approximate formulas:
\begin{eqnarray}
  u_{|\x_0} & = & u_{|\x} - \delta k_{uug}(\x) \times (u_{|\x} - g_{|\x}) \nonumber \\
  g_{|\x_0} & = & g_{|\x} - \delta k_{ggr}(\x) \times (g_{|\x} - r_{|\x}) \nonumber \\
  r_{|\x_0} & = & r_{|\x} - \delta k_{rri}(\x) \times (r_{|\x} - i_{|\x}) \nonumber \\
  i_{|\x_0} & = & i_{|\x} - \delta k_{iri}(\x) \times (r_{|\x} - i_{|\x}) \nonumber \\
  z_{|\x_0} & = & z_{|\x} - \delta k_{ziz}(\x) \times (z_{|\x} - i_{|\x}) 
\end{eqnarray}
Indeed, the error resulting from using the local color: $\delta k(\x)
\times {\rm col}_{|\x}$ instead of the color at the focal plane
reference location: $\delta k(x) \times {\rm col}_{|\x_0}$, as
requested by the grid definition, is smaller than 0.001 mag.

\onllongtab{19}{
\begin{landscape}
\begin{longtable}{cccccccccccccccccc}
\caption{Tertiary catalog for the SNLS field D1.\label{tab:d1_tertiaries}} \\
\hline
\hline
RA & DEC   
           & $\gme$ & $\sigma_{\gme}$ & $N_{\gme}$ & $\delta k_{ggr}$ 
           & $\rme$ & $\sigma_{\rme}$ & $N_{\rme}$ & $\delta k_{rri}$ 
           & $\ime$ & $\sigma_{\ime}$ & $N_{\ime}$ & $\delta k_{iri}$ 
           & $\zme$ & $\sigma_{\zme}$ & $N_{\zme}$ & $\delta k_{ziz}$ \\
(J2000)    & (J2000) 
           &         &                 &           &                        
           &         &                 &           &                        
           &         &                 &           &                        
           &         &                 &           &                        \\
\hline
$35.99905$ & $-4.30274$ & $20.282$ & $ 0.009$ & $3$ & $ 0.033$ & $18.768$ & $ 0.005$ & $6$ & $ 0.037$ & $17.716$ & $ 0.005$ & $7$ & $ 0.024$ & $17.263$ & $ 0.011$ & $3$ & $ 0.053$ \\
$36.00073$ & $-4.44487$ & $20.086$ & $ 0.003$ & $19$ & $ 0.029$ & $18.640$ & $ 0.003$ & $23$ & $ 0.033$ & $17.483$ & $ 0.002$ & $22$ & $ 0.019$ & $16.974$ & $ 0.005$ & $17$ & $ 0.043$ \\
$36.00134$ & $-4.61800$ & $17.520$ & $ 0.003$ & $17$ & $ 0.021$ & $16.727$ & $ 0.003$ & $16$ & $ 0.032$ & $16.312$ & $ 0.005$ & $5$ & $ 0.020$ & $16.124$ & $ 0.005$ & $14$ & $ 0.034$ \\
$36.00196$ & $-4.26749$ & --    &  --    & --  &  --     & $20.921$ & $ 0.010$ & $15$ & $ 0.037$ & $20.483$ & $ 0.008$ & $20$ & $ 0.024$ & --    &  --    & --  &  --     \\
$36.00311$ & $-4.16810$ & $20.873$ & $ 0.004$ & $25$ & $ 0.029$ & $20.122$ & $ 0.004$ & $22$ & $ 0.043$ & $19.706$ & $ 0.004$ & $30$ & $ 0.041$ & --    &  --    & --  &  --     \\
$36.00338$ & $-4.45709$ & $20.845$ & $ 0.004$ & $26$ & $ 0.029$ & $19.361$ & $ 0.003$ & $29$ & $ 0.033$ & $18.399$ & $ 0.002$ & $33$ & $ 0.019$ & $17.976$ & $ 0.005$ & $19$ & $ 0.043$ \\
$36.00416$ & $-4.00506$ & $20.662$ & $ 0.008$ & $5$ & $ 0.040$ & $19.356$ & $ 0.004$ & $10$ & $ 0.040$ & $18.657$ & $ 0.003$ & $15$ & $ 0.050$ & $18.292$ & $ 0.006$ & $18$ & $ 0.037$ \\
$36.00428$ & $-4.65547$ & $18.647$ & $ 0.003$ & $21$ & $ 0.021$ & $18.066$ & $ 0.002$ & $28$ & $ 0.032$ & $17.719$ & $ 0.002$ & $28$ & $ 0.020$ & $17.540$ & $ 0.004$ & $19$ & $ 0.034$ \\
$36.00460$ & $-4.85283$ & $19.996$ & $ 0.003$ & $23$ & $ 0.025$ & $19.556$ & $ 0.003$ & $29$ & $ 0.044$ & $19.267$ & $ 0.003$ & $31$ & $ 0.039$ & --    &  --    & --  &  --     \\
$36.00562$ & $-4.42695$ & --    &  --    & --  &  --     & $20.583$ & $ 0.004$ & $31$ & $ 0.033$ & $19.343$ & $ 0.002$ & $34$ & $ 0.019$ & $18.817$ & $ 0.006$ & $19$ & $ 0.043$ \\
$36.00588$ & $-4.81196$ & $20.736$ & $ 0.004$ & $23$ & $ 0.024$ & $19.257$ & $ 0.002$ & $29$ & $ 0.041$ & $17.770$ & $ 0.002$ & $32$ & $ 0.034$ & $17.107$ & $ 0.004$ & $19$ & $ 0.043$ \\
$36.00604$ & $-4.07668$ & --    &  --    & --  &  --     & $19.870$ & $ 0.003$ & $29$ & $ 0.040$ & $18.397$ & $ 0.002$ & $32$ & $ 0.048$ & $17.717$ & $ 0.004$ & $19$ & $ 0.042$ \\
$36.00673$ & $-4.77046$ & $20.013$ & $ 0.003$ & $26$ & $ 0.024$ & $19.509$ & $ 0.003$ & $29$ & $ 0.041$ & $19.229$ & $ 0.002$ & $34$ & $ 0.034$ & --    &  --    & --  &  --     \\
$36.00686$ & $-4.27621$ & $19.126$ & $ 0.002$ & $27$ & $ 0.033$ & $18.310$ & $ 0.002$ & $29$ & $ 0.037$ & $17.818$ & $ 0.002$ & $34$ & $ 0.024$ & $17.599$ & $ 0.004$ & $19$ & $ 0.053$ \\
$36.00716$ & $-4.53652$ & $18.682$ & $ 0.002$ & $27$ & $ 0.020$ & $17.831$ & $ 0.002$ & $24$ & $ 0.024$ & $17.397$ & $ 0.002$ & $29$ & $ 0.021$ & $17.187$ & $ 0.004$ & $18$ & $ 0.028$ \\
$36.00852$ & $-4.96286$ & --    &  --    & --  &  --     & --    &  --    & --  &  --     & $20.346$ & $ 0.005$ & $32$ & $ 0.042$ & --    &  --    & --  &  --     \\
$36.00870$ & $-4.32838$ & $20.176$ & $ 0.003$ & $28$ & $ 0.033$ & $18.770$ & $ 0.002$ & $32$ & $ 0.037$ & $17.700$ & $ 0.002$ & $35$ & $ 0.024$ & $17.251$ & $ 0.004$ & $19$ & $ 0.053$ \\
$36.00882$ & $-4.41460$ & --    &  --    & --  &  --     & --    &  --    & --  &  --     & $20.204$ & $ 0.003$ & $34$ & $ 0.024$ & --    &  --    & --  &  --     \\
$36.00887$ & $-4.80380$ & --    &  --    & --  &  --     & $20.871$ & $ 0.004$ & $31$ & $ 0.041$ & $20.660$ & $ 0.005$ & $33$ & $ 0.034$ & --    &  --    & --  &  --     \\
$36.00898$ & $-4.54691$ & $18.720$ & $ 0.002$ & $28$ & $ 0.020$ & $17.312$ & $ 0.002$ & $32$ & $ 0.024$ & $16.325$ & $ 0.003$ & $16$ & $ 0.021$ & $15.868$ & $ 0.004$ & $16$ & $ 0.028$ \\
$36.00898$ & $-4.97885$ & --    &  --    & --  &  --     & $20.983$ & $ 0.006$ & $25$ & $ 0.040$ & $19.822$ & $ 0.004$ & $32$ & $ 0.042$ & --    &  --    & --  &  --     \\
$36.00906$ & $-4.87077$ & --    &  --    & --  &  --     & --    &  --    & --  &  --     & $19.991$ & $ 0.003$ & $34$ & $ 0.039$ & --    &  --    & --  &  --     \\
$36.00914$ & $-4.56453$ & --    &  --    & --  &  --     & $20.394$ & $ 0.003$ & $32$ & $ 0.024$ & $18.824$ & $ 0.002$ & $35$ & $ 0.021$ & $18.135$ & $ 0.004$ & $19$ & $ 0.028$ \\
$36.00947$ & $-4.05131$ & --    &  --    & --  &  --     & --    &  --    & --  &  --     & $19.738$ & $ 0.003$ & $35$ & $ 0.050$ & --    &  --    & --  &  --     \\
$36.01037$ & $-4.09691$ & $17.649$ & $ 0.002$ & $27$ & $ 0.033$ & $16.753$ & $ 0.002$ & $27$ & $ 0.040$ & $16.270$ & $ 0.003$ & $11$ & $ 0.048$ & $16.018$ & $ 0.004$ & $18$ & $ 0.042$ \\
$36.01078$ & $-4.60296$ & --    &  --    & --  &  --     & --    &  --    & --  &  --     & $20.328$ & $ 0.003$ & $35$ & $ 0.020$ & --    &  --    & --  &  --     \\
$36.01084$ & $-4.01824$ & $17.405$ & $ 0.002$ & $26$ & $ 0.040$ & $16.364$ & $ 0.002$ & $21$ & $ 0.040$ & $15.844$ & $ 0.005$ & $5$ & $ 0.050$ & $15.601$ & $ 0.004$ & $15$ & $ 0.037$ \\
$36.01106$ & $-4.39612$ & $19.801$ & $ 0.002$ & $27$ & $ 0.034$ & $18.398$ & $ 0.002$ & $32$ & $ 0.035$ & $17.334$ & $ 0.002$ & $36$ & $ 0.024$ & $16.885$ & $ 0.004$ & $19$ & $ 0.049$ \\
$36.01133$ & $-4.17180$ & $19.583$ & $ 0.002$ & $27$ & $ 0.029$ & $18.207$ & $ 0.002$ & $32$ & $ 0.043$ & $17.435$ & $ 0.002$ & $35$ & $ 0.041$ & $17.060$ & $ 0.004$ & $18$ & $ 0.041$ \\
$36.01161$ & $-4.16873$ & $18.587$ & $ 0.002$ & $27$ & $ 0.029$ & $17.464$ & $ 0.002$ & $32$ & $ 0.043$ & $16.863$ & $ 0.002$ & $33$ & $ 0.041$ & $16.565$ & $ 0.004$ & $19$ & $ 0.041$ \\
$36.01239$ & $-4.74109$ & --    &  --    & --  &  --     & $20.891$ & $ 0.004$ & $29$ & $ 0.035$ & $20.212$ & $ 0.004$ & $28$ & $ 0.021$ & --    &  --    & --  &  --     \\
$36.01324$ & $-4.83961$ & $17.036$ & $ 0.003$ & $24$ & $ 0.025$ & $16.306$ & $ 0.002$ & $19$ & $ 0.044$ & $15.899$ & $ 0.005$ & $4$ & $ 0.039$ & $15.719$ & $ 0.004$ & $16$ & $ 0.034$ \\
$36.01386$ & $-4.79779$ & --    &  --    & --  &  --     & --    &  --    & --  &  --     & $20.566$ & $ 0.004$ & $32$ & $ 0.034$ & --    &  --    & --  &  --     \\
$36.01520$ & $-4.90408$ & --    &  --    & --  &  --     & --    &  --    & --  &  --     & $20.887$ & $ 0.005$ & $31$ & $ 0.039$ & --    &  --    & --  &  --     \\
$36.01573$ & $-4.94662$ & $19.364$ & $ 0.002$ & $28$ & $ 0.030$ & $17.900$ & $ 0.002$ & $32$ & $ 0.040$ & $16.556$ & $ 0.002$ & $21$ & $ 0.042$ & $15.914$ & $ 0.003$ & $19$ & $ 0.030$ \\
$36.01626$ & $-4.79657$ & $19.247$ & $ 0.002$ & $28$ & $ 0.024$ & $18.490$ & $ 0.002$ & $32$ & $ 0.041$ & $18.055$ & $ 0.002$ & $35$ & $ 0.034$ & $17.843$ & $ 0.004$ & $19$ & $ 0.043$ \\
$36.01638$ & $-4.38600$ & --    &  --    & --  &  --     & $20.899$ & $ 0.003$ & $32$ & $ 0.035$ & $20.430$ & $ 0.003$ & $34$ & $ 0.024$ & --    &  --    & --  &  --     \\
\ldots & \ldots & 
                                  \ldots & \ldots & \ldots & \ldots &
                                  \ldots & \ldots & \ldots & \ldots &
                                  \ldots & \ldots & \ldots & \ldots &
                                  \ldots & \ldots & \ldots & \ldots \\ 
\hline\end{longtable}\end{landscape}
}
\onllongtab{20}{
\begin{landscape}
\begin{longtable}{cccccccccccccccccc}
\caption{Tertiary catalog for the SNLS field D2.\label{tab:d2_tertiaries}} \\
\hline
\hline
RA & DEC   
           & $\gme$ & $\sigma_{\gme}$ & $N_{\gme}$ & $\delta k_{ggr}$ 
           & $\rme$ & $\sigma_{\rme}$ & $N_{\rme}$ & $\delta k_{rri}$ 
           & $\ime$ & $\sigma_{\ime}$ & $N_{\ime}$ & $\delta k_{iri}$ 
           & $\zme$ & $\sigma_{\zme}$ & $N_{\zme}$ & $\delta k_{ziz}$ \\
(J2000)    & (J2000) 
           &         &                 &           &                        
           &         &                 &           &                        
           &         &                 &           &                        
           &         &                 &           &                        \\
\hline
$149.62000$ & $2.32533$ & --    &  --    & --  &  --     & --    &  --    & --  &  --     & $19.279$ & $ 0.005$ & $12$ & $ 0.024$ & $18.541$ & $ 0.013$ & $7$ & $ 0.049$ \\
$149.62034$ & $2.47045$ & $20.461$ & $ 0.007$ & $7$ & $ 0.029$ & $20.018$ & $ 0.006$ & $9$ & $ 0.043$ & $19.753$ & $ 0.007$ & $10$ & $ 0.041$ & --    &  --    & --  &  --     \\
$149.62099$ & $2.12191$ & --    &  --    & --  &  --     & $19.902$ & $ 0.005$ & $14$ & $ 0.024$ & $18.885$ & $ 0.004$ & $15$ & $ 0.021$ & $18.428$ & $ 0.011$ & $8$ & $ 0.034$ \\
$149.62127$ & $2.28208$ & $20.950$ & $ 0.006$ & $13$ & $ 0.034$ & $19.544$ & $ 0.003$ & $21$ & $ 0.035$ & $18.108$ & $ 0.003$ & $20$ & $ 0.019$ & $17.509$ & $ 0.007$ & $9$ & $ 0.043$ \\
$149.62128$ & $2.05220$ & --    &  --    & --  &  --     & $20.943$ & $ 0.009$ & $14$ & $ 0.032$ & $19.227$ & $ 0.004$ & $14$ & $ 0.020$ & $18.485$ & $ 0.012$ & $7$ & $ 0.034$ \\
$149.62135$ & $2.18374$ & $18.549$ & $ 0.004$ & $9$ & $ 0.020$ & $17.867$ & $ 0.003$ & $18$ & $ 0.024$ & $17.464$ & $ 0.003$ & $16$ & $ 0.021$ & $17.347$ & $ 0.007$ & $8$ & $ 0.028$ \\
$149.62140$ & $2.51753$ & --    &  --    & --  &  --     & $20.898$ & $ 0.008$ & $14$ & $ 0.043$ & $19.744$ & $ 0.005$ & $15$ & $ 0.041$ & --    &  --    & --  &  --     \\
$149.62149$ & $2.67409$ & --    &  --    & --  &  --     & $20.668$ & $ 0.007$ & $14$ & $ 0.040$ & $20.041$ & $ 0.007$ & $14$ & $ 0.050$ & --    &  --    & --  &  --     \\
$149.62175$ & $2.68895$ & --    &  --    & --  &  --     & --    &  --    & --  &  --     & $20.835$ & $ 0.012$ & $9$ & $ 0.050$ & --    &  --    & --  &  --     \\
$149.62177$ & $2.20355$ & $18.592$ & $ 0.004$ & $11$ & $ 0.029$ & $18.000$ & $ 0.003$ & $17$ & $ 0.033$ & $17.642$ & $ 0.003$ & $18$ & $ 0.019$ & $17.531$ & $ 0.006$ & $9$ & $ 0.028$ \\
$149.62230$ & $2.57420$ & --    &  --    & --  &  --     & $20.297$ & $ 0.005$ & $19$ & $ 0.040$ & $19.631$ & $ 0.004$ & $18$ & $ 0.048$ & --    &  --    & --  &  --     \\
$149.62263$ & $1.92858$ & $19.642$ & $ 0.005$ & $9$ & $ 0.024$ & $18.292$ & $ 0.003$ & $15$ & $ 0.041$ & $17.462$ & $ 0.003$ & $14$ & $ 0.034$ & $17.085$ & $ 0.007$ & $8$ & $ 0.043$ \\
$149.62275$ & $2.51222$ & --    &  --    & --  &  --     & $20.832$ & $ 0.008$ & $16$ & $ 0.043$ & $18.985$ & $ 0.003$ & $21$ & $ 0.041$ & $18.223$ & $ 0.007$ & $10$ & $ 0.041$ \\
$149.62294$ & $2.11823$ & $17.441$ & $ 0.003$ & $13$ & $ 0.021$ & $16.822$ & $ 0.003$ & $17$ & $ 0.032$ & $16.465$ & $ 0.004$ & $8$ & $ 0.020$ & $16.305$ & $ 0.006$ & $9$ & $ 0.034$ \\
$149.62313$ & $2.40304$ & $18.341$ & $ 0.004$ & $13$ & $ 0.033$ & $17.832$ & $ 0.003$ & $19$ & $ 0.037$ & --    &  --    & --  &  --     & $17.417$ & $ 0.016$ & $2$ & $ 0.053$ \\
$149.62320$ & $2.50133$ & $20.344$ & $ 0.005$ & $10$ & $ 0.029$ & $19.667$ & $ 0.004$ & $18$ & $ 0.043$ & $19.234$ & $ 0.003$ & $22$ & $ 0.041$ & --    &  --    & --  &  --     \\
$149.62353$ & $2.50588$ & --    &  --    & --  &  --     & $20.410$ & $ 0.006$ & $16$ & $ 0.043$ & $19.174$ & $ 0.003$ & $21$ & $ 0.041$ & $18.641$ & $ 0.008$ & $10$ & $ 0.041$ \\
$149.62382$ & $2.64884$ & $18.973$ & $ 0.004$ & $12$ & $ 0.040$ & $17.916$ & $ 0.003$ & $21$ & $ 0.040$ & $17.353$ & $ 0.003$ & $18$ & $ 0.050$ & $17.090$ & $ 0.006$ & $8$ & $ 0.037$ \\
$149.62427$ & $2.34592$ & --    &  --    & --  &  --     & --    &  --    & --  &  --     & $20.848$ & $ 0.007$ & $20$ & $ 0.024$ & --    &  --    & --  &  --     \\
$149.62444$ & $2.16455$ & --    &  --    & --  &  --     & --    &  --    & --  &  --     & $20.803$ & $ 0.007$ & $21$ & $ 0.021$ & --    &  --    & --  &  --     \\
$149.62516$ & $2.13512$ & $17.438$ & $ 0.003$ & $13$ & $ 0.020$ & $16.817$ & $ 0.003$ & $18$ & $ 0.024$ & $16.478$ & $ 0.004$ & $10$ & $ 0.021$ & $16.320$ & $ 0.005$ & $10$ & $ 0.028$ \\
$149.62525$ & $2.17352$ & $20.459$ & $ 0.005$ & $12$ & $ 0.020$ & $19.001$ & $ 0.003$ & $21$ & $ 0.024$ & $17.972$ & $ 0.002$ & $23$ & $ 0.021$ & $17.539$ & $ 0.006$ & $10$ & $ 0.028$ \\
$149.62570$ & $2.50842$ & $20.549$ & $ 0.005$ & $12$ & $ 0.029$ & $20.088$ & $ 0.004$ & $19$ & $ 0.043$ & $19.771$ & $ 0.004$ & $21$ & $ 0.041$ & --    &  --    & --  &  --     \\
$149.62603$ & $2.44159$ & --    &  --    & --  &  --     & --    &  --    & --  &  --     & $20.360$ & $ 0.005$ & $21$ & $ 0.024$ & --    &  --    & --  &  --     \\
$149.62679$ & $1.83224$ & $20.165$ & $ 0.005$ & $11$ & $ 0.025$ & $18.777$ & $ 0.003$ & $21$ & $ 0.044$ & $17.656$ & $ 0.002$ & $22$ & $ 0.039$ & $17.149$ & $ 0.005$ & $10$ & $ 0.034$ \\
$149.62693$ & $2.67197$ & --    &  --    & --  &  --     & $20.867$ & $ 0.006$ & $16$ & $ 0.040$ & $19.072$ & $ 0.003$ & $19$ & $ 0.050$ & $18.245$ & $ 0.007$ & $9$ & $ 0.037$ \\
$149.62720$ & $2.43592$ & --    &  --    & --  &  --     & $20.383$ & $ 0.004$ & $21$ & $ 0.037$ & $19.929$ & $ 0.004$ & $24$ & $ 0.024$ & --    &  --    & --  &  --     \\
$149.62724$ & $2.00441$ & $16.690$ & $ 0.004$ & $10$ & $ 0.021$ & $16.129$ & $ 0.005$ & $7$ & $ 0.035$ & $15.799$ & $ 0.007$ & $3$ & $ 0.021$ & $15.665$ & $ 0.005$ & $9$ & $ 0.037$ \\
$149.62735$ & $2.66318$ & --    &  --    & --  &  --     & --    &  --    & --  &  --     & $20.590$ & $ 0.006$ & $21$ & $ 0.050$ & --    &  --    & --  &  --     \\
$149.62741$ & $1.79608$ & $20.114$ & $ 0.004$ & $13$ & $ 0.025$ & $18.731$ & $ 0.003$ & $21$ & $ 0.044$ & $17.766$ & $ 0.003$ & $21$ & $ 0.039$ & $17.321$ & $ 0.005$ & $10$ & $ 0.034$ \\
$149.62757$ & $2.49767$ & $19.662$ & $ 0.004$ & $10$ & $ 0.029$ & $18.302$ & $ 0.003$ & $17$ & $ 0.043$ & $17.008$ & $ 0.003$ & $18$ & $ 0.041$ & $16.388$ & $ 0.006$ & $9$ & $ 0.041$ \\
$149.62773$ & $1.86667$ & --    &  --    & --  &  --     & $20.874$ & $ 0.006$ & $18$ & $ 0.044$ & $19.450$ & $ 0.003$ & $23$ & $ 0.039$ & $18.855$ & $ 0.008$ & $10$ & $ 0.034$ \\
$149.62796$ & $2.67539$ & --    &  --    & --  &  --     & --    &  --    & --  &  --     & $20.927$ & $ 0.008$ & $18$ & $ 0.050$ & --    &  --    & --  &  --     \\
$149.62824$ & $1.91106$ & $19.654$ & $ 0.004$ & $12$ & $ 0.024$ & $18.785$ & $ 0.003$ & $21$ & $ 0.041$ & $18.308$ & $ 0.002$ & $24$ & $ 0.034$ & $18.105$ & $ 0.006$ & $10$ & $ 0.043$ \\
$149.62848$ & $1.73792$ & $20.954$ & $ 0.006$ & $11$ & $ 0.030$ & $19.472$ & $ 0.003$ & $21$ & $ 0.040$ & $18.398$ & $ 0.003$ & $21$ & $ 0.042$ & $17.921$ & $ 0.006$ & $10$ & $ 0.030$ \\
$149.62865$ & $2.51922$ & $19.958$ & $ 0.004$ & $13$ & $ 0.029$ & $18.476$ & $ 0.003$ & $21$ & $ 0.043$ & $17.228$ & $ 0.002$ & $23$ & $ 0.041$ & $16.675$ & $ 0.005$ & $10$ & $ 0.041$ \\
$149.62915$ & $2.64270$ & $20.923$ & $ 0.005$ & $13$ & $ 0.040$ & $19.525$ & $ 0.003$ & $21$ & $ 0.040$ & $18.742$ & $ 0.003$ & $25$ & $ 0.050$ & $18.387$ & $ 0.007$ & $10$ & $ 0.037$ \\
\ldots & \ldots & 
                                  \ldots & \ldots & \ldots & \ldots &
                                  \ldots & \ldots & \ldots & \ldots &
                                  \ldots & \ldots & \ldots & \ldots &
                                  \ldots & \ldots & \ldots & \ldots \\ 
\hline\end{longtable}\end{landscape}
}
\onllongtab{21}{
\begin{landscape}
\begin{longtable}{cccccccccccccccccc}
\caption{Tertiary catalog for the SNLS field D3.\label{tab:d3_tertiaries}} \\
\hline
\hline
RA & DEC   
           & $\gme$ & $\sigma_{\gme}$ & $N_{\gme}$ & $\delta k_{ggr}$ 
           & $\rme$ & $\sigma_{\rme}$ & $N_{\rme}$ & $\delta k_{rri}$ 
           & $\ime$ & $\sigma_{\ime}$ & $N_{\ime}$ & $\delta k_{iri}$ 
           & $\zme$ & $\sigma_{\zme}$ & $N_{\zme}$ & $\delta k_{ziz}$ \\
(J2000)    & (J2000) 
           &         &                 &           &                        
           &         &                 &           &                        
           &         &                 &           &                        
           &         &                 &           &                        \\
\hline
$214.04057$ & $53.08957$ & --    &  --    & --  &  --     & --    &  --    & --  &  --     & $20.073$ & $ 0.005$ & $30$ & $ 0.048$ & --    &  --    & --  &  --     \\
$214.04316$ & $52.79265$ & $20.378$ & $ 0.004$ & $19$ & $ 0.034$ & $18.946$ & $ 0.002$ & $28$ & $ 0.035$ & $17.279$ & $ 0.002$ & $29$ & $ 0.024$ & $16.593$ & $ 0.006$ & $8$ & $ 0.049$ \\
$214.04573$ & $52.83204$ & $19.926$ & $ 0.003$ & $27$ & $ 0.034$ & $18.673$ & $ 0.002$ & $32$ & $ 0.035$ & $17.985$ & $ 0.002$ & $39$ & $ 0.024$ & $17.677$ & $ 0.007$ & $10$ & $ 0.049$ \\
$214.04591$ & $53.05735$ & --    &  --    & --  &  --     & $20.777$ & $ 0.004$ & $31$ & $ 0.040$ & $20.161$ & $ 0.004$ & $35$ & $ 0.048$ & --    &  --    & --  &  --     \\
$214.04705$ & $52.63229$ & --    &  --    & --  &  --     & $20.749$ & $ 0.005$ & $28$ & $ 0.024$ & $19.037$ & $ 0.003$ & $32$ & $ 0.021$ & $18.310$ & $ 0.009$ & $9$ & $ 0.028$ \\
$214.04802$ & $52.66266$ & $17.936$ & $ 0.003$ & $25$ & $ 0.020$ & $16.498$ & $ 0.003$ & $15$ & $ 0.024$ & --    &  --    & --  &  --     & $15.027$ & $ 0.008$ & $4$ & $ 0.028$ \\
$214.04907$ & $52.59444$ & $20.944$ & $ 0.004$ & $25$ & $ 0.020$ & $19.554$ & $ 0.003$ & $29$ & $ 0.024$ & $18.505$ & $ 0.002$ & $34$ & $ 0.020$ & $18.030$ & $ 0.007$ & $10$ & $ 0.034$ \\
$214.04999$ & $52.47645$ & $18.327$ & $ 0.004$ & $14$ & $ 0.021$ & $17.435$ & $ 0.003$ & $21$ & $ 0.035$ & $16.947$ & $ 0.003$ & $18$ & $ 0.021$ & $16.714$ & $ 0.007$ & $8$ & $ 0.037$ \\
$214.05001$ & $53.17555$ & $20.731$ & $ 0.004$ & $16$ & $ 0.040$ & $20.247$ & $ 0.004$ & $17$ & $ 0.040$ & $19.964$ & $ 0.005$ & $27$ & $ 0.050$ & --    &  --    & --  &  --     \\
$214.05245$ & $52.27888$ & --    &  --    & --  &  --     & $17.808$ & $ 0.004$ & $7$ & $ 0.044$ & $16.947$ & $ 0.004$ & $8$ & $ 0.039$ & $16.541$ & $ 0.009$ & $4$ & $ 0.034$ \\
$214.05296$ & $52.83856$ & --    &  --    & --  &  --     & $20.885$ & $ 0.003$ & $33$ & $ 0.035$ & $19.240$ & $ 0.002$ & $39$ & $ 0.024$ & $18.604$ & $ 0.008$ & $10$ & $ 0.049$ \\
$214.05366$ & $53.04250$ & $17.383$ & $ 0.003$ & $25$ & $ 0.033$ & $15.995$ & $ 0.004$ & $10$ & $ 0.040$ & --    &  --    & --  &  --     & $14.824$ & $ 0.010$ & $3$ & $ 0.042$ \\
$214.05610$ & $53.12693$ & --    &  --    & --  &  --     & $20.476$ & $ 0.003$ & $34$ & $ 0.040$ & $19.918$ & $ 0.003$ & $39$ & $ 0.050$ & --    &  --    & --  &  --     \\
$214.05647$ & $52.36586$ & --    &  --    & --  &  --     & --    &  --    & --  &  --     & $20.635$ & $ 0.007$ & $29$ & $ 0.034$ & --    &  --    & --  &  --     \\
$214.05660$ & $52.68815$ & $17.613$ & $ 0.002$ & $27$ & $ 0.029$ & $17.017$ & $ 0.002$ & $34$ & $ 0.033$ & $16.667$ & $ 0.002$ & $28$ & $ 0.019$ & $16.524$ & $ 0.005$ & $10$ & $ 0.043$ \\
$214.05815$ & $52.41449$ & $19.833$ & $ 0.008$ & $3$ & $ 0.024$ & $19.148$ & $ 0.004$ & $10$ & $ 0.041$ & $18.760$ & $ 0.003$ & $24$ & $ 0.034$ & $18.572$ & $ 0.010$ & $8$ & $ 0.043$ \\
$214.05909$ & $53.10057$ & $18.543$ & $ 0.002$ & $28$ & $ 0.040$ & $18.069$ & $ 0.002$ & $34$ & $ 0.040$ & $17.790$ & $ 0.002$ & $39$ & $ 0.050$ & $17.664$ & $ 0.006$ & $10$ & $ 0.037$ \\
$214.05940$ & $52.22233$ & $18.895$ & $ 0.004$ & $13$ & $ 0.030$ & $17.517$ & $ 0.003$ & $16$ & $ 0.040$ & $16.711$ & $ 0.003$ & $11$ & $ 0.042$ & $16.323$ & $ 0.006$ & $8$ & $ 0.030$ \\
$214.05970$ & $52.91299$ & $19.717$ & $ 0.003$ & $28$ & $ 0.033$ & $18.259$ & $ 0.002$ & $31$ & $ 0.037$ & $16.948$ & $ 0.002$ & $35$ & $ 0.024$ & $16.382$ & $ 0.005$ & $10$ & $ 0.053$ \\
$214.05972$ & $52.66918$ & --    &  --    & --  &  --     & $20.921$ & $ 0.003$ & $34$ & $ 0.024$ & $19.399$ & $ 0.002$ & $39$ & $ 0.021$ & $18.794$ & $ 0.008$ & $10$ & $ 0.028$ \\
$214.06035$ & $52.53434$ & --    &  --    & --  &  --     & $20.948$ & $ 0.004$ & $34$ & $ 0.032$ & $19.155$ & $ 0.002$ & $39$ & $ 0.020$ & $18.365$ & $ 0.007$ & $10$ & $ 0.034$ \\
$214.06214$ & $52.63167$ & $20.342$ & $ 0.003$ & $28$ & $ 0.020$ & $18.985$ & $ 0.002$ & $34$ & $ 0.024$ & $17.480$ & $ 0.002$ & $39$ & $ 0.021$ & $16.827$ & $ 0.005$ & $10$ & $ 0.028$ \\
$214.06233$ & $52.39881$ & --    &  --    & --  &  --     & --    &  --    & --  &  --     & $20.400$ & $ 0.004$ & $38$ & $ 0.034$ & --    &  --    & --  &  --     \\
$214.06299$ & $52.80484$ & $18.096$ & $ 0.002$ & $28$ & $ 0.034$ & $17.526$ & $ 0.002$ & $34$ & $ 0.035$ & $17.185$ & $ 0.002$ & $37$ & $ 0.024$ & $17.044$ & $ 0.005$ & $10$ & $ 0.049$ \\
$214.06355$ & $52.48590$ & --    &  --    & --  &  --     & $19.753$ & $ 0.002$ & $33$ & $ 0.035$ & $18.366$ & $ 0.002$ & $39$ & $ 0.021$ & $17.755$ & $ 0.006$ & $10$ & $ 0.037$ \\
$214.06365$ & $52.44070$ & --    &  --    & --  &  --     & --    &  --    & --  &  --     & $20.976$ & $ 0.007$ & $32$ & $ 0.021$ & --    &  --    & --  &  --     \\
$214.06534$ & $52.51261$ & $19.890$ & $ 0.003$ & $28$ & $ 0.021$ & $18.477$ & $ 0.002$ & $34$ & $ 0.035$ & $17.470$ & $ 0.002$ & $39$ & $ 0.021$ & $16.996$ & $ 0.005$ & $10$ & $ 0.037$ \\
$214.06552$ & $52.33159$ & --    &  --    & --  &  --     & --    &  --    & --  &  --     & $20.095$ & $ 0.003$ & $39$ & $ 0.039$ & --    &  --    & --  &  --     \\
$214.06559$ & $52.31804$ & $17.934$ & $ 0.002$ & $28$ & $ 0.025$ & $17.156$ & $ 0.002$ & $34$ & $ 0.044$ & $16.696$ & $ 0.002$ & $31$ & $ 0.039$ & $16.464$ & $ 0.005$ & $10$ & $ 0.034$ \\
$214.06638$ & $52.55993$ & $20.364$ & $ 0.003$ & $28$ & $ 0.021$ & $18.905$ & $ 0.002$ & $34$ & $ 0.032$ & $17.751$ & $ 0.002$ & $39$ & $ 0.020$ & $17.220$ & $ 0.005$ & $10$ & $ 0.034$ \\
$214.06685$ & $52.30431$ & --    &  --    & --  &  --     & $19.681$ & $ 0.002$ & $33$ & $ 0.044$ & $18.612$ & $ 0.002$ & $39$ & $ 0.039$ & $18.100$ & $ 0.006$ & $10$ & $ 0.034$ \\
$214.06686$ & $53.11036$ & --    &  --    & --  &  --     & --    &  --    & --  &  --     & $20.936$ & $ 0.005$ & $39$ & $ 0.050$ & --    &  --    & --  &  --     \\
$214.06701$ & $52.63149$ & $20.834$ & $ 0.003$ & $28$ & $ 0.020$ & $19.408$ & $ 0.002$ & $34$ & $ 0.024$ & $18.093$ & $ 0.002$ & $39$ & $ 0.021$ & $17.508$ & $ 0.006$ & $10$ & $ 0.028$ \\
$214.06710$ & $52.59462$ & $17.697$ & $ 0.002$ & $27$ & $ 0.021$ & $16.940$ & $ 0.002$ & $33$ & $ 0.032$ & $16.522$ & $ 0.002$ & $22$ & $ 0.021$ & $16.308$ & $ 0.005$ & $10$ & $ 0.034$ \\
$214.06754$ & $52.89887$ & $19.008$ & $ 0.002$ & $28$ & $ 0.033$ & $18.300$ & $ 0.002$ & $32$ & $ 0.037$ & $17.905$ & $ 0.002$ & $38$ & $ 0.024$ & $17.749$ & $ 0.007$ & $8$ & $ 0.053$ \\
$214.06777$ & $53.05638$ & --    &  --    & --  &  --     & --    &  --    & --  &  --     & $20.478$ & $ 0.004$ & $39$ & $ 0.048$ & --    &  --    & --  &  --     \\
$214.06783$ & $52.32591$ & --    &  --    & --  &  --     & $20.036$ & $ 0.002$ & $34$ & $ 0.044$ & $19.287$ & $ 0.002$ & $39$ & $ 0.039$ & $18.921$ & $ 0.009$ & $9$ & $ 0.034$ \\
\ldots & \ldots & 
                                  \ldots & \ldots & \ldots & \ldots &
                                  \ldots & \ldots & \ldots & \ldots &
                                  \ldots & \ldots & \ldots & \ldots &
                                  \ldots & \ldots & \ldots & \ldots \\ 
\hline\end{longtable}\end{landscape}
}
\onllongtab{22}{
\begin{landscape}
\begin{longtable}{cccccccccccccccccc}
\caption{Tertiary catalog for the SNLS field D4.\label{tab:d4_tertiaries}} \\
\hline
\hline
RA & DEC   
           & $\gme$ & $\sigma_{\gme}$ & $N_{\gme}$ & $\delta k_{ggr}$ 
           & $\rme$ & $\sigma_{\rme}$ & $N_{\rme}$ & $\delta k_{rri}$ 
           & $\ime$ & $\sigma_{\ime}$ & $N_{\ime}$ & $\delta k_{iri}$ 
           & $\zme$ & $\sigma_{\zme}$ & $N_{\zme}$ & $\delta k_{ziz}$ \\
(J2000)    & (J2000) 
           &         &                 &           &                        
           &         &                 &           &                        
           &         &                 &           &                        
           &         &                 &           &                        \\
\hline
$333.35788$ & $-17.79348$ & --    &  --    & --  &  --     & $20.685$ & $ 0.009$ & $14$ & $ 0.024$ & $19.508$ & $ 0.006$ & $18$ & $ 0.021$ & $18.979$ & $ 0.015$ & $10$ & $ 0.028$ \\
$333.35796$ & $-17.84145$ & --    &  --    & --  &  --     & $20.541$ & $ 0.009$ & $11$ & $ 0.032$ & $20.123$ & $ 0.009$ & $14$ & $ 0.020$ & --    &  --    & --  &  --     \\
$333.35919$ & $-17.64129$ & --    &  --    & --  &  --     & $19.762$ & $ 0.003$ & $31$ & $ 0.035$ & $18.954$ & $ 0.003$ & $26$ & $ 0.024$ & $18.648$ & $ 0.009$ & $16$ & $ 0.049$ \\
$333.35945$ & $-17.58542$ & --    &  --    & --  &  --     & --    &  --    & --  &  --     & $20.406$ & $ 0.007$ & $23$ & $ 0.024$ & --    &  --    & --  &  --     \\
$333.35955$ & $-17.45383$ & --    &  --    & --  &  --     & --    &  --    & --  &  --     & $20.456$ & $ 0.010$ & $17$ & $ 0.041$ & --    &  --    & --  &  --     \\
$333.35968$ & $-17.52410$ & --    &  --    & --  &  --     & $20.662$ & $ 0.006$ & $30$ & $ 0.037$ & $20.179$ & $ 0.006$ & $23$ & $ 0.024$ & --    &  --    & --  &  --     \\
$333.35984$ & $-17.39324$ & $20.821$ & $ 0.004$ & $28$ & $ 0.033$ & $19.977$ & $ 0.004$ & $27$ & $ 0.040$ & $19.466$ & $ 0.005$ & $24$ & $ 0.048$ & --    &  --    & --  &  --     \\
$333.36014$ & $-17.76888$ & $17.215$ & $ 0.002$ & $28$ & $ 0.020$ & $16.382$ & $ 0.003$ & $13$ & $ 0.024$ & $15.938$ & $ 0.006$ & $3$ & $ 0.021$ & $15.737$ & $ 0.004$ & $15$ & $ 0.028$ \\
$333.36015$ & $-18.06026$ & --    &  --    & --  &  --     & --    &  --    & --  &  --     & $20.166$ & $ 0.007$ & $21$ & $ 0.034$ & --    &  --    & --  &  --     \\
$333.36124$ & $-17.66393$ & $19.921$ & $ 0.003$ & $29$ & $ 0.029$ & $18.450$ & $ 0.002$ & $28$ & $ 0.033$ & $17.234$ & $ 0.002$ & $28$ & $ 0.019$ & $16.704$ & $ 0.004$ & $16$ & $ 0.043$ \\
$333.36128$ & $-17.38790$ & $19.486$ & $ 0.003$ & $26$ & $ 0.033$ & $18.828$ & $ 0.002$ & $32$ & $ 0.040$ & $18.428$ & $ 0.003$ & $26$ & $ 0.048$ & $18.273$ & $ 0.007$ & $16$ & $ 0.042$ \\
$333.36139$ & $-17.30919$ & $19.850$ & $ 0.003$ & $26$ & $ 0.040$ & $19.126$ & $ 0.003$ & $28$ & $ 0.040$ & $18.737$ & $ 0.003$ & $25$ & $ 0.050$ & $18.567$ & $ 0.008$ & $16$ & $ 0.037$ \\
$333.36183$ & $-17.76437$ & $17.665$ & $ 0.002$ & $28$ & $ 0.020$ & $17.086$ & $ 0.002$ & $29$ & $ 0.024$ & $16.776$ & $ 0.002$ & $22$ & $ 0.021$ & $16.649$ & $ 0.004$ & $17$ & $ 0.028$ \\
$333.36198$ & $-18.15114$ & $20.146$ & $ 0.003$ & $22$ & $ 0.025$ & $19.971$ & $ 0.004$ & $23$ & $ 0.044$ & $19.892$ & $ 0.005$ & $23$ & $ 0.042$ & --    &  --    & --  &  --     \\
$333.36212$ & $-17.52031$ & --    &  --    & --  &  --     & --    &  --    & --  &  --     & $20.728$ & $ 0.008$ & $20$ & $ 0.024$ & --    &  --    & --  &  --     \\
$333.36233$ & $-17.63395$ & $17.770$ & $ 0.002$ & $29$ & $ 0.034$ & $16.802$ & $ 0.002$ & $33$ & $ 0.035$ & $16.309$ & $ 0.003$ & $16$ & $ 0.024$ & $16.097$ & $ 0.004$ & $19$ & $ 0.049$ \\
$333.36247$ & $-17.26413$ & --    &  --    & --  &  --     & $20.044$ & $ 0.004$ & $28$ & $ 0.040$ & $18.830$ & $ 0.003$ & $28$ & $ 0.050$ & $18.273$ & $ 0.007$ & $16$ & $ 0.037$ \\
$333.36252$ & $-17.69630$ & $20.351$ & $ 0.003$ & $29$ & $ 0.029$ & $19.056$ & $ 0.002$ & $34$ & $ 0.033$ & $18.344$ & $ 0.002$ & $29$ & $ 0.019$ & $18.028$ & $ 0.005$ & $18$ & $ 0.043$ \\
$333.36290$ & $-17.68844$ & $19.345$ & $ 0.002$ & $30$ & $ 0.029$ & $17.842$ & $ 0.002$ & $33$ & $ 0.033$ & $16.563$ & $ 0.002$ & $22$ & $ 0.019$ & $16.033$ & $ 0.004$ & $18$ & $ 0.043$ \\
$333.36309$ & $-17.75397$ & $19.149$ & $ 0.002$ & $29$ & $ 0.020$ & $18.453$ & $ 0.002$ & $34$ & $ 0.024$ & $18.079$ & $ 0.002$ & $29$ & $ 0.021$ & $17.912$ & $ 0.005$ & $19$ & $ 0.028$ \\
$333.36337$ & $-17.84771$ & --    &  --    & --  &  --     & $20.525$ & $ 0.004$ & $33$ & $ 0.032$ & $20.149$ & $ 0.005$ & $28$ & $ 0.020$ & --    &  --    & --  &  --     \\
$333.36352$ & $-18.03406$ & --    &  --    & --  &  --     & $20.913$ & $ 0.006$ & $28$ & $ 0.041$ & $19.717$ & $ 0.004$ & $29$ & $ 0.034$ & --    &  --    & --  &  --     \\
$333.36386$ & $-18.14651$ & $20.709$ & $ 0.004$ & $24$ & $ 0.025$ & $19.197$ & $ 0.003$ & $29$ & $ 0.044$ & $17.800$ & $ 0.002$ & $27$ & $ 0.042$ & $17.157$ & $ 0.005$ & $14$ & $ 0.034$ \\
$333.36395$ & $-17.43115$ & --    &  --    & --  &  --     & --    &  --    & --  &  --     & $19.502$ & $ 0.004$ & $27$ & $ 0.041$ & $18.721$ & $ 0.008$ & $16$ & $ 0.041$ \\
$333.36439$ & $-18.02007$ & --    &  --    & --  &  --     & $20.318$ & $ 0.004$ & $30$ & $ 0.041$ & $19.057$ & $ 0.003$ & $29$ & $ 0.034$ & $18.484$ & $ 0.006$ & $17$ & $ 0.043$ \\
$333.36442$ & $-17.45408$ & --    &  --    & --  &  --     & --    &  --    & --  &  --     & $20.022$ & $ 0.006$ & $20$ & $ 0.041$ & --    &  --    & --  &  --     \\
$333.36459$ & $-17.21264$ & $18.992$ & $ 0.004$ & $10$ & $ 0.040$ & $17.581$ & $ 0.003$ & $13$ & $ 0.040$ & $16.582$ & $ 0.004$ & $8$ & $ 0.050$ & $16.100$ & $ 0.010$ & $4$ & $ 0.037$ \\
$333.36467$ & $-17.54705$ & $20.560$ & $ 0.003$ & $30$ & $ 0.033$ & $19.276$ & $ 0.002$ & $34$ & $ 0.037$ & $18.598$ & $ 0.002$ & $29$ & $ 0.024$ & $18.300$ & $ 0.006$ & $18$ & $ 0.053$ \\
$333.36483$ & $-18.15027$ & --    &  --    & --  &  --     & --    &  --    & --  &  --     & $20.736$ & $ 0.008$ & $21$ & $ 0.042$ & --    &  --    & --  &  --     \\
$333.36490$ & $-18.03335$ & --    &  --    & --  &  --     & --    &  --    & --  &  --     & $20.708$ & $ 0.007$ & $25$ & $ 0.034$ & --    &  --    & --  &  --     \\
$333.36497$ & $-17.41420$ & --    &  --    & --  &  --     & --    &  --    & --  &  --     & $20.862$ & $ 0.009$ & $18$ & $ 0.041$ & --    &  --    & --  &  --     \\
$333.36550$ & $-17.51455$ & $19.162$ & $ 0.002$ & $30$ & $ 0.033$ & $18.401$ & $ 0.002$ & $34$ & $ 0.037$ & $17.962$ & $ 0.002$ & $29$ & $ 0.024$ & $17.773$ & $ 0.005$ & $19$ & $ 0.053$ \\
$333.36579$ & $-17.95646$ & --    &  --    & --  &  --     & --    &  --    & --  &  --     & $20.198$ & $ 0.005$ & $27$ & $ 0.021$ & --    &  --    & --  &  --     \\
$333.36613$ & $-17.77279$ & $17.218$ & $ 0.002$ & $29$ & $ 0.020$ & $16.556$ & $ 0.002$ & $27$ & $ 0.024$ & $16.205$ & $ 0.003$ & $10$ & $ 0.021$ & $16.045$ & $ 0.004$ & $19$ & $ 0.028$ \\
$333.36655$ & $-17.51977$ & $20.637$ & $ 0.003$ & $30$ & $ 0.033$ & $19.208$ & $ 0.002$ & $34$ & $ 0.037$ & $17.913$ & $ 0.002$ & $29$ & $ 0.024$ & $17.375$ & $ 0.004$ & $19$ & $ 0.053$ \\
$333.36671$ & $-17.89921$ & $19.272$ & $ 0.002$ & $30$ & $ 0.021$ & $18.311$ & $ 0.002$ & $34$ & $ 0.035$ & $17.827$ & $ 0.002$ & $29$ & $ 0.021$ & $17.578$ & $ 0.004$ & $19$ & $ 0.037$ \\
$333.36710$ & $-17.47462$ & $17.946$ & $ 0.002$ & $27$ & $ 0.029$ & $17.193$ & $ 0.002$ & $31$ & $ 0.043$ & $16.767$ & $ 0.002$ & $23$ & $ 0.024$ & $16.558$ & $ 0.004$ & $16$ & $ 0.041$ \\
\ldots & \ldots & 
                                  \ldots & \ldots & \ldots & \ldots &
                                  \ldots & \ldots & \ldots & \ldots &
                                  \ldots & \ldots & \ldots & \ldots &
                                  \ldots & \ldots & \ldots & \ldots \\ 
\hline\end{longtable}\end{landscape}
}

\end{appendix}

\begin{appendix}
\section{Calibration of the $u_M$-band Data}
\label{sec:u_band_magnitudes}

The DEEP survey $\ume$-band data have also been analyzed. This
calibration is less robust however, given the difficulties inherent to
the calibration of near-UV data, and also given the small number of
$\ume$-band epochs. Indeed, the $\ume$-band dataset not being formally
part of the SNLS dataset, and the exposures are not time sequenced. In
this section, we list the main results obtained on the $\ume$-band
data.

\begin{figure*}
\centering
\mbox{\subfigure[$\delta zp_{u,u-g}(\x)$]{\includegraphics[width=0.45\linewidth]{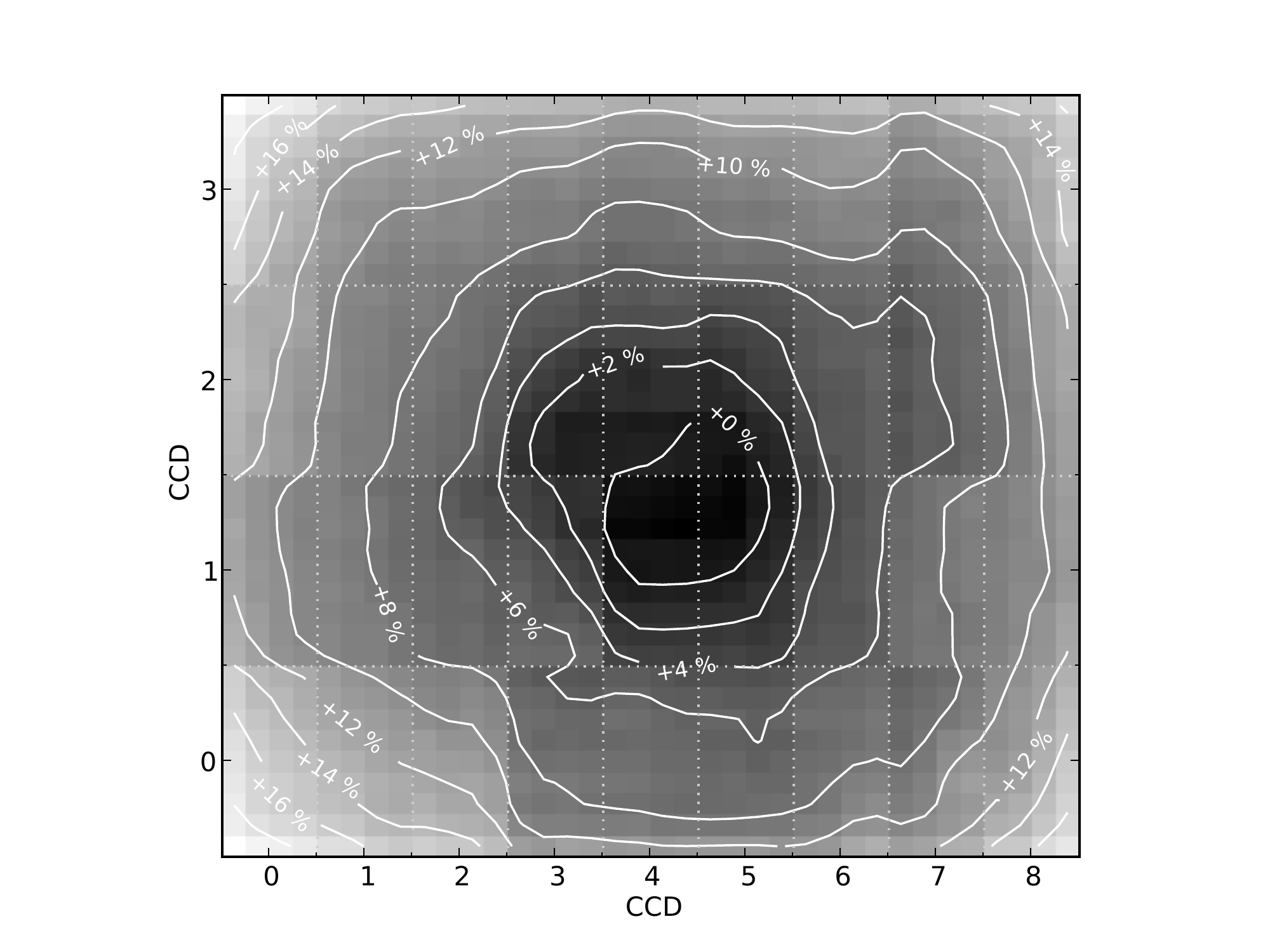}}\quad
      \subfigure[$\delta k_{u,u-g}(\x)$] {\includegraphics[width=0.45\linewidth]{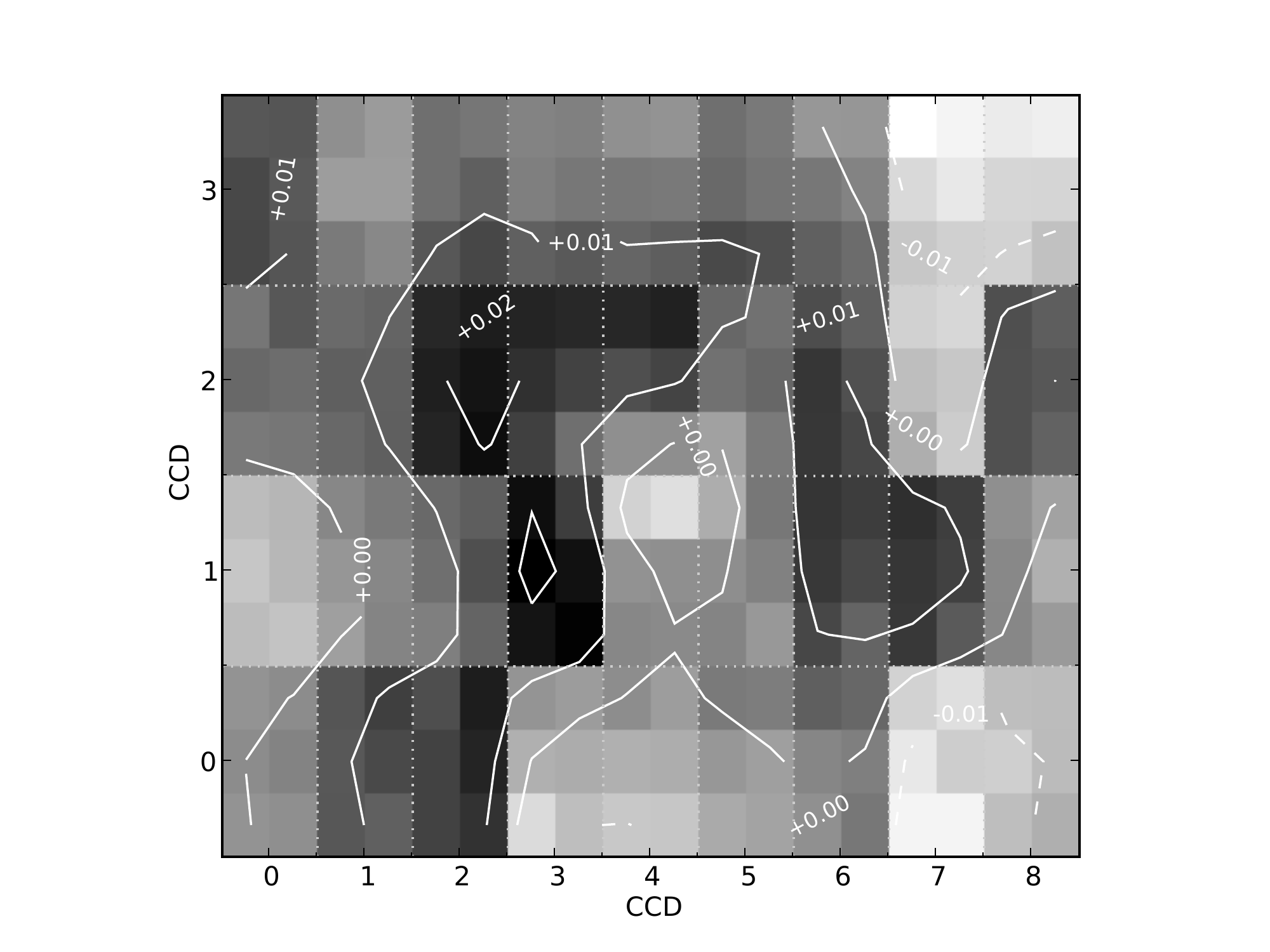}}}
\caption{$\ume$-band grid maps. As can be seen, the $\ume$ filter
  seems more uniform than the other band filters. Sharp variations can
  be observed in the $\delta k(\x)$ maps. This is due to the fact that
  the blue edge of the $\ume$-filter is determined by the quantum
  efficiency of each CCD. Hence, the passband variations.\label{fig:u_band_grid_dzp}}
\end{figure*}

The uniformity studies described in \S \ref{sec:the_photometric_grids}
have also been performed on the $\ume$-band data. Figure
\ref{fig:u_band_grid_dzp} presents one of the $\delta zp(\x)$ and $\delta
k(\x)$ maps obtained in semester 2005B. We note that the $\ume$-band
filter seems to be much more uniform than the other band filters. Also
there seem to be sharp passband variations between the various
CCDs. This is due to the fact that the blue edge of the MegaCam
$\ume$-band is determined by the CCD quantum efficiencies. 

The $\ume$-band Landolt observations have been analyzed with the same
procedure as the one described in this paper. The Landolt-to-MegaCam
color transformations were modeled using a piecewise-linear function,
with a break at $(U-B)_0 \sim 0.4$:
\begin{align*}
  u_{ADU|\x_0} & = & U - k_u \times (X-1) &+& \alpha_u \times (U-B) \\
              &   &                      & & {\rm if\ U-B < (U-B)_0} \\
  u_{ADU|\x_0} & = & U - k_u \times (X-1) & + & \alpha_u \times (U-B)_0 \\
             & &                         & + & \beta_u \times \left[(U-B) - (U-B)_0\right] \\
             & &                         &   & {\rm if\ U-B > (U-B)_0} \\
\end{align*}
We found:
\begin{eqnarray}
  \alpha_u & = & -0.2450 \pm 0.0108 \nonumber \\
  \beta_u  & = & -0.2787 \pm 0.0070 \nonumber \\
  k_u      & = & -0.0758 \pm 0.0087
\end{eqnarray}

The zero-points determined along with the global parameters
$\alpha_u$, $\beta_u$ and $k_u$ were applied to the tertiary
instrumental magnitudes which where then averaged to give final
calibrated magnitudes. Due to the smaller number of epochs, the night
selection was much less robust. We just remove the ``pathological
nights'' that display an absorption larger than 0.1 mag.

\begin{figure*}
\centering
\mbox{\subfigure[$\ume-\gme\ vs. \gme-\rme$]{\includegraphics[width=0.45\linewidth]{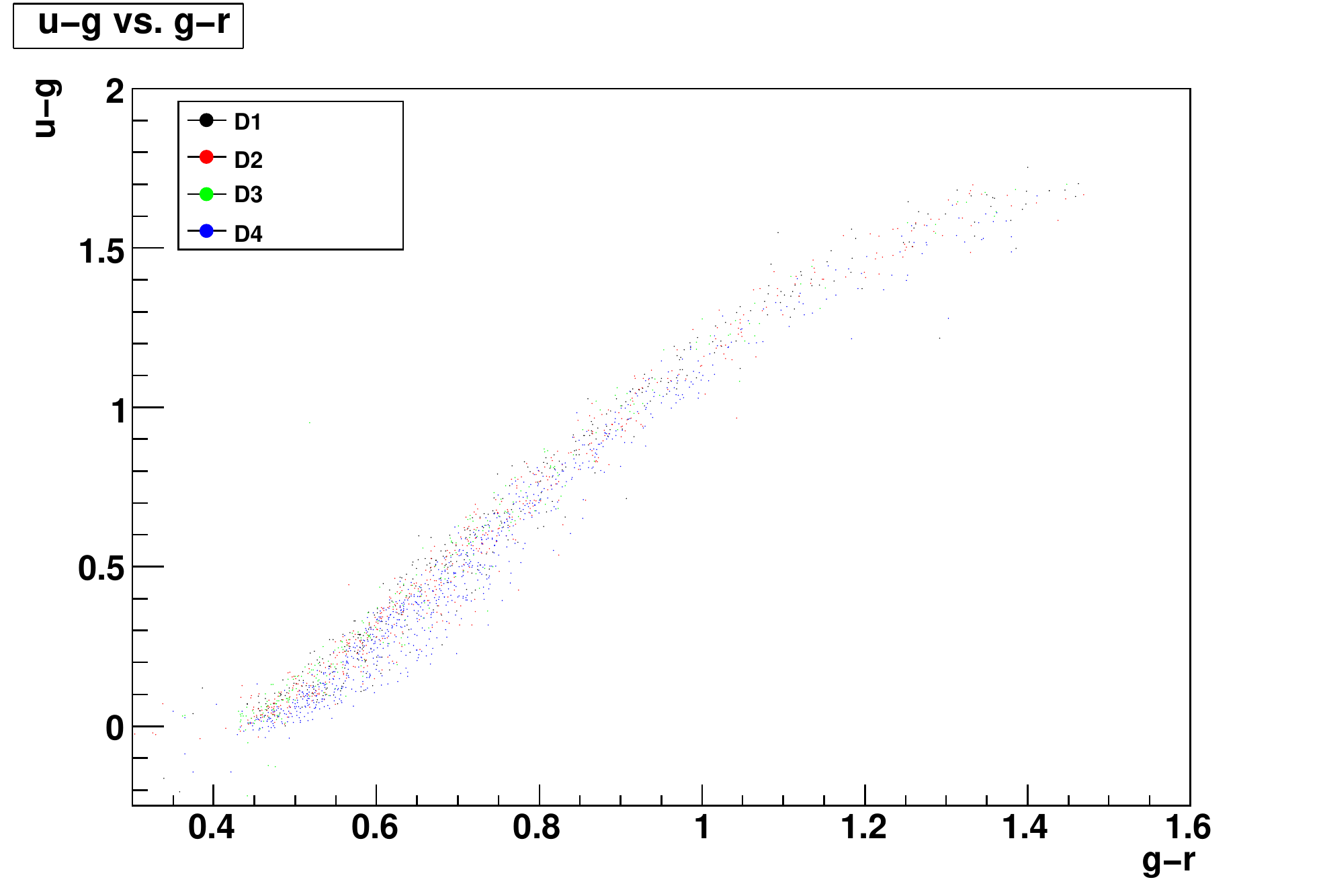}}
      \subfigure[$\ume-\gme\ vs. \gme-\rme$ (profile)]{\includegraphics[width=0.45\linewidth]{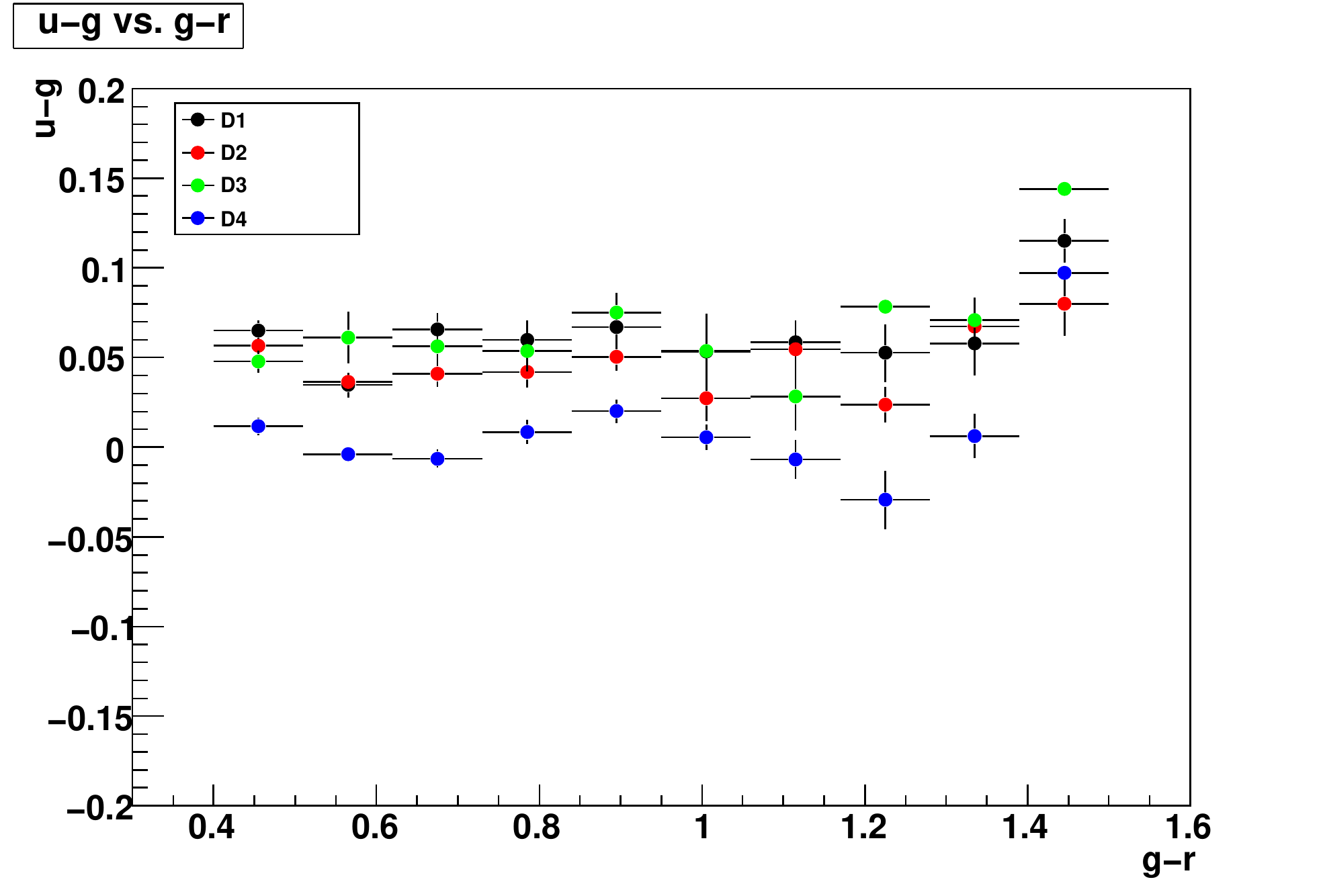}}}
\caption{Left: $\ume - \gme$ versus $\gme - \rme$ color-color plot of
  the tertiary standards.  Right: a zoom on the $0.4 < \ume - \gme <
  1.5$ region of the color-color plot, after subtraction of the mean
  slope. We see that the $\ume - \gme$ colors of the three D1, D2 and
  D3 fields agree within one percent. The origin of the systematic
  shift which affect the D4 colors is not clearly identified.
  \label{fig:u_band_color_color_plots}}
\end{figure*}

Figure \ref{fig:u_band_color_color_plots} presents the $\ume-\gme$
versus $\gme-\rme$ color-color diagram for all four SNLS fields. As
can be seen, the $\ume-\gme$ colors of the three D1, D2 and D3 fields
agree within one percent. On the other hand, the stellar locus of the
D4 field is about 4\% away from that of the other fields. The reason
for this is not clear yet. In particular, D4 is affected by the same
Galactic absorption as, for exampled, D1 ($E(B-V) \sim 0.027$ for
both). Note that we observe the same phenomenon --although much
weaker-- on the $\rme-\ime$ vs. $\gme-\rme$ diagram (figure
\ref{fig:color_color_plots}).

The $\ume$-band magnitudes determined in this section can be retrieved
in electronic form at the CDS\footnote{{\tt http://cdsweb.u-strasbg.fr/cgi-bin/qcat?J/A+A/}}. Aside from the main
catalogs, described in the previous section, we have released similar
catalogs containing the subset of tertiary standards with $\ume$-band
magnitudes.

As for the main catalogs, we report the Local Natural Magnitudes of
each star. The flux interpretation of those magnitudes relies on the
MegaCam $\ume$-band magnitude of \bdtruc, which can be derived as
described in \S
\ref{sec:flux_interpretation_of_megacam_magnitudes}. Using the color
transformations determine above, we can compute a first order estimate
of the $\ume$-band magnitude of \bdtruc:
\begin{equation*}
  u_{\bdtruc} = 9.7688 \pm 0.0027 \ + \Delta \ume
\end{equation*}
This estimate should be corrected by an offset $\Delta u$, which would
account for the fact that the actual $\ume$-band magnitude of
\bdtruc\ which would be observed by MegaCam does not necessarily
follow exactly the linear Landolt-to-MegaCam color law. Such an offset
may be computed using synthetic photometry of \bdtruc\ as described in
\S \ref{sec:flux_interpretation_of_megacam_magnitudes}. However, given
the fact that we have not extensively tested the instrument and the
Mauna Kea atmospheric extinction in the $\ume$-band, we prefer not to
publish any number.

Clearly, the $\ume$-band calibration of the CFHTLS survey will be
considerably improved by (1) adding more epochs and (2) observing a
set of fundamental standards directly with MegaCam. This is the goal
of the MAPC program, already discussed in this paper.

\end{appendix}

\begin{appendix}
\section{MegaCam-to-SDSS Color Transformations}
\label{sec:comparison_with_sdss}

Two CFHTLS DEEP fields, D2 and D3 were also observed by the SDSS
collaboration. While the intercalibration of the MegaCam and SDSS
surveys deserves its own paper, it is possible to use the observations
of D2 and D3 to derive the color transformations between the SDSS and
MegaCam magnitudes. We used the SDSS DR6 catalogs and determined the
color transformations between the SDSS and MegaCam transformations
{\em at the focal plane center}. We found the color transformations to
be linear, and equal to:
\begin{eqnarray}
  u_{|\x_0} - u_{SDSS} & = & -0.211 (\pm 0.004) \times (u_{SDSS} - g_{SDSS}) + {\Delta zp_u}\nonumber \\
  g_{|\x_0} - g_{SDSS} & = & -0.155 (\pm 0.003) \times (g_{SDSS} - r_{SDSS}) + {\Delta zp_g}\nonumber \\
  r_{|\x_0} - r_{SDSS} & = & -0.030 (\pm 0.004) \times (r_{SDSS} - i_{SDSS}) + {\Delta zp_r}\nonumber \\
  i_{|\x_0} - i_{SDSS} & = & -0.102 (\pm 0.005) \times (r_{SDSS} - i_{SDSS}) + {\Delta zp_i}\nonumber \\
  z_{|\x_0} - z_{SDSS} & = & +0.036 (\pm 0.008) \times (i_{SDSS} - z_{SDSS}) + {\Delta zp_z}
\end{eqnarray}
As one can expect, there are non-zero offsets between the SDSS 2.5-m
and the SNLS system ($\Delta u \sim -0.56$, $\Delta g \sim +0.11$,
$\Delta r \sim -0.14$, $\Delta i \sim -0.32$ and $\Delta z \sim
-0.43$). The precise measurement of these offsets along with the
associated uncertainties will be the subject of a later paper.

\end{appendix}

\end{document}